\documentclass[12pt,a4paper,wide]{article}

\usepackage{graphicx}
\usepackage{amsfonts}
\usepackage{amsmath}
\usepackage{hyperref}
\usepackage{psfrag,subfigure,epsfig}
\usepackage{ytableau}
\usepackage{tikz}
\usetikzlibrary{calc}
\usepackage{color, colortbl}
\usepackage{pdflscape}

\usepackage[numbers,square,comma, compress]{natbib}

\newcommand{\weyl}{\xi}

\ytableausetup{boxsize=0.6em}

\newcommand{\myarrow}[1][-45]{%
  \mathrel{%
    \text{$
     \begin{tikzpicture}[baseline = -0.5ex]
       \node[inner sep=0pt,outer sep=0pt,rotate = #1] (a) at (0,0)  {$\xleftarrow{-\widehat{\alpha}_0}$};
    \end{tikzpicture}
    $}%
  }%
}%

\csname @addtoreset\endcsname{equation}{section}

\definecolor{Gray}{gray}{0.9}

\usepackage[lmargin=1.7 cm, rmargin=1.7 cm, tmargin=3.0cm, bmargin=3.0cm]{geometry}

\usepackage{float}

\title{
\vspace{-1.0cm}
\bf{Five-Brane Webs and Highest Weight Representations}\\[25pt]}
\author{\large \textsc{Brice~Bastian\footnote{\tt b.bastian@ipnl.in2p3.fr} ~and~ Stefan~Hohenegger\footnote{\tt s.hohenegger@ipnl.in2p3.fr}}}
\date{}

\begin{document}

\maketitle

\begin{center}
\renewcommand{\thefootnote}{\fnsymbol{footnote}}\vspace{-0.5cm}
${}^{\footnotemark[1]\footnotemark[2]}$ Universit\'e de Lyon\\
UMR 5822, CNRS/IN2P3, Institut de Physique Nucl\'eaire de Lyon\\ 4 rue Enrico Fermi, 69622 Villeurbanne Cedex, \rm FRANCE\\[0.4cm]
${}^{\footnotemark[2]}$ Fields, Gravity \& Strings, CTPU\\
Institute for Basic Sciences, Daejeon 34047 \rm KOREA\\[2.5cm]
\end{center}

\begin{abstract}
\noindent 
We consider BPS-counting functions $\mathcal{Z}_{N,M}$ of $N$ parallel M5-branes probing a transverse $\mathbb{Z}_M$ orbifold geometry. These brane web configurations can be dualised into a class of toric non-compact Calabi-Yau threefolds which have the structure of an elliptic fibration over (affine) $A_{N-1}$. We make this symmetry of $\mathcal{Z}_{N,M}$ manifest in particular regions of the parameter space of the setup: we argue that for specific choices of the deformation parameters, the supercharges of the system acquire particular holonomy charges which lead to infinitely many cancellations among states contributing to the partition function. The resulting (simplified) $\mathcal{Z}_{N,M}$ can be written as a sum over weights forming a single irreducible representation of the Lie algebra $\mathfrak{a}_{N-1}$ (or its affine counterpart). We show this behaviour explicitly for an extensive list of examples for specific values of $(N,M)$ and generalise the arising pattern for generic configurations. Finally, for a particular compact M5-brane setup we use this form of the partition function to make the duality $N\leftrightarrow M$ manifest.

\end{abstract}

\newpage
\setcounter{tocdepth}{2}
\tableofcontents

\section{Introduction}
Six dimensional superconformal field theories (SCFTs) along with their compactifications to lower dimensions have attracted a lot of attention in recent years: on the one hand, the dynamics of these theories display very rich structures which are interesting to explore in their own right. On the other hand, these SCFTs have seen numerous applications in string- and field theories. Indeed, the fact that many of them can be engineered from string or M-theory through various brane constructions (see for example \cite{Haghighat:2013gba,Haghighat:2013tka,Hohenegger:2013ala,Haghighat:2015coa,Hohenegger:2015cba,Haghighat:2015ega,Ahmed:2017hfr} for recent work on theories constructed from parallel M5-branes (with M2-branes stretched between them)), has allowed to identify interesting structures in the latter and has provided an invaluable window into their inner workings \cite{Indirect20}. Similarly, from the point of view of field theory, the recent years have brought to light interesting new dualities: for example, different types of compactifications of six dimensional SCFTs lead to various lower dimensional theories. The connection to a common higher dimensional parent theory gives rise to relations between certain quantities computed in these theories. The first example of this phenomenon was discussed in \cite{Gaiotto,Dijkgraaf:2009pc}, relating the partition functions of four dimensional gauge theories to conformal blocks in Liouville theory. Since then, multiple other examples of this type have been found.

Describing these SCFT's, however, using traditional tools in field theory, is typically rather difficult, since in general no Lagrangian description is known. Therefore, different methods -- many of them inspired by their relation to string-theory -- have been developped. In particular, considering compactifications of F-theory \cite{Vafa:1996xn} on elliptically fibered Calabi-Yau threefolds, a classification \cite{Heckman:2013pva,DelZotto:2014hpa,Heckman:2014qba,Haghighat:2014vxa,Heckman:2015bfa} (see also \cite{Choi:2017vtd} for recent work in this direction) of six-dimensional SCFTs has been proposed. Those theories with $\mathcal{N}=(2,0)$ supersymmetry allow an ADE classification and can be realised within type II string theory compactified on a $\mathbb{R}^4/\Gamma$ singularity, with $\Gamma$ a discrete ADE subgroup of $SU(2)$. In the case of an $A$-type orbifold (\emph{i.e.} $\Gamma=\mathbb{Z}_N$) these theories have a dual description in terms of $N$ parallel M5-branes probing a transverse $\mathbb{R}^4_\perp$ space. 

In this paper we study the $A_{N}$ symmetry in a series of mass-deformed theories that are described by $N$ parallel M5-branes (separated along $\mathbb{R}$ or $\mathbb{S}^1$) that probe a transverse $\mathbb{R}_\perp/\mathbb{Z}_M$ singularity. The BPS partition functions $\mathcal{Z}_{N,M}$ of this system have been computed explicitly in \cite{Haghighat:2013gba} for $M=1$ and in \cite{Haghighat:2013tka,Hohenegger:2013ala} for generic $M\in \mathbb{Z}$. There are various techniques to obtain $\mathcal{Z}_{N,M}$, which exploit different dual descriptions of the M-brane setup: 
\begin{itemize}
\item For general $N,M$ one can associate a toric Calabi-Yau threefold\footnote{In the case that the M5-branes are separated along $\mathbb{R}$ (called the non-compact setup in this work), the Calabi-Yau is an elliptic fibration over $A_{N-1}$ while in the case that $\mathbb{R}$ is compactified to $\mathbb{S}^1$ (called the compact brane setup in this work), the latter is replaced by affine $\widehat{A}_{N-1}$.} $X_{N,M}$ to the M-brane setup whose topological string partition function captures $\mathcal{Z}_{N,M}$.
\item The M-brane setup is dual to a $(p,q)$ 5-brane web in type II string theory \cite{Aharony:1997bh}. The Nekrasov partition function on the world-volume of the D5-branes corresponds to $\mathcal{Z}_{N,M}$.
\item Considering BPS M2-branes stretched between the M5-branes, the intersection of the two has been dubbed \emph{M-string} in \cite{Haghighat:2013gba}. The partition function of the latter is computed by a $\mathcal{N}=(0,2)$ sigma model, whose elliptic genus was shown in \cite{Haghighat:2013gba} to capture $\mathcal{Z}_{N,M}$.
\end{itemize}
Besides the mass parameter $m$, the partition function $\mathcal{Z}_{N,M}$ needs to be regularised by the introduction of two deformation parameters $\epsilon_{1,2}$, which (from the perspective of the dual gauge theory) correspond to the introduction of the $\Omega$-background \cite{Moore:1997dj,Lossev:1997bz}. For generic values of $m,\epsilon_{1,2}$, the M-string world-sheet theory is described by a sigma model with $\mathcal{N}=(2,0)$ supersymmetry. However, it was remarked in  \cite{Haghighat:2013gba} that $m=\pm\tfrac{\epsilon_{1}-\epsilon_2}{2}$ the supersymmetry is enhanced to $\mathcal{N}=(2,2)$ leading to $\mathcal{Z}_{N,M}(m=\pm\tfrac{\epsilon_{1}-\epsilon_2}{2})=1$ (after a suitable normalisation).

In this paper we generalise this observation to make the $A_{N-1}$ (or affine $\widehat{A}_{N-1}$) symmetry of the partition function $\mathcal{Z}_{N,M}$ manifest and organise it according to irreducible (integrable) representations of the associated Lie algebra $\mathfrak{a}_{N-1}$ (or affine $\widehat{\mathfrak{a}}_{N-1}$) for certain choices of the deformation parameters: for simplicity, we consider the unrefined partition functions (\emph{i.e.} we choose $\epsilon_1=-\epsilon_2=\epsilon$) and consider the case $m=n\epsilon$ with $n\in\mathbb{N}$. While the former enhances the supersymmetry to $\mathcal{N}=(0,4)$, the latter choice does not change the superconformal algebra on the M-string world-sheet. However, nevertheless, the partition function $\mathcal{Z}_{N,M}(m=n\epsilon,\epsilon)$ simplifies dramatically due to the fact that the corresponding supercharges obtain a non-trivial holonomy structure. This allows for infinitely many cancellations of different BPS-states contributing to the partition function, thus dramatically simplifying $\mathcal{Z}_{N,M}(m=n\epsilon,\epsilon)$: indeed, by studying a series of examples, we show that in the case of a non-compact brane configuration (\emph{i.e.} in the cases where the M5-branes are separated along non-compact $\mathbb{R}$), the partition function becomes a polynomial of order $Mn^2$ in $Q_{f_a}=e^{-t_{f_a}}$, where $t_{f_a}$ is the distance between the M5-branes (in suitable units). Similarly, also the partition functions of the compact brane configurations simplify (although their dependence on the $Q_{f_a}$ remains non-polynomial).

Moreover, since the choice $m=n\epsilon$ is fully compatible with all symmetries of the elliptic fibration $X_{N,M}$, notably $A_{N-1}$ (or affine $\widehat{A}_{N-1}$ in the case of a compact brane configuration), the latter are manifestly visible in $\mathcal{Z}_{N,M}(m=n\epsilon,\epsilon)$. Indeed, from the perspective of the Calabi-Yau manifold $X_{N,M}$ the $t_{f_a}$ can be written as integrals of the K\"ahler form over a set of $\mathbb{P}^1$'s that can be identified with the simple positive roots of the Lie algebra $\mathfrak{a}_{N-1}$ (or affine $\widehat{\mathfrak{a}}_{N-1}$) (see \emph{e.g.} \cite{Hohenegger:2015cba,Shabbir:2016nhp} for recent applications). Using this identification, specifically for the choice $m=n\epsilon$ we show in a large series of examples that $\mathcal{Z}_{N,M}(m=n\epsilon,\epsilon)$ can be written as a sum over weights that form a single irreducible (or integrable in the affine case) representation of the Lie algebra $\mathfrak{a}_{N-1}$ (affine $\widehat{\mathfrak{a}}_{N-1}$). In the basis of the fundamental weights, the highest weight of these representations is given by $[Mn^2,\ldots,Mn^2]$. Furthermore, each summand in the sum over weights is a quotient of Jacobi theta functions transforming with a well-defined index under an $SL(2,\mathbb{Z})$ symmetry corresponding to the elliptic fibration of $X_{N,M}$. Based on an extensive list of examples of different brane configurations (and choices for $n\in\mathbb{N}$) we find a pattern for all these symmetries that allows us to formulate precise conjectures for generic values $(N,M)$ and $n$.

Finally, the compact M-brane configurations (\emph{i.e.} where the M5-branes separated along $\mathbb{S}^1$ rather than $\mathbb{R}$) enjoy a duality upon exchanging $M\leftrightarrow N$ as can be seen directly from the web diagram of $X_{N,M}$. For the simplest\footnote{We expect that similar results hold true for generic values $(N,M)$.} such configuration (\emph{i.e.} $N=2=M$) we show explicitly that the partition function can be written as a double sum over integrable representations of affine $\widehat{\mathfrak{a}}_{N-1}$ and $\widehat{\mathfrak{a}}_{M-1}$ respectively. The latter not only makes the algebraic structures but also the duality manifest. Since compact brane setups of the type $(N,M)$ capture \cite{Bhardwaj:2015oru,Hohenegger:2015btj,Hohenegger:2016eqy} a class of little string theories (see \cite{Witten:1995zh,Aspinwall:1997ye,Seiberg:1997zk,Intriligator:1997dh,Hanany:1997gh,Brunner:1997gf} for various different approaches as well as \cite{Aharony:1999ks,Kutasov:2001uf} for reviews) with $\mathcal{N}=(1,0)$ supersymmetry we expect that these findings will turn out useful for the further study of little string theories  in general, in particular their symmetries and dualities (see \emph{e.g.} \cite{Harvey:2014cva} for a recent application).

The outline of this paper is as follows. In section \ref{Sect:braneConfiguration} we describe the M-theory brane setup probing a transverse orbifold geometry. We introduce all necessary parameters to describe the configurations and discuss different approaches in the literature to compute the BPS counting functions $\mathcal{Z}_{N,M}$. Finally, we also discuss the supersymmetry preserved by these configurations (from the point of view of the M-string world-sheet theory) specifically focusing on their holonomy charges as a function of the deformation parameters $(m,\epsilon_1=-\epsilon_2=\epsilon)$. In section \ref{Sect:PartitionFunctions} the expression for the topological string partition function is introduced. We furthermore motivate the choice $m=n\epsilon$ of deformation parameters by exhibit explicitly cancellations in $\mathcal{Z}_{N,M}$. In section \ref{Sect:ExamplesNonCompact} we present specific examples of non-compact brane setups and rewrite the corresponding partitions functions as sums over Weyl orbits of weights forming specific irreducible representations of $\mathfrak{a}_{N-1}\cong\mathfrak{sl}(N,\mathbb{C})$. In section \ref{Sect:CompactBraneConfigs} we repeat a similar analysis for certain compact brane configuration and rewrite them in a similar manner as sums of Weyl orbits of weights forming integrable representation of the affine Lie algebra $\widehat{\mathfrak{a}}_{N-1}\cong\widehat{\mathfrak{sl}}(N,\mathbb{C})$. Based on the examples of the previous two sections, in section \ref{Sect:GenericConfiguration} we give a general expression for the compact partition functions $\mathcal{Z}_{N,M}$ (for generic $(N,M)$) as a sum over integrable representations of $\widehat{\mathfrak{a}}_{N-1}$. The non-compact partition functions in turn are obtained by an appropriate decompactification limit. Finally section~\ref{Sect:Conclusions} contains our conclusions. Several supplementary computations as well as additional information on simple and affine Lie algebras and their representations are relegated to 5 appendices.
\section{M-Brane Configurations and Calabi-Yau Manifolds}\label{Sect:braneConfiguration}
In this paper we consider theories which can be described through particular BPS configurations of M-branes. In the following subsection we provide a review of these M-brane webs and relate them to a class of toric Calabi-Yau threefolds in section~\ref{Sect:ToricCY}.

\subsection{M-Brane Webs}
In the following we describe configurations of parallel M5-branes with M2-branes stretched between them. Depending on whether the M5-branes are separated along $\mathbb{S}^1$ or $\mathbb{R}$, we call these configurations either compact or non-compact.
\subsubsection{Non-Compact Brane Webs}
We first discuss non-compact brane webs in M-theory compactified on $\mathbb{T}^2\times \mathbb{R}^4_{||}\times \mathbb{R}\times \mathbb{R}^4_{\perp}$ (with coordinates $x^0,\ldots, x^{10}$) and consider a configuration of $N$ M5- and $K$ M2-branes as shown in table~\ref{Tab:MbraneConfig}.
\begin{table}[h!]
\centering
\begin{tabular}{ccc|cccc|c|cccc}
&&&&&&&&&&&\\[-12pt]
 & \multicolumn{2}{c|}{$\mathbb{S}_0^1 \times \mathbb{S}_1^1 $} & \multicolumn{4}{c|}{$\mathbb{R}^4_{||}$} & $\mathbb{R}$ & \multicolumn{4}{c}{$\mathbb{R}_{\perp}^4$} \\[2pt] \cline{2-12}
&&&&&&&&&&&\\[-12pt]
 & $x^0$ & $x^1$ & $x^2$ & $x^3$ & $x^4$ & $x^5$ & $x^6$ & $x^7$ & $x^8$ & $x^9$ & $x^{10}$ \\  &&&&&&&&&&&\\[-12pt]\hline \hline
\multicolumn{1}{c|}{M5-branes} & = & = & = & = & = & = &  &  &  &  &  \\[2pt] \hline
\multicolumn{1}{c|}{M2-branes} & = & = &  &  &  &  & = &  &  &  &  \\[2pt]
\end{tabular}
\caption{Non-compact BPS configuration of M5- and M2-branes.}
\label{Tab:MbraneConfig}
\end{table}
Here the M5-branes are spread out along the $x^6$ direction and we denote their positions $\text{\bf a}_a$ with $a=1,\ldots, N$ (such that ${\bf a}_a<{\bf a}_b$ for $a<b$). For explicit computations we introduce the $N-1$ distances between adjacent M5-branes as
\begin{align}
&t_{f_i}={\bf a}_{a+1}-{\bf a}_a\,,&&\forall a=1,\ldots, N-1\,.
\end{align}
which typically appear in the form of 
\begin{align}
&Q_{f_a}=e^{-t_{f_a}/R_0}\,,&&\forall a=1,\ldots, N-1\,.
\end{align}
Furthermore, we also denote the $t_{f_{a}}$ collectively as ${\bf t}=(t_{f_1},\ldots, t_{f_N-1})$. The M2-branes are stretched between adjacent M5-branes and their two-dimensional intersections have been termed \emph{M-strings} in \cite{Haghighat:2013gba}. Furthermore, denoting the radius of $\mathbb{S}_0^1$ and $\mathbb{S}_1^1$ by $R_0$ and $R_1$ respectively (\emph{i.e.} $x^0\sim x^0+2\pi R_0$ and $x^1\sim x^1+2\pi R_1$) we  introduce the parameter
\begin{align}
&\tau:= i R_0/R_1&&\text{and} &&Q_\tau=e^{2\pi i\tau}\,.\label{DefTau}
\end{align}

\subsubsection{Compact Brane Webs}
By arranging the M5-branes on a circle rather than on $\mathbb{R}$, we obtain compact M-brane configurations. Specifically, we replace the $\mathbb{R}$ along direction $x^6$ by $\mathbb{S}^1_6$ with radius $R_6$ (\emph{i.e.} $x^6\sim x^6+2\pi R_6$), as shown in table~\ref{Tab:CompactMbraneConfig}.
\begin{table}[h!]
\centering
\begin{tabular}{ccc|cccc|c|cccc}
&&&&&&&&&&&\\[-12pt]
 & \multicolumn{2}{c|}{$\mathbb{S}_0^1 \times \mathbb{S}_1^1 $} & \multicolumn{4}{c|}{$\mathbb{R}^4_{||}$} & $\mathbb{S}_6^1$ & \multicolumn{4}{c}{$\mathbb{R}_{\perp}^4$} \\[2pt] \cline{2-12}
&&&&&&&&&&&\\[-12pt]
 & $x^0$ & $x^1$ & $x^2$ & $x^3$ & $x^4$ & $x^5$ & $x^6$ & $x^7$ & $x^8$ & $x^9$ & $x^{10}$ \\  &&&&&&&&&&&\\[-12pt]\hline \hline
\multicolumn{1}{c|}{M5-branes} & = & = & = & = & = & = &  &  &  &  &  \\[2pt] \hline
\multicolumn{1}{c|}{M2-branes} & = & = &  &  &  &  & = &  &  &  &  \\[2pt] 
\end{tabular}
\caption{Compact BPS configuration of M5- and M2-branes.}
\label{Tab:CompactMbraneConfig}
\end{table}
As before, we denote the $N$ positions of the M5-branes on $\mathbb{S}_6^1$ by ${\bf a}_a$ (with $a=1,\ldots, N$) which satisfy the relation
\begin{align}
0\leq \text{\bf a}_1\leq {\bf a}_2\leq \ldots\leq {\bf a}_N\leq 2\pi R_6\,,
\end{align}
and introduce the $N$ distance between adjacent branes as
\begin{align}
t_{f_a}=\left\{\begin{array}{lcl} \text{\bf a}_{a+1}-\text{\bf a}_a & \text{for} & a=1,\ldots, N-1\,, \\ 2\pi R_6-(\text{\bf a}_N-\text{\bf a}_1) & \text{for} & a=N\,. \end{array}\right.\label{M5braneDistance}
\end{align}
As in the non-compact case, we also introduce
\begin{align}
&Q_{f_a}=e^{- t_{f_a}/R_0}\,,&&\forall a=1,\ldots, N\,,
\end{align}
along with the parameter\footnote{We use the definition (\ref{DefTau}) also in the compact case.}
\begin{align}
&\rho:=i R_6/R_0&&\text{and} &&Q_\rho=e^{2\pi i\rho}\,.\label{DefRho}
\end{align}
Notice the following relation 
\begin{align}
&\rho=\frac{i}{2\pi}\sum_{a=1}^N \tfrac{t_{f_a}}{R_0}&&\text{and} &&Q_\rho=Q_{f_1} Q_{f_2}\ldots Q_{f_N}\,.
\end{align}
With this notation, the non-compact brane configurations are obtained in the limit $\rho\to 0$.

\subsubsection{Deformation Parameters}
Computing the partition functions for the brane configurations introduced above, the latter are typically divergent. To circumvent this problem, one can introduce various deformation parameters \cite{Haghighat:2013gba}. Indeed, the underlying geometries allow for two different types of $U(1)$ twists. Upon introducing the complex coordinates for $\mathbb{R}_{||}$ and $\mathbb{R}_{\perp}$
\begin{align}
&z^1=x^2+ix^3\,,&&z^2=x^4+ix^5\,,&&w^1=x^7+ix^8\,,&&w^2=x^9+ix^{10}\,.
\end{align}
we can define
\begin{itemize}
\item $\epsilon$-deformation:\\
As we go around the compact $x^0$-direction (\emph{i.e.} the circle $\mathbb{S}_0^1$) we can twist by
\begin{align}
U(1)_{\epsilon_1} \times U(1)_{\epsilon_2}\,:\,  (z_1,z_2) &\longrightarrow (e^{2 \pi i \epsilon_1}z_1,e^{2 \pi i \epsilon_2}z_2)\,, \nonumber\\
 (w_1,w_2) &\longrightarrow (e^{-\frac{\epsilon_1+ \epsilon_2}{2}}w_1, e^{-\frac{\epsilon_1+ \epsilon_2}{2}}w_2)\,.\label{DefEps1Eps2}
\end{align}
From the point of view of supersymmetric gauge theories which can be associated with the brane configurations described above (see \cite{Haghighat:2013gba,Hohenegger:2013ala}) this deformation introduces the $\Omega$-background \cite{Moore:1997dj,Lossev:1997bz} allowing to compute the partition functions in an efficient manner.
\item mass deformation:\\
As we go around the compact $x^1$-direction (\emph{i.e.} the circle $\mathbb{S}^1_1$) we can twist by:
\begin{equation}
U(1)_m : (w_1,w_2) \longrightarrow (e^{2 \pi i m}w_1,e^{-2 \pi i m}w_2)\,.\label{DefMass}
\end{equation}
As we shall briefly discuss further below, from the perspective of the gauge theories (that are engineered from a dual type II setup), this deformation parameter corresponds to a mass for certain hypermultiplet fields.
\end{itemize}
The action of the deformation parameters $\epsilon_{1,2}$ and $m$ can be schematically represented in table~\ref{Tab:DeformedMbraneConfig}.
\begin{table}[h!]
\centering
\begin{tabular}{ccc|cccc|c|cccc}
&&&&&&&&&&&\\[-12pt]
 & \multicolumn{2}{c|}{$\mathbb{S}_0^1 \times \mathbb{S}_1^1 $} & \multicolumn{4}{c|}{$\mathbb{R}^4_{||}$} & $\mathbb{R}$ or $\mathbb{S}_6^1$ & \multicolumn{4}{c}{$\mathbb{R}_{\perp}^4$} \\[2pt] \cline{2-12}
&&&&&&&&&&&\\[-12pt]
 & $x^0$ & $x^1$ & $x^2$ & $x^3$ & $x^4$ & $x^5$ & $x^6$ & $x^7$ & $x^8$ & $x^9$ & $x^{10}$ \\  &&&&&&&&&&&\\[-12pt]\hline \hline
\multicolumn{1}{c|}{M5-branes} & = & = & = & = & = & = &  &  &  &  &  \\[2pt] \hline
\multicolumn{1}{c|}{M2-branes} & = & = &  &  &  &  & = &  &  &  &  \\[2pt] \hline\hline
\multicolumn{1}{c|}{$\epsilon_1$} &  &  & $\bullet$ & $\bullet$ &  &  &  & $\bullet$ & $\bullet$ & $\bullet$ & $\bullet$ \\ \hline
\multicolumn{1}{c|}{$\epsilon_2$} &  &  &  &  & $\bullet$ & $\bullet$ &  & $\bullet$ & $\bullet$ & $\bullet$ & $\bullet$ \\ \hline
\multicolumn{1}{c|}{$m$} &  &  &  &  &  &  &  & $\bullet$ & $\bullet$ & $\bullet$ & $\bullet$ 
\end{tabular}
\caption{$U(1)$ twists alowing for deformations of the BPS configurations of M5- and M2-branes.}
\label{Tab:DeformedMbraneConfig}
\end{table}
The former regularise divergences in the partition function coming from contributions of the non-compact dimensions while at the same time breaking part of the supersymmetries, as we shall discuss in sections~\ref{Sect:PartitionFunctions} and \ref{Sect:SUSYEnhance} respectively. Finally, we remark that in the later sections of this paper, the parameters $\epsilon_{1,2}$ and $m$ appear through
\begin{align}
&Q_m=e^{2\pi im}\,,&&q=e^{2 \pi i\epsilon_1}\,,&&t=e^{-2 \pi i\epsilon_2}\,.
\end{align}

\subsection{Orbifolds of M-brane webs}
\subsubsection{Orbifold Action and Brane Web Parameters}
A generalisation of the above M-brane configurations has been discussed in \cite{Haghighat:2013tka} (see also \cite{Hohenegger:2013ala}). Indeed, upon considering M5-branes probing an orbifold geometry (rather than $\mathbb{R}^4_\perp$), the positions of the M2-branes can be separated in the transverse direction.  Specifically, we generalise $\mathbb{R}^4_\perp$ to an Asymptotically Locally Euclidean space of type $A_{M-1}$ (which we denote by ALE$_{A_{M-1}}$) for $M\in\mathbb{N}$, which can be obtain as the following orbifold
\begin{align}
\text{ALE}_{A_{M-1}}=\mathbb{R}^4_\perp/\mathbb{Z}_M\,,&&\text{with} &&\mathbb{Z}_M:\,\left\{\begin{array}{l}w_1\longmapsto e^{\frac{2\pi in}{M}}\,w_1 \\ w_2\longmapsto e^{-\frac{2\pi i n}{M}}\,w_2 \end{array}\right.&&\text{for} &&n=0,\dots,M-1\,.\label{ActionOrbifold}
\end{align}
As explained in \cite{Haghighat:2013tka}, the twists (\ref{DefEps1Eps2}) and (\ref{DefMass}), which introduce the deformation parameters $\epsilon_{1,2}$ and $m$, are compatible with the $\text{ALE}_{A_{M-1}}$ geometry. Indeed, when viewed as an $\mathbb{S}^1$ fibration over $\mathbb{R}^3$, the latter posses two distinct $U(1)$ isometries related to the fiber and base respectively. Therefore, the generalised M-brane configuration (including the deformation parameters $\epsilon_{1,2}$ and $m$) can be represented by table~\ref{Tab:DeformedMbraneConfigOrbifold}, 
\begin{table}[h!]
\centering
\begin{tabular}{ccc|cccc|c|cccc}
&&&&&&&&&&&\\[-12pt]
 & \multicolumn{2}{c|}{$\mathbb{S}_0^1 \times \mathbb{S}_1^1 $} & \multicolumn{4}{c|}{$\mathbb{R}^4_{||}$} & $\mathbb{R}$ or $\mathbb{S}_6^1$ & \multicolumn{4}{c}{$\text{ALE}_{A_{M-1}}$} \\[2pt] \cline{2-12}
&&&&&&&&&&&\\[-12pt]
 & $x^0$ & $x^1$ & $x^2$ & $x^3$ & $x^4$ & $x^5$ & $x^6$ & $x^7$ & $x^8$ & $x^9$ & $x^{10}$ \\  &&&&&&&&&&&\\[-12pt]\hline \hline
\multicolumn{1}{c|}{M5-branes} & = & = & = & = & = & = &  &  &  &  &  \\[2pt] \hline
\multicolumn{1}{c|}{M2-branes} & = & = &  &  &  &  & = &  &  &  &  \\[2pt] \hline\hline
\multicolumn{1}{c|}{$\epsilon_1$} &  &  & $\bullet$ & $\bullet$ &  &  &  & $\bullet$ & $\bullet$ & $\bullet$ & $\bullet$ \\ \hline
\multicolumn{1}{c|}{$\epsilon_2$} &  &  &  &  & $\bullet$ & $\bullet$ &  & $\bullet$ & $\bullet$ & $\bullet$ & $\bullet$ \\ \hline
\multicolumn{1}{c|}{$m$} &  &  &  &  &  &  &  & $\bullet$ & $\bullet$ & $\bullet$ & $\bullet$ 
\end{tabular}
\caption{BPS configurations of M5-branes probing a transverse orbifold geometry with M2-branes stretched between them.}
\label{Tab:DeformedMbraneConfigOrbifold}
\end{table}
where we again allowed for the possibility of arranging the M5-branes along the $x^6$-direction either on $\mathbb{R}$ or on $\mathbb{S}_6^1$.

As in the case $M=1$, the distances between the M5-branes along the direction $x^6$ give rise to $N$ parameters $t_{f_a}$ for $a=1,\ldots,N$ (see eq.~\ref{M5braneDistance}). The case $|t_{f_N}|<\infty$ corresponds to a compact brane configuration (\emph{i.e.} the direction $x^6$ is compactified on $\mathbb{S}_6^1$ with finite radius $R_6$), while the limit $|t_{f_N}|\to \infty$ corresponds to a non-compact brane configuration (\emph{i.e.} the direction $x^6$ is non-compact). As explained in \cite{Haghighat:2013tka}, besides the $(t_{f_a},m,\epsilon_{1,2})$, the orbifolded configuration allows for another set of parameters, corresponding to the expectation values $T_i$ (for $i=1,\ldots,M$) of the M-theory three-form along $\mathbb{S}_1^1\times \mathcal{C}_i$, where $\mathcal{C}_i$ is a basis of the 2-cycles of $\text{ALE}_{A_{M-1}}$. In later computations, these parameters typically appear in the form
\begin{align}
&\bar{Q}_i=e^{-T_i}\,,&&\forall i=1,\ldots,M\,.
\end{align}
Furthermore, the parameters $\tau$ and $\rho$ (see (\ref{DefTau}) and (\ref{DefRho}) respectively) are in this duality frame given by
\begin{align}
&\tau=\frac{i}{2\pi}\sum_{i=1}^M T_i\,,&&\text{and}&&\rho=\frac{i}{2\pi}\sum_{a=1}^N \tfrac{t_{f_a}}{R_0}\,,\label{DefRhoTau}
\end{align}
which is equivalent to
\begin{align}
&Q_\tau=\bar{Q}_1\ldots \bar{Q}_M\,,&&\text{and}&&Q_\rho=Q_{f_1}\ldots Q_{f_N}\,.
\end{align}
The full orbifolded M-brane configuration is finally parametrised by $(t_{f_1},\ldots,t_{f_N},T_1,\ldots,T_M,m,\epsilon_{1,2})$, which we denote more compactly by $(\mathbf{t},\mathbf{T},m,\epsilon_{1,2})$.

\subsubsection{Type II Description}\label{Sect:TypeII5brane}
The parameters introduced in the above M-brane configurations can be given a more geometric interpretation when dualising to the corresponding type II picture. Indeed, upon reducing the orbifold M-theory configuration along $\mathbb{S}_1^1$, it can be dualised into a web of intersecting D5- and NS5-branes as shown in table~\ref{Tab:DeformedD5NS5}, where we represented the $\text{ALE}_{A_{M-1}}$ space as a (particular limit of a) fibration of $\mathbb{S}^1_7$ over $\mathbb{R}^3_\perp$ (see \cite{Haghighat:2013tka}  for more details).
\begin{table}[h!]
\centering
\begin{tabular}{cc|cccc|c|c|ccc}
&&&&&&&&&\\[-12pt]
 & $\mathbb{S}_0^1$ & \multicolumn{4}{c|}{$\mathbb{R}^4_{||}$} & $\mathbb{R}$ or $\mathbb{S}_6^1$ & $\mathbb{S}^1_7$ & \multicolumn{3}{c}{$\mathbb{R}_\perp$} \\[2pt] \cline{2-11}
&&&&&&&&&\\[-12pt]
 & $x^0$ & $x^2$ & $x^3$ & $x^4$ & $x^5$ & $x^6$ & $x^7$ & $x^8$ & $x^9$ & $x^{10}$ \\  &&&&&&&&&\\[-12pt]\hline \hline
\multicolumn{1}{c|}{D5-branes} & = & = & = & = & = &  & = &  &  &  \\[2pt] \hline
\multicolumn{1}{c|}{NS5-branes} & = & =  & =  & =  & =  & = &  &  &  &  \\[2pt] 
\end{tabular}
\caption{Configuration of intersecting D5- and NS5-branes.}
\label{Tab:DeformedD5NS5}
\end{table}

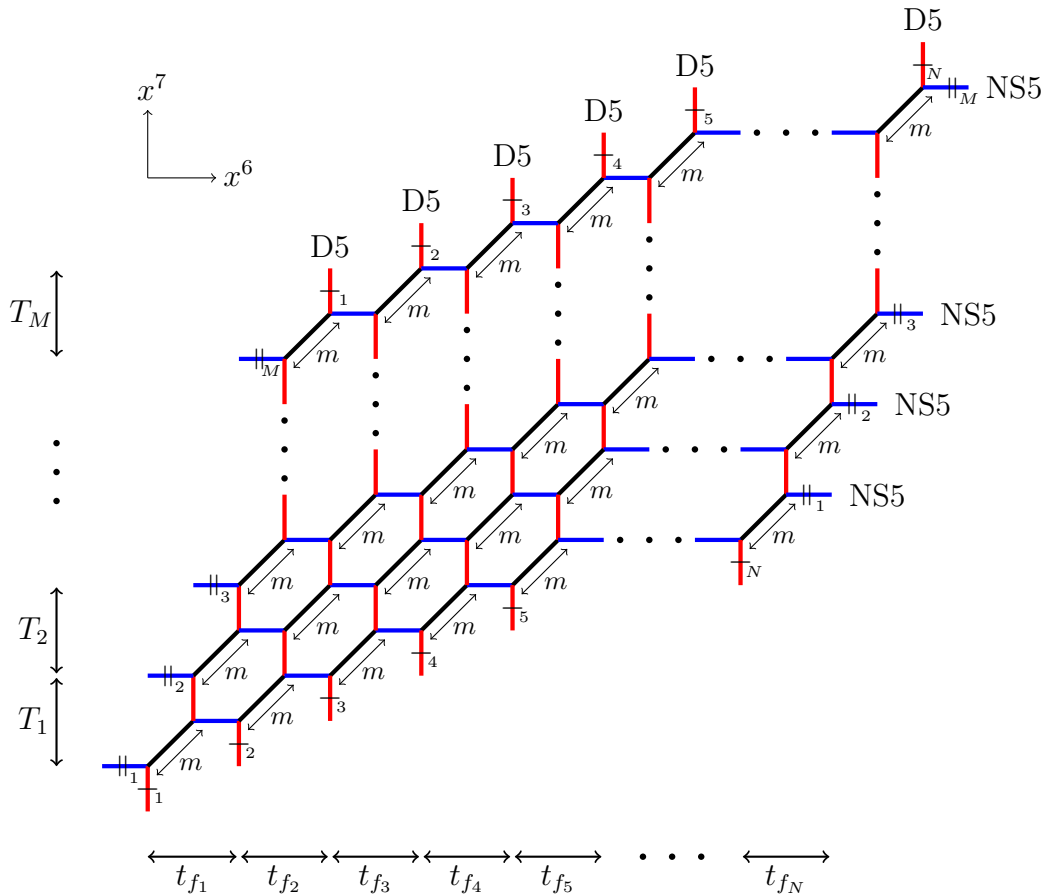
\begin{figure}[htb]
\begin{center}
\begin{tikzpicture}[scale = 0.60]
\draw[ultra thick,blue] (-6,0) -- (-5,0);
\draw[ultra thick,red] (-5,-1) -- (-5,0);
\draw[ultra thick] (-5,0) -- (-4,1);
\draw[ultra thick,blue] (-4,1) -- (-3,1);
\draw[ultra thick,red] (-4,1) -- (-4,2);
\draw[ultra thick,red] (-3,1) -- (-3,0);
\draw[ultra thick] (-3,1) -- (-2,2);
\draw[ultra thick] (-4,2) -- (-3,3);
\draw[ultra thick,blue] (-5,2) -- (-4,2);
\draw[ultra thick,red] (-3,3) -- (-3,4);
\draw[ultra thick,blue] (-3,3) -- (-2,3);
\draw[ultra thick,red] (-2,2) -- (-2,3);
\draw[ultra thick,blue] (-2,2) -- (-1,2);
\draw[ultra thick] (-2,3) -- (-1,4);
\draw[ultra thick] (-1,2) -- (0,3);
\draw[ultra thick,red] (-1,2) -- (-1,1);
\draw[ultra thick,blue] (-4,4) -- (-3,4);
\draw[ultra thick] (-3,4) -- (-2,5);
\draw[ultra thick,blue] (-2,5) -- (-1,5);
\draw[ultra thick,blue] (-1,4) -- (0,4);
\draw[ultra thick,blue] (0,3) -- (1,3);
\draw[ultra thick,red] (-2,5) -- (-2,6);
\draw[ultra thick,red] (-1,4) -- (-1,5);
\draw[ultra thick,red] (0,3) -- (0,4);
\draw[ultra thick] (1,3) -- (2,4);
\draw[ultra thick] (0,4) -- (1,5);
\draw[ultra thick] (-1,5) -- (0,6);
\draw[ultra thick,blue] (2,4) -- (3,4);
\draw[ultra thick,blue] (1,5) -- (2,5);
\draw[ultra thick,blue] (0,6) -- (1,6);
\draw[ultra thick,red] (1,3) -- (1,2);
\draw[ultra thick,red] (2,4) -- (2,5);
\draw[ultra thick,red] (1,5) -- (1,6);
\draw[ultra thick,red] (0,6) -- (0,7);
\draw[ultra thick,blue] (-1,10) -- (0,10);
\node[rotate=90] at (0,8) {{\Huge$\ldots$}};
\node[rotate=90] at (2,9) {{\Huge$\ldots$}};
\node[rotate=90] at (4,10) {{\Huge$\ldots$}};
\node[rotate=90] at (6,11) {{\Huge$\ldots$}};
\node[rotate=90] at (-2,7) {{\Huge$\ldots$}};
\draw[ultra thick,red] (0,9) -- (0,10);
\draw[ultra thick] (0,10) -- (1,11);
\draw[ultra thick,red] (1,11) -- (1,12);
\draw[ultra thick,red] (2,11) -- (2,10);
\draw[ultra thick,blue] (1,11) -- (2,11);
\draw[ultra thick] (2,11) -- (3,12);
\draw[ultra thick,red] (3,12) -- (3,13);
\draw[ultra thick,red] (4,12) -- (4,11);
\draw[ultra thick,blue] (3,12) -- (4,12);
\draw[ultra thick] (4,12) -- (5,13);
\draw[ultra thick,blue] (5,13) -- (6,13);
\draw[ultra thick,red] (5,13) -- (5,14);
\draw[ultra thick] (6,13) -- (7,14);
\draw[ultra thick,red] (6,13) -- (6,12);
\draw[ultra thick,blue] (7,14) -- (8,14);
\draw[ultra thick,red] (7,14) -- (7,15);
\draw[ultra thick] (1,6) -- (2,7);
\draw[ultra thick,red] (2,7) -- (2,8);
\draw[ultra thick,red] (3,7) -- (3,6);
\draw[ultra thick,blue] (2,7) -- (3,7);
\draw[ultra thick] (3,7) -- (4,8);
\draw[ultra thick,red] (4,8) -- (4,9);
\draw[ultra thick,red] (5,8) -- (5,7);
\draw[ultra thick,blue] (4,8) -- (5,8);
\draw[ultra thick,blue] (3,6) -- (4,6);
\draw[ultra thick,blue] (5,7) -- (6,7);
\draw[ultra thick] (5,8) -- (6,9);
\draw[ultra thick,red] (6,9) -- (6,10);
\draw[ultra thick] (2,5) -- (3,6) ;
\draw[ultra thick] (4,6) -- (5,7) ;
\draw[ultra thick,blue] (6,9) -- (7,9);
\node at (8,9) {{\Huge $\ldots$}};
\draw[ultra thick] (3,4) -- (4,5);
\draw[ultra thick,red] (3,4) -- (3,3);
\draw[ultra thick,red] (4,5) -- (4,6);
\draw[ultra thick,blue] (4,5) -- (5,5);
\node at (7,7) {{\Huge $\ldots$}};
\draw[ultra thick,blue] (10,14) -- (11,14);
\draw[ultra thick,red] (11,14) -- (11,13);
\draw[ultra thick] (11,14) -- (12,15);
\node at (9,14) {{\Huge $\ldots$}};
\draw[ultra thick,red] (12,15) -- (12,16);
\draw[ultra thick,blue] (12,15) -- (13,15);
\node[rotate=90] at (11,12) {{\Huge $\ldots$}};
\node at (6,5) {{\Huge $\ldots$}};
\draw[ultra thick] (10,9) -- (11,10);
\draw[ultra thick,blue] (9,9) -- (10,9);
\draw[ultra thick,red] (11,10) -- (11,11);
\draw[ultra thick,blue] (11,10) -- (12,10);
\draw[ultra thick,red] (10,9) -- (10,8);
\draw[ultra thick] (9,7) -- (10,8);
\draw[ultra thick,blue] (8,7) -- (9,7);
\draw[ultra thick,red] (9,6) -- (9,7);
\draw[ultra thick,blue] (10,8) -- (11,8);
\draw[ultra thick] (8,5) -- (9,6);
\draw[ultra thick,blue] (9,6) -- (10,6);
\draw[ultra thick,blue] (7,5) -- (8,5);
\draw[ultra thick,red] (8,4) -- (8,5);
\draw[ultra thick,blue] (-3,9) -- (-2,9);
\draw[ultra thick] (-2,9) -- (-1,10);
\draw[ultra thick,red] (-1,10) -- (-1,11);
\draw[ultra thick,red] (-2,8) -- (-2,9);
\node[rotate=90] at (-5.5,0) {$=$};
\node at (-5.3,-0.25) {{\tiny$1$}};
\node[rotate=90] at (-4.5,2) {$=$};
\node at (-4.3,1.75) {{\tiny$2$}};
\node[rotate=90] at (-3.5,4) {$=$};
\node at (-3.3,3.75) {{\tiny$3$}};
\node[rotate=90] at (-2.5,9) {$=$};
\node at (-2.3,8.75) {{\tiny$M$}};
\node[rotate=90] at (9.5,6) {$=$};
\node at (9.75,5.75) {{\tiny$1$}};
\node[rotate=90] at (10.5,8) {$=$};
\node at (10.75,7.75) {{\tiny$2$}};
\node[rotate=90] at (11.5,10) {$=$};
\node at (11.75,9.75) {{\tiny$3$}};
\node[rotate=90] at (12.7,15) {$=$};
\node at (13,14.75) {{\tiny$M$}};
\node at (-5,-0.5) {$-$};
\node at (-4.75,-0.7) {{\tiny $1$}};
\node at (-3,0.5) {$-$};
\node at (-2.75,0.3) {{\tiny $2$}};
\node at (-1,1.5) {$-$};
\node at (-0.75,1.3) {{\tiny $3$}};
\node at (1,2.5) {$-$};
\node at (1.25,2.3) {{\tiny $4$}};
\node at (3,3.5) {$-$};
\node at (3.25,3.3) {{\tiny $5$}};
\node at (8,4.5) {$-$};
\node at (8.3,4.3) {{\tiny $N$}};
\node at (-1,10.5) {$-$};
\node at (-0.7,10.3) {{\tiny $1$}};
\node at (1,11.5) {$-$};
\node at (1.3,11.3) {{\tiny $2$}};
\node at (3,12.5) {$-$};
\node at (3.3,12.3) {{\tiny $3$}};
\node at (5,13.5) {$-$};
\node at (5.3,13.3) {{\tiny $4$}};
\node at (7,14.5) {$-$};
\node at (7.3,14.3) {{\tiny $5$}};
\node at (12,15.5) {$-$};
\node at (12.25,15.3) {{\tiny $N$}};
\draw[thick, <->] (-5,-2) -- (-3.05,-2);
\node at (-4,-2.5) {$t_{f_1}$};
\draw[thick, <->] (-2.95,-2) -- (-1.05,-2);
\node at (-2,-2.5) {$t_{f_2}$};
\draw[thick, <->] (-0.95,-2) -- (0.95,-2);
\node at (0,-2.5) {$t_{f_3}$};
\draw[thick, <->] (1.05,-2) -- (2.95,-2);
\node at (2,-2.5) {$t_{f_4}$};
\draw[thick, <->] (3.05,-2) -- (4.95,-2);
\node at (4,-2.5) {$t_{f_5}$};
\node at (6.5,-2) {{\Huge $\ldots$}};
\draw[thick, <->] (8.05,-2) -- (9.95,-2);
\node at (9,-2.5) {$t_{f_N}$};
\draw[thick, <->] (-7,0) -- (-7,1.95);
\node at (-7.5,1) {$T_1$};
\draw[thick, <->] (-7,2.05) -- (-7,3.95);
\node at (-7.5,3) {$T_2$};
\node[rotate=90] at (-7,6.5) {{\Huge $\ldots$ }};
\draw[thick, <->] (-7,9.05) -- (-7,10.95);
\node at (-7.55,10) {$T_M$};
\draw[<->] (-4.8,-0.2) -- (-3.8,0.8);
\node at (-4.05,0.05) {\footnotesize $m$};
\draw[<->,xshift=2cm,yshift=1cm] (-4.8,-0.2) -- (-3.8,0.8);
\node at (-2.05,1.05) {\footnotesize $m$};
\draw[<->,xshift=4cm,yshift=2cm] (-4.8,-0.2) -- (-3.8,0.8);
\node at (-0.05,2.05) {\footnotesize $m$};
\draw[<->,xshift=6cm,yshift=3cm] (-4.8,-0.2) -- (-3.8,0.8);
\node at (1.95,3.05) {\footnotesize $m$};
\draw[<->,xshift=8cm,yshift=4cm] (-4.8,-0.2) -- (-3.8,0.8);
\node at (3.95,4.05) {\footnotesize $m$};
\draw[<->,xshift=13cm,yshift=5cm] (-4.8,-0.2) -- (-3.8,0.8);
\node at (8.95,5.05) {\footnotesize $m$};
\draw[<->,xshift=1cm,yshift=2cm] (-4.8,-0.2) -- (-3.8,0.8);
\node at (-3.05,2.05) {\footnotesize $m$};
\draw[<->,xshift=3cm,yshift=3cm] (-4.8,-0.2) -- (-3.8,0.8);
\node at (-1.05,3.05) {\footnotesize $m$};
\draw[<->,xshift=5cm,yshift=4cm] (-4.8,-0.2) -- (-3.8,0.8);
\node at (0.95,4.05) {\footnotesize $m$};
\draw[<->,xshift=7cm,yshift=5cm] (-4.8,-0.2) -- (-3.8,0.8);
\node at (2.95,5.05) {\footnotesize $m$};
\draw[<->,xshift=9cm,yshift=6cm] (-4.8,-0.2) -- (-3.8,0.8);
\node at (4.95,6.05) {\footnotesize $m$};
\draw[<->,xshift=14cm,yshift=7cm] (-4.8,-0.2) -- (-3.8,0.8);
\node at (9.95,7.05) {\footnotesize $m$};
\draw[<->,xshift=2cm,yshift=4cm] (-4.8,-0.2) -- (-3.8,0.8);
\node at (-2.05,4.05) {\footnotesize $m$};
\draw[<->,xshift=4cm,yshift=5cm] (-4.8,-0.2) -- (-3.8,0.8);
\node at (-0.05,5.05) {\footnotesize $m$};
\draw[<->,xshift=6cm,yshift=6cm] (-4.8,-0.2) -- (-3.8,0.8);
\node at (1.95,6.05) {\footnotesize $m$};
\draw[<->,xshift=8cm,yshift=7cm] (-4.8,-0.2) -- (-3.8,0.8);
\node at (3.95,7.05) {\footnotesize $m$};
\draw[<->,xshift=10cm,yshift=8cm] (-4.8,-0.2) -- (-3.8,0.8);
\node at (5.95,8.05) {\footnotesize $m$};
\draw[<->,xshift=15cm,yshift=9cm] (-4.8,-0.2) -- (-3.8,0.8);
\node at (10.95,9.05) {\footnotesize $m$};
\draw[<->,xshift=3cm,yshift=9cm] (-4.8,-0.2) -- (-3.8,0.8);
\node at (-1.05,9.05) {\footnotesize $m$};
\draw[<->,xshift=5cm,yshift=10cm] (-4.8,-0.2) -- (-3.8,0.8);
\node at (0.95,10.05) {\footnotesize $m$};
\draw[<->,xshift=7cm,yshift=11cm] (-4.8,-0.2) -- (-3.8,0.8);
\node at (2.95,11.05) {\footnotesize $m$};
\draw[<->,xshift=9cm,yshift=12cm] (-4.8,-0.2) -- (-3.8,0.8);
\node at (4.95,12.05) {\footnotesize $m$};
\draw[<->,xshift=11cm,yshift=13cm] (-4.8,-0.2) -- (-3.8,0.8);
\node at (6.95,13.05) {\footnotesize $m$};
\draw[<->,xshift=16cm,yshift=14cm] (-4.8,-0.2) -- (-3.8,0.8);
\node at (11.95,14.05) {\footnotesize $m$};
\node at (-1,11.5) {D5};
\node at (1,12.5) {D5};
\node at (3,13.5) {D5};
\node at (5,14.5) {D5};
\node at (7,15.5) {D5};
\node at (12,16.5) {D5};
\node at (14,15) {NS5};
\node at (13,10) {NS5};
\node at (12,8) {NS5};
\node at (11,6) {NS5};
\draw[->] (-5,13) -- (-3.5,13);
\node at (-3,13.1) {$x^6$};
\draw[->] (-5,13) -- (-5,14.5);
\node at (-4.9,15) {$x^7$};
\end{tikzpicture}
\end{center}
\caption{\sl Configuration of intersecting D5-branes (red) and NS5-branes (blue). The deformation parameter $m$ introduced by the $(1,1)$-branes (black) is chosen to be the same throughout the diagram. For $|t_{f_N}|<\infty$, the D5-branes are arranged on a circle (compact case), while the non-compact case corresponds to the limit $t_{f_N}\to \infty$.}
\label{Fig:Webpq}
\end{figure}

\noindent
While the parameters $\epsilon_{1,2}$ can be introduced in the same fashion as in the M-theory case, the parameter $m$ can no longer be interpreted as a $U(1)$ deformation (since the corresponding circle $\mathbb{S}_1^1$ is no longer present). The latter is introduced by giving mass $m$ to the bifundamental hypermultiplets corresponding to strings stretched between the D5- and NS5-branes. At the level of the brane web, it corresponds to a deformation with $(1,1)$ branes in the $(x^6,x^7)$-plane, as shown in figure~\ref{Fig:Webpq}. This figure also shows the remaining parameters $(\mathbf{t},\mathbf{T})$ as the distances of the D5- and NS5-branes in the $x^6$ and $x^7$ direction respectively.\footnote{For latter convenience, we adopt the convention that the $T_i$ are counted in units of $R_0$.}
As discussed in \cite{Haghighat:2013tka,Hohenegger:2016eqy,Hohenegger:2016yuv} choosing the deformation parameter $m$ to be the same for all intersections of D5-NS5-branes is not the most general case since a generic such brane web has $NM+2$ independent parameters. In the following, however, we focus on this simpler case, where all mass deformations are the same (as indicated in Figure~\ref{Fig:Webpq}).

\subsubsection{Toric Calabi-Yau Manifolds}\label{Sect:ToricCY}
There is a further description of the theories introduced above. Indeed, as explained in \cite{Haghighat:2013gba,Haghighat:2013tka,Hohenegger:2013ala}, one can associate a toric non-compact Calabi-Yau 3-fold (CY3fold) $X_{N,M}$ with the 5-brane web. More precisely, the web diagram shown in Figure~\ref{Fig:Webpq} can be interpreted as the dual of the Newton polygon which encodes how $X_{N,M}$ is constructed from $\mathbb{C}^3$ patches.

A generic $X_{N,M}$ can be described as a $\mathbb{Z}_N\times \mathbb{Z}_M$ orbifold of $X_{1,1}$. The latter is a Calabi-Yau threefold that resembles the geometry of the resolved conifold at certain boundary-regions of its moduli space (\emph{i.e.} upon sending $\tau,\rho\to \infty$).\footnote{Orbifolds of the latter have for example been studied in \cite{Aganagic:1999fe}.} More importantly, $X_{N,M}$ has the structure of a double elliptic fibration: it can be understood as an elliptic fibration over the affine $A_{N-1}$ space, which (as already mentioned) itself is an elliptic fibration. The two elliptic parameters are $\rho$ and $\tau$, which were introduced in (\ref{DefRhoTau}). The remaining parameters $(t_{f_1},\ldots,t_{f_{N-1}})$, $(T_1,\ldots,T_{M-1})$ as well as $m$ correspond to further K\"ahler parameters of $X_{N,M}$. We shall further elaborate on the interpretation of the parameters $\epsilon_{1,2}$ from the point of view of the Calabi-Yau manifold once we discuss the topological partition function on $X_{N,M}$ in section~\ref{Sect:PartitionFunctions}.

The double elliptic fibration structure of $X_{N,M}$ corresponds to the presence of two $SL(2,\mathbb{Z})$ symmetries which act separately on the modular parameters $\tau$ and $\rho$. Particularly for the case $M=1$ we have the following action on the various parameters~\cite{Hohenegger:2015btj}
\begin{align}
&SL(2,\mathbb{Z})_\tau\,: &&(\tau,\rho,m,t_{f_1},\ldots,t_{f_{N-1}},\epsilon_1,\epsilon_2)\longrightarrow \left(\tfrac{a\tau+b}{c\tau+b},\rho,\tfrac{m}{c\tau+d},t_{f_1},\ldots,t_{f_{N-1}},\tfrac{\epsilon_1}{c\tau+d},\tfrac{\epsilon_2}{c\tau+d}\right)\,,\nonumber\\
&SL(2,\mathbb{Z})_\rho\,: &&(\tau,\rho,m,t_{f_1},\ldots,t_{f_{N-1}},\epsilon_1,\epsilon_2)\longrightarrow \left(\tau,\tfrac{a\rho+b}{c\rho+b},\tfrac{m}{c\tau+d},\tfrac{t_{f_1}}{c\rho+d},\ldots,\tfrac{t_{f_{N-1}}}{c\rho+d},\tfrac{\epsilon_1}{c\rho+d},\tfrac{\epsilon_2}{c\rho+d}\right)\,,\label{SL2Actions}
\end{align}
where $\left(\begin{array}{cc}a & b \\ c & d\end{array}\right)\in SL(2,\mathbb{Z})$, \emph{i.e.} $a,b,c,d\in\mathbb{Z}$ and $ad-bc=1$.

\subsection{Supersymmetry}\label{Sect:SUSYEnhance}
In order to discuss the amount of supersymmetry preserved by the M-brane configurations described above, we adopt the point of view of the M-string~\cite{Haghighat:2013gba}: for a configuration of parallel M5-branes probing a flat $\mathbb{R}^4_{\perp}$ with M2-branes stretched between them (\emph{i.e.} configurations with $M=1$), the M-string preserves $\mathcal{N}=(4,4)$ supersymmetry with R-symmetry group $Spin_R(4)$. The latter acts on the space $\mathbb{R}^4_\perp$ transverse to the M5-branes. The supercharges \cite{Haghighat:2013gba} transform as the representations
\begin{align}
(\mathbf{2},\mathbf{1},\mathbf{2},\mathbf{1})_+\oplus (\mathbf{1},\mathbf{2},\mathbf{1},\mathbf{2})_-\,,
\end{align}
under $Spin(4)\times Spin_R(4)\times Spin(1,1)$, where $Spin(4)\times Spin(1,1) $ is the Lorentz group on the M5 world-volume (with $Spin(1,1)$ the Lorentz-group on the world-volume of the M-string) and the $\pm$ subscript denotes the chirality with respect to $Spin(1,1)$. As was explained in \cite{Haghighat:2013gba}, upon introducing the simple roots of $Spin(8)\supset Spin(4)_R\times Spin(4)$
\begin{align}
&u_1=e_1-e_2\,,&&u_2=e_2-e_3\,,&&u_3=e_3-e_4\,,&&u_4=e_3+e_4\,,
\end{align}
the weight vectors of the preserved supercharges are
\begin{align}
&(\mathbf{2},\mathbf{1},\mathbf{2},\mathbf{1})_+\,:\,\left\{\frac{e_1+e_2+e_3+e_4}{2}\,,\frac{e_1+e_2-e_3-e_4}{2}\,,-\frac{e_1+e_2-e_3-e_4}{2}\,,-\frac{e_1+e_2+e_3+e_4}{2}\right\}\,,\nonumber\\
&(\mathbf{1},\mathbf{2},\mathbf{1},\mathbf{2})_-\,:\,\left\{\frac{e_1-e_2+e_3-e_4}{2}\,,\frac{e_1-e_2-e_3+e_4}{2}\,,-\frac{e_1-e_2+e_3-e_4}{2}\,,-\frac{e_1-e_2-e_3+e_4}{2}\right\}\,.\nonumber
\end{align}
Furthermore, as discussed in \cite{Haghighat:2013tka}, the orbifold action (\ref{ActionOrbifold}) is not compatible with all 8 supercharges and indeed only $(\mathbf{2},\mathbf{1},\mathbf{2},\mathbf{1})_+$ (\emph{i.e.} the supercharge with positive chirality) is invariant. Therefore, for configurations with $M>1$, supersymmetry is broken to $\mathcal{N}=(4,0)$. The latter is in general further reduced by the deformations (\ref{DefEps1Eps2}): while the mass deformation (\ref{DefMass}) (which acts in a similar manner on $\mathbb{R}^4_\perp$ as the $\mathbb{Z}_M$ orbifold (\ref{ActionOrbifold})) breaks the same supercharges as the orbifold action (and leaves invariant all of $(\mathbf{2},\mathbf{1},\mathbf{2},\mathbf{1})_+$), the $\epsilon$-deformation in general\footnote{In the unrefined case (\emph{i.e.} for $\epsilon_1=-\epsilon_2$), in fact all supercharges $(\mathbf{2},\mathbf{1},\mathbf{2},\mathbf{1})_+$ remain invariant, such that the supersymmetry remains $\mathcal{N}=(4,0)$.} only leaves the supercharges corresponding to 
\begin{align}
&\frac{e_1+e_2+e_3+e_4}{2}\,,&&\text{and} &&-\frac{e_1+e_2+e_3+e_4}{2}\,,
\end{align}
invariant. It therefore reduces the supersymmetry to $\mathcal{N}=(2,0)$.

\section{Partition Functions}\label{Sect:PartitionFunctions}
\subsection{Compact and Non-Compact M-brane Configurations}
An important quantity to describe the different M-brane configurations introduced above is the partition function $\mathcal{Z}_{N,M}$ that counts BPS states. The latter can be weighted by fugacities related to the various symmetries described above. Concretely, the partition functions can be computed in various different manners, as explained in \cite{Haghighat:2013gba,Haghighat:2013tka,Hohenegger:2013ala}
\begin{itemize}
\item Topological string partition function\\
The partition function $\mathcal{Z}_{N,M}$ is captured by the (refined) topological string partition function on the toric Calabi-Yau threefold $X_{N,M}$. The latter can efficiently be computed using the (refined) topological vertex \cite{Aganagic:2003db, Hollowood:2003cv,Iqbal:2007ii}
\item M-string partition function\\
$\mathcal{Z}_{N,M}$ can also be computed as the M-string partition function. For configurations $(N,1)$ (\emph{i.e.} for $M=1$) it was shown in \cite{Haghighat:2013gba} that the latter can be obtained as the elliptic genus of a sigma model with $\mathcal{N}=(2,0)$ supersymmetry whose target space is a product of $\text{Hilb}[\mathbb{C}^2]$, the Hilbert scheme of points in $\mathbb{R}^4$. This result was generalised in \cite{Hohenegger:2013ala} to the case $M>1$ where it was shown that $\mathcal{Z}_{N,M}$ can be computed as the elliptic genus of a sigma model with $\mathcal{N}=(2,0)$ supersymmetry whose target space is given by $\mathcal{M}(r,k)$, the moduli spaces of $U(r)$ instantons of charge $k$. 
\item Nekrasov instanton calculus\\
The partition function can also be obtained from the 5-dimensional gauge theory that lives on the world-volume of the D5-branes in the type II brane-web description (see section~\ref{Sect:TypeII5brane}). The non-perturbative partition function of the latter can be computed using Nekrasov's instanton calculus on the $\Omega$-background \cite{Nekrasov:2002qd}.  
\item BPS scattering amplitudes in type II string theory\\
As discussed in \cite{Hohenegger:2013ala}, certain of the partition functions $\mathcal{Z}_{N,M}$ can also be obtained from a specific class of higher derivative scattering amplitudes in type II string theory.
\end{itemize}

\noindent
Using either of these approaches, the partition function for a compact (\emph{i.e.} $Q_\rho\neq 0$) brane configuration $(N,M)$ can be written in the following manner  \cite{Haghighat:2013gba,Haghighat:2013tka,Hohenegger:2013ala}
\begin{align}
\mathcal{Z}_{N,M}(\mathbf{T},&\mathbf{t},m,\epsilon_1,\epsilon_2)=\,W_M(\emptyset)^N\sum_{\alpha_i^{(a)}}Q_\rho^{\sum_{i=1}^M|\alpha_i^{(N)}|}\left(\prod_{a=1}^NQ_{f_a}^{\sum_{i=1}^M\left(|\alpha_i^{(a)}|-|\alpha_i^{(N)}|\right)}\right)\,\left(\prod_{a=1}^N\prod_{i=1}^M\frac{\vartheta_{\alpha_i^{(a+1)}\alpha_i^{(a)}}(Q_m;\tau)}{\vartheta_{\alpha_i^{(a)}\alpha_i^{(a)}}(\sqrt{t/q};\tau)}\right)\nonumber\\
&\times \left(\prod_{1\leq i<j\leq M}\prod_{a=1}^N\frac{\vartheta_{\alpha_i^{(a)}\alpha_j^{(a+1)}}(Q_{ij}Q_m^{-1};\tau)\,\vartheta_{\alpha_i^{(a+1)}\alpha_j^{(b)}}(Q_{ij}Q_m;\tau)}{\vartheta_{\alpha_i^{(a)}\alpha_j^{(a)}}(Q_{ij}\sqrt{t/q};\tau)\,\vartheta_{\alpha_i^{(a)}\alpha_j^{(a)}}(Q_{ij}\sqrt{q/t};\tau)}\right)\bigg|_{\alpha_i^{(1)}=\alpha_i^{(N+1)}}\,,\label{GenDefPartFct}
\end{align}
where $\alpha_i^{(a)}$ are $NM$ integer partitions (with size $|\alpha_a^{(i)}|$) and $\alpha_i^{N+1}=\alpha_i^{(1)}$ and
\begin{align}
&Q_{ij}=\bar{Q}_{i}\bar{Q}_{i+1}\ldots \bar{Q}_{j-1}\,,&&\text{for} &&1\leq i<j\leq M\,.
\end{align}
Furthermore, for two integer partitions $\mu=(\mu_1,\ldots,\mu_{\ell_1})$ and $\nu=(\nu_1,\ldots,\nu_{\ell_2})$ of length $\ell_{1,2}$ respectively, we have
\begin{align}
\vartheta_{\mu\nu}(x;\tau)=\prod_{(i,j)\in\mu}\vartheta\left(x^{-1}q^{-\mu_i+j-\tfrac{1}{2}}t^{-\nu_j^t+i-\tfrac{1}{2}};\tau\right)\,\prod_{(i,j)\in\nu}\vartheta\left(x^{-1}q^{\nu_i-j+\tfrac{1}{2}}t^{\mu^t_j-i+\tfrac{1}{2}};\tau\right)\,.
\end{align}
Here $(i,j)$ denotes the position of a given box in the Young diagram of the partitions $\mu$ and $\nu$ respectively, $\mu^t$ denotes the transposed partition of $\mu$ and 
\begin{align}
\vartheta(x;\tau)=\frac{i\theta_1(\tau,x)}{Q_\tau^{1/8}\prod_{k=1}^\infty(1-Q_\tau^k)}\,,
\end{align}
where $\theta_1(\tau;x)$ (for $x=e^{2\pi iz}$) is the Jacobi theta-function
\begin{equation}
\theta_1(\tau;z) = 2Q_\tau^{1/8} \sin(\pi z) \prod_{n=1}^{\infty}(1-Q_\tau^n)(1-xQ_\tau^n)(1-x^{-1}Q_\tau^n)\,.
\label{jacobitheta}
\end{equation}
Finally, the factor $W_M(\emptyset)$ in (\ref{GenDefPartFct}) is defined as
\begin{align}
W_M(\emptyset;\mathbf{T},m,\epsilon_1,\epsilon_2)=\lim_{\rho\to i\infty} \mathcal{Z}_{1,M}(\mathbf{T},\rho,m,\epsilon_1,\epsilon_2)\,,
\end{align}
and we also introduce the normalised partition function
{\allowdisplaybreaks\begin{align}
\widetilde{\mathcal{Z}}_{N,M}(\mathbf{T},\mathbf{t},m,\epsilon_1,\epsilon_2)=&\,\frac{\mathcal{Z}_{N,M}(\mathbf{T},\mathbf{t},m,\epsilon_1,\epsilon_2)}{W_M(\emptyset)^N}\,.
\label{completepartitionNorm}
\end{align}}
The latter was related in \cite{Haghighat:2013tka,Hohenegger:2015cba,Bhardwaj:2015oru,Hohenegger:2015btj} to an $U(N)^M$ gauge theory (which is dual to an $U(M)^N$ gauge theory), as well as (five-dimensional) little string theory. For the explicit computations in the remainder of this work it is more convenient to rewrite the partiton function in the following form:
{\allowdisplaybreaks\begin{align}
\mathcal{Z}_{N,M}(\mathbf{T},\mathbf{t},m,\epsilon_1,\epsilon_2)=W_M(\emptyset)^N &\displaystyle \sum_{\alpha_i^{(a)}} \Bigg( \prod_{a=1}^N (-Q_{f_a})^{\sum_{i=1}^M \vert \alpha_i^{(a)} \vert} \Bigg) \left( \prod_{a=1}^N \prod_{k=1}^M \prod_{(i,j) \in \alpha_{k}^{(a)}} \frac{\theta_1(\tau;z_{k,ij}^{(a+1)}) \theta_1(\tau;v_{k,ij}^{(a-1)})}{\theta_1(\tau;u_{k,ij}^{(a)}) \theta_1(\tau;w_{k,ij}^{(a)})} \right)\nonumber \\
&
\times \prod_{a=1}^N \prod_{1 \leq k < l \leq M}  \Bigg( \prod_{(i,j) \in \alpha_{k}^{(a)} } \frac{\theta_1(\tau;z_{l,ij}^{(a+1)}+\tilde{T}_{kl}) \theta_1(\tau;v_{l,ij}^{(a-1)}-\tilde{T}_{kl})}{\theta_1(\tau;u_{l,ij}^{(a)}+\tilde{T}_{kl}) \theta_1(\tau;w_{l,ij}^{(a)}+\tilde{T}_{kl})} \Bigg) \nonumber \\
&
\times \Bigg( \prod_{(i,j) \in \alpha_{l}^{(a)} } \frac{\theta_1(\tau;z_{k,ij}^{(a+1)}-\tilde{T}_{kl}) \theta_1(\tau;v_{k,ij}^{(a-1)}+\tilde{T}_{kl})}{\theta_1(\tau;u_{k,ij}^{(a)}-\tilde{T}_{kl}) \theta_1(\tau;w_{k,ij}^{(a)}-\tilde{T}_{kl})} \Bigg)\,.
\label{completepartition}
\end{align}}
Here we introduced
\begin{align}
&\tilde{T}_i=\frac{i}{2\pi}\,T_i\,,&&\text{and} &&\tilde{T}_{kl}=\tilde{T}_k+\tilde{T}_{k+1}+\ldots+\tilde{T}_{l-1}\,,&&\text{for}&&\begin{array}{l}k,l=1,\ldots, M \\ k\leq l\end{array}\label{IntroTilde}
\end{align}
and the arguments of the Jacobi-theta functions in (\ref{completepartition}) are given by:
\begin{align}
z_{k,ij}^{(a)}&=-m+\epsilon_{1}(\alpha_{k,i}^{(a)}-j+\tfrac{1}{2})-\epsilon_{2}((\alpha^{(a+1)}_{k,j})^t-i+\tfrac{1}{2})\,,\nonumber\\
v^{(a)}_{k,ij}&=-m-\epsilon_{1}(\alpha_{k,i}^{(a)}-j+\tfrac{1}{2})+\epsilon_{2}((\alpha^{(a-1)}_{k,j})^t-i+\tfrac{1}{2})\,,\nonumber\\
w^{(a)}_{k,ij}&=\epsilon_{1}(\alpha_{k,i}^{(a)}-j+1)-\epsilon_{2}((\alpha^{(a)}_{k,j})^t-i)\,,\nonumber\\
u^{(a)}_{k,ij}&=\epsilon_{1}(\alpha_{k,i}^{(a)}-j)-\epsilon_{2}((\alpha^{(a)}_{k,j})^t-i+1)
\label{arg}
\end{align}
Specifically, for $M=1$ we have the following expression
\begin{align}
\mathcal{Z}_{N,1}(\tau,t_{f_{1}},\ldots,t_{f_N},m,\epsilon_1,\epsilon_2)&=\sum_{\nu_{1},\mathellipsis, \nu_{N}}\left(\prod_{a=1}^{N}(-Q_{f_{a}})^{|\nu_{a}|}\right)\,\prod_{a=1}^{N}\prod_{(i,j)\in \nu_{(a)}}\frac{\theta_{1}(\tau;z^{(a)}_{ij})\,\theta_{1}(\tau;v^{(a)}_{ij})}{\theta_1(\tau;w^{(a)}_{ij})\theta_1(\tau;u^{(a)}_{ij})}\,,\label{DefPartFctComp}
\end{align}
where we introduced the following shorthand notation for the arguments of the Jacobi theta-functions
\begin{align} 
z_{ij}^{(a)}&=-m+\epsilon_{1}(\nu_{a,i}-j+\tfrac{1}{2})-\epsilon_{2}(\nu^{t}_{a+1,j}-i+\tfrac{1}{2})\,,\nonumber\\
v^{(a)}_{ij}&=-m-\epsilon_{1}(\nu_{a,i}-j+\tfrac{1}{2})+\epsilon_{2}(\nu^{t}_{a-1,j}-i+\tfrac{1}{2})\,,\nonumber\\
w^{(a)}_{ij}&=\epsilon_{1}(\nu_{a,i}-j+1)-\epsilon_{2}(\nu^{t}_{a,j}-i)\,,\nonumber\\
u^{(a)}_{ij}&=\epsilon_{1}(\nu_{a,i}-j)-\epsilon_{2}(\nu^{t}_{a,j}-i+1)\,.\hspace{3cm}\text{for} \hspace{0.5cm}a=1,\ldots, N\,.\label{arguments}
\end{align}

\noindent
The partition function for non-compact brane webs (which we denote $\mathcal{Z}^{\text{line}}_{N,M}$) can be obtained from (\ref{GenDefPartFct}) through the limit $Q_\rho\to 0$ (\emph{i.e.} $\rho\to i\infty$):
\begin{align}
\mathcal{Z}^{\text{line}}_{N,M}(\mathbf{T},\mathbf{t},m,\epsilon_1,\epsilon_2)=&\,W_M(\emptyset)^N\sum_{\alpha_i^{(a)}}Q_\rho^{\sum_{i=1}^M|\alpha_i^{(N)}|}\left(\prod_{a=1}^NQ_{f_a}^{\sum_{i=1}^M\left(|\alpha_i^{(a)}|-|\alpha_i^{(N)}|\right)}\right)\,\left(\prod_{a=1}^N\prod_{i=1}^M\frac{\vartheta_{\alpha_i^{(a+1)}\alpha_i^{(a)}}(Q_m;\tau)}{\vartheta_{\alpha_i^{(a)}\alpha_i^{(a)}}(\sqrt{t/q};\tau)}\right)\nonumber\\
&\times \left(\prod_{1\leq i<j\leq M}\prod_{a=1}^N\frac{\vartheta_{\alpha_i^{(a)}\alpha_j^{(a+1)}}(Q_{ij}Q_m^{-1};\tau)\,\vartheta_{\alpha_i^{(a+1)}\alpha_j^{(b)}}(Q_{ij}Q_m;\tau)}{\vartheta_{\alpha_i^{(a)}\alpha_j^{(a)}}(Q_{ij}\sqrt{t/q};\tau)\,\vartheta_{\alpha_i^{(a)}\alpha_j^{(a)}}(Q_{ij}\sqrt{q/t};\tau)}\right)\bigg|_{\alpha_i^{(0)}=\alpha_i^{(N)}=\emptyset}\,,\label{GenDefPartFctNon}
\end{align}
where $\mathbf{t}=\{t_{f_1},\ldots,t_{f_{N-1}}\}$ and $\mathbf{T}=\{T_{1},\ldots,T_{M}\}$. Specifically for $M=1$ we have
\begin{align}
\mathcal{Z}^{\text{line}}_{N,1}(\tau,t_{f_{1}},\ldots,t_{f_{N-1}},m,\epsilon_1,\epsilon_2)&=\sum_{{\nu_{1},\mathellipsis, \nu_{N-1}}\atop{\nu_0=\nu_N=\emptyset}}\left(\prod_{a=1}^{N-1}(-Q_{f_{a}})^{|\nu_{a}|}\right)\,\prod_{a=1}^{N-1}\prod_{(i,j)\in \nu_{a}}\frac{\theta_{1}(\tau;z^{a}_{ij})\,\theta_{1}(\tau;v^{a}_{ij})}{\theta_1(\tau;w^{a}_{ij})\theta_1(\tau;u^{a}_{ij})}\,,\label{DefZline}
\end{align}
where the arguments $(z_{ij}^{(a)},v_{ij}^{(a)},w_{ij}^{(a)},u_{ij}^{(a)})$ for $a=1,\ldots,N-1$ are the same as in (\ref{arguments}).

\subsection{Particular Values of the Deformation Parameters}\label{Sect:ParticularValues}
Viewed as a BPS counting function (\ref{completepartition}) (and its non-compact counterpart (\ref{GenDefPartFctNon})) depend on the fugacities $(\mathbf{T},\mathbf{t},m,\epsilon_1,\epsilon_2)$ that refine various symmetries associated with the $(N,M)$ brane-web. We can summarise the latter in the following table
\begin{center}
\begin{tabular}{c|c|c}
{\bf parameter} & {\bf symmetry, compact case} & {\bf symmetry, non-compact case}\\[2pt]\hline
&&\\[-10pt]
$\mathbf{T}=\{\tau,T_1,\ldots T_{M-1}\}$ & $SL(2,\mathbb{Z})_\tau$ & $SL(2,\mathbb{Z})_\tau$\\[4pt]\hline
&&\\[-10pt]
$\mathbf{t}$ & $\widehat{A}_{N-1}$ & $A_{N-1}$\\[4pt]\hline
&&\\[-10pt]
$m$ & $U(1)_{m}$ & $U(1)_{m}$\\[4pt]\hline
&&\\[-10pt]
$\epsilon_1\,,\epsilon_2$ & $U(1)_{\epsilon_1}\times U(1)_{\epsilon_2}$ & $U(1)_{\epsilon_1}\times U(1)_{\epsilon_2}$\\[4pt]
\end{tabular}
\end{center}
Here $SL(2,\mathbb{Z})_\tau$ is a generalisation of (\ref{SL2Actions}) to the case $M>1$ 
\begin{align}
&(\tau,T_1,\ldots,T_{M-1},\rho,m,t_{f_1},\ldots,t_{f_{N-1}},\epsilon_1,\epsilon_2)\longrightarrow \left(\tfrac{a\tau+b}{c\tau+b},\tfrac{T_1}{c\tau+d},\ldots,\tfrac{T_{M-1}}{c\tau+d},\rho,\tfrac{m}{c\tau+d},t_{f_1},\ldots,t_{f_{N-1}},\tfrac{\epsilon_1}{c\tau+d},\tfrac{\epsilon_2}{c\tau+d}\right)\,.\label{GenSl2tau}
\end{align}
From the point of view of the Calabi-Yau manifold $X_{N,M}$ (described in section~\ref{Sect:ToricCY}), the $\mathbf{t}$ are K\"ahler parameters associated with the structure of an elliptic fibration over (affine) $A_{N-1}$. From the point of view of the M-brane web, the connection of the $\mathbf{t}$ to (affine) $A_{N-1}$ seems less clear, since the former correspond to the distances of the M5-branes along the (non-)compact $x^6$ direction. However, as remarked in \emph{e.g.} \cite{Hohenegger:2015cba}, the structure of the M5-branes along this direction can be interpreted as the Dynkin diagram of $\mathfrak{a}_{N-1}$ (or its affine extension $\widehat{\mathfrak{a}}_{N-1}$) and the $Q_{f_a}$ can be linked to the roots of these algebras respectively. Indeed, we will explain this connection in more detail in the following sections, when considering explicit examples of the partition functions $\mathcal{Z}_{N,M}$. Finally, we notice that in the compact case, the roles of $\mathbf{T}$ and $\mathbf{t}$ can be exchanged upon replacing $(N,M)\longrightarrow (M,N)$. In the above table the parameters $\mathbf{t}$ have been singled out since we have decided to write $\mathcal{Z}_{N,M}$ in (\ref{GenDefPartFct}) as a power series expansion in $Q_{f_a}$ (rather than $\bar{Q}_i$).\footnote{From the point of view of the (refined) topological vertex (which was used to compute the topological string partition function $\mathcal{Z}_{N,M}$), this corresponds to a particular choice of the preferred direction of the vertex. In the current case, the latter has been chosen horizontally with respect to Figure~\ref{Fig:Webpq}.}

Written as a function of all parameters mentioned above $\mathcal{Z}_{N,M}$ is rather complicated and very difficult to analyse. In this paper we therefore consider particular values for some of the parameters, such that $\mathcal{Z}_{N,M}$ simplifies and the various symmetries can be made more manifest. First, for simplicity, we choose to work in the unrefined case, \emph{i.e.} we set
\begin{align}
\epsilon_1=-\epsilon_2=\epsilon\,,\label{UnrefinedChoice}
\end{align}
which (as mentioned in section~\ref{Sect:SUSYEnhance}) leads to an enhancement of supersymmetry to $\mathcal{N}=(4,0)$. Furthermore, (\ref{UnrefinedChoice}) is fully compatible with the symmetries $SL(2,\mathbb{Z})_\tau$ as well as $A_{N-1}$ (or $\widehat{A}_{N-1}$).

In order to further define regions in the parameter space in which the partition function simplifies, we first consider the case $M=1$. In this case, the $Spin(8)$ holonomy charges corresponding to the deformations (\ref{DefEps1Eps2}) and (\ref{DefMass}) read
\begin{align}
(\epsilon,-\epsilon,m,-m)\,,
\end{align}
where we recall that the first two entries (depending on $\epsilon$) correspond to a holonomy with respect to $\mathbb{S}^1_0$ and the last two (depending on $m$) with respect to $\mathbb{S}^1_1$. For generic values of $\epsilon$ and $m$ (in particular for $m/\epsilon\in \mathbb{R}\slash \mathbb{N}$) there is no cancellation between the corresponding holonomy phases. Phrased differently, there is no mixing between states with distinct charges under $U(1)_m$ and $U(1)_{\epsilon_1}\times U(1)_{\epsilon_2}$ in the partition function. However, if we choose
\begin{align}
&m=n\epsilon\,,&&\text{with} && n \in \mathbb{N}\,,\label{choicemasseps}
\end{align}
the holonomy charges are no longer linear independent over $\mathbb{Z}$ and thus holonomy phases may cancel when we go multiple times around the circle $\mathbb{S}^1_0$. In this way, there may be non-trivial cancellations between the contributions of states with distinct charges under $U(1)_m$ and $U(1)_{\epsilon_1}\times U(1)_{\epsilon_2}$ in the partition function $\mathcal{Z}_{N,1}$ leading to possible simplifications of $\mathcal{Z}_{N,1}$.\footnote{Notice that for $n=1$, in addition to $(\mathbf{2},\mathbf{1},\mathbf{2},\mathbf{1})_+$ the (anti-chiral) supercharges with the weight vectors $\tfrac{e_1-e_2-e_3+e_4}{2}$ and $-\tfrac{e_1-e_2-e_3+e_4}{2}$ remain unbroken, thus leading to an enhancement of supersymmetry. This fact was already remarked in \cite{Haghighat:2013gba} for the more generic case $m=\tfrac{\epsilon_1-\epsilon_2}{2}$.} For $M>1$, the same effect appears (at least) in the untwisted sector of the orbifold, such that we expect similar simplifications. Finally, we also remark that the choice (\ref{choicemasseps}) is still compatible with $SL(2,\mathbb{Z})_\tau$ as well as $A_{N-1}$ (or $\widehat{A}_{N-1}$). Therefore, we can analyse the simplified partition functions $\mathcal{Z}_{N,M}(\mathbf{T},\mathbf{t},m=n\epsilon,\epsilon,-\epsilon)$ with respect to these symmetries and write them in a fashion that makes them manifest. 

Explicitly, at the level of the partition function, the reason for the above mentioned simplifications is the following: when choosing the parameters
\begin{align}
&\epsilon_1=-\epsilon_2=\epsilon \,,&&\text{and} &&m=n\epsilon\,,&&\text{for} && n \in \mathbb{N}\,,\label{parchoice}
\end{align}
the arguments (\ref{arg}) of the theta-functions in (\ref{completepartition}) take the following form
\begin{align}
&z_{k,ij}^{(a)}=\epsilon(\alpha_{k,i}^{(a)}+(\alpha^{(a+1)}_{k,j})^t-i-j+1-n)\,,&&v^{(a)}_{k,ij}=-\epsilon(\alpha_{k,i}^{(a)}+(\alpha^{(a-1)}_{k,j})^t-i-j+1+n)\,,\nonumber\\
&w^{(a)}_{k,ij}=u^{(a)}_{k,ij}=\epsilon(\alpha_{k,i}^{(a)}+(\alpha^{(a)}_{k,j})^t-i-j+1)\,.
\label{simparg}
\end{align}
For specific partitions $\alpha^{(a)}_k$ these combinations may become zero even for generic $\epsilon$, thereby (with $\theta_1(\tau;0)=0$) leading to a vanishing contribution to the partition function. We also notice that for (\ref{parchoice}) in general $w_{k,ij}^{(a)}\neq 0\neq u_{k,ij}^{(a)}$: indeed, the coordinates $(i,j)$ of the boxes in a given Young diagram are bounded from above by $(\alpha_{k,j}^{(a)})^t$ and $\alpha_{k,i}^{(a)}$ respectively, so $w_{a,ij}^{(k)},u_{a,ij}^{(k)}\geq 1$ as can be seen from (\ref{simparg}). Therefore, there are no divergences coming from the denominator of (\ref{completepartition}) and $\mathcal{Z}_{N,M}(\mathbf{T},\mathbf{t},m=n\epsilon,\epsilon,-\epsilon)$ is well defined for $n\in \mathbb{N}$.

In the following we discuss specific examples of partition functions with the choice of parameters (\ref{parchoice}) and analyse their symmetries. 
\section{Examples: Non-Compact Brane Configurations}\label{Sect:ExamplesNonCompact}
\subsection{Configuration $(N,M)=(2,1)$}\label{Sect:Con21NC}
\subsubsection{Choice $\epsilon_1=-\epsilon_2=\epsilon$ and $m=\epsilon$}
We start with the non-compact configuration $(N,M)=(2,1)$ for which the partition function (\ref{DefZline}) is a sum over a single partition $\nu_1$. For the choice of the deformation parameters $\epsilon_1=-\epsilon_2=m=\epsilon$ we can show that the only integer partitions $\nu_1$ contributing to the partition function $\mathcal{Z}_{2,1}^{\text{line}}(t_{f_1},m=\epsilon,\epsilon,-\epsilon)$ are in fact $\nu_1=\emptyset$ and $\nu_1=\ydiagram{1}$. To see this, we recall that $\theta_1(\tau;0)=0$ such that only those partitions $\nu_1$ contribute for which (see (\ref{arguments}) for the definitions of $z_{ij}^{(a)}$ and $v_{ij}^{(a)}$)
\begin{align}
&z_{ij}^{(1)}\neq0\,,&&\text{and} && v_{ij}^{(1)}\neq 0\,,&&\forall (i,j)\in \nu_1\,.\label{VanishCond}
\end{align} 
Starting from a generic partition $\nu_1=(\nu_{1,1}\,,\nu_{1,2}\,,\ldots\,,\nu_{1,\ell})$ of length $\ell$, the condition (\ref{VanishCond}) can be checked explicitly. In particular, we can consider the following two particular boxes:
\begin{itemize}
\item the last box in the second row (\emph{i.e.} $(i,j)=(2,\nu_{1,2})$):
\begin{align}
&\parbox{1.35cm}{\ydiagram{7, 5, 4,4,3,1} *[*(black)]{0,5+1,0,0,0,0}} 
\end{align}
For this particular box we have $v_{2,\nu_{1,2}}^{(1)}=\epsilon(\nu_{2}-2-\nu_2+2)=0$, such that all partitions with $\ell\geq 2$ violate (\ref{VanishCond}) and therefore do not contribute to the partition function (\ref{DefZline}). 
\item $(\nu_1-1)$th box in the first row (\emph{i.e.} $(i,j)=(1,\nu_{1,1}-1)$)\\
Due to the previous constraint the only remaining partitions correspond to Young diagrams with a single row: 
\begin{align}
&\parbox{1.35cm}{\ydiagram{7} *[*(black)]{5+1}} 
\end{align}
For this particular box we have $z_{1,\nu_{1,1}-1}^{(1)}=\epsilon(\nu_{1,1}-1-(\nu_{1,1}-1))=0$, such that all partitions with $\nu_{1,1}\geq 2$ violate (\ref{VanishCond}) and do not  contribute to the partition function (\ref{DefZline}). 
\end{itemize}
Combining these two constraints we find that the only possible choices are $\nu_1=\emptyset$ or $\nu_1=\ydiagram{1}$ and the partition function therefore is 
\begin{align}
\mathcal{Z}^{\text{line}}_{2,1}(\tau,t_{f_{1},m=\epsilon},\epsilon,-\epsilon)&=\sum_{\nu\in\{\emptyset,{\tiny \ydiagram{1}}\}}(-Q_{f_{a}})^{|\nu|}\,\prod_{(i,j)\in \nu}\frac{\theta_{1}(\tau;z^{(1)}_{ij})\,\theta_{1}(\tau;v^{(1)}_{ij})}{\theta_1(\tau;w^{(1)}_{ij})\theta_1(\tau;u^{(1)}_{ij})}
=1-Q_{f_1}\frac{\theta_1(\tau;-\epsilon)\theta_1(\tau;-\epsilon)}{\theta_1(\tau;\epsilon)^2}\nonumber\\
&=1-Q_{f_1}\,.\label{PartF1N2}
\end{align}
Notice that the right hand side is independent of $\tau$ and $\epsilon$ and only depends linearly on $Q_{f_1}$.

The partition function (\ref{PartF1N2}) can be rewritten in fashion that makes an $\mathfrak{a}_1$ symmetry manifest. Indeed, upon identifying 
\begin{align}
&Q_{f_1}=e^{-\alpha_1}\,,\label{DefQrootA1}
\end{align}
where $\alpha_1$ is the simple root of $\mathfrak{a}_1$ we can write
\begin{align}
\mathcal{Z}_{2,1}^{\text{line}}(\tau,t_{f_1},m=\epsilon,\epsilon,-\epsilon)=\prod_{\alpha\in\Delta_+(\mathfrak{a}_1)}(1-e^{-\alpha})^{\text{mult}(\alpha)}\,,\label{ProdRootA1}
\end{align}
with $\text{mult}(\alpha_1)=1$ and $\Delta_+(\mathfrak{a}_1)$ the space of positive roots of $\mathfrak{a}_1$. Using the Weyl character formula, we can rewrite the product (\ref{ProdRootA1}) as a sum over the Weyl group $\mathcal{W}(\mathfrak{a}_1)\cong \mathbb{Z}_2$ of $\mathfrak{a}_1$
\begin{align}
\mathcal{Z}_{2,1}^{\text{line}}(\tau,t_{f_1},m=\epsilon,\epsilon,-\epsilon)=\sum_{w\in\mathcal{W}(\mathfrak{a}_1)}(-1)^{\ell(w)}\,e^{w(\weyl )-\weyl }\,,\label{A1WeylSum}
\end{align}
where $\weyl $ is the Weyl vector of $\mathfrak{a}_1$ and $\ell(w)$ is the length of $w\in\mathcal{W}(\mathfrak{a}_1)\cong \mathbb{Z}_2$, \emph{i.e.} the number of Weyl reflections that $w$ is decomposed of. 

While the re-writings (\ref{ProdRootA1}) and (\ref{A1WeylSum}) seem trivial (due to the fact that the root space of $\mathfrak{a}_1$ is one-dimensional, \emph{i.e.} $\Delta_+(\mathfrak{a}_1)=\{\alpha_1\}$), we shall see that both equations can be directly generalised for other choices $m=n\epsilon$ (with $n>1$) and also $N>2$ (as we shall discuss in section~\ref{Sect:NonCompactN3}).
\subsubsection{Choice $\epsilon_1=-\epsilon_2$ and $m=n\epsilon$ for $n>1$}\label{Sect:N21ng1}
For the cases $n>1$ we can repeat the above analysis to find all partitions that yield a non-vanishing contribution to the partition function (\ref{DefZline}). In doing so, we find a generic pattern, which can be summarised as follows:\footnote{This pattern has explicitly been checked up to $n=11$ and we conjecture it to hold for generic $n\in\mathbb{N}$.} only those partitions $\nu_1=(\nu_{1,1}\,,\nu_{1,2}\,,\ldots\,,\nu_{1,\ell})$ with
\begin{align}
&\ell\leq n\,,&&\text{and} &&\nu_{1,a}\leq n\,,&&\forall a=1,\ldots,n\,,
\end{align}
satisfy (\ref{VanishCond}). As a consequence, we propose that the partition function is a polynomial in $Q_{f_1}$ and can be written as the finite sum
\begin{align}
\mathcal{Z}_{2,1}^{\text{line}}(\tau,t_{f_1},m=n\epsilon,\epsilon,-\epsilon)=\sum_{k=0}^{n^2}(-1)^k\,c^{(n)}_{k}(\tau,\epsilon)\,Q_{f_{1}}^k\,.\label{ExpZline2}
\end{align}
For $n>1$ the coefficients $c_{k}^{(n)}$ depend explicitly on $\tau$ and $\epsilon$ and have the property
\begin{align}
&c_{k}^{(n)}(\tau,\epsilon)=(-1)^n\,c_{n^2-k}^{(n)}(\tau,\epsilon)\,.\label{MirroringN2}
\end{align}
Explicit expressions for the first few $c_{k}^{(n)}$ with the condition\footnote{If $k$ does not satisfy (\ref{CondkCof}), the corresponding coefficient is determined by (\ref{MirroringN2}).}
\begin{align}
k\leq \left\{\begin{array}{ccl}\frac{n^2}{2}+1 & \ldots & n\text{ even} \\ \frac{n^2+1}{2} & \ldots & n\text{ odd}\end{array}\right.\label{CondkCof}
\end{align}
are given by (we recall that relations (\ref{Coef11}) -- (\ref{Coef1fin}) have in fact been checked explicitly up to $n=11$):
{\allowdisplaybreaks
\begin{align}
c_{0}^{(n)}&=1\,,\label{Coef11}\\
c_{1}^{(n)}&=\frac{\theta(n)^2}{\theta(1)^2}\,,\\
c_{2}^{(n)}&=2\,\frac{\theta(n-1)\theta^2(n)\theta(n+1)}{\theta(1)^2\theta(2)^2}\,,\\
c_{3}^{(n)}&=\frac{\theta(n-1)^2\theta(n)^2\theta(n+1)^2}{\theta(1)^4\theta(3)^2}+2\,\frac{\theta(n-2)\theta(n-1)\theta(n)^2\theta(n+1)\theta(n+2)}{\theta(1)^2\theta(2)^2\theta(3)^2}\,,\\
c_{4}^{(n)}&=\frac{\theta(n-1)^2\theta(n)^4\theta(n+1)^2}{\theta(1)^2\theta(2)^4\theta(3)^2}+2\,\frac{\theta(n-2)\theta(n-1)^2\theta(n)^2\theta(n+1)^2\theta(n+2)}{\theta(1)^4\theta(2)^2\theta(4)^2}\nonumber\\
&\hspace{0.5cm}+2\,\frac{\theta(n-3)\theta(n-2)\theta(n-1)\theta(n)^2\theta(n+1)\theta(n+2)\theta(n+3)}{\theta(1)^2\theta(2)^2\theta(3)^2\theta(4)^2}\,,\\
c_{5}^{(n)}&=\frac{\theta(n-2)^2\theta(n-1)^2\theta(n)^2\theta(n+1)^2\theta(n+2)^2}{\theta(1)^4\theta(2)^4\theta(5)^2}+2\,\frac{\theta(n-2)\theta(n-1)^2\theta(n)^4\theta(n+1)^2\theta(n+2)}{\theta(1)^4\theta(2)^2\theta(3)^2\theta(4)^2}\nonumber\\
&\hspace{0.5cm}+2\,\frac{\theta(n-3)\theta(n-2)\theta(n-1)^2\theta(n)^2\theta(n+1)^2\theta(n+2)\theta(n+3)}{\theta(1)^4\theta(2)^2\theta(3)^2\theta(5)^2}\nonumber\\
&\hspace{0.5cm}+2\,\frac{\theta(n-4)\theta(n-3)\theta(n-2)\theta(n-1)\theta(n)^2\theta(n+1)\theta(n+2)\theta(n+3)\theta(n+4)}{\theta(1)^2\theta(2)^2\theta(3)^2\theta(4)^2\theta(5)^2}\,,
%
\label{Coef1fin}
\end{align}}
where for simplicity we have introduced the shorthand notation
\begin{align}
&\theta(\ell):=\theta_1(\tau,\ell\epsilon)\,,&&\forall \ell\in\mathbb{N}\,.
\end{align}
While not constant (as in the case of $n=1$), the coefficients $c_{k}^{(n)}$ display a clear pattern, which we propose to hold for generic $(k,n)$ satisfying (\ref{CondkCof}): every coefficient itself can be written in the form
\begin{align}
c_{k}^{(n)}(\tau,\epsilon)=\sum_{\mu(k,n)=(\mu_1,\ldots,\mu_\ell)\atop{|\mu|=k+1}}\,c(\mu)\,\frac{\theta(n)^{\mu_1}\prod_{a=2}^\ell\left(\theta(n-a+1)\theta(n+a-1)\right)^{\mu_a}}{f(\mu;\epsilon,\tau)}\,,\label{Coefsckn}
\end{align}
where the sum is over partitions $\mu(k,n)=\left(\mu_1(k,n)\,,\mu_2(k,n)\,,\ldots\,,\mu_\ell(k,n)\right)$ of length $\ell$ (with $0\leq \ell\leq k+1$) and $f(\mu)$ is a product of theta functions
\begin{align}
&f(\mu;\epsilon,\tau)=\prod_{i=1}^k\theta(i)^{r_i}\,,
\end{align}
with $r_i\in \mathbb{N}_{\text{even}}$ and $r_i\leq 4$ that satisfy
\begin{align}
&\sum_{i=1}^k r_i=2k\,,&&\text{and} &&\mu_1n^2+2\sum_{a=2}^\ell\left[n^2+(a-1)^2\right]\mu_a-\sum_{j=1}^k\,r_j\,j^2=2k(n^2-k)\,.\label{ConditionPadding}
\end{align}
Here the first condition states that the number of $\theta_1$-functions in the numerator and denominator of (\ref{Coefsckn}) is the same, while the second condition ensures that each coefficient $c_{k}^{(n)}(\tau,\epsilon)$ transforms in an appropriate manner under $SL(2,\mathbb{Z})_\tau$ transformations (see (\ref{SL2Actions})). Specifically, we have
\begin{align}
c^{(n)}_{k}\left(-1/\tau,\epsilon/\tau\right)=e^{\frac{2\pi i\epsilon^2}{\tau}\,k(n^2-k)}\,c^{(n)}_{k}(\tau,\epsilon)\,.
\end{align}
Thus, we can assign an index under $SL(2,\mathbb{Z})_\tau$ to each of the $c^{(n)}_{k}$
\begin{align}
\mathcal{I}_\tau(c^{(n)}_{k})=k(n^2-k)\,.
\end{align}
Finally, the $c(\mu)$ in (\ref{Coefsckn}) are numerical coefficients which take values $c(\mu)\in \{0,1,2\}$.

While the expressions for $c_{k}^{(n)}(\tau,\epsilon)$ in (\ref{Coefsckn}) are rather complicated, they are essentially determined by specifying all partitions $\mu(n,k)$ for which $c(\mu)\neq 0$. These can be obtained from the partitions $\mu(n,k-1)$ in an algorithmic fashion by increasing one of the $\mu_a(k-1,n)$ by either $1$ or $2$. The precise relation (along with explicit examples up to $k=5$) is explained in appendix~\ref{App:RecursionN2} and can be summarised by the fact that there is an operator $R_+$ such that
\begin{align}
c_{k}^{(n)}(\tau,\epsilon)=R_+\,c_{k-1}^{(n)}(\tau,\epsilon)\,.\label{OperationRp}
\end{align}
Schematically, the action of $R_+$ can be represented graphically in the following manner
\begin{center}
\scalebox{1.4}{\parbox{10cm}{\begin{tikzpicture}
\draw[->] (-4,0) -- (4,0);
\draw[ultra thick] (0,-0.25) -- (0,0.25);
\node[red] at (-3,0) {$\bullet$};
\node[red] at (-2,0) {$\bullet$};
\node[red] at (-1,0) {$\bullet$};
\node[red] at (0,0) {$\bullet$};
\node[red] at (1,0) {$\bullet$};
\node[red] at (2,0) {$\bullet$};
\node[red] at (3,0) {$\bullet$};
%
\node at (-3,0.4) {\tiny $c^{(n)}_{0}$};
\node at (-2,0.4) {\tiny $\ldots$};
\node[rotate=90] at (-1,0.75) {\tiny $c^{(n)}_{n^2/2-1}$};
\node[rotate=90] at (0,0.75) {\tiny $c^{(n)}_{n^2/2}$};
\node[rotate=90] at (1,0.75) {\tiny $c^{(n)}_{n^2/2-1}$};
\node at (2,0.4) {\tiny $\ldots$};
\node at (3,0.4) {\tiny $c^{(n)}_{0}$};
\draw[ultra thick, ->, blue] (-2.9,-0.1) to [out=315,in=180] (-2.5,-0.3) to [out=0,in=225] (-2.1,-0.1);
\node[blue] at (-2.5,-0.5) {\tiny $R_+$};
\draw[ultra thick, ->, blue, xshift=1cm] (-2.9,-0.1) to [out=315,in=180] (-2.5,-0.3) to [out=0,in=225] (-2.1,-0.1);
\node[blue] at (-1.5,-0.5) {\tiny $R_+$};
\draw[ultra thick, ->, blue, xshift=2cm] (-2.9,-0.1) to [out=315,in=180] (-2.5,-0.3) to [out=0,in=225] (-2.1,-0.1);
\node[blue] at (-0.5,-0.5) {\tiny $R_+$};
\draw[ultra thick, ->, blue, xshift=3cm] (-2.9,-0.1) to [out=315,in=180] (-2.5,-0.3) to [out=0,in=225] (-2.1,-0.1);
\node[blue] at (0.5,-0.5) {\tiny $R_+$};
\draw[ultra thick, ->, blue, xshift=4cm] (-2.9,-0.1) to [out=315,in=180] (-2.5,-0.3) to [out=0,in=225] (-2.1,-0.1);
\node[blue] at (1.5,-0.5) {\tiny $R_+$};
\draw[ultra thick, ->, blue, xshift=5cm] (-2.9,-0.1) to [out=315,in=180] (-2.5,-0.3) to [out=0,in=225] (-2.1,-0.1);
\node[blue] at (2.5,-0.5) {\tiny $R_+$};
\node at (5.5,0) {\footnotesize $n$ even};
\end{tikzpicture}}}\\[30pt]
\scalebox{1.4}{\parbox{10cm}{\begin{tikzpicture}
\draw[->] (-4,0) -- (4,0);
\draw[ultra thick] (0,-0.25) -- (0,0.25);
\node at (3,0) {{\tiny $|$}};
\node at (2,0) {{\tiny $|$}};
\node at (1,0) {{\tiny $|$}};
\node at (-1,0) {{\tiny $|$}};
\node at (-2,0) {{\tiny $|$}};
\node at (-3,0) {{\tiny $|$}};
\node[red] at (-3.5,0) {$\bullet$};
\node[red] at (-2.5,0) {$\bullet$};
\node[red] at (-1.5,0) {$\bullet$};
\node[red] at (-0.5,0) {$\bullet$};
\node[red] at (0.5,0) {$\bullet$};
\node[red] at (1.5,0) {$\bullet$};
\node[red] at (2.5,0) {$\bullet$};
\node[red] at (3.5,0) {$\bullet$};
\node at (-3.5,0.4) {\tiny $c^{(n)}_{0}$};
\node at (-2.5,0.4) {\tiny $\ldots$};
\node[rotate=90] at (-1.5,0.8) {\tiny $c^{(n)}_{(n^2-3)/2}$};
\node[rotate=90] at (-0.5,0.8) {\tiny $c^{(n)}_{(n^2-1)/2}$};
\node[rotate=90] at (0.5,0.85) {\tiny $-c^{(n)}_{(n^2-1)/2}$};
\node[rotate=90] at (1.5,0.85) {\tiny $-c^{(n)}_{(n^2-3)/2}$};
\node at (2.5,0.4) {\tiny $\ldots$};
\node at (3.5,0.45) {\tiny $-c^{(n)}_{0}$};
\draw[ultra thick, ->, blue] (-3.4,-0.1) to [out=315,in=180] (-3,-0.3) to [out=0,in=225] (-2.6,-0.1);
\node[blue] at (-3,-0.5) {\tiny $R_+$};
\draw[ultra thick, ->, blue, xshift=1cm] (-3.4,-0.1) to [out=315,in=180] (-3,-0.3) to [out=0,in=225] (-2.6,-0.1);
\node[blue] at (-2,-0.5) {\tiny $R_+$};
\draw[ultra thick, ->, blue, xshift=2cm] (-3.4,-0.1) to [out=315,in=180] (-3,-0.3) to [out=0,in=225] (-2.6,-0.1);
\node[blue] at (-1,-0.5) {\tiny $R_+$};
\draw[ultra thick, ->, blue, xshift=3cm] (-3.4,-0.1) to [out=315,in=180] (-3,-0.3) to [out=0,in=225] (-2.6,-0.1);
\node[blue] at (0,-0.5) {\tiny $R_+$};
\draw[ultra thick, ->, blue, xshift=4cm] (-3.4,-0.1) to [out=315,in=180] (-3,-0.3) to [out=0,in=225] (-2.6,-0.1);
\node[blue] at (1,-0.5) {\tiny $R_+$};
\draw[ultra thick, ->, blue, xshift=5cm] (-3.4,-0.1) to [out=315,in=180] (-3,-0.3) to [out=0,in=225] (-2.6,-0.1);
\node[blue] at (2,-0.5) {\tiny $R_+$};
\draw[ultra thick, ->, blue, xshift=6cm] (-3.4,-0.1) to [out=315,in=180] (-3,-0.3) to [out=0,in=225] (-2.6,-0.1);
\node[blue] at (3,-0.5) {\tiny $R_+$};
\node at (5.5,0) {\footnotesize $n$ odd};
\end{tikzpicture}}}
\end{center}
which also reflects the symmetry (\ref{MirroringN2}). These graphical representations are reminiscent of the highest-weight representation $\Gamma_{n^2}$ of $\mathfrak{sl}(2,\mathbb{C})$ where one can move between the various points (which represent certain one-dimensional functional spaces of theta-quotients) with the help of raising and lowering operators. In fact, we can make this connection more precise by writing
\begin{align}
\mathcal{Z}_{2,1}^{\text{line}}(\tau,t_{f_1},m=n\epsilon,\epsilon,-\epsilon)=e^{-n^2\weyl} \sum_{\lambda=[c]\in P^+_{n^2}}(-1)^c\,\phi^n_{[c]}(\tau,\epsilon)\,\mathcal{O}^n_\lambda(t_{f_1})\label{N2gennForm}
\end{align}
where $e^{-n^2\weyl}=Q_{f_{1}}^{n^2/2}$ and $\xi=\alpha_1/2=t_{f_1}/2$ can be identified with the Weyl vector of $\mathfrak{a}_1$ following the identification (\ref{DefQrootA1}). Furthermore, the sum is over all elements of the fundamental Weyl chamber of the representation $\Gamma_{n^2}$ which are labelled by their weights $\lambda=[c]$, \emph{i.e.}
\begin{align}
P^+_{n^2}=\left\{\begin{array}{lcl}\{[2c]|c=0,\ldots,n^2/2\}=\{[0],[2],[4],\ldots,[n^2]\} & \ldots & n\text{ even}\,,\\[8pt] \{[2c+1]|c=0,\ldots,(n^2-1)/2\}=\{[1],[3],[5],\ldots,[n^2]\} & \ldots & n\text{ odd}\,.\end{array}\right.
\end{align}
while we have for the coefficients
\begin{align}
&\phi^n_{[k]}=c^{(n)}_{(n^2-k)/2}\,,&&\text{for} &&[k]\in P^+_{n^2}\,.\label{Phin1M1}
\end{align}
Finally, the $\mathcal{O}_\lambda^n$ in (\ref{N2gennForm}) can be understood as the (normalised) orbits of  $\lambda\in P^+_{n^2}$ under the Weyl group $\mathcal{W}(\mathfrak{a}_1)\cong \mathbb{Z}_2$ of $\mathfrak{a}_1$, \emph{i.e.}
\begin{align}
\mathcal{O}_{\lambda=[c]}^n(t_{f_1})=d_\lambda\,\sum_{w\in \mathcal{W}(\mathfrak{a}_1)}(-1)^{n\ell(w)}\,e^{w(\lambda)}=d_\lambda\left(Q_{f_1}^{\frac{n^2-c}{2}}+(-1)^nQ_{f_1}^{\frac{n^2+c}{2}}\right)\,,\label{WeylOrbM1}
\end{align}
where we have used the identification (\ref{DefQrootA1}) and the normalisation factor is given by
\begin{align}
d_{\lambda=[c]}=\frac{|\text{Orb}_\lambda(\mathcal{W}(\mathfrak{a}_1))|}{|\mathcal{W}(\mathfrak{a}_1)|}=\left\{\begin{array}{lcl}\tfrac{1}{2} & \ldots & c=0\,, \\[6pt] 1 &\ldots & \text{else}\,.\end{array}\right.\label{NormWeylOrb2}
\end{align}
Here $|\text{Orb}_\lambda(\mathcal{W}(\mathfrak{a}_1))|$ is the order of the orbit of $\lambda$ under the Weyl group of $\mathfrak{a}_1$ and $|\mathcal{W}(\mathfrak{a}_1)|=|\mathbb{Z}_2|=2$.

To summarise, we propose that $\mathcal{Z}_{2,1}^{\text{line}}(\tau,t_{f_1},m=n\epsilon,\epsilon,-\epsilon)$ can be written as a sum over weights of $\mathfrak{sl}(2,\mathbb{C})$, whose representatives fall into the fundamental Weyl chamber of the irreducible representation $\Gamma_{n^2}$. As we shall see in the following, this pattern continues to hold for the partition functions of other non-compact M-brane configurations $(N,1)$ for $N>2$.
\subsection{Configuration $(N,M)=(3,1)$}\label{Sect:NonCompactN3}
\subsubsection{Case $\epsilon_1=-\epsilon_2=\epsilon$ and $m=\epsilon$}
The case $(N,M)=(3,1)$ for the choice $m=\epsilon$ is analysed in detail in appendix~\ref{App:NM31m1}. Summarising the results, as above only finitely many partitions contribute to $\mathcal{Z}^{\text{line}}_{N,1}$ in (\ref{DefZline}) which are given in the following table
\begin{center}
\begin{tabular}{|c|c|c|}\hline
&&\\[-10pt]
$\nu_1$ & $\nu_2$ & $\left(\prod_{a=1}^{2}(-Q_{f_{a}})^{|\nu_{a}|}\right)\,\prod_{a=1}^{2}\prod_{(i,j)\in \nu_{a}}\frac{\theta_{1}(\tau;z^{a}_{ij})\,\theta_{1}(\tau;v^{a}_{ij})}{\theta_1(\tau;w^{a}_{ij})\theta_1(\tau;u^{a}_{ij})}$ \\[8pt]\hline\hline
&&\\[-10pt]
$\emptyset$ & $\emptyset$ & $1$ \\[4pt]\hline\hline
&&\\[-10pt]
$\parbox{0.3cm}{\ydiagram{1}}$ & $\emptyset$ & $-Q_{f_1}$ \\[4pt]\hline
&&\\[-10pt]
$\emptyset$ & $\parbox{0.3cm}{\ydiagram{1}}$ & $-Q_{f_2}$ \\[4pt]\hline\hline
&&\\[-10pt]
$\parbox{0.6cm}{\ydiagram{2}}$ & $\parbox{0.3cm}{\ydiagram{1}}$ & $Q_{f_1}^2Q_{f_2}$ \\[4pt]\hline
&&\\[-10pt]
$\parbox{0.3cm}{\ydiagram{1}}$ & $\parbox{0.3cm}{\ydiagram{1,1}}$ & $Q_{f_1}Q_{f_2}^2$ \\[4pt]\hline\hline
&&\\[-10pt]
$\parbox{0.6cm}{\ydiagram{2}}$ & $\parbox{0.3cm}{\ydiagram{1,1}}$ & $-Q_{f_1}^2Q_{f_2}^2$ \\[4pt]\hline
\end{tabular}
\end{center}
Combining these expressions, we find for the partition function
\begin{align}
\mathcal{Z}^{\text{line}}_{3,1}(\tau,t_{f_{1}},t_{f_2},m=\epsilon,\epsilon,-\epsilon)&=1-Q_{f_1}-Q_{f_2}+Q_{f_1}^2Q_{f_2}+Q_{f_1}Q_{f_2}^2-Q_{f_1}^2Q_{f_2}^2\nonumber\\
&=(1-Q_{f_1})(1-Q_{f_2})(1-Q_{f_1}Q_{f_2})\,.\label{PartF1N3}
\end{align}
Notice that this result is independent of $\tau$ and $\epsilon$ and only depends on $Q_{f_{1,2}}$ in a polynomial fashion. Moreover, the partition function (\ref{PartF1N3}) can be rewritten in a fashion that makes an $\mathfrak{a}_2$ symmetry manifest. Indeed, upon defining
\begin{align}
&Q_{f_1}=e^{-\alpha_1}\,,&&Q_{f_2}=e^{-\alpha_2}\,,&&\text{with} &&\alpha_{1,2}\in\Delta_+(\mathfrak{a}_2)\,,\label{DefQrootA2}
\end{align}
where $\Delta_+(\mathfrak{a}_2)$ denotes the simple positive roots of $\mathfrak{a}_2$, we can write
\begin{align}
\mathcal{Z}_{3,1}^{\text{line}}(\tau,t_{f_1},t_{f_2},m=\epsilon,\epsilon,-\epsilon)=\prod_{\alpha\in\Delta_+(\mathfrak{a}_2)}(1-e^{-\alpha})^{\text{mult}(\alpha)}\,.\label{ProdRootA2}
\end{align}
Here we have used the fact that $\text{mult}(\alpha_1)=\text{mult}(\alpha_2)=1$. Using the Weyl character formula, we can rewrite the product (\ref{ProdRootA2}) as an orbit of the Weyl group
\begin{align}
\mathcal{Z}_{3,1}^{\text{line}}(\tau,t_{f_1},t_{f_2},m=\epsilon,\epsilon,-\epsilon)=\sum_{w\in\mathcal{W}(\mathfrak{a}_2)}(-1)^{\ell(w)}\,e^{w(\weyl )-\weyl }\,,\label{A2WeylSum}
\end{align}
where $\weyl =\alpha_1+\alpha_2$ is the Weyl vector and $\ell(w)$ is the length of $w\in\mathcal{W}(\mathfrak{a}_2)$, \emph{i.e.} the number of Weyl reflections that $w$ is decomposed of: the Weyl reflections of $\mathfrak{a}_2$ are defined as $s_i:\hspace{0.2cm}\gamma\longrightarrow s_i(\gamma)=\gamma-\langle\gamma,\alpha_i^\vee\rangle\,\alpha_i$ for $i=1,2$, where $\alpha_i^\vee$ are the co-roots associated with $\alpha_{1,2}$, \emph{i.e.} $\alpha_i^\vee=\tfrac{2\alpha_i}{(\alpha_i,\alpha_i)}$. They are subject to the relations $s_1^2=s_2^2=(s_1s_2)^3=0$. With this notation we can check (\ref{A2WeylSum}) by working out all non-equivalent Weyl reflections 
\begin{center}
\begin{tabular}{c|c|c}
$w\in\mathcal{W}(\mathfrak{a}_2)$ & $w(\weyl )-\weyl $ & $\ell(w)$\\\hline
$1$ & $0$ & $0$ \\
$s_1$ & $-\alpha_1$ & $1$ \\
$s_2$ & $-\alpha_2$ & $1$ \\
$s_1s_2$ & $-2\alpha_1-\alpha_2$ & $2$ \\
$s_2s_1$ & $-\alpha_1-2\alpha_2$ & $2$ \\
$s_1s_2s_1$ & $-2\alpha_1-2\alpha_2$ & $3$ 
\end{tabular}
\end{center}
Therefore, (using (\ref{DefQrootA2})), we have 
\begin{align}
\mathcal{Z}_{3,1}^{\text{line}}(\tau,t_{f_1},t_{f_2},m=\epsilon,\epsilon,-\epsilon)=\sum_{w\in\mathcal{W}(\mathfrak{a}_2)}(-1)^{\ell(w)}\,e^{w(\weyl )-\weyl }=1-Q_{f_1}-Q_{f_2}+Q_{f_1}^2Q_{f_2}+Q_{f_1}Q_{f_2}^2-Q_{f_1}^2Q_{f_2}^2\,,\label{WeylSuma21}
\end{align}
which indeed matches (\ref{PartF1N3}). Thus the partition function $\mathcal{Z}_{3,1}^{\text{line}}(\tau,m=\epsilon,t_{f_1},t_{f_2},\epsilon,-\epsilon)$ can be written in the form of a single Weyl-orbit of $\mathcal{W}(\mathfrak{a}_2)$. 

In view of generalising (\ref{A2WeylSum}) to the cases $m=n\epsilon$ for $n>1$, we prefer to write the action of the Weyl group $\mathcal{W}(\mathfrak{a}_2)\cong S_3$ in a slightly different and more intuitive manner. To this end we introduce the simple weights $(L_1,L_2,L_3)$ that span the dual of the Cartan subalgebra $\mathfrak{h}^*_{\mathfrak{a}_2}$ (as explained in appendix~\ref{App:Irreps}) and identify
\begin{align}
&t_{f_a}=L_a-L_{a+1} &&\forall a=1,2\,,
\end{align}
which is compatible with (\ref{DefQrootA2}). Furthermore, we introduce
\begin{align}
&x_r:=e^{L_{r}}\,,&&\forall r=1,2,3\,,\label{Defxr}
\end{align}
such that
\begin{align}
&Q_{f_1}=x_2/x_1\,,&&\text{and} &&Q_{f_2}=x_3/x_2\,.
\end{align}
We note that the $x_{r=1,2,3}$ are not independent, but satisfy $x_1x_2x_3=1$ due to the constraint $L_1+L_2+L_3=0$ (see (\ref{DefSl3Dual})). Using the latter condition, we can write  (\ref{WeylSuma21}) in the following fashion
\begin{align}
\mathcal{Z}_{3,1}^{\text{line}}(\tau,t_{f_1},t_{f_2},m=\epsilon,\epsilon,-\epsilon)&=x_2x_3^2\left(x_1^2 x_2-x_1^2 x_3-x_1 x_2^2+x_1 x_3^2+x_2^2 x_3-x_2 x_3^2\right)\nonumber\\
&=e^{-\weyl }\sum_{\sigma\in S_3}\text{sign}(\sigma)\,x_{\sigma(1)}^2\,x_{\sigma(2)}^1\,x_{\sigma(3)}^{0}\,,\label{ExpS3Weyl}
\end{align}
where $e^{-\weyl }=e^{L_3-L_1}=Q_{f_1}Q_{f_2}=x_2x_3^2$ (and $e^{\weyl }=e^{2L_1+L_2}=x_1^2x_2$). The action of the Weyl group in (\ref{ExpS3Weyl}) can also be illustrated graphically by arranging all terms in the following weight diagram:
\begin{align}
\parbox{11.2cm}{\begin{tikzpicture}
\draw (-2,0) -- (-1,1.732) -- (1,1.732) -- (2,0) -- (1,-1.732) -- (-1,-1.732) -- (-2,0);
\draw[fill=black] (1,-1.732) circle (0.15);
\draw[fill=black] (-1,-1.732) circle (0.15);
\draw[fill=black] (2,0) circle (0.15);
\draw[fill=black] (1,1.732) circle (0.15);
\draw[fill=black] (-1,1.732) circle (0.15);
\draw[fill=black] (-2,0) circle (0.15);
\draw (-2,0) -- (2,0);
\draw (-1,1.732) -- (1,-1.732);
\draw (1,1.732) -- (-1,-1.732);
%
%
%
\node at (3.1,0) {\footnotesize $e^{\weyl }=x_1^2x_2$};
\node[blue] at (2.9,-0.4) {\footnotesize $+1$};
\node at (-3.7,0) {\footnotesize $e^\weyl \,Q_{f_1}^2Q_{f_2}^2 =x_2x_3^2$};
\node[blue] at (-3.5,-0.4) {\footnotesize $-1$};
\node at (2.35,2) {\footnotesize $e^{\weyl }\,Q_{f_1}=x_1x_2^2$};
\node[blue] at (2.4,1.6) {\footnotesize $-1$};
\node at (-2.7,2.1) {\footnotesize $e^{\weyl }\,Q_{f_1}^2 Q_{f_2}=x_2^2 x_3$};
\node[blue] at (-2.6,1.7) {\footnotesize $+1$};
\node at (2.35,-2) {\footnotesize $e^{\weyl }\,Q_{f_2}= x_1^2 x_3$};
\node[blue] at (2.4,-2.45) {\footnotesize $-1$};
\node at (-2.7,-2) {\footnotesize $e^{\weyl }\,Q_{f_1}Q_{f_2}^2=x_1x_3^2$};
\node[blue] at (-2.5,-2.45) {\footnotesize $+1$};
\draw[red] (0,0) circle (0.15);
\draw[dashed] (0,0) -- (0,2.3);
\node at (0,1) {$\times$};
\node at (0.3,1.2) {\footnotesize $L_2$};
\draw[dashed] (0,0) -- (2.165,-1.25);
\node at (0.866,-0.5) {$+$};
\node at (0.9,-0.85) {\footnotesize $L_1$};
\draw[dashed] (0,0) -- (-2.165,-1.25);
\node at (-0.866,-0.5) {$+$};
\node at (-0.9,-0.85) {\footnotesize $L_3$};
\end{tikzpicture}} 
\end{align}
where the blue numbers represent the factor $\text{sign}(\sigma)$ in (\ref{ExpS3Weyl}). This picture indeed illustrates the $S_3\cong \mathcal{W}(\mathfrak{a}_2)$ symmetry inherent in $\mathcal{Z}_{3,1}^{\text{line}}(\tau,t_{f_1},t_{f_2},m=\epsilon,\epsilon,-\epsilon)$. 

Finally, before continuing with further examples with $m=n\epsilon$ for $n>1$ there are two comments we would like to make
\begin{itemize}
\item The prefactor $e^{-\weyl }$ in (\ref{ExpS3Weyl}) simply serves to arrange the various terms in the expansion of $\mathcal{Z}_{3,1}^{\text{line}}(\tau,t_{f_1},t_{f_2},m=\epsilon,\epsilon,-\epsilon)$ to be concentric with respect to the origin of the weight lattice spanned by $(L_1,L_2,L_3)$.   
\item We can also add the 'central point' $e^{-\weyl }$ (marked by a red circle in the above figure) to the partition function $\mathcal{Z}_{3,1}^{\text{line}}(\tau,t_{f_1},t_{f_2},m=\epsilon,\epsilon,-\epsilon)$ in (\ref{ExpS3Weyl}) since
\begin{align}
e^{-\weyl }\sum_{\sigma\in S_3}\text{sign}(\sigma)\,x_{\sigma(1)}^0\,x_{\sigma(2)}^0\,x_{\sigma(3)}^0=e^{-\weyl }(1-1-1+1+1-1)=0\,.\label{Vanishn1}
\end{align}
Therefore, we can write the partition function in the more suggestive form
\begin{align}
\mathcal{Z}_{3,1}^{\text{line}}(\tau,t_{f_1},t_{f_2},m=\epsilon,\epsilon,-\epsilon)=e^{-\weyl }\sum_{\lambda\in P^+_{1,1}}\sum_{w\in \mathcal{W}(\mathfrak{a}_2)}(-1)^{\ell(w)}\,e^{w(\lambda)}\label{FormIrrep11}
\end{align}
where $P^+_{1,1}$ is the fundamental Weyl chamber of the irreducible representation $\Gamma_{1,1}$ of $\mathfrak{a}_2$
\begin{align}
P^+_{1,1}=\{0,\weyl \}=\left\{0,2L_1+L_2\right\}\,.
\end{align}
As we shall discuss in the following, the form (\ref{FormIrrep11}) can be generalised to the cases $m=n\epsilon$ for $n>1$.
\end{itemize}
\subsubsection{Case $\epsilon_1=-\epsilon_2$ and $m=2\epsilon$} 
\label{31n}
Generalising the discussion of the previous subsection to the case $\epsilon_1=-\epsilon_2=\epsilon$ and $m=2\epsilon$ we find again specific conditions for the partitions $\nu_{1,2}$ in (\ref{DefZline}) to yield a non-vanishing contribution to the partition function $\mathcal{Z}^{\text{line}}_{3,1}(\tau,t_{f_{1}},t_{f_2},m=2\epsilon,\epsilon,-\epsilon)$. As a consequence, the latter is again polynomial in $Q_{f_1}$ and $Q_{f_2}$ with highest powers $Q_{f_1}^{8}$ and $Q_{f_2}^{8}$ respectively. However, the coefficient of each term in this polynomial is no longer a constant (\emph{i.e.} $\pm 1$), but rather a quotient of Jacobi theta functions, \emph{i.e.} schematically
\begin{align}
\mathcal{Z}_{3,1}^{\text{line}}(\tau,t_{f_1},t_{f_2},m=2\epsilon,\epsilon,-\epsilon)= \displaystyle \sum_{i,j=1}^{8} Q_{f_1}^iQ_{f_2}^j \prod_{r} \frac{\theta(a_r(i,j) \epsilon)}{\theta(b_r(i,j) \epsilon)} \,, \quad a_r, b_r \in \mathbb{Z}\,,\label{Part3n2}
\end{align}
where the integers $a_r(i,j)$ and $b_r(i,j)$ implicitly depend on $i,j$. However, as we shall discuss presently, this expressions can still be written in a manner that makes the action of $\mathfrak{a}_2$ manifest. To this end, we group together all terms corresponding to a given quotient of theta functions, however, rather than using the variables $Q_{f_i}$, we use the variables $x_r$ as introduced in (\ref{Defxr}). In terms of the monomials $Q_{f_1}^i Q_{f_2}^i$ we have
\begin{align}
&Q_{f_1}^iQ_{f_2}^j = e^{-iL_1 + (i-j)L_2 + jL_3}=x_1^{-i}x_2^{i-j}x_3^{j}\,,&&\text{for} &&0\leq i,j\leq 2n^2\,.\label{RepartQx}
\end{align}
The relation $L_1+L_2+L_3=0$ then implies $x_1x_2x_3=1$, which allows us to a generic monomial $Q_{f_1}^i Q_{f_2}^j$ as a polynomial of $x_{1,2,3}$ with only positive powers. Specifically, for $n=2$ we find :
\begin{align}
\mathcal{Z}_{3,1}^{\text{line}}(\tau,t_{f_1},t_{f_2},m=2\epsilon,\epsilon,-\epsilon)= x_2^4x_3^8  \big\{& \phi^2_{[4,4]}\,(x_1^8x_2^4 + x_1^8x_3^4 + x_2^8x_3^4 + x_2^8x_1^4 + x_3^8x_1^4 + x_3^8x_2^4) \nonumber \\
+ &\phi^2_{[2,5]}\,(x_1^7x_2^5+ x_1^7x_3^5 + x_2^7x_3^5 + x_2^7x_1^5 + x_3^7x_1^5 + x_3^7x_2^5) \nonumber\\
+&\phi^2_{[5,2]}\,(x_1^7x_2^2+ x_1^7x_3^2 + x_2^7x_3^2 + x_2^7x_1^2 + x_3^7x_1^2 + x_3^7x_2^2) \nonumber \\ 
+& \phi^2_{[0,6]}\,(x_1^6x_2^6+ x_1^6x_3^6 + x_2^6x_3^6) + \phi^2_{[6,0]}\,(x_1^6 + x_2^6 + x_3^6) \nonumber\\
+& \phi^2_{[3,3]}\,(x_1^6x_2^3 + x_1^6x_3^3 + x_2^6x_3^3 + x_2^6x_1^3 + x_3^6x_1^3 + x_3^6x_2^3) \nonumber  \\
+& \phi^2_{[1,4]}\, (x_1^5x_2^4 + x_1^5x_3^4 + x_2^5x_3^4 + x_2^5x_1^4 + x_3^5x_1^4 + x_3^5x_2^4)\nonumber\\
+&\phi^2_{[4,1]}\,(x_1^5x_2 + x_1^5x_3 + x_1^5x_3 + x_2^5x_1 + x_3^5x_1 + x_3^5x_2) \nonumber \\ 
 +& \phi^2_{[2,2]}\,(x_1^4x_2^2 + x_1^4x_3^2 + x_2^4x_3^2 + x_2^4x_1^2 + x_3^4x_1^2 + x_3^4x_2^2) \nonumber\\
+& \phi^2_{[0,3]}\,(x_1^3x_2^3+ x_1^3x_3^3 + x_2^3x_3^3) +\phi^2_{[3,0]} \,(x_1^3 + x_2^3 + x_3^3) \nonumber \\
+& \phi^2_{[1,1]} \,(x_1^2x_2 + x_1^2x_3 + x_2^2x_3 + x_2^2x_1 + x_3^2x_1 + x_3^2x_2)\nonumber\\
+& \phi^2_{[0,0]}  \big\}\,, \label{Part312}
\end{align}
where the factors $\phi^2_{[c_1,c_2]}(\tau,\epsilon)$ depend on $\tau$ and $\epsilon$ and are given as follows
{\allowdisplaybreaks
\begin{align}
&\phi^2_{[4,4]}(\tau,\epsilon)=1\,,&&\phi^2_{[2,2]}(\tau,\epsilon)=\frac{-\theta_1(\tau;3\epsilon)^2+\theta_1(\tau;\epsilon)^2\theta_1(\tau;5\epsilon)}{\theta_1(\tau;\epsilon)^3}\,,\nonumber\\
&\phi^2_{[5,2]}(\tau,\epsilon)=\phi^2_{[2,5]}(\tau,\epsilon)=\frac{\theta_1(\tau;2\epsilon)^2}{\theta_1(\tau;\epsilon)^2}\,,&&\phi^2_{[3,0]}(\tau,\epsilon)=\phi^2_{[0,3]}(\tau,\epsilon)=-2\,\frac{\theta_1(\tau;2\epsilon)\theta_1(\tau;3\epsilon)\theta_1(\tau;4\epsilon)}{\theta_1(\tau;\epsilon)^3}\,,\nonumber\\
&\phi^2_{[6,0]}(\tau,\epsilon)=\phi^2_{[0,6]}(\tau,\epsilon)=2\,\frac{\theta_1(\tau;3\epsilon)}{\theta_1(\tau;\epsilon)}\,,&&\phi^2_{[1,1]}(\tau,\epsilon)=-\frac{\theta_1(\tau;\epsilon)\theta_1(\tau;4\epsilon)^2+\theta_1(\tau;2\epsilon)^2\theta_1(\tau;5\epsilon)}{\theta_1(\epsilon)^3}\,,\nonumber\\
&\phi^2_{[3,3]}(\tau,\epsilon)=\frac{\theta_1(\tau;2\epsilon)^2\theta_1(\tau;3\epsilon)}{\theta_1(\tau;\epsilon)^3}\,,&&\phi^2_{[0,0]}(\tau,\epsilon)=6\,\frac{\theta_1(\tau;3\epsilon)\theta_1(\tau;5\epsilon)}{\theta_1(\tau;\epsilon)^2}\,,\nonumber\\
&\phi^2_{[4,1]}(\tau,\epsilon)=\phi^2_{[1,4]}(\tau,\epsilon)=\frac{\theta_1(\tau;2\epsilon)\theta_1(\tau;4\epsilon)}{\theta_1(\tau;\epsilon)^2}\,.
\label{Pfac}
\end{align}}
The subscripts\footnote{The superscript has been added as a reminder of the fact that we are dealing with the case $n=2$.} are chosen in such a way to make an action of the Weyl group $\mathcal{W}(\mathfrak{a}_2)\cong S_3$ of $\mathfrak{sl}(3,\mathbb{C})$ on $\mathcal{Z}_{3,1}^{\text{line}}(\tau,t_{f_1},t_{f_2},m=2\epsilon,\epsilon,-\epsilon)$ (along the lines of (\ref{FormIrrep11}) for $n=1$) visible. They can be identified with the Dynkin labels of the irreducible representation $\Gamma_{4,4}$, as we shall explain in the following: as in the case of $n=1$ (see eq.~(\ref{ExpS3Weyl})), the Weyl group $\mathcal{W}(\mathfrak{a}_2)$  acts as a permutation of the powers of a given monomial of the $x_{1,2,3}$:
\begin{align}
&s_{\sigma}(x_1^ix_2^jx_3^k)=x_{\sigma(1)}^ix_{\sigma(2)}^jx_{\sigma(3)}^k \,,&&\text{for} && \sigma \in S_3\,,
\end{align}
which allows us to describe all monomials multiplying a given $\phi^2_{[c_1,c_2]}$ as the Weyl orbit of a single element. To describe the latter, we introduce the fundamental weights of $\mathfrak{a}_2$
\begin{align}
&\omega_1= L_1\,,&& \text{and} && \omega_2=L_1 + L_2\,,
\end{align}
which serve as a basis for the weight lattice of $\mathfrak{a}_2$ and span the fundamental Weyl chamber. Concretely, every weight vector can be written as
\begin{align}
&\lambda= c_1 \omega_1 + c_2 \omega_2 \,,&&\text{for} && c_1,c_2 \in \mathbb{Z}\,.
\end{align}
For example the Weyl vector is given by $\weyl = \omega_1 + \omega_2$. 

In order to illustrate the structure of the partition function $\mathcal{Z}_{3,1}^{\text{line}}(\tau,t_{f_1},t_{f_2},m=2\epsilon,\epsilon,-\epsilon)$ graphically, we can represent each term in (\ref{Part312}) in the weight lattice of $\mathfrak{a}_2$
\begin{center}
\begin{tikzpicture}[scale=1.68]
\draw[fill=black] (-2,3.464) circle (0.1);
\node at (-2.35,3.65) {\footnotesize $\phi^2_{[4,4]}$};
\draw[fill=black] (-1,3.464) circle (0.1);
\node at (-1,3.8) {\footnotesize $\phi^2_{[5,2]}$};
\draw[fill=black] (0,3.464) circle (0.1);
\node at (0,3.8) {\footnotesize $\phi^2_{[6,0]}$};
\draw[fill=black] (1,3.464) circle (0.1);
\node at (1,3.8) {\footnotesize $\phi^2_{[2,5]}$};
\draw[fill=black] (2,3.464) circle (0.1);
\node at (2.35,3.65) {\footnotesize $\phi^2_{[4,4]}$};
\draw[fill=black] (-2.5,2.598) circle (0.1);
\node at (-2.85,2.8) {\footnotesize $\phi^2_{[2,5]}$};
\draw[fill=black] (-1.5,2.598) circle (0.1);
\node at (-1.875,2.775) {\footnotesize $\phi^2_{[3,3]}$};
\draw[fill=black] (-0.5,2.598) circle (0.1);
\node at (-0.9,2.775) {\footnotesize $\phi^2_{[4,1]}$};
\draw[fill=black] (0.5,2.598) circle (0.1);
\node at (0.9,2.775) {\footnotesize $\phi^2_{[1,4]}$};
\draw[fill=black] (1.5,2.598) circle (0.1);
\node at (1.875,2.775) {\footnotesize $\phi^2_{[3,3]}$};
\draw[fill=black] (2.5,2.598) circle (0.1);
\node at (2.85,2.8) {\footnotesize $\phi^2_{[5,2]}$};
\draw[fill=black] (-3,1.732) circle (0.1);
\node at (-3.35,1.95) {\footnotesize $\phi^2_{[0,6]}$};
\draw[fill=black] (-2,1.732) circle (0.1);
\node at (-2.4,1.9) {\footnotesize $\phi^2_{[1,4]}$};
\draw[fill=black] (-1,1.732) circle (0.1);
\node at (-1.4,1.9) {\footnotesize $\phi^2_{[2,2]}$};
\draw[fill=black] (0,1.732) circle (0.1);
\node at (0.4,1.9) {\footnotesize $\phi^2_{[3,0]}$};
\draw[fill=black] (1,1.732) circle (0.1);
\node at (1.4,1.9) {\footnotesize $\phi^2_{[2,2]}$};
\draw[fill=black] (2,1.732) circle (0.1);
\node at (2.4,1.9) {\footnotesize $\phi^2_{[4,1]}$};
\draw[fill=black] (3,1.732) circle (0.1);
\node at (3.4,1.6) {\footnotesize $\phi^2_{[0,6]}$};
\draw[fill=black] (-3.5,0.866) circle (0.1);
\node at (-3.85,1.05) {\footnotesize $\phi^{2}_{[5,2]}$};
\draw[fill=black] (-2.5,0.866) circle (0.1);
\node at (-2.9,1.02) {\footnotesize $\phi^2_{[4,1]}$};
\draw[fill=black] (-1.5,0.866) circle (0.1);
\node at (-1.85,1.02) {\footnotesize $\phi^2_{[0,3]}$};
\draw[fill=black] (-0.5,0.866) circle (0.1);
\node at (-0.9,1.02) {\footnotesize $\phi^2_{[1,1]}$};
\draw[fill=black] (0.5,0.866) circle (0.1);
\node at (0.1,1.02) {\footnotesize $\phi^2_{[1,1]}$};
\draw[fill=black] (1.5,0.866) circle (0.1);
\node at (1.175,1.02) {\footnotesize $\phi^2_{[0,3]}$};
\draw[fill=black] (2.5,0.866) circle (0.1);
\node at (2.9,1.02) {\footnotesize $\phi^2_{[1,4]}$};
\draw[fill=black] (3.5,0.866) circle (0.1);
\node at (3.85,1.05) {\footnotesize $\phi^{2}_{[2,5]}$};
\draw[fill=black] (-4,0) circle (0.1);
\node at (-4.4,0) {\footnotesize $\phi^{2}_{[4,4]}$};
\draw[fill=black] (-3,0) circle (0.1);
\node at (-3.4,-0.19) {\footnotesize $\phi^{2}_{[3,3]}$};
\draw[fill=black] (-2,0) circle (0.1);
\node at (-2.375,0.2) {\footnotesize $\phi^2_{[2,2]}$};
\draw[fill=black] (-1,0) circle (0.1);
\node at (-1.4,-0.19) {\footnotesize $\phi^2_{[1,1]}$};
\draw[fill=black] (0,0) circle (0.1);
\node at (-0.35,0.175) {\footnotesize $\phi^2_{[0,0]}$};
\draw[fill=black] (1,0) circle (0.1);
\node at (1.38,-0.175) {\footnotesize $\phi^2_{[1,1]}$};
\draw[fill=black] (2,0) circle (0.1);
\node at (2.4,0.15) {\footnotesize $\phi^2_{[2,2]}$};
\draw[fill=black] (3,0) circle (0.1);
\node at (3.425,-0.2) {\footnotesize $\phi^{2}_{[3,3]}$};
\draw[fill=black] (4,0) circle (0.1);
\node at (4.4,0) {\footnotesize $\phi^{2}_{[4,4]}$};
\draw[fill=black] (-3.5,-0.866) circle (0.1);
\node at (-3.85,-1.05) {\footnotesize $\phi^{2}_{[2,5]}$};
\draw[fill=black] (-2.5,-0.866) circle (0.1);
\node at (-2.9,-1.02) {\footnotesize $\phi^2_{[1,4]}$};
\draw[fill=black] (-1.5,-0.866) circle (0.1);
\node at (-1.9,-1.02) {\footnotesize $\phi^2_{[3,0]}$};
\draw[fill=black] (-0.5,-0.866) circle (0.1);
\node at (-0.9,-1.02) {\footnotesize $\phi^2_{[1,1]}$};
\draw[fill=black] (0.5,-0.866) circle (0.1);
\node at (0.1,-1.02) {\footnotesize $\phi^2_{[1,1]}$};
\draw[fill=black] (1.5,-0.866) circle (0.1);
\node at (1.15,-1.02) {\footnotesize $\phi^2_{[3,0]}$};
\draw[fill=black] (2.5,-0.866) circle (0.1);
\node at (2.9,-1.02) {\footnotesize $\phi^2_{[4,1]}$};
\draw[fill=black] (3.5,-0.866) circle (0.1);
\node at (3.85,-1.05) {\footnotesize $\phi^{2}_{[5,2]}$};
\draw[fill=black] (-3,-1.732) circle (0.1);
\node at (-3.35,-1.95) {\footnotesize $\phi^2_{[6,0]}$};
\draw[fill=black] (-2,-1.732) circle (0.1);
\node at (-2.4,-1.9) {\footnotesize $\phi^2_{[4,1]}$};
\draw[fill=black] (-1,-1.732) circle (0.1);
\node at (-1.4,-1.9) {\footnotesize $\phi^2_{[2,2]}$};
\draw[fill=black] (0,-1.732) circle (0.1);
\node at (-0.4,-1.9) {\footnotesize $\phi^2_{[0,3]}$};
\draw[fill=black] (1,-1.732) circle (0.1);
\node at (1.4,-1.9) {\footnotesize $\phi^2_{[2,2]}$};
\draw[fill=black] (2,-1.732) circle (0.1);
\node at (2.4,-1.9) {\footnotesize $\phi^2_{(1,4]}$};
\draw[fill=black] (3,-1.732) circle (0.1);
\node at (3.4,-1.6) {\footnotesize $\phi^2_{[6,0]}$};
\draw[fill=black] (-2.5,-2.598) circle (0.1);
\node at (-2.85,-2.8) {\footnotesize $\phi^2_{[5,2]}$};
\draw[fill=black] (-1.5,-2.598) circle (0.1);
\node at (-1.9,-2.775) {\footnotesize $\phi^2_{[3,3]}$};
\draw[fill=black] (-0.5,-2.598) circle (0.1);
\node at (-0.9,-2.775) {\footnotesize $\phi^2_{[1,4]}$};
\draw[fill=black] (0.5,-2.598) circle (0.1);
\node at (0.9,-2.775) {\footnotesize $\phi^2_{[4,1]}$};
\draw[fill=black] (1.5,-2.598) circle (0.1);
\node at (1.9,-2.775) {\footnotesize $\phi^2_{[3,3]}$};
\draw[fill=black] (2.5,-2.598) circle (0.1);
\node at (2.85,-2.8) {\footnotesize $\phi^2_{[2,5]}$};
\draw[fill=black] (-2,-3.464) circle (0.1);
\node at (-2.35,-3.65) {\footnotesize $\phi^2_{[4,4]}$};
\draw[fill=black] (-1,-3.464) circle (0.1);
\node at (-1,-3.8) {\footnotesize $\phi^2_{[2,5]}$};
\draw[fill=black] (0,-3.464) circle (0.1);
\node at (0,-3.8) {\footnotesize $\phi^2_{[0,6]}$};
\draw[fill=black] (1,-3.464) circle (0.1);
\node at (1,-3.8) {\footnotesize $\phi^2_{[5,2]}$};
\draw[fill=black] (2,-3.464) circle (0.1);
\node at (2.35,-3.65) {\footnotesize $\phi^2_{[4,4]}$};
\draw (-2,3.464) -- (2,3.464);
\draw (-2.5,2.598) -- (2.5,2.598);
\draw (-3,1.732) -- (3,1.732);
\draw (-3.5,0.866) -- (3.5,0.866);
\draw (-4,0) -- (4,0);
\draw (-3.5,-0.866) -- (3.5,-0.866);
\draw (-3,-1.732) -- (3,-1.732);
\draw (-2.5,-2.598) -- (2.5,-2.598);
\draw (-2,-3.464) -- (2,-3.464);
\draw (-2,-3.464) -- (-4,0);
\draw (-1,-3.464) -- (-3.5,0.866);
\draw (0,-3.464) -- (-3,1.732);
\draw (1,-3.464) -- (-2.5,2.598);
\draw (2,-3.464) -- (-2,3.464);
\draw (2.5,-2.598) -- (-1,3.464);
\draw (3,-1.732) -- (0,3.464);
\draw (3.5,-0.866) -- (1,3.464);
\draw (4,0) -- (2,3.464);
\draw (-4,0) -- (-2,3.464);
\draw (-3.5,-0.866) -- (-1,3.464);
\draw (-3,-1.732) -- (0,3.464);
\draw (-2.5,-2.598) -- (1,3.464);
\draw (-2,-3.464) -- (2,3.464);
\draw (-1,-3.464) -- (2.5,2.598);
\draw (0,-3.464) -- (3,1.732);
\draw (1,-3.464) -- (3.5,0.866);
\draw (2,-3.464) -- (4,0);
\draw[dashed, red, ultra thick] (0,0) -- (4.25,2.45);
\draw[dashed, red, ultra thick] (0,0) -- (4.25,-2.45);
\draw[->, ultra thick] (0,0) -- (0.5,0.288);
\node at (0.5,0.45) {\footnotesize $\omega_2$};
\draw[->, ultra thick] (0,0) -- (0.5,-0.288);
\node at (0.5,-0.45) {\footnotesize $\omega_1$};
\draw[->,xshift=-4.7cm, yshift=3.5cm] (0,0) -- (0.5,-0.288);
\node at (-4,3.1) {\footnotesize $L_1$};
\draw[->,xshift=-4.7cm, yshift=3.5cm] (0,0) -- (-0.5,-0.288);
\node at (-5.35,3.1) {\footnotesize $L_3$};
\draw[->,xshift=-4.7cm, yshift=3.5cm] (0,0) -- (0,0.577);
\node at (-4.7,4.25) {\footnotesize $L_2$};
\end{tikzpicture}
\end{center}
where we have also indicated the fundamental Weyl chamber (spanned by the fundamental weights $\omega_{1,2}$) and attributed the factors $\phi^2_{[c_1,c_2]}$ accordingly. Comparing with the irreducible representations of $\mathfrak{sl}(3,\mathbb{C})$ (see appendix~\ref{App:IrrepsSL3} for a review), we can write the partition function (\ref{Part312}) as a sum over the Weyl orbits of the 13 representatives in the fundamental Weyl chamber of the irreducible representation $\Gamma_{4,4}$. Concretely, we have
\begin{align}
&\mathcal{Z}_{3,1}^{\text{line}}(\tau,t_{f_1},t_{f_2},m=2\epsilon,\epsilon,-\epsilon)= e^{-4\weyl }  \sum_{\lambda=[c_1,c_2] \in P^+_{4,4}} \, (-1)^{c_1+c_2}\,\phi_{[c_1,c_2]}^{2}(\tau,\epsilon) \,\mathcal{O}^2_{\lambda}(t_{f_1},t_{f_2})\,,\label{PartFctGam44}
\end{align}
where the individual (normalised) Weyl orbits are labelled by the Dynkin labels $[c_1,c_2]$ and are given as
\begin{align}
\mathcal{O}^2_{\lambda=[c_1,c_2]}(t_{f_1},t_{f_2})= d_\lambda\sum_{w \in \mathcal{W}(\mathfrak{a}_2)}  e^{w(\lambda)}=d_\lambda\sum_{\sigma\in S_3}\,x_{\sigma(1)}^{c_1+c_2}\,x_{\sigma(2)}^{c_2}\,x_{\sigma(3)}^{0}\,,\label{N3Orbitn2}
\end{align}
and the normalisation factor is given by
\begin{align}
d_{\lambda=[c_1,c_2]}=\frac{|\text{Orb}_\lambda(\mathcal{W}(\mathfrak{a}_2))|}{|\mathcal{W}(\mathfrak{a}_2)|}=\left\{\begin{array}{lcl}\tfrac{1}{6} & \ldots & c_1=c_2=0 \\[6pt] \tfrac{1}{2} & \ldots & \begin{array}{l}c_1=0 \text{ or }c_2=0\text{ and } \\ \,[c_1,c_2]\neq [0,0]  \end{array} \\[12pt] 1 &\ldots & \text{else}\end{array}\right.\label{NormWeylOrb}
\end{align}
where $|\text{Orb}_\lambda(\mathcal{W}(\mathfrak{a}_2))|$ is the order of the orbit of $\lambda$ under the Weyl group of $\mathfrak{a}_2$ and $|\mathcal{W}(\mathfrak{a}_2)|=|S_3|=6$. Finally the following weights of $\Gamma_{4,4}$ are in the fundamental Weyl chamber 
\begin{align}
P^+_{4,4}=\{[0,0]\,,[1,1]\,,[3,0]\,,[0,3]\,,[2,2]\,,[4,1]\,,[1,4]\,,[3,3]\,,[6,0]\,,[0,6]\,,[5,2]\,,[2,5]\,,[4,4]\}\,.
\end{align}
For example we have explicitly\footnote{In order to make contact with the $Q_{f_1}$ and $Q_{f_2}$ we recall that upon using (\ref{RepartQx}), a given monomial in (\ref{Part3n2}) can be written in the form $Q_{f_1}^iQ_{f_2}^j = e^{(j-2i)\omega_1 + (i-2j)\omega_2}$.}
\begin{equation}
e^{-4\weyl }\,\mathcal{O}^2_{[3,0]}(t_{f_1},t_{f_2})=\frac{1}{2}\,x_2^4x_3^8(2x_1^3+2x_2^2+2x_3^3)=Q_{f_1}^4Q_{f_2}^4(Q_{f_1}^2Q_{f_2}+Q_{f_1}^{-1}Q_{f_2}^1+Q_{f_1}^{-1}Q_{f_2}^{-2})\,.
\end{equation}
Here the factors $2$ (which cancel $d_{[3,0]}=\tfrac{1}{2}$) are due to the fact that \emph{e.g.} $x_1^3x_2^0x_3^0=x_1^3x_3^0x_2^0$, such that $x_1^3$ is invariant under two elements of $S_3$. Notice also $|\text{Orb}_{[3,0]}(\mathcal{W}(\mathfrak{a}_2))|=3$. 
 
Before further generalising this discussion to generic $m=n\epsilon$ for $n\in\mathbb{N}$, there are a few comments we would like to make
\begin{itemize}
\item Comparing (\ref{PartFctGam44}) to (\ref{FormIrrep11}), both are structurally very similar in the sense that they are sums over Weyl orbits whose representatives are in the fundamental Weyl chamber of a certain irreducible representation of $\mathfrak{sl}(3,\mathbb{C})$. However, in the case of (\ref{PartFctGam44}), each orbit is still multiplied by a non-trivial function which depends on $\tau$ and $\epsilon$. Another difference is the fact that the terms in each orbit in (\ref{N3Orbitn2}) come with the same relativ sign due to the absence of $(-1)^{\ell(w)}$ which is present in (\ref{FormIrrep11}).
\item The arguments of the theta functions of the individual $\phi^2_{[c_1,c_2]}$ are related to the Dynkin labels $[c_1,c_2]$. Indeed, recall that the $\phi^2_{[c_1,c_2]}$ are quotients of Jacobi-theta functions, schematically
\begin{align}
&\phi^2_{[c_1,c_2]}(\tau,\epsilon) =\prod_r \frac{\theta_1(\tau;a_r\epsilon)}{\theta_1(\tau;b_r\epsilon)} &&\text{with} && a_r, b_r \in \mathbb{N}
\end{align}
Each such quotient has a well-defined index $\mathcal{I}_\tau$ under the action of $SL(2,\mathbb{Z})_\tau$ (which was introduced in (\ref{SL2Actions})\footnote{Notice that $SL(2,\mathbb{Z})_\tau$ remains a symmetry of the partition function even after the identification $\epsilon_1=-\epsilon_2=\epsilon$ and $m=2\epsilon$.})
\begin{align}
&(\tau,\epsilon)\longrightarrow \left(\frac{a\tau+b}{c\tau+d}\,,\frac{\epsilon}{c\tau+d}\right)\,,&&\text{with} &&\left(\begin{array}{cc}a & b \\ c & d\end{array}\right)\in SL(2,\mathbb{Z})_\tau\,.\label{DefActSL2tau}
\end{align}
Specifically, $\mathcal{I}_\tau$ is given as
\begin{equation}
\mathcal{I}_\tau(\phi^2_{\lambda=[c_1,c_2]})=\frac{1}{2} \displaystyle \sum_r (a_r^2 -b_r^2)\,,
\end{equation}
which is related to the weight $\lambda=[c_1,c_2]$ of $\phi^2_{[c_1,c_2]}$ through \begin{equation}
\mathcal{I}_\tau(\phi^2_{\lambda=[c_1,c_2]})=(4\weyl ,4\weyl ) - (\lambda , \lambda)=16 -\frac{1}{3} (c_1^2 + c_1c_2 + c_2^2)\,. \label{len212}
\end{equation}
Here $(.,.)$ stands for the inner product in the basis $(\omega_1,\omega_2)$. 
\end{itemize}

\subsubsection{Case $\epsilon_1=-\epsilon_2$ and $m=n\epsilon$ for generic $n\in\mathbb{N}$} 
\label{31n2}
The results of the previous two subsections show an emergent pattern which can be generalised directly and which we conjecture\footnote{We have indeed verified the results further up to $n=6$.} to hold for generic $n\in \mathbb{N}$: for $n$ a (finite) integer, only a finite number of partitions $\nu_{1,2}$ can contribute to the partition function (\ref{completepartition}). Therefore $\mathcal{Z}_{3,1}^{\text{line}}(\tau,t_{f_1},t_{f_2},m=n\epsilon,\epsilon,-\epsilon)$ is polynomial in the parameters $Q_{f_1}$ and $Q_{f_2}$ with the highest powers $Q_{f_1}^{2n^2}$ and $Q_{f_2}^{2n^2}$. Each monomial $Q_{f_1}^i Q_{f_2}^j$ is multiplied by a quotient of Jacobi-theta functions that depend on $\tau$ and $\epsilon$. Specifically, we can write in a similar fashion as in (\ref{Part3n2}) 
\begin{align}
\mathcal{Z}_{3,1}^{\text{line}}(\tau,m=n\epsilon,t_{f_1},t_{f_2},\epsilon,-\epsilon)= \displaystyle \sum_{i,j=1}^{2n^2} Q_{f_1}^iQ_{f_2}^j \prod_{r} \frac{\theta(a_r(i,j) \epsilon)}{\theta(b_r(i,j) \epsilon)} \,, \quad a_r, b_r \in \mathbb{Z}\,,\label{Part3ngen}
\end{align}
Using the same notation as in the previous subsection, we propose that we can re-write the partition function in the following manner
\begin{align}
&\mathcal{Z}_{3,1}^{\text{line}}(\tau,t_{f_1},t_{f_2},m=n\epsilon,\epsilon,-\epsilon)= e^{-n^2\weyl }  \sum_{\lambda=[c_1,c_2] \in P^+_{n^2,n^2}}\,(-1)^{c_1+c_2} \, \phi_{[c_1,c_2]}^{n}(\tau,\epsilon) \,\mathcal{O}^n_{\lambda}(t_{f_1},t_{f_2})\,.\label{PartFctGamnn}
\end{align}
Here $\mathcal{O}^n_{\lambda}(t_{f_1},t_{f_2})$ denotes the following normalised orbits of the Weyl group $\mathcal{W}(\mathfrak{a}_2)\cong S_3$ (with $d_\lambda$ defined in (\ref{NormWeylOrb}))
\begin{align}
\mathcal{O}^n_{\lambda=[c_1,c_2]}(t_{f_1},t_{f_2})=d_\lambda\,\sum_{w \in \mathcal{W}(\mathfrak{a}_2)}  (-1)^{n\,\ell(w)}\,e^{w(\lambda)}=d_\lambda\,\sum_{\sigma\in S_3}(\text{sign}(\sigma))^n\,x_{\sigma(1)}^{c_1+c_2}\,x_{\sigma(2)}^{c_2}\,x_{\sigma(3)}^{0}\,,\label{N3Orbitnn}
\end{align}
while the representatives $\lambda$ fit into the irreducible representation $\Gamma_{n^2,n^2}$ of $\mathfrak{sl}(3,\mathbb{C})$ (see appendix~\ref{App:IrrepsSL3} for further information and notation) and are chosen from the fundamental Weyl chamber, \emph{i.e.}
\begin{align}
P^+_{n^2,n^2}&=\{[r-s,r+2s]\big| r=0,\ldots, n^2\,\,\text{and}\,\,s=0,\ldots,\text{min}(r,n^2-r)\}\nonumber\\
&\hspace{0.5cm}\cup \{[r+2s,r-s]\big| r=1,\ldots, n^2\,\,\text{and}\,\,s=1,\ldots,\text{min}(r,n^2-r)\}\nonumber\\
&=\{[0,0]\,,\,[1,1]\,,\,[3,0]\,,\,[0,3]\,,\,[2,2]\,,\ldots,[n^2-2,n^2+1]\,,\,[n^2+1,n^2-1]\,,\,[n^2,n^2]\}\,.
\end{align}
Finally, the $\phi^n_{[c_1,c_2]}(\tau,\epsilon)$ are quotients of theta functions and the first few of them are given explicitly in appendix~\ref{App:Coeffs31}. These expressions are compatible with (\ref{Pfac}): notice in particular the appearance of the numerical overall factors $2$ for the weights $[k,0]$ and $[0,k]$ (for $k\in \mathbb{N}$) or $6$ for the weight $[0,0]$, \emph{e.g.} 
\begin{align}
&\phi^n_{[n^2-4,n^2-1]}\big|_{n=2}=\phi^2_{[0,3]}=-2\,\frac{\theta(1)\theta(3)\theta(4)}{\theta(1)^3}&&\text{or}&&\phi^n_{[n^2-4,n^2-4]}\big|_{n=2}=\phi^2_{[0,0]}=6\,\frac{\theta(1)\theta(5)}{\theta(1)^2}\,,
\end{align}
which agree with (\ref{Pfac}) and compensate the factor $d_\lambda$ for the cases $|\text{Orb}_\lambda(\mathcal{W}(\mathfrak{a}_2))|<6$, in order to avoid overcounting. Furthermore, just as in the case $n=2$ in (\ref{Pfac}), the functions $\phi^n_{[c_1,c_2]}$ can be assigned an index under the $SL(2,\mathbb{Z})_\tau$ action defined in (\ref{DefActSL2tau}) 
\begin{equation}
\mathcal{I}(\phi^n_{\lambda=[c_1,c_2]})=(n^2\weyl  , n^2 \weyl )-(\lambda,\lambda)=n^4 - \frac{1}{3}\,(c_1^2 + c_1c_2 + c_2^2)\,.\label{Index21}
\end{equation} 
The structure of (\ref{PartFctGamnn}) can be made more transparent by arranging all terms on the weight lattice of $\mathfrak{a}_2$ as shown in figure \ref{Fig:WeightGenn}. Here the red lines indicate the fundamental Weyl chamber and we have attached the coefficients for each weight respectively. In this way the symmetry under the Weyl group is made manifest.
\begin{figure}[h!tbp]
\begin{center}
\rotatebox{90}{\scalebox{0.97}{\parbox{22.5cm}{\begin{tikzpicture}[scale=1.75]
\draw[fill=black] (-2,3.464) circle (0.1);
\draw[fill=black] (-1.5,3.464) circle (0.1);
\draw[fill=black] (-1,3.464) circle (0.1);
\node at (-0.5,3.464) {$\cdots$};
\node at (0.5,3.464) {$\cdots$};
\draw[fill=black] (1,3.464) circle (0.1);
\draw[fill=black] (1.5,3.464) circle (0.1);
\draw[fill=black] (2,3.464) circle (0.1);
\draw[fill=black] (-2.25,3.031) circle (0.1);
\draw[fill=black] (-1.75,3.031) circle (0.1);
\draw[fill=black] (-1.25,3.031) circle (0.1);
\node at (-0.75,3.031) {$\cdots$};
\node at (0.75,3.031) {$\cdots$};
\draw[fill=black] (1.25,3.031) circle (0.1);
\draw[fill=black] (1.75,3.031) circle (0.1);
\draw[fill=black] (2.25,3.031) circle (0.1);
\draw[fill=black] (-2.5,2.598) circle (0.1);
\draw[fill=black] (-2,2.598) circle (0.1);
\draw[fill=black] (-1.5,2.598) circle (0.1);
\draw[fill=black] (1.5,2.598) circle (0.1);
\draw[fill=black] (2,2.598) circle (0.1);
\draw[fill=black] (2.5,2.598) circle (0.1);
\node[rotate=60] at (-2.75,2.165) {$\cdots$};
\node[rotate=60] at (-2.25,2.165) {$\cdots$};
\node[rotate=300] at (2.25,2.165) {$\cdots$};
\node[rotate=300] at (2.75,2.165) {$\cdots$};
\node[rotate=90] at (0,1.732) {$\cdots$};
\node[rotate=60] at (-3.25,1.299) {$\cdots$};
\node[rotate=300] at (3.25,1.299) {$\cdots$};
\draw[fill=black] (-3.5,0.866) circle (0.1);
\node[rotate=60] at (-3,0.866) {$\cdots$};
\node[rotate=330] at (-1.5,0.866) {$\cdots$};
\draw[fill=black] (-0.5,0.866) circle (0.1);
\draw[fill=black] (0,0.866) circle (0.1);
\draw[fill=black] (0.5,0.866) circle (0.1);
\node[rotate=300] at (3,0.866) {$\cdots$};
\draw[fill=black] (3.5,0.866) circle (0.1);
\draw[fill=black] (-3.75,0.433) circle (0.1);
\draw[fill=black] (-3.25,0.433) circle (0.1);
\draw[fill=black] (-0.75,0.433) circle (0.1);
\draw[fill=black] (-0.25,0.433) circle (0.1);
\draw[fill=black] (0.25,0.433) circle (0.1);
\draw[fill=black] (0.75,0.433) circle (0.1);
\draw[fill=black] (3.25,0.433) circle (0.1);
\draw[fill=black] (3.75,0.433) circle (0.1);
\draw[fill=black] (-4,0) circle (0.1);
\draw[fill=black] (-3.5,0) circle (0.1);
\draw[fill=black] (-3,0) circle (0.1);
\node at (-2,0) {$\cdots$};
\draw[fill=black] (-1,0) circle (0.1);
\draw[fill=black] (-0.5,0) circle (0.1);
\draw[fill=black] (0,0) circle (0.1);
\draw[fill=black] (0.5,0) circle (0.1);
\draw[fill=black] (1,0) circle (0.1);
\node at (2,0) {$\cdots$};
\draw[fill=black] (3,0) circle (0.1);
\draw[fill=black] (3.5,0) circle (0.1);
\draw[fill=black] (4,0) circle (0.1);
\draw[fill=black] (-3.75,-0.433) circle (0.1);
\draw[fill=black] (-3.25,-0.433) circle (0.1);
\draw[fill=black] (-0.75,-0.433) circle (0.1);
\draw[fill=black] (-0.25,-0.433) circle (0.1);
\draw[fill=black] (0.25,-0.433) circle (0.1);
\draw[fill=black] (0.75,-0.433) circle (0.1);
\draw[fill=black] (3.25,-0.433) circle (0.1);
\draw[fill=black] (3.75,-0.433) circle (0.1);
\draw[fill=black] (-3.5,-0.866) circle (0.1);
\node[rotate=300] at (-3,-0.866) {$\cdots$};
\node[rotate=30] at (-1.5,-0.866) {$\cdots$};
\draw[fill=black] (-0.5,-0.866) circle (0.1);
\draw[fill=black] (0,-0.866) circle (0.1);
\draw[fill=black] (0.5,-0.866) circle (0.1);
\node[rotate=60] at (3,-0.866) {$\cdots$};
\draw[fill=black] (3.5,-0.866) circle (0.1);
\node[rotate=300] at (-3.25,-1.299) {$\cdots$};
\node[rotate=60] at (3.25,-1.299) {$\cdots$};
\node[rotate=90] at (0,-1.732) {$\cdots$};
\node[rotate=300] at (-2.75,-2.165) {$\cdots$};
\node[rotate=300] at (-2.25,-2.165) {$\cdots$};
\node[rotate=60] at (2.25,-2.165) {$\cdots$};
\node[rotate=60] at (2.75,-2.165) {$\cdots$};
\draw[fill=black] (-2.5,-2.598) circle (0.1);
\draw[fill=black] (-2,-2.598) circle (0.1);
\draw[fill=black] (-1.5,-2.598) circle (0.1);
\draw[fill=black] (1.5,-2.598) circle (0.1);
\draw[fill=black] (2,-2.598) circle (0.1);
\draw[fill=black] (2.5,-2.598) circle (0.1);
\draw[fill=black] (-2.25,-3.031) circle (0.1);
\draw[fill=black] (-1.75,-3.031) circle (0.1);
\draw[fill=black] (-1.25,-3.031) circle (0.1);
\node at (-0.75,-3.031) {$\cdots$};
\node at (0.75,-3.031) {$\cdots$};
\draw[fill=black] (1.25,-3.031) circle (0.1);
\draw[fill=black] (1.75,-3.031) circle (0.1);
\draw[fill=black] (2.25,-3.031) circle (0.1);
\draw[fill=black] (-2,-3.464) circle (0.1);
\draw[fill=black] (-1.5,-3.464) circle (0.1);
\draw[fill=black] (-1,-3.464) circle (0.1);
\node at (-0.5,-3.464) {$\cdots$};
\node at (0.5,-3.464) {$\cdots$};
\draw[fill=black] (1,-3.464) circle (0.1);
\draw[fill=black] (1.5,-3.464) circle (0.1);
\draw[fill=black] (2,-3.464) circle (0.1);
\draw[style={line width=1pt}] (2,3.464) -- (0.75,3.464);
\draw[style={line width=1pt}] (2.25,3.031) -- (1,3.031);
\draw[style={line width=1pt}] (2.5,2.598) -- (1.25,2.598);
\draw[style={line width=1pt}] (2,3.464) -- (2.625,2.381);
\draw[style={line width=1pt}] (1.5,3.464) -- (2.125,2.381);
\draw[style={line width=1pt}] (1,3.464) -- (1.625,2.381);
\draw[style={line width=1pt}] (2,3.464) -- (1.375,2.381);
\draw[style={line width=1pt}] (1.5,3.464) -- (1.125,2.814);
\draw[style={line width=1pt}] (2.25,3.031) -- (1.875,2.381);
\draw[style={line width=1pt}] (-2,3.464) -- (-0.75,3.464);
\draw[style={line width=1pt}] (-2.25,3.031) -- (-1,3.031);
\draw[style={line width=1pt}] (-2.5,2.598) -- (-1.25,2.598);
\draw[style={line width=1pt}] (-2,3.464) -- (-2.625,2.381);
\draw[style={line width=1pt}] (-1.5,3.464) -- (-2.125,2.381);
\draw[style={line width=1pt}] (-1,3.464) -- (-1.625,2.381);
\draw[style={line width=1pt}] (-2,3.464) -- (-1.375,2.381);
\draw[style={line width=1pt}] (-1.5,3.464) -- (-1.125,2.814);
\draw[style={line width=1pt}] (-2.25,3.031) -- (-1.875,2.381);
\draw[style={line width=1pt}] (4,0) -- (2.75,0);
\draw[style={line width=1pt}] (3.75,-0.433) -- (3,-0.433);
\draw[style={line width=1pt}] (3.75,0.433) -- (3,0.433);
\draw[style={line width=1pt}] (4,0) -- (3.375,-1.082);
\draw[style={line width=1pt}] (3.75,0.433) -- (3.125,-0.6495);
\draw[style={line width=1pt}] (3.5,0.866) -- (2.875,-0.2165);
\draw[style={line width=1pt}] (4,0) -- (3.375,1.082);
\draw[style={line width=1pt}] (3.75,-0.433) -- (3.125,0.6495);
\draw[style={line width=1pt}] (3.5,-0.866) -- (2.875,0.2165);
\draw[style={line width=1pt}] (-1.25,0) -- (1.25,0);
\draw[style={line width=1pt}] (-1,-0.433) -- (1,-0.433);
\draw[style={line width=1pt}] (-1,0.433) -- (1,0.433);
\draw[style={line width=1pt}] (-0.75,-0.866) -- (0.75,-0.866);
\draw[style={line width=1pt}] (-0.75,0.866) -- (0.75,0.866);
\draw[style={line width=1pt}] (-1.125,-0.216) -- (-0.375,1.082);
\draw[style={line width=1pt}] (-0.875,-0.6495) -- (0.125,1.082);
\draw[style={line width=1pt}] (-0.625,-1.082) -- (0.625,1.082);
\draw[style={line width=1pt}] (-0.125,-1.082) -- (0.875,0.6495);
\draw[style={line width=1pt}] (0.375,-1.082) -- (1.125,0.2165);
\draw[style={line width=1pt}] (-1.125,0.216) -- (-0.375,-1.082);
\draw[style={line width=1pt}] (-0.875,0.6495) -- (0.125,-1.082);
\draw[style={line width=1pt}] (-0.625,1.082) -- (0.625,-1.082);
\draw[style={line width=1pt}] (-0.125,1.082) -- (0.875,-0.6495);
\draw[style={line width=1pt}] (1.125,-0.216) -- (0.375,1.082);
\draw[style={line width=1pt}] (-4,0) -- (-2.75,0);
\draw[style={line width=1pt}] (-3.75,-0.433) -- (-3,-0.433);
\draw[style={line width=1pt}] (-3.75,0.433) -- (-3,0.433);
\draw[style={line width=1pt}] (-4,0) -- (-3.375,-1.082);
\draw[style={line width=1pt}] (-3.75,0.433) -- (-3.125,-0.6495);
\draw[style={line width=1pt}] (-3.5,0.866) -- (-2.875,-0.2165);
\draw[style={line width=1pt}] (-4,0) -- (-3.375,1.082);
\draw[style={line width=1pt}] (-3.75,-0.433) -- (-3.125,0.6495);
\draw[style={line width=1pt}] (-3.5,-0.866) -- (-2.875,0.2165);
\draw[style={line width=1pt}] (2,-3.464) -- (0.75,-3.464);
\draw[style={line width=1pt}] (2.25,-3.031) -- (1,-3.031);
\draw[style={line width=1pt}] (2.5,-2.598) -- (1.25,-2.598);
\draw[style={line width=1pt}] (2,-3.464) -- (2.625,-2.381);
\draw[style={line width=1pt}] (1.5,-3.464) -- (2.125,-2.381);
\draw[style={line width=1pt}] (1,-3.464) -- (1.625,-2.381);
\draw[style={line width=1pt}] (2,-3.464) -- (1.375,-2.381);
\draw[style={line width=1pt}] (1.5,-3.464) -- (1.125,-2.814);
\draw[style={line width=1pt}] (2.25,-3.031) -- (1.875,-2.381);
\draw[style={line width=1pt}] (-2,-3.464) -- (-0.75,-3.464);
\draw[style={line width=1pt}] (-2.25,-3.031) -- (-1,-3.031);
\draw[style={line width=1pt}] (-2.5,-2.598) -- (-1.25,-2.598);
\draw[style={line width=1pt}] (-2,-3.464) -- (-2.625,-2.381);
\draw[style={line width=1pt}] (-1.5,-3.464) -- (-2.125,-2.381);
\draw[style={line width=1pt}] (-1,-3.464) -- (-1.625,-2.381);
\draw[style={line width=1pt}] (-2,-3.464) -- (-1.375,-2.381);
\draw[style={line width=1pt}] (-1.5,-3.464) -- (-1.125,-2.814);
\draw[style={line width=1pt}] (-2.25,-3.031) -- (-1.875,-2.381);
\node at (-2.2,-3.7) {\tiny $\phi^n_{[n^2,n^2]}$};
\node at (-2.2,3.7) {\tiny $\phi^n_{[n^2,n^2]}$};
\node at (4.125,0.866) {\tiny $\phi^n_{[n^2-4,n^2+2]}$};
\node at (4.4,0.433) {\tiny $\phi^n_{[n^2-2,n^2+1]}$};
\node at (4.475,0) {\tiny $\phi^n_{[n^2,n^2]}$};
\node at (4.4,-0.433) {\tiny $\phi^n_{[n^2+1,n^2-2]}$};
\node at (4.125,-0.866) {\tiny $\phi^n_{[n^2+2,n^2-4]}$};
\node at (2.725,0.58) {\tiny $\phi^n_{[n^2-3,n^2]}$};
\node at (2.4,0.17) {\tiny $\phi^n_{[n^2-2,n^2-2]}$};
\node at (2.725,-0.58) {\tiny $\phi^n_{[n^2,n^2-3]}$};
\draw[dashed, <-] (3.325,-0.08) to [out=200,in=10] (2.9,-0.35) to [out=180,in=45] (2,-0.5) to [out=210,in=90] (1.9,-0.75); 
\node at (2.2,-0.9) {\tiny $\phi^n_{[n^2-1,n^2-1]}$};
\node at (1.25,0.433) {\tiny $\phi^n_{[0,3]}$};
\node at (1.35,0.13) {\tiny $\phi^n_{[2,2]}$};
\node at (-1.35,-0.13) {\tiny $\phi^n_{[2,2]}$};
\node at (1.25,-0.433) {\tiny $\phi^n_{[3,0]}$};
\node at (-4.475,0) {\tiny $\phi^n_{[n^2,n^2]}$};
\node at (2.3,-3.7) {\tiny $\phi^n_{[n^2,n^2]}$};
\node at (2.3,3.7) {\tiny $\phi^n_{[n^2,n^2]}$};
\node[rotate=270] at (-1.5,-4.1) {\tiny $\phi^n_{[n^2-2,n^2+1]}$};
\node[rotate=270] at (-1,-4.1) {\tiny $\phi^n_{[n^2-4,n^2+2]}$};
\node[rotate=270] at (1.5,-4.1) {\tiny $\phi^n_{[n^2+1,n^2-2]}$};
\node[rotate=270] at (1,-4.1) {\tiny $\phi^n_{[n^2+2,n^2-4]}$};
\node at (-0.75,-2.9) {\tiny $\phi^n_{[n^2-3,n^2]}$};
\node[rotate=90] at (-1.5,-1.9) {\tiny $\phi^n_{[n^2-2,n^2-2]}$};
\node[rotate=90] at (-2,-2) {\tiny $\phi^n_{[n^2,n^2-3]}$};
\node at (0,-2.4) {\tiny $\phi^n_{[n^2-1,n^2-1]}$};
\draw[dashed,->] (-0.3,-2.55) to [out=220,in=45] (-1.4,-2.85) to [out=225,in=20] (-1.6,-2.95);
\node at (0.75,-2.9) {\tiny $\phi^n_{[n^2,n^2-3]}$};
\node[rotate=270] at (1.5,-1.9) {\tiny $\phi^n_{[n^2-2,n^2-2]}$};
\node[rotate=270] at (2,-2) {\tiny $\phi^n_{[n^2-3,n^2]}$};
\draw[dashed,->] (0.3,-2.55) to [out=320,in=135] (1.4,-2.85) to [out=315,in=110] (1.6,-2.95);
\node at (-0.75,2.9) {\tiny $\phi^n_{[n^2,n^2-3]}$};
\node[rotate=270] at (-1.5,1.9) {\tiny $\phi^n_{[n^2-2,n^2-2]}$};
\node[rotate=270] at (1.5,1.9) {\tiny $\phi^n_{[n^2-2,n^2-2]}$};
\node[rotate=270] at (-2,2) {\tiny $\phi^n_{[n^2-3,n^2]}$};
\node at (0,2.4) {\tiny $\phi^n_{[n^2-1,n^2-1]}$};
\draw[dashed,->] (-0.3,2.55) to [out=130,in=315] (-1.4,2.85) to [out=135,in=340] (-1.6,2.95);
\node at (0.75,2.9) {\tiny $\phi^n_{[n^2,n^2-3]}$};
\node[rotate=270] at (2,2) {\tiny $\phi^n_{[n^2-3,n^2]}$};
\draw[dashed,->] (0.3,2.55) to [out=50,in=225] (1.4,2.85) to [out=45,in=200] (1.6,2.95);
\node[rotate=90] at (-1.5,4.1) {\tiny $\phi^n_{[n^2+1,n^2-2]}$};
\node[rotate=90] at (-1,4.1) {\tiny $\phi^n_{[n^2+2,n^2-4]}$};
\node[rotate=90] at (1.5,4.1) {\tiny $\phi^n_{[n^2-2,n^2+1]}$};
\node[rotate=90] at (1,4.1) {\tiny $\phi^n_{[n^2-4,n^2+2]}$};
\node at (-2.85,-3) {\tiny $\phi^n_{[n^2+1,n^2-2]}$};
\node at (-3.1,-2.6) {\tiny $\phi^n_{[n^2+2,n^2-4]}$};
\node at (2.85,-3) {\tiny $\phi^n_{[n^2-2,n^2+1]}$};
\node at (3.1,-2.6) {\tiny $\phi^n_{[n^2-4,n^2+2]}$};
\node at (2.85,3) {\tiny $\phi^n_{[n^2+1,n^2-2]}$};
\node at (3.05,2.65) {\tiny $\phi^n_{[n^2+2,n^2-4]}$};
\node at (-2.85,3) {\tiny $\phi^n_{[n^2-2,n^2+1]}$};
\node at (-3.1,2.65) {\tiny $\phi^n_{[n^2-4,n^2+2]}$};
\node at (-4.125,0.866) {\tiny $\phi^n_{[n^2+2,n^2-4]}$};
\node at (-4.4,0.433) {\tiny $\phi^n_{[n^2+1,n^2-2]}$};
\node at (-4.4,-0.433) {\tiny $\phi^n_{[n^2-2,n^2+1]}$};
\node at (-4.125,-0.866) {\tiny $\phi^n_{[n^2-4,n^2+2]}$};
\node at (-2.725,0.58) {\tiny $\phi^n_{[n^2,n^2-3]}$};
\node at (-2.4,0.17) {\tiny $\phi^n_{[n^2-2,n^2-2]}$};
\node at (-2.725,-0.58) {\tiny $\phi^n_{[n^2-3,n^2]}$};
\draw[dashed, <-] (-3.325,-0.08) to [out=340,in=170] (-2.9,-0.35) to [out=0,in=135] (-2,-0.5) to [out=340,in=90] (-1.9,-0.75); 
\node at (-2.2,-0.9) {\tiny $\phi^n_{[n^2-1,n^2-1]}$};
\node at (-1.1,0.6) {\tiny $\phi^n_{[0,3]}$};
\node at (-1.1,-0.6) {\tiny $\phi^n_{[3,0]}$};
\node at (-0.85,-1) {\tiny $\phi^n_{[2,2]}$};
\node at (-0.8,1) {\tiny $\phi^n_{[2,2]}$};
\node at (0.85,-1) {\tiny $\phi^n_{[2,2]}$};
\draw[ultra thick,blue,dashed] (3.3,1.6) -- (3.025,2.076) -- (2.332,1.676) -- (2.607,1.2) -- (3.3,1.6);
\draw[ultra thick,blue,->] (3.025,2.076) to [out=45,in=180] (5.5,3.7);
\draw[ultra thick,blue,->] (3.3,1.6) to [out=0,in=180] (5.5,1.5);
\node at (8.7,3.8) {for $n$ even};
\draw[fill=black] (6.938,4.041) circle (0.1);
\draw[fill=black] (6.438,4.041) circle (0.1);
\draw[fill=black] (6.688,3.608) circle (0.1);
\draw[fill=black] (6.188,3.608) circle (0.1);
\draw[style={line width=1pt}] (7.063,3.824) -- (6.813,4.258);
\draw[style={line width=1pt}] (6.313,4.258) -- (6.813,3.391);
\draw[style={line width=1pt}] (6.063,3.825) -- (6.313,3.391);
\draw[style={line width=1pt}] (6.938,4.041) -- (6.563,3.391);
\draw[style={line width=1pt}] (6.938,4.041) -- (6.188,4.041);
\draw[style={line width=1pt}] (6.563,4.258) -- (6.063,3.391);
\draw[style={line width=1pt}] (6.938,3.608) -- (5.938,3.608);
\node[rotate=300] at (6.688,4.474) {$\cdots$};
\node[rotate=300] at (7.188,3.608) {$\cdots$};
\node[rotate=300] at (6.188,4.474) {$\cdots$};
\node[rotate=300] at (6.938,3.175) {$\cdots$};
\node[rotate=300] at (6.438,3.175) {$\cdots$};
\node[rotate=300] at (5.938,4.041) {$\cdots$};
\node[rotate=30] at (5.639,3.291) {$\cdots$};
\node at (7.45,4.0) {\tiny $\phi^n_{[0,3n^2/2]}$};
\node at (7.3,3.4) {\tiny $\phi^n_{[1,3n^2/2-2]}$};
\node[rotate=30] at (5.85,3.2) {\tiny $\phi^n_{[0,3n^2/2-3]}$};
\node at (5.85,4.3) {\tiny $\phi^n_{[3n^2/2-2,1]}$};
\draw[red, ultra thick, dashed] (5.9134,3.45) -- (7.3,4.25);
\node at (8.7,1.5) {for $n$ odd};
\draw[fill=black] (6.813,1.958) circle (0.1);
\draw[fill=black] (7.063,1.525) circle (0.1);
\draw[fill=black] (6.563,1.525) circle (0.1);
\draw[fill=black] (6.063,1.525) circle (0.1);
\draw[fill=black] (6.313,1.091) circle (0.1);
\draw[style={line width=1pt}] (7.188,1.308) -- (6.688,2.174);
\draw[style={line width=1pt}] (6.688,1.308) -- (6.438,1.741);
\draw[style={line width=1pt}] (6.438,0.875) -- (5.938,1.741);
\draw[style={line width=1pt}] (7.063,1.525) -- (5.813,1.525);
\draw[style={line width=1pt}] (6.813,1.958) -- (6.188,0.875);
\node[rotate=300] at (6.563,2.391) {$\cdots$};
\node[rotate=300] at (7.313,1.091) {$\cdots$};
\node[rotate=300] at (6.313,1.958) {$\cdots$};
\node[rotate=300] at (6.813,1.091) {$\cdots$};
\node[rotate=300] at (5.813,1.958) {$\cdots$};
\node[rotate=300] at (6.563,0.658) {$\cdots$};
\node[rotate=30] at (5.639,0.991) {$\cdots$};
\node at (7.6,1.725) {\tiny $\phi^n_{[1,(3n^2-1)/2]}$};
\node at (7.4,2.1) {\tiny $\phi^n_{[(3n^2-1)/2,1]}$};
\node[rotate=-60] at (7.025,1.05) {\tiny $\phi^n_{[0,3n^2/2-1]}$};
\node[rotate=-60] at (6.45,0.5) {\tiny $\phi^n_{[2,3n^2/2-2]}$};
\node at (5.4,1.7) {\tiny $\phi^n_{[3n^2/2-2,2]}$};
\draw[red, ultra thick, dashed] (5.9134,1.15) -- (7.3,1.95);
\draw[ultra thick,blue,dashed] (3.3,-1.6) -- (3.025,-2.076) -- (2.332,-1.676) -- (2.607,-1.2) -- (3.3,-1.6);
\draw[ultra thick,blue,->] (3.025,-2.076) to [out=-45,in=180] (5.5,-3.7);
\draw[ultra thick,blue,->] (3.3,-1.6) to [out=0,in=180] (5.5,-1.5);
\node at (8.7,-3.8) {for $n$ even};
\draw[fill=black] (6.938,-4.041) circle (0.1);
\draw[fill=black] (6.438,-4.041) circle (0.1);
\draw[fill=black] (6.688,-3.608) circle (0.1);
\draw[fill=black] (6.188,-3.608) circle (0.1);
\draw[style={line width=1pt}] (7.063,-3.824) -- (6.813,-4.258);
\draw[style={line width=1pt}] (6.313,-4.258) -- (6.813,-3.391);
\draw[style={line width=1pt}] (6.063,-3.825) -- (6.313,-3.391);
\draw[style={line width=1pt}] (6.938,-4.041) -- (6.563,-3.391);
\draw[style={line width=1pt}] (6.938,-4.041) -- (6.188,-4.041);
\draw[style={line width=1pt}] (6.563,-4.258) -- (6.063,-3.391);
\draw[style={line width=1pt}] (6.938,-3.608) -- (5.938,-3.608);
\node[rotate=60] at (6.688,-4.474) {$\cdots$};
\node[rotate=60] at (7.188,-3.608) {$\cdots$};
\node[rotate=60] at (6.188,-4.474) {$\cdots$};
\node[rotate=60] at (6.938,-3.175) {$\cdots$};
\node[rotate=60] at (6.438,-3.175) {$\cdots$};
\node[rotate=60] at (5.938,-4.041) {$\cdots$};
\node[rotate=330] at (5.639,-3.291) {$\cdots$};
\node at (7.45,-4.0) {\tiny $\phi^n_{[3n^2/2,0]}$};
\node at (7.3,-3.4) {\tiny $\phi^n_{[3n^2/2-2,1]}$};
\node[rotate=300] at (6.05,-3.05) {\tiny $\phi^n_{[3n^2/2-3,0]}$};
\node[rotate=60] at (6,-4.5) {\tiny $\phi^n_{[1,3n^2/2-2]}$};
\draw[red, ultra thick, dashed] (5.9134,-3.45) -- (7.3,-4.25);
\node at (8.7,-1.5) {for $n$ odd};
\draw[fill=black] (6.813,-1.958) circle (0.1);
\draw[fill=black] (7.063,-1.525) circle (0.1);
\draw[fill=black] (6.563,-1.525) circle (0.1);
\draw[fill=black] (6.063,-1.525) circle (0.1);
\draw[fill=black] (6.313,-1.091) circle (0.1);
\draw[style={line width=1pt}] (7.188,-1.308) -- (6.688,-2.174);
\draw[style={line width=1pt}] (6.688,-1.308) -- (6.438,-1.741);
\draw[style={line width=1pt}] (6.438,-0.875) -- (5.938,-1.741);
\draw[style={line width=1pt}] (7.063,-1.525) -- (5.813,-1.525);
\draw[style={line width=1pt}] (6.813,-1.958) -- (6.188,-0.875);
\node[rotate=60] at (6.563,-2.391) {$\cdots$};
\node[rotate=60] at (7.313,-1.091) {$\cdots$};
\node[rotate=60] at (6.313,-1.958) {$\cdots$};
\node[rotate=60] at (6.813,-1.091) {$\cdots$};
\node[rotate=60] at (5.813,-1.958) {$\cdots$};
\node[rotate=60] at (6.563,-0.658) {$\cdots$};
\node[rotate=-30] at (5.639,-0.991) {$\cdots$};
\node at (7.6,-1.725) {\tiny $\phi^n_{[(3n^2-1)/2,1]}$};
\node at (7.4,-2.1) {\tiny $\phi^n_{[1,(3n^2-1)/2]}$};
\node[rotate=60] at (7.025,-1.05) {\tiny $\phi^n_{[3n^2/2-1,0]}$};
\node[rotate=60] at (6.45,-0.5) {\tiny $\phi^n_{[3n^2/2-2,2]}$};
\node at (5.4,-1.7) {\tiny $\phi^n_{[2,3n^2/2-2]}$};
\draw[red, ultra thick, dashed] (5.9134,-1.15) -- (7.3,-1.95);
%
\draw[red, ultra thick, dashed] (0,0) -- (3.464,2);
\draw[red, ultra thick, dashed] (0,0) -- (3.464,-2);
\node at (0.85,1) {\tiny $\phi^n_{[2,2]}$};
\node at (1.1,1.5) {\tiny $\phi^n_{[1,1]}$};
\draw[dashed,->] (0.9,1.4) to [out=190,in=30] (-0.2,1.4) to [out=240,in=90] (-0.25,0.65); 
\draw[dashed,->] (1,1.4) to [out=200,in=60] (0.3,1.1) to [out=210,in=90] (0.25,0.65); 
\draw[dashed,->] (1.1,1.4) to [out=300,in=30] (1.1,0.9) to [out=200,in=90] (0.5,0.15); 
\node at (-1.1,-1.5) {\tiny $\phi^n_{[1,1]}$};
\draw[dashed,->] (-0.9,-1.4) to [out=10,in=210] (0.2,-1.4) to [out=60,in=270] (0.25,-0.65); 
\draw[dashed,->] (-1,-1.4) to [out=20,in=240] (-0.3,-1.1) to [out=80,in=270] (-0.25,-0.65); 
\draw[dashed,->] (-1.1,-1.4) to [out=150,in=210] (-1.1,-0.9) to [out=20,in=270] (-0.5,-0.15); 
\node at (-1.6,0.5) {\tiny $\phi^n_{[0,0]}$};
\draw[dashed,->] (-1.4,0.35) to [out=340,in=170] (-0.45,0.2) to [out=0,in=160] (-0.125,0.125);
\node[rotate=90] at (0,1.25) {\tiny $\phi^n_{[3,0]}$};
\node[rotate=270] at (0,-1.25) {\tiny $\phi^n_{[0,3]}$};
\draw[->] (-4.2,3.8) -- (-4.2,4.017);
\node at (-4.2,4.125) {\tiny $L_2$};
\draw[->] (-4.2,3.8) -- (-4.013,3.692);
\node at (-3.9,3.65) {\tiny $L_1$};
\draw[->] (-4.2,3.8) -- (-4.387,3.692);
\node at (-4.5,3.65) {\tiny $L_3$};
\end{tikzpicture}}}}
\end{center}
\caption{Representation of $\mathcal{Z}_{3,1}^{\text{line}}(\tau,m=n\epsilon,t_{f_1},t_{f_2},\epsilon,-\epsilon)$ on the weight lattice of $\mathfrak{a}_2$.}
\label{Fig:WeightGenn}
\end{figure}
We notice, however, that for $n$ odd, the weights $[c_1,c_2]$ for $c_1=0$ or $c_2=0$ do not contribute to the partition function. Indeed, in these cases we have 
\begin{align}
&\mathcal{O}^n_{[c,0]}(t_{f_1},t_{f_2})=0=\mathcal{O}^n_{[0,c]}(t_{f_1},t_{f_2})\,,&&\text{for} &&\begin{array}{l} c\in\mathbb{N} \\ n\in \mathbb{N}_{\text{odd}} \end{array}
\end{align}
due to the sign factors $\text{sign}(\sigma)$ in the definition (\ref{N3Orbitnn}).\footnote{The vanishing is due to the same mechanism which leads to (\ref{Vanishn1}) for $n=1$.}

In order to further elucidate the connection between $\mathcal{Z}_{3,1}^{\text{line}}(\tau,t_{f_1},t_{f_2},m=n\epsilon,\epsilon,-\epsilon)$ and the irreducible representation $\Gamma_{n^2,n^2}$ of $\mathfrak{a}_2$, we remark another property of the $\phi^n_{\lambda}(\tau,\epsilon)$ in (\ref{PhinGen}). As explained in appendix~\ref{App:IrrepsSL3}, the weight diagram of the representation $\Gamma_{n^2,n^2}$ is made from concentric hexagons whose weight spaces share the same multiplicity. Thus, one would expect that the quotients of the theta-functions $\phi^n_\lambda$ are elements of a vector space of functions whose dimension corresponds to the latter multiplicity. Concretely, we expect
\begin{center}
\begin{tabular}{r|l}
multiplicity & weights\\\hline
&\\[-10pt]
1 & $\left\{[n^2-2r,n^2+r]|r=0,\ldots,\lfloor \tfrac{n^2}{2}\rfloor\right\}\cup \left\{[n^2+r,n^2-2r]|r=1,\ldots,\lfloor \tfrac{n^2}{2}\rfloor\right\}$\\[8pt]
&\\[-10pt]
2 & \parbox{14cm}{
$\left\{[n^2-1-2r,n^2-1+r]|r=0,\ldots,\lfloor \tfrac{n^2-1}{2}\rfloor\right\} \cup$\\
${}$\hspace{1cm}$ \left\{[n^2-1+r,n^2-1-2r]|r=1,\ldots,\lfloor \tfrac{n^2-1}{2}\rfloor\right\}$
}\\[24pt]
&\\[-10pt]
3 & \parbox{14cm}{
$\left\{[n^2-2-2r,n^2-2+r]|r=0,\ldots,\lfloor \tfrac{n^2-2}{2}\rfloor\right\} \cup$\\
${}$\hspace{1cm}$ \left\{[n^2-2+r,n^2-2-2r]|r=1,\ldots,\lfloor \tfrac{n^2-2}{2}\rfloor\right\}$
}\\[24pt]
&\\[-10pt]
$\vdots$ & $\vdots$\\[4pt]
&\\[-10pt]
$k$ & \parbox{14cm}{
$\left\{[n^2-(k-1)-2r,n^2-(k-1)+r]|r=0,\ldots,\lfloor \tfrac{n^2-(k-1)}{2}\rfloor\right\} \cup$\\
${}$\hspace{1cm}$ \left\{[n^2-(k-1)+r,n^2-(k-1)-2r]|r=1,\ldots,\lfloor \tfrac{n^2-(k-1)}{2}\rfloor\right\}$
}\\[24pt]
\end{tabular}
\end{center} 
Comparing with the explicit expressions (\ref{PhinGen}), we find that the functions $\phi^n_\lambda$ with weights $\lambda=[c_1,c_2]$ that are expected to be of multiplicity $k\in \mathbb{N}$ according to the above table, are indeed linear combinations of theta-quotients of the following type:
{\allowdisplaybreaks\begin{align}
\mathcal{S}_{k}=\bigg\{&\frac{\theta(n-p)^{a^{(1)}_{p}}\theta(n-p+1)^{a^{(1)}_{p-1}}\ldots \theta(n)^{a^{(1)}_0}\theta(n+p-1)^{a^{(1)}_{p-1}}\theta(n+p)^{a^{(1)}_{p}}}{\theta(1)^{b^{(1)}_1}\theta(2)^{b^{(1)}_2}\ldots\theta(p-1)^{b^{(1)}_{p-1}}}\,,\nonumber\\
&\frac{\theta(n+2-p))^{a^{(2)}_{2-p}}\theta(n+4-p)^{a^{(2)}_{4-p}}\theta(n+5-p)^{a^{(2)}_{5-p}}\ldots \theta(n+p-4)^{a^{(2)}_{p-4}}\theta(n+p-3)^{a^{(2)}_{p-3}}}{\theta(1)^{b^{(2)}_1}\theta(2)^{b^{(2)}_2}\ldots\theta(p-2)^{b^{(2)}_{p-2}}}\nonumber\\
&\hspace{1cm}+\frac{\theta(n+3-p))^{a^{(2)}_{p-3}}\theta(n+4-p)^{a^{(2)}_{p-4}}\ldots \theta(n+p-5)^{a^{(2)}_{5-p}}\theta(n+p-4)^{a^{(2)}_{4-p}}\theta(n+p-2)^{a^{(2)}_{2-p}}}{\theta(1)^{b^{(2)}_1}\theta(2)^{b^{(2)}_2}\ldots\theta(p-2)^{b^{(2)}_{p-2}}}\,,\nonumber\\
&\frac{\theta(n+2-p)^{a^{(3)}_{2-p}}\theta(n+5-p)^{a^{(3)}_{5-p}}\theta(n+6-p)^{a^{(3)}_{6-p}}\ldots \theta(n+p-3)^{a^{(3)}_{p-3}}\theta(n+p-4)^{a^{(3)}_{p-4}}}{\theta(1)^{b^{(3)}_1}\theta(2)^{b^{(3)}_2}\ldots \theta(p-3)^{b^{(3)}_{p-3}}}\nonumber\\
&\hspace{1cm}+\frac{\theta(n+4-p)^{a^{(3)}_{p-4}}\theta(n+3-p)^{a^{(3)}_{p-3}}\ldots \theta(n+p-6)^{a^{(3)}_{6-p}}\theta(n+p-5)^{a^{(3)}_{5-p}}\theta(n+p-2)^{a^{(3)}_{2-p}}}{\theta(1)^{b^{(3)}_1}\theta(2)^{b^{(3)}_2}\ldots \theta(p-3)^{b^{(3)}_{p-3}}}\,,\nonumber\\
&,\ldots\,,\nonumber\\
&\frac{\theta(n+2-p)^{a^{(k)}_{2-p}}\theta(n+2+k-p)^{a^{(k)}_{2+k-p}}\theta(n+3+k-p)^{a^{(k)}_{3+k-p}}\ldots \theta(n+p-k-1)^{a^{(k)}_{p-k-1}}}{\theta(1)^{b^{(k)}_1}\theta(2)^{b^{(k)}_2}\ldots \theta(p-k)^{b^{(k)}_{p-k}}}\nonumber\\
&\hspace{1cm}+\frac{\theta(n+k+1-p)^{a^{(k)}_{p-k-1}}\ldots\theta(n+p-k-3)^{a^{(k)}_{3+k-p}}\theta(n+p-k-2)^{a^{(k)}_{2+k-p}}\theta(n+p-2)^{a^{(k)}_{2-p}}}{\theta(1)^{b^{(k)}_1}\theta(2)^{b^{(k)}_2}\ldots \theta(p-k)^{b^{(k)}_{p-k}}}\bigg\}\,,\label{SpaceFunctionsMult}
\end{align}}
where $p=2n^2+1-c_1-c_2$ and $a^{(r)}_{i},b^{(r)}_j\in\mathbb{N}$. Thus, according to the grouping in (\ref{SpaceFunctionsMult}), the $\phi^n_{\lambda}$ are indeed elements of a space of functions $\mathcal{S}_k$ whose dimension $k$ matches the expected mutliplicity. This is a further indication that the partition functions $\mathcal{Z}_{3,1}^{\text{line}}(\tau,t_{f_1},t_{f_2},m=n\epsilon,\epsilon,-\epsilon)$ can be arranged according to the irreducible representation $\Gamma_{n^2,n^2}$ of $\mathfrak{sl}(3,\mathbb{C}) $ for $n\in \mathbb{N}$.

Finally, we would like to comment on the relation between $\mathcal{Z}_{3,1}^{\text{line}}(\tau,m=n\epsilon,t_{f_1},t_{f_2},\epsilon,-\epsilon)$ and $\mathcal{Z}_{2,1}^{\text{line}}(\tau,m=n\epsilon,t_{f_1},\epsilon,-\epsilon)$ from the point of view of the representation theory of $\mathfrak{sl}(3,\mathbb{C})$ and $\mathfrak{sl}(2,\mathbb{C})$ respectively. Indeed, starting from the highest weight $[n^2,n^2]$ of $\Gamma_{n^2,n^2}$ and acting with only a single root produces a highest weight representation of $\mathfrak{sl}(2,\mathbb{C})$. Indeed, considering the functions
\begin{align}
\left\{\phi^n_{[n^2,n^2]},\phi^n_{[n^2+1,n^2-2]},\phi^n_{[n^2+2,n^2-4]},\ldots,\phi^n_{\left[n^2+\lfloor\tfrac{n^2}{2}\rfloor,n^2-2\lfloor\tfrac{n^2}{2}\rfloor\right]}\right\}
\end{align}
they exactly correspond to the $c_r^{(n)}$ defined in (\ref{Coef11}) -- (\ref{Coef1fin}), that appear in the expansion of $\mathcal{Z}_{2,1}^{\text{line}}(\tau,t_{f_1},m=n\epsilon,\epsilon,-\epsilon)$ in (\ref{ExpZline2}) and which we already argued in section~\ref{Sect:N21ng1} follow the irreducible representations $\Gamma_{n^2}$ of $\mathfrak{sl}(2,\mathbb{C})$. From a physical perspective, acting with only a single root on the highest weight $[n^2,n^2]$ amounts to setting $Q_{f_2}\rightarrow 0$ and thus reducing the M-brane web configuration $(N,M)=(3,1)$ to $(2,1)$ by decoupling one of the M5-branes.

\subsection{Configurations $(N,M)=(4,1)$ and $(5,1)$}
\begin{figure}[H]
\begin{center}
      \includegraphics[scale=0.7]{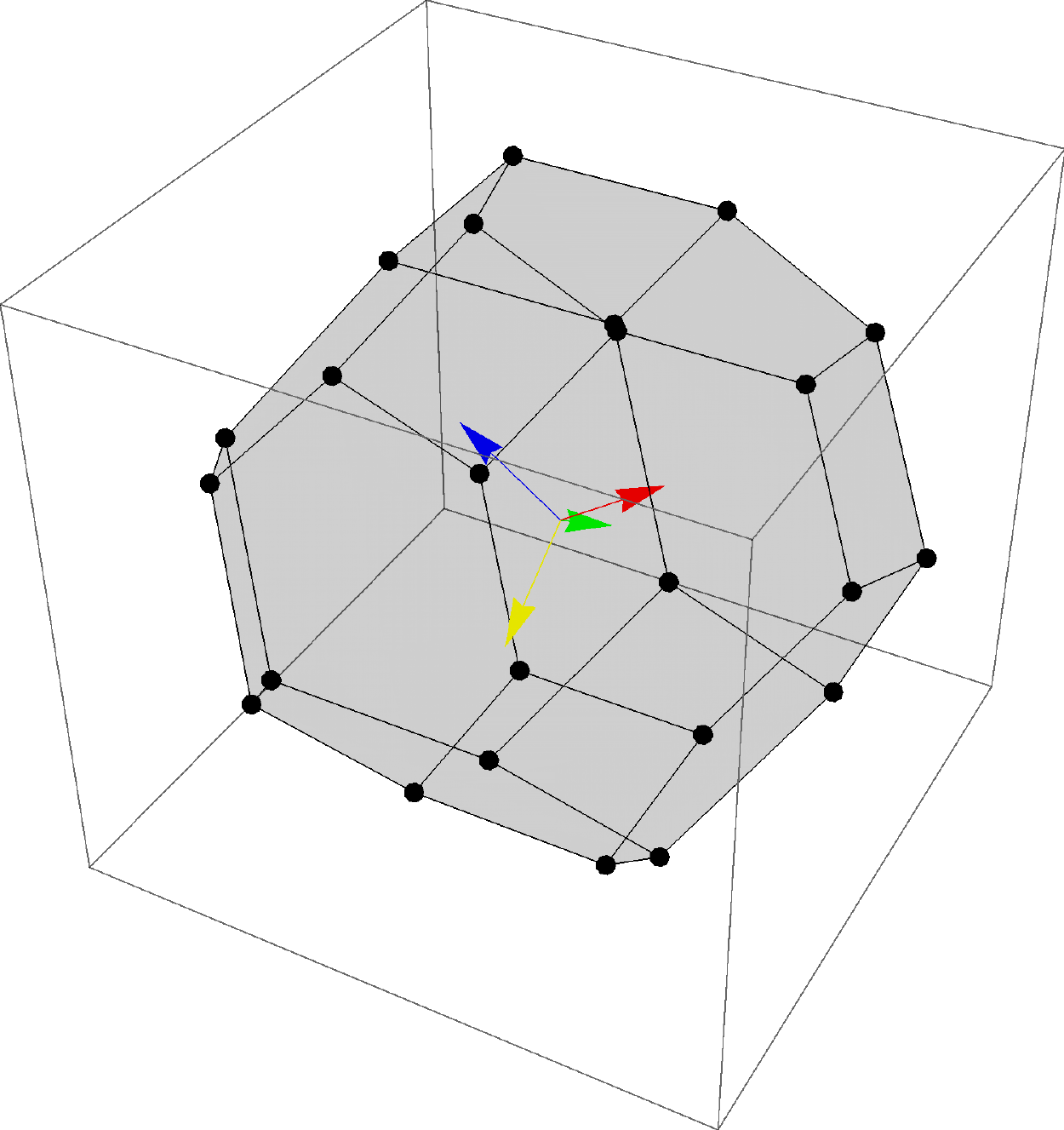}
      \caption{Structure of $\mathcal{Z}_{4,1}^{\text{line}}$ with the simple weights $L_1$ (red), $L_2$ (blue), $L_3$ (green) and $L_4$ (yellow).}
      \label{Fig:A3PartFct}
   \end{center}
\end{figure}

We can repeat the above analysis for $(N,M)=(4,1)$ and $(5,1)$. For simplicity, we restrict ourselves to the case $m=\epsilon$ In the former case, the partition function (\ref{DefZline}) contains a sum over non-trivial partitions $(\nu_1,\nu_2,\nu_3)$ and the relevant contributions are given by
\begin{align}
\mathcal{Z}^{\text{line}}_{4,1}(\tau,t_{f_1},t_{f_2},t_{f_3},m=\epsilon,\epsilon,-\epsilon)&=\sum_{\nu_1,\nu_2,\nu_3}f_{(4,1)}^{(\nu_1,\nu_2,\nu_3)}(\tau,t_{f_1},t_{f_2},t_{f_3},\epsilon)\nonumber\\
&=\sum_{\nu_1,\nu_2,\nu_3}\left(\prod_{a=1}^{3}(-Q_{f_{a}})^{|\nu_{a}|}\right)\,\prod_{(i,j)\in \nu_{a}}\frac{\theta_{1}(\tau;z^{a}_{ij})\,\theta_{1}(\tau;v^{a}_{ij})}{\theta_1(\tau;w^{a}_{ij})\theta_1(\tau;u^{a}_{ij})}
\end{align}
which are tabulated as follows
\begin{center}
\begin{tabular}{|c|c|c|c|}\hline
&&&\\[-10pt]
$\nu_1$ & $\nu_2$ & $\nu_3$ & $f_{(4,1)}^{(\nu_1,\nu_2,\nu_3)}$ \\[8pt]\hline\hline
&&&\\[-10pt]
$\emptyset$ & $\emptyset$ & $\emptyset$ &$1$ \\[4pt]\hline\hline
&&&\\[-10pt]
$\parbox{0.3cm}{\ydiagram{1}}$ & $\emptyset$ & $\emptyset$ & $-Q_{f_1}$ \\[4pt]\hline
&&&\\[-10pt]
$\emptyset$ & $\parbox{0.3cm}{\ydiagram{1}}$ & $\emptyset$ & $-Q_{f_2}$ \\[4pt]\hline
&&&\\[-10pt]
$\emptyset$ & $\emptyset$ &$\parbox{0.3cm}{\ydiagram{1}}$ & $-Q_{f_3}$ \\[4pt]\hline\hline
&&&\\[-10pt]
$\parbox{0.3cm}{\ydiagram{1}}$ & $\emptyset$ & $\parbox{0.3cm}{\ydiagram{1}}$ & $Q_{f_1}Q_{f_3}$ \\[4pt]\hline\hline
&&&\\[-10pt]
$\parbox{0.6cm}{\ydiagram{2}}$ & $\parbox{0.3cm}{\ydiagram{1}}$ & $\emptyset$ & $Q_{f_1}^2Q_{f_2}$ \\[4pt]\hline
&&&\\[-10pt]
$\parbox{0.3cm}{\ydiagram{1}}$ & $\parbox{0.6cm}{\ydiagram{2}}$ &  $\emptyset$ & $Q_{f_1}Q_{f_2}^2$ \\[4pt]\hline
&&&\\[-10pt]
$\emptyset$ & $\parbox{0.6cm}{\ydiagram{2}}$ & $\parbox{0.3cm}{\ydiagram{1}}$ & $Q_{f_2}^2Q_{f_3}$ \\[4pt]\hline
&&&\\[-10pt]
$\emptyset$ & $\parbox{0.3cm}{\ydiagram{1}}$ & $\parbox{0.3cm}{\ydiagram{1,1}}$ & $Q_{f_2}Q_{f_3}^2$ \\[4pt]\hline\hline
&&&\\[-10pt]
$\emptyset$ & $\parbox{0.6cm}{\ydiagram{2}}$ & $\parbox{0.3cm}{\ydiagram{1,1}}$ & $-Q_{f_2}^2Q_{f_3}^2$ \\[4pt]\hline
&&&\\[-10pt]
$\parbox{0.6cm}{\ydiagram{2}}$ & $\parbox{0.3cm}{\ydiagram{1,1}}$ & $\emptyset$ & $-Q_{f_1}^2Q_{f_2}^2$ \\[4pt]\hline\hline
&&&\\[-10pt]
$\parbox{0.6cm}{\ydiagram{2}}$ & $\parbox{0.3cm}{\ydiagram{1}}$ & $\parbox{0.3cm}{\ydiagram{1,1}}$ & $-Q_{f_1}^2Q_{f_2}Q_{f_3}^2$ \\[4pt]\hline
&&&\\[-10pt]
$\parbox{0.3cm}{\ydiagram{1}}$ & $\parbox{0.6cm}{\ydiagram{2,1}}$ & $\parbox{0.3cm}{\ydiagram{1}}$ & $-Q_{f_1}Q_{f_2}^3Q_{f_3}$ \\[4pt]\hline
\end{tabular}
\hspace{3cm}
\begin{tabular}{|c|c|c|c|}\hline
&&&\\[-10pt]
$\nu_1$ & $\nu_2$ & $\nu_3$ & $f_{(4,1)}^{(\nu_1,\nu_2,\nu_3)}$ \\[8pt]\hline\hline
&&&\\[-10pt]
$\parbox{0.3cm}{\ydiagram{1}}$ & $\parbox{0.3cm}{\ydiagram{1,1}}$ & $\parbox{0.3cm}{\ydiagram{1,1,1}}$ & $-Q_{f_1}^2Q_{f_2}^2Q_{f_3}^3$ \\[8pt]\hline
&&&\\[-10pt]
$\parbox{0.9cm}{\ydiagram{3}}$ & $\parbox{0.6cm}{\ydiagram{2}}$ & $\parbox{0.3cm}{\ydiagram{1}}$ & $-Q_{f_1}^3Q_{f_2}^2Q_{f_3}$ \\[4pt]\hline\hline
&&&\\[-10pt]
$\parbox{0.3cm}{\ydiagram{1}}$ & $\parbox{0.6cm}{\ydiagram{2,1}}$ & $\parbox{0.3cm}{\ydiagram{1,1,1}}$ & $Q_{f_1}Q_{f_2}^3Q_{f_3}^3$ \\[8pt]\hline
&&&\\[-10pt]
$\parbox{0.6cm}{\ydiagram{2}}$ & $\parbox{0.3cm}{\ydiagram{1,1}}$ & $\parbox{0.3cm}{\ydiagram{1,1,1}}$ & $Q_{f_1}^2Q_{f_2}^2Q_{f_3}^3$ \\[8pt]\hline
&&&\\[-10pt]
$\parbox{0.9cm}{\ydiagram{3}}$ & $\parbox{0.6cm}{\ydiagram{2}}$ & $\parbox{0.3cm}{\ydiagram{1,1}}$ & $Q_{f_1}^3Q_{f_2}^2Q_{f_3}^2$ \\[8pt]\hline
&&&\\[-10pt]
$\parbox{0.9cm}{\ydiagram{3}}$ & $\parbox{0.6cm}{\ydiagram{2,1}}$ & $\parbox{0.3cm}{\ydiagram{1}}$ & $Q_{f_1}^3Q_{f_2}^3Q_{f_3}$ \\[4pt]\hline\hline
&&&\\[-10pt]
&&&\\[-10pt]
$\parbox{0.6cm}{\ydiagram{2}}$ & $\parbox{0.6cm}{\ydiagram{2,2}}$ & $\parbox{0.3cm}{\ydiagram{1,1}}$ & $Q_{f_1}^2Q_{f_2}^4Q_{f_3}^2$ \\[8pt]\hline\hline
&&&\\[-10pt]
$\parbox{0.6cm}{\ydiagram{2}}$ & $\parbox{0.6cm}{\ydiagram{2,2}}$ & $\parbox{0.3cm}{\ydiagram{1,1,1}}$ & $-Q_{f_1}^2Q_{f_2}^4Q_{f_3}^3$ \\[8pt]\hline
&&&\\[-10pt]
$\parbox{0.9cm}{\ydiagram{3}}$ & $\parbox{0.6cm}{\ydiagram{2,1}}$ & $\parbox{0.3cm}{\ydiagram{1,1,1}}$ & $-Q_{f_1}^3Q_{f_2}^3Q_{f_3}^3$ \\[8pt]\hline
&&&\\[-10pt]
$\parbox{0.9cm}{\ydiagram{3}}$ & $\parbox{0.6cm}{\ydiagram{2,2}}$ & $\parbox{0.3cm}{\ydiagram{1,1}}$ & $-Q_{f_1}^3Q_{f_2}^4Q_{f_3}^2$ \\[4pt]\hline\hline
&&&\\[-10pt]
$\parbox{0.9cm}{\ydiagram{3}}$ & $\parbox{0.6cm}{\ydiagram{2,2}}$ & $\parbox{0.3cm}{\ydiagram{1,1,1}}$ & $Q_{f_1}^3Q_{f_2}^4Q_{f_3}^3$ \\[8pt]\hline
\end{tabular}
\end{center}
Combining all these contributions, we find for the partition function
\begin{align}
\mathcal{Z}^{\text{line}}_{4,1}(\tau,t_{f_{1}},t_{f_2},f_{f_3},&m=\epsilon,\epsilon,-\epsilon)\nonumber\\
&=(1-Q_{f_1})(1-Q_{f_2})(1-Q_{f_3})(1-Q_{f_1}Q_{f_2})(1-Q_{f_2}Q_{f_3})(1-Q_{f_1}Q_{f_2}Q_{f_3})\,,\label{PartF1N4}
\end{align}
which is polynomial in $Q_{f_{1,2,3}}$ and invariant under the exchange $Q_{f_1}\leftrightarrow Q_{f_3}$. By making the following identifications with the simple roots of $\mathfrak{sl}(4,\mathbb{C})$ 
\begin{equation}
Q_{f_1}=e^{-\alpha_1} \,, \quad Q_{f_2}=e^{-\alpha_2} \,, \quad Q_{f_3}=e^{-\alpha_3} \,, 
\end{equation}
we can write (\ref{PartF1N4}) 
\begin{equation}
\mathcal{Z}^{\text{line}}_{X_{4,1}}(\tau,m=\epsilon,t_{f_{1}},t_{f_2},f_{f_3},\epsilon,-\epsilon)
= \prod_{\alpha \in \Delta_+(\mathfrak{a}_3)} (1-e^{-\alpha})^{\text{mult}(\alpha)}\,.
\end{equation}
As before, this can be rewritten, using the Weyl denominator formula, as a sum over the Weyl group for the corresponding root lattice
\begin{equation}
\mathcal{Z}^{\text{line}}_{X_{4,1}}(\tau,m=\epsilon,t_{f_{1}},t_{f_2},f_{f_3},\epsilon,-\epsilon)
= \sum_{w \in \mathcal{W}(\mathfrak{a}_3)} e^{w(\xi) - \xi}
\end{equation}
where $\xi=\frac{3}{2} \alpha_1 + 2 \alpha_2 + \frac{3}{2} \alpha_3$ is the Weyl vector for $ \mathfrak{sl}(4,\mathbb{C}) $. In a similar fashion as in the previous section we can give a graphical representation of the partition function by arranging its various terms on the weight lattice of $\mathfrak{a}_3$ (see figure~\ref{Fig:A3PartFct}). This presentation of the partition function indeed resembles a highest weight representation of $\mathfrak{a}_3\cong \mathfrak{sl}(4,\mathbb{C})$. We have also performed checks for $n>1$: in all cases the partition function still has the structure of irreducible $\mathfrak{sl}(4,\mathbb{C})$ representations.

In the case $(N,M)=(5,1)$, the partition function is a sum over four partitions $(\nu_1,\nu_2,\nu_3,\nu_4)$. Analysing the individual contributions, we find that the partition function can be written as
\begin{align}
\mathcal{Z}^{\text{line}}_{5,1}(\tau,&t_{f_{1}},t_{f_2},f_{f_3},t_{f_4},m=\epsilon,\epsilon,-\epsilon)=
(1-Q_{f_1})(1-Q_{f_2})(1-Q_{f_3})(1-Q_{f_4})(1-Q_{f_1}Q_{f_2})\nonumber\\
&\times (1-Q_{f_2}Q_{f_3})(1-Q_{f_3}Q_{f_4})(1-Q_{f_1}Q_{f_2}Q_{f_3})(1-Q_{f_2}Q_{f_3}Q_{f_4})(1-Q_{f_1}Q_{f_2}Q_{f_3}Q_{f_4})
\nonumber \\
&=\prod_{\alpha \in \Delta_{+}(\mathfrak{a}_4)}(1-e^{-\alpha})
\,.\label{PartF1N5}
\end{align}
where we used 
\begin{equation}
Q_{f_i}=e^{-\alpha_i}\,, \quad i=1,2,3,4
\end{equation}
As in the previous cases this can be rewritten as
\begin{equation}
\mathcal{Z}^{\text{line}}_{5,1}(\tau,t_{f_{1}},t_{f_2},f_{f_3},t_{f_4},m=\epsilon,\epsilon,-\epsilon)=\sum_{w \in \mathcal{W}(\mathfrak{a}_4)} e^{w(\xi) - \xi}
\end{equation}
where $\xi=2\alpha_1+3\alpha_2+3\alpha_3+2\alpha_4$ is the Weyl vector of $\mathfrak{a}_4$.

Comparing (\ref{PartF1N2}), (\ref{PartF1N3}), (\ref{PartF1N4}) and (\ref{PartF1N5}) we conjecture the following pattern
\begin{align}
&\mathcal{Z}^{\text{line}}_{N,1}(\tau,t_{f_{1}},\ldots,t_{f_N{-1}},m=\epsilon,\epsilon,-\epsilon)=\prod_{I=1}^{N-1}\left[\prod_{a=1}^{I}\left(1-\prod_{b=a}^{N-1-I+a}Q_{f_{b}}\right)\right]\nonumber\\
&=
\left[\prod_{a=1}^{N-1}(1-Q_{f_a})\right]\,\left[\prod_{b=1}^{N-2}(1-Q_{f_b}Q_{f_{b+1}})\right]\ldots\left[\prod_{c=1}^2(1-Q_{f_c}Q_{f_{c+1}}\ldots Q_{f_{c+N-3}})\right](1-Q_{f_1}\ldots Q_{f_{N-1}})\,.
\end{align}
which is independent of $\tau$ and $\epsilon$.
\subsection{Configuration $(N,M)=(2,2)$}\label{Sect:Config22NC}
After discussing examples of partition functions for non-compact configurations $(N,M)$ with $M=1$, we can generalise the analysis to cases with $M>1$. We recall that the latter correspond to brane configurations with M5-branes probing a transverse $\mathbb{Z}_M$ orbifold background (\emph{i.e.} an $\text{ALE}_{A_{M-1}}$-space). The simplest such configuration is $(N,M)=(2,2)$, \emph{i.e.} two M5-branes probing a transverse ALE$_{A_1}$ space.
\subsubsection{Choice $\epsilon_1=-\epsilon_2=\epsilon$ and $m=\epsilon$}
We begin with the specific choice $\epsilon_1=-\epsilon_2=m=\epsilon$ for the deformation parameters. Analysing all integer partitions that may contribute to (\ref{GenDefPartFctNon}) in this case, we find 
\begin{align}
\mathcal{Z}^{\text{line}}_{2,2}(\tau,T_{1},t_{f_1},m=\epsilon, \epsilon,-\epsilon)&=
Q_{f_1} \left[Q_{f_1}^{-1}\,\phi^{1,2}_{[2]}(\tau,T_1,\epsilon)  -  \phi^{1,2}_{[0]}(\tau,T_1,\epsilon) + Q_{f_1}\,\phi^{1,2}_{[2]}(\tau,T_1,\epsilon)  \right]\,,
\label{(2,2)}
\end{align}
where $\tilde{T}_1$ was defined in (\ref{IntroTilde}) and
\begin{align}
&\phi^{n=1,M=2}_{[2]}(\tau,T_1,\epsilon) =1\,,&&\text{and} &&\phi^{n=1,M=2}_{[0]}(\tau,T_1,\epsilon) =2\,\frac{\theta_1(\tau; \tilde{T}_1-\epsilon)\theta_1(\tau;\tilde{T}_1+\epsilon)}{\theta_1(\tau;\tilde{T}_1)^2}\,.
\end{align}
Here we have added an additional superscript $M$ in order to distinguish the coefficients from their counterparts with $M=1$ defined in (\ref{Phin1M1}). Moreover, similar to the configuration $(N,M)=(2,1)$, we can write (\ref{(2,2)}) as a sum over Weyl orbits of representatives in the fundamental Weyl chamber of the irreducible representation $\Gamma_2$ of $\mathfrak{sl}(2,\mathbb{C})$. Indeed, similar to (\ref{N2gennForm}), we can write
\begin{align}
\mathcal{Z}_{2,2}^{\text{line}}(\tau,T_1,t_{f_1},m=\epsilon,\epsilon,-\epsilon)=e^{-2\weyl} \sum_{\lambda=[c]\in P^+_{2}}(-1)^{1-\frac{c}{2}}\phi^{1,2}_{\lambda}(\tau,T_1,\epsilon)\,\mathcal{O}^{1,2}_\lambda(t_{f_1})\,,
\end{align}
where $e^{-2\weyl}=Q_{f_1}$ and $P^+_2$ corresponds to the $\mathfrak{sl}(2,\mathbb{C})$ weights in the fundamental Weyl chamber of $\Gamma_2$, \emph{i.e.} $P^+_2=\{[0]\,,[2]\}$. Furthermore, we have the following definition of the Weyl orbits
\begin{align}
\mathcal{O}_{\lambda=[c]}^{n=1,M=2}(t_{f_1})=d_\lambda\,\sum_{w\in \mathcal{W}(\mathfrak{a}_1)}\,e^{w(\lambda)}=d_\lambda\left(Q_{f_1}^{-\frac{c}{2}}+Q_{f_1}^{\frac{c}{2}}\right)\,,\label{WeylOrbitM2}
\end{align}
where $d_\lambda$ was defined in (\ref{NormWeylOrb}). Comparing $\mathcal{O}_{\lambda=[c]}^{n=1,M=2}(t_{f_1})$ to its counterpart for $M=1$ and $n=1$ defined in  (\ref{WeylOrbM1}), we notice that there is no relative sign between the two factors due to the absence of the factor $(-1)^{\ell(w)}$. We can represent $\mathcal{Z}_{2,2}^{\text{line}}(\tau,t_{f_1},m=\epsilon,\epsilon,-\epsilon)$ schematically in the following weight diagram
\begin{center}
\begin{tikzpicture}

\draw (-1,0) -- (1,0);

\draw (-1,-0.2) -- (-1,0.2);
\draw(0,-0.2) -- (0,0.2);
\draw (1,-0.2) -- (1,0.2);

\node[red] at (-1,0) {$\bullet$};
\node[red] at (0,0) {$\bullet$};
\node[red] at (1,0) {$\bullet$};
\node at (-1,0.5) {\footnotesize{$\phi^{1,2}_{[2]}$}};
\node at (0,-0.6) {\footnotesize{$\phi^{1,2}_{[0]}$}};
\node at (1,0.5) {\footnotesize{$\phi^{1,2}_{[2]}$}};

\end{tikzpicture}
\end{center}
The coefficient functions $\phi^{1,2}_{\lambda}$ transform in a particular manner under modular transformations with respect to $\tau$ (generalising the action of $SL(2,\mathbb{Z})$ to the case $M>1$ as in (\ref{GenSl2tau})). Specifically, we have
\begin{equation}
\phi^{1,2}_{[c]}\left(-\tfrac{1}{\tau},\tfrac{\tilde{T}_1}{\tau},\tfrac{\epsilon}{\tau}\right)=e^{-\frac{2\pi i\,\epsilon^2\,\mathcal{I}_\tau}{\tau}}\,\phi^{1,2}_{[c]}(\tau,\tilde{T}_1,\epsilon)\,,\label{Trafo22}
\end{equation}
where for $\lambda=[c]$ we have
\begin{equation}
\mathcal{I}_\tau\left(\phi^{1,2}_{\lambda=[c]}\right)=(2\weyl  , 2\weyl  ) - (\lambda,\lambda)= 1- \frac{1}{4} c^2\,,\label{GenIndex2}
\end{equation}
where $(.,.)$ denotes the inner product in the fundamental weight basis $\{\omega_1\}$ of $\mathfrak{a}_1$. Generalising (\ref{Index21}) we call $\mathcal{I}_\tau$ the index of $\phi^{1,2}_{\lambda}$ under $SL(2,\mathbb{Z})_\tau$. We point out in particular that the phase-factor (\ref{Trafo22}) is independent of $T_1$ and only depends on $\epsilon$.

Before generalising the above discussion to cases $m=n\epsilon$ for $n\in\mathbb{N}$, we would like to make a further remark: in section~\ref{Sect:ParticularValues} we argued that the simplification of the partition function $\mathcal{Z}_{N,M=1}$ for $m=n\epsilon$ and $\epsilon_1=-\epsilon_2=\epsilon$ is due to the fact that the $Spin(8)$ holonomy charges are no longer linear independent over $\mathcal{Z}$. Therefore, there are possible cancellation among contributions with different charges with respect to $U(1)_m$ and $U(1)_{\epsilon_1}\times U(1)_{\epsilon_2}$. For $M>1$, the same simplifications take place in the untwisted sector of the orbifold action (\ref{ActionOrbifold}), leading to similar simplifications of the partition function, as is indeed showcased in (\ref{(2,2)}). However, along the same line of reasoning, identifying $\tilde{T}_1=k\epsilon$ for $k\in\mathbb{N}$, should lead to further cancellations among different contributions in the partition function. Indeed, setting $\tilde{T}_1=\epsilon$ in (\ref{(2,2)}) we get $\phi^{1,2}_{[0]}(\tau,\epsilon,\epsilon)=0$, such that
\begin{align}
\mathcal{Z}^{\text{line}}_{2,2}(\tau,\tilde{T}_1=\epsilon,t_{f_1}, \epsilon,-\epsilon)= 1+ Q_{f_1}^2\,.
\end{align}
This choice of parameters is still compatible with the $SL(2,\mathbb{Z})_\tau$ transformation (\ref{Trafo22}).

\subsubsection{Choice $\epsilon_1=-\epsilon_2=\epsilon$ and $m=n \epsilon$ for $n>1$}
Generalising the discussion of the previous subsubsection for $m=n\epsilon$ with $n>1$ the partition function can schematically be written in the following form:\footnote{We have checked this expression explicitly up to $n=10$ and conjecture that it holds in general.}
\begin{align}
\mathcal{Z}^{\text{line}}_{2,2}(\tau,T_1,t_{f_1},m=n\epsilon,\epsilon,-\epsilon) =\displaystyle \sum_{i=1}^{2n^2}Q_{f_1}^i \prod_{r} \frac{\theta_1(\tau;a_{1,r} \epsilon)\theta_1(\tau;a_{2,r} \epsilon + \tilde{T}_1)}{\theta_1(\tau;b_{1,r} \epsilon)\theta_1(\tau;b_{2,r} \epsilon + \tilde{T}_1)}\,.
\label{(2,2)2}
\end{align}
Analogously to the previous cases we propose that the partition function (\ref{(2,2)2}) can be written by summing the Weyl orbits for the weights in the fudamental Weyl chamber $P_{2n^2}^+$ of the irreducible representation $\Gamma_{2n^2}$ of $\mathfrak{sl}(2,\mathbb{C})$
\begin{equation}
\mathcal{Z}_{2,2}^{\text{line}}(\tau,T_1,t_{f_1},m=n\epsilon,\epsilon,-\epsilon)= e^{-2n^2\weyl } \sum_{\lambda \in P^+_{2n^2,2n^2}} (-1)^{\frac{2n^2-c}{2}}\, \phi^{n,M=2}_{[c_1]}(\tau,T_1,\epsilon)\,\mathcal{O}_{\lambda}^{n,M=2}(t_{f_1})
\end{equation}
where $e^{-2n^2\weyl}=Q_{f_1}^{n^2}$, $P^+_{2n^2}=\{[2k]|k=0,\ldots,n^2\}$ and the Weyl orbits $\mathcal{O}_{\lambda}^{n,M=2}(t_{f_1})$ are defined as
\begin{align}
\mathcal{O}_{\lambda=[c]}^{n=1,M=2}(t_{f_1})=d_\lambda\,\sum_{w\in \mathcal{W}(\mathfrak{a}_1)}\,(-1)^{Mnl(w)}\,e^{w(\lambda)}=d_\lambda \left(Q_{f_1}^{-\frac{c}{2}}+Q_{f_1}^{\frac{c}{2}}\right)\,,
\end{align}
which is equivalent to (\ref{WeylOrbitM2}) since $(-1)^{Mnl(w)}=1$ for $n\in\mathbb{Z}$. Furthermore, the first few coefficient functions $\phi^{n,M=2}_{[c_1]}(\tau,T_1,\epsilon)$ are given by (for $\lambda=[c]$ with $c\geq 0$)
{\allowdisplaybreaks
\begin{align}
&\phi^{n,2}_{[2n^2]}=1\,,\nonumber\\
&\phi^{n,2}_{[2(n^2-1)]}=2\frac{\theta(n)^2\theta(\tilde{T}_1-n)\theta(\tilde{T}_1+n)}{\theta(1)^2\theta(\tilde{T}_1)^2}\,,\nonumber\\
&\phi^{n,2}_{[2(n^2-2)]}=\frac{\theta(n)^4\theta(\tilde{T}_1-n)^2\theta(n+\tilde{T}_1)^2}{\theta(1)^4\theta(\tilde{T}_1-1)^2\theta(\tilde{T}_1+1)^2}\nonumber\\
&\hspace{0.5cm}+2\frac{\theta(n-1)\theta(n)^2\theta(n+1)\theta(\tilde{T}_1-n-1)\theta(\tilde{T}_1-n)\theta(\tilde{T}_1+n-1)\theta(\tilde{T}_1+n)}{\theta(1)^2\theta(2)^2\theta(\tilde{T}_1)^2\theta(\tilde{T}_1+1)^2}\nonumber\\
&\hspace{0.5cm}+2\frac{\theta(n-1)\theta(n)^2\theta(n+1)\theta(\tilde{T}_1-n)\theta(\tilde{T}_1-n+1)\theta(\tilde{T}_1+n)\theta(\tilde{T_1}+n+1)}{\theta(1)^2\theta(2)^2\theta(\tilde{T}_1)^2\theta(\tilde{T}_1+1)^2}\nonumber\\
%
\end{align}}
Generalising (\ref{GenIndex2}) and using the notation (\ref{(2,2)2}), the index of the theta ratios is 
\begin{equation}
\mathcal{I}_\tau(\phi^{n,2}_{[c]})= \sum_r (a_{1,r}^2 + a_{2,r}^2 -b_{1,r}^2 -b_{2,r}^2)=(2n^2 \weyl  ,2n^2 \weyl ) - (\lambda, \lambda)=n^4-\frac{1}{4} c^2\,.
\end{equation}
Finally, as for the case $n=1$, there are additional cancellations in the partition function once we set $\tilde{T}_1=k \epsilon$ (with $k\in \mathbb{N}$) to be a(n integer) multiple of $\epsilon$. Notice, however, when $k<n$ the partition function $\mathcal{Z}_{2,2}^{\text{line}}(\tau,T_1=k\epsilon,t_{f_1},m=n\epsilon,\epsilon,-\epsilon)$ appears to diverge due to the fact that theta-functions in the denominator vanish. The choice $k=n$ provides the simplest expression in the sense that certain $\phi^{n,M=2}_{\lambda}$ vanish. Schematically, the vanishing coefficient functions can be shown in the following weight diagram of $\mathfrak{sl}(2,\mathbb{C})$:
\begin{center}
\begin{tikzpicture}
\draw[<-] (-8,0) -- (-6,0);
\draw[dashed] (-6,0) -- (-4,0);
\draw (-4,0) -- (-3,0);
\draw[dashed] (-3,0) -- (-1,0);
\draw (-1,0) -- (1,0);
\draw[dashed] (1,0) -- (3,0);
\draw (3,0) -- (4,0);
\draw[dashed] (4,0) -- (6,0);
\draw[->] (6,0) -- (8,0);
\draw (-7,-0.2) -- (-7,0.2);
\draw(-6,-0.2) -- (-6,0.2);
\draw[dashed]  (-5,-0.2) -- (-5,0.2);
\draw (-4,-0.2) -- (-4,0.2);
\draw (-3,-0.2) -- (-3,0.2);
\draw[dashed] (-2,-0.2) -- (-2,0.2);
\draw (-1,-0.2) -- (-1,0.2);
\draw[thick,black] (0,-0.2) -- (0,0.2);
\draw (1,-0.2) -- (1,0.2);
\draw[dashed] (2,-0.2) -- (2,0.2);
\draw (3,-0.2) -- (3,0.2);
\draw (4,-0.2) -- (4,0.2);
\draw[dashed] (5,-0.2) -- (5,0.2);
\draw (6,-0.2) -- (6,0.2);
\draw (7,-0.2) -- (7,0.2);
\node[red] at (-7,0) {$\bullet$};
\node[red] at (-3,0) {$\bullet$};
\node[red] at (-2,0) {$\bullet$};
\node[red] at (-1,0) {$\bullet$};
\draw[red] (0,0) circle (0.1);
\node[red] at (1,0) {$\bullet$};
\node[red] at (2,0) {$\bullet$};
\node[red] at (3,0) {$\bullet$};
\node[red] at (7,0) {$\bullet$};
\node at (-7,0.5) {\footnotesize{$\phi^{n,2}_{[2n^2]}$}};
\node[rotate=270] at (-6,-1.4) {\footnotesize{$\phi^{n,2}_{[2(n^2-1)]}\rightarrow 0$}};
\node[rotate=270] at (-4,-1.6) {\footnotesize{$\phi^{n,2}_{[2(n^2-n+1)]}\rightarrow 0$}};
\node[rotate=90] at (-3,1.1) {\footnotesize{$\phi^{n,2}_{[2(n^2-n)]}$}};
\node at (-1,0.5) {\footnotesize{$\phi^{n,2}_{[2]}$}};
\node at (0,-0.5) {\footnotesize{$\phi^{n,2}_{[0]}$}};
\draw[->] (0,-0.9) -- (1,-2.2);
\node at (1,-2.5) {\footnotesize $=0$};
\node[rotate=308,scale=1.2] at (0.65,-1.45) {\tiny for $n$ odd};
\draw[->] (0,-0.9) -- (-1,-2.2);
\node[rotate=52,scale=1.2] at (-0.65,-1.45) {\tiny for $n$ even};
\node at (-1,-2.5) {\footnotesize $\neq0$};
\node at (1,0.5) {\footnotesize{$\phi^{n,2}_{[2]}$}};
\node[rotate=90] at (3,1.1) {\footnotesize{$\phi^{n,2}_{[2(n^2-n)]}$}};
\node[rotate=270] at (4,-1.6) {\footnotesize{$\phi^{n,2}_{[2(n^2-n+1)]}\rightarrow 0$}};
\node[rotate=270] at (6,-1.4) {\footnotesize{$\phi^{n,2}_{[2(n^2-1)]}\rightarrow 0$}};
\node at (7,0.5) {\footnotesize{$\phi^{n,2}_{[2n^2]}$}};
\end{tikzpicture}
\end{center}
The vanishing theta-quotients correspond to the following powers of $Q_{f_1}$ in the partition function: $Q_{f_1},Q_{f_1}^2,\dots,Q_{f_1}^n,Q_{f_1}^{2n^2-n},Q_{f_1}^{2n^2-n+1},\dots,Q_{f_1}^{2n^2-1}$, while for odd $n$ also the power $Q_{f_1}^{n^2}$ is vanishing as well. 
\subsection{Configuration $(N,M)=(3,2)$}
We can analyse the configuration $(N,M)=(3,2)$ in a similar fashion. The latter corresponds to a brane web with 3 M5-branes probing a transverse ALE$_{A_1}$ space. 
\subsubsection{Choice $\epsilon_1=-\epsilon_2=\epsilon$ and $m=\epsilon$}
We again begin with the case $m=\epsilon$. In order to write the partition function, we use the same notation as in section~\ref{Sect:NonCompactN3}. In particular we use the variables $x_{1,2,3}$ as defined in (\ref{Defxr}) to write
\begin{align}
\mathcal{Z}_{3,2}^{\text{line}}(\tau,T_1,t_{f_1},t_{f_2},m=\epsilon,\epsilon,-\epsilon)= x_2^2x_3^4 \big\{ &\phi^{1,2}_{[2,2]}(\tau,T_1,\epsilon)\,(x_1^4x_2^2 + x_1^4x_3^2 + x_2^4x_3^2 + x_2^4x_1^2 + x_3^4x_1^2 + x_3^4x_2^2) \nonumber \\
+ &\phi^{1,2}_{[3,0]}(\tau,T_1,\epsilon)\, (x_1^3+x_2^3+x_3^3)\nonumber\\
+& \phi_{[0,3]}^{1,2}(\tau,T_1,\epsilon)\,(x_1^3x_2^3+x_2^3x_3^3+x_1^3x_3^3) \nonumber \\
+& \phi^{1,2}_{[1,1]}(\tau,T_1,\epsilon)\,( x_1^2x_2 + x_1^2x_3 + x_2^2x_3 + x_2^2x_1 + x_3^2x_1 + x_3^2x_2  )  \nonumber \\
+&\phi^{1,2}_{[0,0]} \big\}\label{PartFct32n1}
\end{align}
where $x_2^2x_3^4=e^{-2\weyl}=Q_{f_1}^2 Q_{f_2}^2$ with $\weyl $ the Weyl vector of $ \mathfrak{a}_2 $ and the $\phi^{1,2}_{[c_1,c_2]}(\tau,T_1,\epsilon)$ are defined as follows
\begin{align} 
&\phi^{1,2}_{[2,2]}=1\,, \quad
&&\phi^{1,2}_{[1,1]}=\frac{\theta_1(\tilde{T}_1+2\epsilon)\theta_1(\tilde{T}_1-\epsilon)^2+\theta_1(\tilde{T}_1-2\epsilon)\theta_1(\tilde{T}_1+\epsilon)^2}{\theta_1(\tilde{T}_1)^3}  \nonumber \\
&\phi^{1,2}_{[0,0]}=-6\frac{\theta_1(\tilde{T}_1-2\epsilon)\theta_1(\tilde{T}_1+2\epsilon)}{\theta_1(\tilde{T_1})^2}\,, \quad
&& \phi^{1,2}_{[3,0]}=\phi^{1,2}_{[0,3]}=-2\frac{\theta_1(\tilde{T}_1-\epsilon)\theta_1(\tilde{T}_1+\epsilon)}{\theta_1(\tilde{T_1})^2}\,.\label{Defphi32n1}
\end{align}
As in section~\ref{Sect:NonCompactN3}, the polynomials in $x_{1,2,3}$ in (\ref{PartFct32n1}) resemble orbits of the Weyl action $\mathcal{W}(\mathfrak{a}_2)\cong S_3$ (and the subscripts in (\ref{Defphi32n1}) correspond to weights of $\mathfrak{a}_2$). More precisely, $\mathcal{Z}_{3,2}^{\text{line}}(\tau,T_1,t_{f_1},t_{f_2},m=\epsilon,\epsilon,-\epsilon)$ can be expressed as a sum over the Weyl orbits of the weights in the fundamental Weyl chamber $P_{2,2}^+$ of the irreducible representation $\Gamma_{2,2}$ of $\mathfrak{sl}(3,\mathbb{C})$
\begin{equation}
\mathcal{Z}_{3,2}^{\text{line}}(\tau,T_1, t_{f_1},t_{f_2}, m=\epsilon, \epsilon, -\epsilon)= e^{-2\weyl } \displaystyle \sum_{\lambda=[c_1,c_2] \in P^+_{2,2}} \phi^{1,2}_{[c_1,c_2]}(\tau,T_1,\epsilon) \mathcal{O}_{\lambda}^1(t_{f_1},t_{f_2})
\end{equation}
where $P^{+}_{2,2}=\{[0,0]\,,[1,1]\,,[0,3]\,,[3,0]\,,[2,2]\}$ and the Weyl orbits are given by
\begin{equation}
\mathcal{O}_{\lambda}^1(t_{f_1},t_{f_2})= d_\lambda\,\displaystyle \sum_{w\in \mathcal{W}}e^{w(\lambda)}
\end{equation}
with $d_\lambda$ defined in (\ref{NormWeylOrb}). As in section~\ref{Sect:Config22NC}, the partition function $\mathcal{Z}_{3,2}^{\text{line}}(\tau,T_1, t_{f_1},t_{f_2}, m=\epsilon, \epsilon, -\epsilon)$ transforms well under $SL(2,\mathbb{Z})_\tau$: following the transformation (\ref{Trafo22}) we have for example 
\begin{equation}
\phi^{1,2}_{[1,1]}\left(-\tfrac{1}{\tau},\tfrac{\epsilon}{\tau},\tfrac{\tilde{T}_1}{\tau}\right)=e^{i \pi (-2\epsilon+\tilde{T}_1)^2}e^{i \pi 2(\epsilon+\tilde{T}_1)^2} e^{-i \pi 3 \tilde{T}_1^2} \phi^{1,2}_{[1,1]}(\tau,\epsilon,\tilde{T}_1)=e^{2 i \pi 3\epsilon^2} \phi^{1,2}_{[1,1]}(\tau,\epsilon,\tilde{T}_1)\,.
\end{equation}
In general we can introduce the index 
\begin{equation}
\mathcal{I}_\tau(\phi^{1,2}_{[c_1,c_2]}) = (2\weyl  , 2 \weyl ) - (\lambda , \lambda)= 4 - \frac{1}{3}(c_1^2 + c_1c_2+c_2^2)\,.\label{relX32}
\end{equation}
As in the previous section the partition function can be further simplified by setting $\tilde{T}_1=2\epsilon$.
We can represent the partition function by the following diagram:
\begin{center}
\begin{tikzpicture}[scale=1.4]
\draw[fill=black] (-1,1.732) circle (0.1);
\node at (-1.2,2) {\footnotesize $\phi_{[2,2]}$};
\draw[red] (0,1.732) circle (0.1);
\node at (0,2.05) {\footnotesize $\phi_{[3,0]}$};
\draw[fill=black] (1,1.732) circle (0.1);
\node at (1.2,2) {\footnotesize $\phi_{[2,2]}$};
\draw[red] (-1.5,0.866) circle (0.1);
\node at (-1.8,1.1) {\footnotesize $\phi_{[0,3]}$};
\draw[fill=black] (-0.5,0.866) circle (0.1);
\node at (-0.85,1.05) {\footnotesize $\phi_{[1,1]}$};
\draw[fill=black] (0.5,0.866) circle (0.1);
\node at (0.95,1.05) {\footnotesize $\phi_{[1,1]}$};
\draw[red] (1.5,0.866) circle (0.1);
\node at (1.9,0.8) {\footnotesize $\phi_{[0,3]}$};
\draw[fill=black] (-2,0) circle (0.1);
\node at (-2.45,0) {\footnotesize $\phi_{[2,2]}$};
\draw[fill=black] (-1,0) circle (0.1);
\node at (-1.4,-0.15) {\footnotesize $\phi_{[1,1]}$};
\draw[fill=black] (0,0) circle (0.1);
\node at (-0.4,0.2) {\footnotesize $\phi_{[0,0]}$};
\draw[fill=black] (1,0) circle (0.1);
\node at (1.45,-0.15) {\footnotesize $\phi_{[1,1]}$};
\draw[fill=black] (2,0) circle (0.1);
\node at (2.45,0) {\footnotesize $\phi_{[2,2]}$};
\draw[red] (-1.5,-0.866) circle (0.1);
\node at (-1.8,-1.1) {\footnotesize $\phi_{[3,0]}$};
\draw[fill=black] (-0.5,-0.866) circle (0.1);
\node at (-0.9,-1.05) {\footnotesize $\phi_{[1,1]}$};
\draw[fill=black] (0.5,-0.866) circle (0.1);
\node at (0.95,-1.05) {\footnotesize $\phi_{[1,1]}$};
\draw[red] (1.5,-0.866) circle (0.1);
\node at (1.9,-0.8) {\footnotesize $\phi_{[3,0]}$};
\draw[fill=black] (-1,-1.732) circle (0.1);
\node at (-1.2,-2) {\footnotesize $\phi_{[2,2]}$};
\draw[red] (0,-1.732) circle (0.1);
\node at (0,-2.05) {\footnotesize $\phi_{[0,3]}$};
\draw[fill=black] (1,-1.732) circle (0.1);
\node at (1.2,-2) {\footnotesize $\phi_{[2,2]}$};
\draw (-1,1.732) -- (1,1.732);
\draw (-1.5,0.866) -- (1.5,0.866);
\draw (-2,0) -- (2,0);
\draw (-1.5,-0.866) -- (1.5,-0.866);
\draw (-1,-1.732) -- (1,-1.732);
\draw (-1,-1.732) -- (-2,0);
\draw (0,-1.732) -- (-1.5,0.866);
\draw (1,-1.732) -- (-1,1.732);
\draw (1.5,-0.866) -- (0,1.732);
\draw (1,1.732) -- (2,0);
\draw (1,-1.732) -- (2,0);
\draw (0,-1.732) -- (1.5,0.866);
\draw (-1,-1.732) -- (1,1.732);
\draw (-1.5,-0.866) -- (0,1.732);
\draw (-1,1.732) -- (-2,0);
\draw[dashed,red] (0,0) -- (3,1.732);
\draw[dashed,red] (0,0) -- (3,-1.732);
\draw[->, ultra thick] (0,0) -- (0.5,0.288);
\node at (0.5,0.45) {\footnotesize $\omega_2$};
\draw[->, ultra thick] (0,0) -- (0.5,-0.288);
\node at (0.5,-0.45) {\footnotesize $\omega_1$};
\end{tikzpicture}
\end{center}
The red circles correspond to the terms that are removed by the simplification.
\subsubsection{Choice $\epsilon_1=-\epsilon_2=\epsilon$ and $m=n\epsilon$ for $n>1$ }
For $n\geq 2$ the number (and size) of all expressions grows very quickly. However, all terms can still be arranged according to irreducible representations of $\mathfrak{a}_2$, as \emph{e.g.} is graphically shown below for $n=2$
\begin{center}
\begin{tikzpicture}[scale=1.4]
\draw[fill=black] (-2,3.464) circle (0.1);
\draw[red] (-1.5,3.464) circle (0.1);
\draw[red] (-1,3.464) circle (0.1);
\draw[fill=black] (-0.5,3.464) circle (0.1);
\draw[fill=black] (0,3.464) circle (0.1);
\draw[fill=black] (0.5,3.464) circle (0.1);
\draw[red] (1,3.464) circle (0.1);
\draw[red] (1.5,3.464) circle (0.1);
\draw[fill=black] (2,3.464) circle (0.1);
\draw[red] (-2.25,3.031) circle (0.1);
\draw[fill=black] (-1.75,3.031) circle (0.1);
\draw[red] (-1.25,3.031) circle (0.1);
\draw[fill=black] (-0.75,3.031) circle (0.1);
\draw[fill=black] (-0.25,3.031) circle (0.1);
\draw[fill=black] (0.25,3.031) circle (0.1);
\draw[fill=black] (0.75,3.031) circle (0.1);
\draw[red] (1.25,3.031) circle (0.1);
\draw[fill=black] (1.75,3.031) circle (0.1);
\draw[red] (2.25,3.031) circle (0.1);
\draw[red] (-2.5,2.598) circle (0.1);
\draw[red] (-2,2.598) circle (0.1);
\draw[fill=black] (-1.5,2.598) circle (0.1);
\draw[fill=black] (-1,2.598) circle (0.1);
\draw[fill=black] (-0.5,2.598) circle (0.1);
\draw[fill=black] (0,2.598) circle (0.1);
\draw[fill=black] (0.5,2.598) circle (0.1);
\draw[fill=black] (1,2.598) circle (0.1);
\draw[fill=black] (1.5,2.598) circle (0.1);
\draw[red] (2,2.598) circle (0.1);
\draw[red] (2.5,2.598) circle (0.1);
\draw[fill=black] (-2.75,2.165) circle (0.1);
\draw[fill=black] (-2.25,2.165) circle (0.1);
\draw[fill=black] (-1.75,2.165) circle (0.1);
\draw[fill=black] (-1.25,2.165) circle (0.1);
\draw[fill=black] (-0.75,2.165) circle (0.1);
\draw[fill=black] (-0.25,2.165) circle (0.1);
\draw[fill=black] (0.25,2.165) circle (0.1);
\draw[fill=black] (0.75,2.165) circle (0.1);
\draw[fill=black] (1.25,2.165) circle (0.1);
\draw[fill=black] (1.75,2.165) circle (0.1);
\draw[fill=black] (2.25,2.165) circle (0.1);
\draw[fill=black] (2.75,2.165) circle (0.1);
\draw[fill=black] (-3,1.732) circle (0.1);
\draw[fill=black] (-2.5,1.732) circle (0.1);
\draw[fill=black] (-2,1.732) circle (0.1);
\draw[fill=black] (-1.5,1.732) circle (0.1);
\draw[fill=black] (-1,1.732) circle (0.1);
\draw[fill=black] (-0.5,1.732) circle (0.1);
\draw[fill=black] (0,1.732) circle (0.1);
\draw[fill=black] (0.5,1.732) circle (0.1);
\draw[fill=black] (1,1.732) circle (0.1);
\draw[fill=black] (1.5,1.732) circle (0.1);
\draw[fill=black] (2,1.732) circle (0.1);
\draw[fill=black] (2.5,1.732) circle (0.1);
\draw[fill=black] (3,1.732) circle (0.1);
\node at (3.5,1.732) {{\footnotesize $\phi^{2,2}_{[0,12]}$}};
\draw[fill=black] (-3.25,1.299) circle (0.1);
\draw[fill=black] (-2.75,1.299) circle (0.1);
\draw[fill=black] (-2.25,1.299) circle (0.1);
\draw[fill=black] (-1.75,1.299) circle (0.1);
\draw[fill=black] (-1.25,1.299) circle (0.1);
\draw[fill=black] (-0.75,1.299) circle (0.1);
\draw[fill=black] (-0.25,1.299) circle (0.1);
\draw[fill=black] (0.25,1.299) circle (0.1);
\draw[fill=black] (0.75,1.299) circle (0.1);
\draw[fill=black] (1.25,1.299) circle (0.1);
\draw[fill=black] (1.75,1.299) circle (0.1);
\draw[fill=black] (2.25,1.299) circle (0.1);
\draw[fill=black] (2.75,1.299) circle (0.1);
\draw[fill=black] (3.25,1.299) circle (0.1);
\node at (3.75,1.299) {{\footnotesize $\phi^{2,2}_{[2,11]}$}};
\draw[red] (-3.5,0.866) circle (0.1);
\draw[fill=black] (-3,0.866) circle (0.1);
\draw[fill=black] (-2.5,0.866) circle (0.1);
\draw[fill=black] (-2,0.866) circle (0.1);
\draw[fill=black] (-1.5,0.866) circle (0.1);
\draw[fill=black] (-1,0.866) circle (0.1);
\draw[fill=black] (-0.5,0.866) circle (0.1);
\draw[fill=black] (0,0.866) circle (0.1);
\draw[fill=black] (0.5,0.866) circle (0.1);
\draw[fill=black] (1,0.866) circle (0.1);
\draw[fill=black] (1.5,0.866) circle (0.1);
\draw[fill=black] (2,0.866) circle (0.1);
\draw[fill=black] (2.5,0.866) circle (0.1);
\draw[fill=black] (3,0.866) circle (0.1);
\draw[red] (3.5,0.866) circle (0.1);
\node at (4,0.866) {{\footnotesize $\phi^{2,2}_{[4,10]}$}};
\draw[red] (-3.75,0.433) circle (0.1);
\draw[red] (-3.25,0.433) circle (0.1);
\draw[fill=black] (-2.75,0.433) circle (0.1);
\draw[fill=black] (-2.25,0.433) circle (0.1);
\draw[fill=black] (-1.75,0.433) circle (0.1);
\draw[fill=black] (-1.25,0.433) circle (0.1);
\draw[fill=black] (-0.75,0.433) circle (0.1);
\draw[fill=black] (-0.25,0.433) circle (0.1);
\draw[fill=black] (0.25,0.433) circle (0.1);
\draw[fill=black] (0.75,0.433) circle (0.1);
\draw[fill=black] (1.25,0.433) circle (0.1);
\draw[fill=black] (1.75,0.433) circle (0.1);
\draw[fill=black] (2.25,0.433) circle (0.1);
\draw[fill=black] (2.75,0.433) circle (0.1);
\draw[red] (3.25,0.433) circle (0.1);
\draw[red] (3.75,0.433) circle (0.1);
\node at (4.25,0.433) {{\footnotesize $\phi^{2,2}_{[6,9]}$}};
\draw[fill=black] (-4,0) circle (0.1);
\draw[fill=black] (-3.5,0) circle (0.1);
\draw[fill=black] (-3,0) circle (0.1);
\draw[fill=black] (-2.5,0) circle (0.1);
\draw[fill=black] (-2,0) circle (0.1);
\draw[fill=black] (-1.5,0) circle (0.1);
\draw[fill=black] (-1,0) circle (0.1);
\draw[fill=black] (-0.5,0) circle (0.1);
\draw[fill=black] (0,0) circle (0.1);
\draw[fill=black] (0.5,0) circle (0.1);
\draw[fill=black] (1,0) circle (0.1);
\draw[fill=black] (1.5,0) circle (0.1);
\draw[fill=black] (2,0) circle (0.1);
\draw[fill=black] (2.5,0) circle (0.1);
\draw[fill=black] (3,0) circle (0.1);
\draw[fill=black] (3.5,0) circle (0.1);
\draw[fill=black] (4,0) circle (0.1);
\node at (4.5,0) {{\footnotesize $\phi^{2,2}_{[8,8]}$}};
\draw[red] (-3.75,-0.433) circle (0.1);
\draw[red] (-3.25,-0.433) circle (0.1);
\draw[fill=black] (-2.75,-0.433) circle (0.1);
\draw[fill=black] (-2.25,-0.433) circle (0.1);
\draw[fill=black] (-1.75,-0.433) circle (0.1);
\draw[fill=black] (-1.25,-0.433) circle (0.1);
\draw[fill=black] (-0.75,-0.433) circle (0.1);
\draw[fill=black] (-0.25,-0.433) circle (0.1);
\draw[fill=black] (0.25,-0.433) circle (0.1);
\draw[fill=black] (0.75,-0.433) circle (0.1);
\draw[fill=black] (1.25,-0.433) circle (0.1);
\draw[fill=black] (1.75,-0.433) circle (0.1);
\draw[fill=black] (2.25,-0.433) circle (0.1);
\draw[fill=black] (2.75,-0.433) circle (0.1);
\draw[red] (3.25,-0.433) circle (0.1);
\draw[red] (3.75,-0.433) circle (0.1);
\node at (4.25,-0.433) {{\footnotesize $\phi^{2,2}_{[9,6]}$}};
\draw[red] (-3.5,-0.866) circle (0.1);
\draw[fill=black] (-3,-0.866) circle (0.1);
\draw[fill=black] (-2.5,-0.866) circle (0.1);
\draw[fill=black] (-2,-0.866) circle (0.1);
\draw[fill=black] (-1.5,-0.866) circle (0.1);
\draw[fill=black] (-1,-0.866) circle (0.1);
\draw[fill=black] (-0.5,-0.866) circle (0.1);
\draw[fill=black] (0,-0.866) circle (0.1);
\draw[fill=black] (0.5,-0.866) circle (0.1);
\draw[fill=black] (1,-0.866) circle (0.1);
\draw[fill=black] (1.5,-0.866) circle (0.1);
\draw[fill=black] (2,-0.866) circle (0.1);
\draw[fill=black] (2.5,-0.866) circle (0.1);
\draw[fill=black] (3,-0.866) circle (0.1);
\draw[red] (3.5,-0.866) circle (0.1);
\node at (4,-0.866) {{\footnotesize $\phi^{2,2}_{[10,4]}$}};
\draw[fill=black] (-3.25,-1.299) circle (0.1);
\draw[fill=black] (-2.75,-1.299) circle (0.1);
\draw[fill=black] (-2.25,-1.299) circle (0.1);
\draw[fill=black] (-1.75,-1.299) circle (0.1);
\draw[fill=black] (-1.25,-1.299) circle (0.1);
\draw[fill=black] (-0.75,-1.299) circle (0.1);
\draw[fill=black] (-0.25,-1.299) circle (0.1);
\draw[fill=black] (0.25,-1.299) circle (0.1);
\draw[fill=black] (0.75,-1.299) circle (0.1);
\draw[fill=black] (1.25,-1.299) circle (0.1);
\draw[fill=black] (1.75,-1.299) circle (0.1);
\draw[fill=black] (2.25,-1.299) circle (0.1);
\draw[fill=black] (2.75,-1.299) circle (0.1);
\draw[fill=black] (3.25,-1.299) circle (0.1);
\node at (3.75,-1.299) {{\footnotesize$\phi^{2,2}_{[11,2]}$}};
\draw[fill=black] (-3,-1.732) circle (0.1);
\draw[fill=black] (-2.5,-1.732) circle (0.1);
\draw[fill=black] (-2,-1.732) circle (0.1);
\draw[fill=black] (-1.5,-1.732) circle (0.1);
\draw[fill=black] (-1,-1.732) circle (0.1);
\draw[fill=black] (-0.5,-1.732) circle (0.1);
\draw[fill=black] (0,-1.732) circle (0.1);
\draw[fill=black] (0.5,-1.732) circle (0.1);
\draw[fill=black] (1,-1.732) circle (0.1);
\draw[fill=black] (1.5,-1.732) circle (0.1);
\draw[fill=black] (2,-1.732) circle (0.1);
\draw[fill=black] (2.5,-1.732) circle (0.1);
\draw[fill=black] (3,-1.732) circle (0.1);
\node at (3.5,-1.732) {{\footnotesize $\phi^{2,2}_{[12,0]}$}};
\draw[fill=black] (-2.75,-2.165) circle (0.1);
\draw[fill=black] (-2.25,-2.165) circle (0.1);
\draw[fill=black] (-1.75,-2.165) circle (0.1);
\draw[fill=black] (-1.25,-2.165) circle (0.1);
\draw[fill=black] (-0.75,-2.165) circle (0.1);
\draw[fill=black] (-0.25,-2.165) circle (0.1);
\draw[fill=black] (0.25,-2.165) circle (0.1);
\draw[fill=black] (0.75,-2.165) circle (0.1);
\draw[fill=black] (1.25,-2.165) circle (0.1);
\draw[fill=black] (1.75,-2.165) circle (0.1);
\draw[fill=black] (2.25,-2.165) circle (0.1);
\draw[fill=black] (2.75,-2.165) circle (0.1);
\draw[red] (-2.5,-2.598) circle (0.1);
\draw[red] (-2,-2.598) circle (0.1);
\draw[fill=black] (-1.5,-2.598) circle (0.1);
\draw[fill=black] (-1,-2.598) circle (0.1);
\draw[fill=black] (-0.5,-2.598) circle (0.1);
\draw[fill=black] (0,-2.598) circle (0.1);
\draw[fill=black] (0.5,-2.598) circle (0.1);
\draw[fill=black] (1,-2.598) circle (0.1);
\draw[fill=black] (1.5,-2.598) circle (0.1);
\draw[red] (2,-2.598) circle (0.1);
\draw[red] (2.5,-2.598) circle (0.1);
\draw[red] (-2.25,-3.031) circle (0.1);
\draw[fill=black] (-1.75,-3.031) circle (0.1);
\draw[red] (-1.25,-3.031) circle (0.1);
\draw[fill=black] (-0.75,-3.031) circle (0.1);
\draw[fill=black] (-0.25,-3.031) circle (0.1);
\draw[fill=black] (0.25,-3.031) circle (0.1);
\draw[fill=black] (0.75,-3.031) circle (0.1);
\draw[red] (1.25,-3.031) circle (0.1);
\draw[fill=black] (1.75,-3.031) circle (0.1);
\draw[red] (2.25,-3.031) circle (0.1);
\draw[fill=black] (-2,-3.464) circle (0.1);
\draw[red] (-1.5,-3.464) circle (0.1);
\draw[red] (-1,-3.464) circle (0.1);
\draw[fill=black] (-0.5,-3.464) circle (0.1);
\draw[fill=black] (0,-3.464) circle (0.1);
\draw[fill=black] (0.5,-3.464) circle (0.1);
\draw[red] (1,-3.464) circle (0.1);
\draw[red] (1.5,-3.464) circle (0.1);
\draw[fill=black] (2,-3.464) circle (0.1);
\draw (-2,3.464) -- (2,3.464);
\draw (-2.25,3.031) -- (2.25,3.031);
\draw (-2.5,2.598) -- (2.5,2.598);
\draw (-2.75,2.165) -- (2.75,2.165);
\draw (-3,1.732) -- (3,1.732);
\draw (-3.25,1.299) -- (3.25,1.299);
\draw (-3.5,0.866) -- (3.5,0.866);
\draw (-3.75,0.433) -- (3.75,0.433);
\draw (-4,0) -- (4,0);
\draw (-2,-3.464) -- (2,-3.464);
\draw (-2.25,-3.031) -- (2.25,-3.031);
\draw (-2.5,-2.598) -- (2.5,-2.598);
\draw (-2.75,-2.165) -- (2.75,-2.165);
\draw (-3,-1.732) -- (3,-1.732);
\draw (-3.25,-1.299) -- (3.25,-1.299);
\draw (-3.5,-0.866) -- (3.5,-0.866);
\draw (-3.75,-0.433) -- (3.75,-0.433);
\draw (2,3.464) -- (4,0);
\draw (1.5,3.464) -- (3.75,-0.433);
\draw (1,3.464) -- (3.5,-0.866);
\draw (0.5,3.464) -- (3.25,-1.299);
\draw (0,3.464) -- (3,-1.732);
\draw (-0.5,3.464) -- (2.75,-2.165);
\draw (-1,3.464) -- (2.5,-2.598);
\draw (-1.5,3.464) -- (2.25,-3.031);
\draw (-2,3.464) -- (2,-3.464);
\draw (-2,-3.464) -- (-4,0);
\draw (-1.5,-3.464) -- (-3.75,0.433);
\draw (-1,-3.464) -- (-3.5,0.866);
\draw (-0.5,-3.464) -- (-3.25,1.299);
\draw (0,-3.464) -- (-3,1.732);
\draw (0.5,-3.464) -- (-2.75,2.165);
\draw (1,-3.464) -- (-2.5,2.598);
\draw (1.5,-3.464) -- (-2.25,3.031);
\draw (-2,3.464) -- (-4,0);
\draw (-1.5,3.464) -- (-3.75,-0.433);
\draw (-1,3.464) -- (-3.5,-0.866);
\draw (-0.5,3.464) -- (-3.25,-1.299);
\draw (-0,3.464) -- (-3,-1.732);
\draw (0.5,3.464) -- (-2.75,-2.165);
\draw (1,3.464) -- (-2.5,-2.598);
\draw (1.5,3.464) -- (-2.25,-3.031);
\draw (2,3.464) -- (-2,-3.464);
\draw (2,-3.464) -- (4,0);
\draw (1.5,-3.464) -- (3.75,0.433);
\draw (1,-3.464) -- (3.5,0.866);
\draw (0.5,-3.464) -- (3.25,1.299);
\draw (0,-3.464) -- (3,1.732);
\draw (-0.5,-3.464) -- (2.75,2.165);
\draw (-1,-3.464) -- (2.5,2.598);
\draw (-1.5,-3.464) -- (2.25,3.031);
\draw[dashed,red] (0,0) -- (4.5,2.598);
\draw[dashed,red] (0,0) -- (4.5,-2.598);
\draw[->, ultra thick] (-5,2.5) -- (-4.5,2.788);
\node at (-4.5,2.95) {\footnotesize $\omega_2$};
\draw[->, ultra thick] (-5,2.5) -- (-4.5,2.212);
\node at (-4.5,2.05) {\footnotesize $\omega_1$};
\end{tikzpicture}
\end{center}
where some of the $\phi^{2,2}_{[c_1,c_2]}$ are given by
\begin{align}
&\phi^{2,2}_{[8,8]}(\tau,T_1,\epsilon)=1 \,, \quad \quad \phi^{2,2}_{[6,9]}=\phi^{2,2}_{[9,6]}=-2\frac{\theta_1(\tau;4\epsilon)^2\theta_1(\tau;4\epsilon-\tilde{T}_1)\theta_1(\tau;4\epsilon+\tilde{T}_1)}{\theta_1(\tau;\epsilon)^2\theta_1(\tau;\tilde{T}_1)^2} \nonumber \\
&\phi^{2,2}_{[4,10]}=\phi^{2,2}_{[10,4]}=\frac{\theta_1(\tau;4\epsilon)^4\theta_1(\tau;4\epsilon-\tilde{T}_1)^2\theta_1(\tau;4\epsilon+\tilde{T}_1)^2}{\theta_1(\tau;\epsilon)^4\theta_1(\tau;\epsilon-\tilde{T}_1)^2\theta_1(\tau;\epsilon+\tilde{T}_1)^2} \nonumber \\
& +\frac{\theta_1(\tau;3\epsilon )\theta_1(\tau; 4\epsilon)^2\theta_1(\tau;5\epsilon )\theta_1(\tau;4\epsilon-\tilde{T}_1 )\theta_1(\tau;5\epsilon-\tilde{T}_1 )\theta_1(\tau;3\epsilon+\tilde{T}_1 )\theta_1(\tau;4\epsilon+ \tilde{T}_1 )}{\theta_1(\tau; \epsilon)\theta_1(\tau;2\epsilon )^2\theta_1(\tau;2\tilde{T}_1 )^2\theta_1(\tau;\epsilon-\tilde{T}_1 )^2} \nonumber \\
& + \frac{\theta_1(\tau;3\epsilon )\theta_1(\tau; 4\epsilon)^2\theta_1(\tau;5\epsilon )\theta_1(\tau;3\epsilon-\tilde{T}_1 )\theta_1(\tau;4\epsilon-\tilde{T}_1 )\theta_1(\tau;4\epsilon+\tilde{T}_1 )\theta_1(\tau;5\epsilon+ \tilde{T}_1 )}{\theta_1(\tau; \epsilon)\theta_1(\tau;2\epsilon )^2\theta_1(\tau;2\tilde{T}_1 )^2\theta_1(\tau;\epsilon+\tilde{T}_1 )^2}
\end{align}
Here again the red circles stand for the terms removed when setting $\tilde{T}_1=4\epsilon$.\footnote{In general, cancellation of this type occur for generic $n$ by setting $\tilde{T}_1=2n\epsilon$}

Based on the above results, for generic $n\geq 2$, we propose that the partition function can then be expressed by summing the Weyl orbits for the weights in the fundamental Weyl chamber $P^+_{2n^2,2n^2}$ of the irreducible representation $\Gamma_{2n^2,2n^2}$
\begin{align}
&\mathcal{Z}_{3,2}^{\text{line}}(\mathbf{T},\mathbf{t},m=n\epsilon,\epsilon,-\epsilon)= e^{-2n^2\weyl } \sum_{\lambda \in P^+_{2n^2,2n^2}} \phi^{n,2}_{[c_1,c_2]}(\tau,\epsilon)\mathcal{O}_{\lambda}^n(t_{f_1},t_{f_2})\,,&&\text{with} &&\mathcal{O}_{\lambda}^n(t_{f_1},t_{f_2})=d_\lambda \displaystyle \sum_{w \in \mathcal{W}} e^{w(\lambda)}\,,\nonumber
\end{align}
and the $SL(2,\mathbb{Z})_\tau$ indices are
\begin{equation}
\mathcal{I}_\tau(\phi^{n,M}_{\lambda[c_1,c_2]})=(2n^2 \weyl  , 2 n^2 \weyl ) - (\lambda , \lambda)= 4n^4 - \frac{1}{3}(c_1^2 + c_1c_2 + c_2^2)
\end{equation}
\section{Examples: Compact Brane Configuration}\label{Sect:CompactBraneConfigs}
After having discussed examples of partition functions of non-compact brane configurations for the particular choice $m=n\epsilon$ (with $n\in\mathbb{N}$), we now consider compact brane configurations. The non-compact case can be recovered in the limit $\prod_{a=1}^nQ_{f_a}=Q_\rho\rightarrow 0$, as we shall discuss in the following.
\subsection{Configuration $(N,M)=(2,1)$}\label{Sect:ConfigCompact21}
\subsubsection{Choice $ \epsilon_1=-\epsilon_2=\epsilon$ and $m=\epsilon$}
We start with the case of two M5-branes, in which case there are two different partitions contributing to (\ref{DefPartFctComp}). To describe the configurations contributing, we introduce the following class of partitions 
\begin{align}
&\mathfrak{m}_n=(n,n-1,n-2,\ldots, 1)\,,&&\overbrace{\parbox{1.6cm}{\vspace{2pt}\ydiagram{6, 5, 4,3,2,1}}}^{n\text{-boxes}}\,,
\end{align}
with length
\begin{align}
|\mathfrak{m}_n|=\sum_{i=1}^ni=\frac{n(n+1)}{2}\,.
\end{align}
We also use the notation $\mathfrak{m}_0=\emptyset$. With this notation, we only get the following three types of contributions to the partition function ($n\in\mathbb{N}$)
\begin{center}
\begin{tabular}{|c|c|c|}\hline
&&\\[-10pt]
$\nu_1$ & $\nu_2$ & $\left(\prod_{a=1}^{2}(-Q_{f_{a}})^{|\nu_{a}|}\right)\,\prod_{a=1}^{2}\prod_{(i,j)\in \nu_{a}}\frac{\theta_{1}(\tau;z^{a}_{ij})\,\theta_{1}(\tau;v^{a}_{ij})}{\theta_1(\tau;w^{a}_{ij})\theta_1(\tau;u^{a}_{ij})}$ \\[8pt]\hline\hline
&&\\[-10pt]
$\emptyset$ & $\emptyset$ &$1$ \\[4pt]\hline
&&\\[-10pt]
$\mathfrak{m}_n=\parbox{1.6cm}{\vspace{2pt}\ydiagram{6, 5, 4,3,2,1}}$ & $\mathfrak{m}_{n-1}=\parbox{1.3cm}{\vspace{2pt}\ydiagram{5, 4,3,2,1}}$ &$(-Q_{f_1})^{\frac{n(n+1)}{2}}(-Q_{f_2})^{\frac{n(n-1)}{2}}$ \\[20pt]\hline
&&\\[-10pt]
$\mathfrak{m}_{n-1}=\parbox{1.3cm}{\vspace{2pt}\ydiagram{5, 4,3,2,1}}$ & $\mathfrak{m}_n=\parbox{1.6cm}{\vspace{2pt}\ydiagram{6, 5, 4,3,2,1}}$ & $(-Q_{f_1})^{\frac{n(n-1)}{2}}(-Q_{f_2})^{\frac{n(n+1)}{2}}$ \\[20pt]\hline
\end{tabular}
\end{center}
Thus, the normalised partition function (\ref{completepartitionNorm}) is 
\begin{align}
\widetilde{\mathcal{Z}}_{2,1}(\tau,m=\epsilon,t_{f_1},t_{f_2},\epsilon,-\epsilon)&=1+\sum_{n=1}^\infty(-1)^{n^2}\left[Q_{f_1}^{\frac{n(n+1)}{2}}Q_{f_2}^{\frac{n(n-1)}{2}}+Q_{f_1}^{\frac{n(n-1)}{2}}Q_{f_2}^{\frac{n(n+1)}{2}}\right]\,.\nonumber
\label{Part21}
\end{align}
This expression can also be written in the form
\begin{align}
\widetilde{\mathcal{Z}}_{2,1}(\tau,t_{f_1},t_{f_2},m=\epsilon,\epsilon,-\epsilon)&=\prod_{k=1}^\infty \mathcal{Z}_{2,1}^{(k)}(\tau,m=\epsilon,t_{f_1},t_{f_2},\epsilon,-\epsilon)\nonumber\\
&=\prod_{k=1}^\infty(1-Q_\rho^k)\,(1-Q_{f_1}Q_\rho^{k-1})\,(1-Q_{f_2}Q_\rho^{k-1})\,,
\end{align}
where $Q_\rho=Q_{f_1}Q_{f_2}$.\footnote{Notice the relation $\mathcal{Z}^{\text{line}}_{3,1}(\tau,t_{f_1},t_{f_2},m=\epsilon,\epsilon,-\epsilon)=\widetilde{\mathcal{Z}}_{2,1}^{(1)}(\tau,t_{f_1},t_{f_2},m=\epsilon,\epsilon,-\epsilon)$ relating compact to the non-compact M-brane configurations.} 
Following the discussion of the non-compact examples, we would like to Identify the K\"ahler parameters $t_{f_1}$ and $t_{f_2}$ with the affine roots $ \widehat{\alpha}_0 $ and $ \widehat{\alpha}_1 $, which are introduced in appendix \ref{AffA1}. This involves choosing which $t_{f_a}$ contains the null root $\delta$ . The final answer does not depend on this choice as the exchange $Q_{f_1} \leftrightarrow Q_{f_2}$ does not change the partition function. Here we choose the following
\begin{align}
&Q_{f_1}=e^{-\widehat{\alpha}_1}\,,&&Q_{f_2}=Q_\rho/Q_{f_1}=e^{\widehat{\alpha}_1-\delta}=e^{-\widehat{\alpha}_0}\,,&&Q_\rho=e^{-\delta}
\label{choice}
\end{align}
and using expression (\ref{PosRootA1p}) for the positive roots of $\widehat{\mathfrak{a}}_1$ we can write
\begin{align}
\widetilde{\mathcal{Z}}_{2,1}(\tau,\mathbf{t},m=\epsilon,\epsilon,-\epsilon)&=\left(\prod_{n=0}^\infty(1-Q_{1}Q_\rho^{n})\right)\left(\prod_{n=1}^\infty(1-Q_\rho^{n}/Q_{1})\right)\left(\prod_{k=1}^\infty(1-Q_\rho^k)\right)=\prod_{\widehat{\alpha}\in\widehat{\Delta}_+(\widehat{\mathfrak{a}}_1)}(1-e^{-\widehat{\alpha}})\,.\label{ProdRootA1p}
\end{align}
Using the affine Weyl denominator formula (\ref{ProdRootA1p}) can be written ass a sum over elements of the affine Weyl group (with $\text{mult}(\widehat{\alpha})=1$ for $\widehat{\alpha}\in\widehat{\Delta}(\widehat{\mathfrak{a}}_1)$)
\begin{equation}
\widetilde{\mathcal{Z}}_{2,1}(\tau,t_{f_1},t_{f_2},m=\epsilon,\epsilon,-\epsilon)= \displaystyle \sum_{w \in \widehat{W}} (-1)^{l(w)} e^{w(\hat{\weyl })-\hat{\weyl }}
\label{Z21}
\end{equation}
where $\widehat{\weyl }=\widehat{\omega}_0+\widehat{\omega}_1=[1,1,0]$ is the affine Weyl vector. We recall the action of the affine Weyl group $\widehat{\mathcal{W}}(\widehat{\mathfrak{a}}_1)$ as given in (\ref{AffWeylActionA1})
\begin{align}
&s_0[c_0,c_1,l]=[-c_0,c_1+2c_0,l-c_0]\,, &&\text{and} &&s_1[c_0,c_1,l]=[c_0+2c_1,-c_1,l]\,.\label{AffineWeylRelections}
\end{align} 
We can work out the first few Weyl reflections to check (\ref{Z21})
\begin{center}
\begin{tabular}{c|c|c|c}
$w\in\widehat{\mathcal{W}}(\widehat{\mathfrak{a}}_1)$ & $w(\widehat{\weyl })-\widehat{\weyl }$ & $\ell(w)$ & grade\\\hline
$1$ & $0$ & $0$ & $0$\\
$s_0$ & $-\widehat{\alpha}_0$ & $1$ & $-1$\\
$s_1$ & $-\widehat{\alpha}_1$ & $1$ & $0$ \\
$s_1s_0$ & $-\widehat{\alpha}_0-3\widehat{\alpha}_1$ & $2$ & $-1$\\
$s_0s_1$ & $-3\widehat{\alpha}_0-\widehat{\alpha}_1$ & $2$ & $-3$ \\
$s_1s_0s_1$ & $-3\widehat{\alpha}_0-6\widehat{\alpha}_1$ & $3$ & $-3$
\end{tabular}
\end{center}
Therefore, using (\ref{Z21}), we have 
\begin{align}
\widetilde{\mathcal{Z}}_{2,1}(\tau,t_{f_1},t_{f_2},m=\epsilon,\epsilon,-\epsilon)&= \displaystyle \sum_{w \in \widehat{W}} (-1)^{l(w)} e^{w(\hat{\weyl })-\hat{\weyl }} \nonumber \\
&=1-Q_{f_1}-Q_{f_2}+Q_{f_1}^3Q_{f_2}+Q_{f_1}^3Q_{f_2} - Q_{f_1}^6Q_{f_2}^3+ \dots\label{Z21compn1}
\end{align}
which matches (\ref{Part21}). While written as a sum of Weyl reflections of $\widehat{\weyl}$, we can also interpret (\ref{Z21compn1}) as a sum over Weyl orbits of weights in the fundamental domain $\widehat{P}^+_{1,1}$ of the highest weight representation $\widehat{\Gamma}_{1,1}$ of $\widehat{\mathfrak{a}}_1$\footnote{This directly generalises the discussion of section~\ref{Sect:Con21NC} to compact M5-brane configurations.}:  following the discussion of appendix \ref{AffA1}, every affine weight of $\widehat{\mathfrak{sl}}(2,\mathbb{C})$ can be decomposed into fundamental weights $(\widehat{\omega}_0,\widehat{\omega}_1)$ as follows
\begin{equation}
\widehat{\lambda}= c_0 \widehat{\omega}_0 + c_1 \widehat{\omega}_1 + l \delta = [c_0,c_1,l] \,, \quad c_0,c_1,l\in \mathbb{Z}\label{WeightBaseComp2}
\end{equation}
such that the affine root $t_{f_1}$ and a generic monomial $Q_{f_1}^iQ_{f_2}^j$ are decomposed as
\begin{align}
&t_{f_1}=-2\omega_0 + 2 \omega_1\,,&&\text{and} &&Q_{f_1}^i Q_{f_2}^j=e^{2(j-i)(\omega_1-\omega_0)}e^{-j\delta}\,.\label{OmegaIntrod}
\end{align} 
Furthermore, in table \ref{AffRep110} in appendix \ref{affrep} we give the the first few grades of the affine representation generated by $\widehat{\weyl }=[1,1,0]$. The affine weights which are colored in red are contained in the Weyl orbit of $\widehat{\weyl }$. To make the connection to the remaining weights even more manifest, we rewrite (\ref{Z21compn1}) in a slightly different manner: we observe that the Weyl-orbit of the weight $\lambda=[1,1,r]$ for $r\in \mathbb{Z}$ can be written as
\begin{align}
e^{-\widehat{\weyl }}\sum_{w\in\widehat{W}}(-1)^{\ell(w)}\,e^{w([1,1,r])}=e^{-\widehat{\weyl }}e^{r\delta}\sum_{w\in\widehat{W}}(-1)^{\ell(w)}\,e^{w([1,1,0])}=e^{-\widehat{\weyl }}e^{r\delta}\sum_{w\in\widehat{W}}(-1)^{\ell(w)}\,e^{w(\widehat{\weyl })}\,,
\end{align}
such that
\begin{align}
\sum_{l=0}^\infty e^{-\widehat{\weyl }}\sum_{w\in\widehat{W}}(-1)^{\ell(w)}\,e^{w([1,1,-l])}=e^{-\widehat{\weyl }}\left(\sum_{l=0}^\infty e^{-l\delta}\right)\sum_{w\in\widehat{W}}(-1)^{\ell(w)}\,e^{w(\widehat{\weyl })}=\frac{e^{-\widehat{\weyl }}}{1-e^{-\delta}}\,\sum_{w\in\widehat{W}}(-1)^{\ell(w)}\,e^{w(\widehat{\weyl })}\,.\nonumber
\end{align}
Therefore, we can write
\begin{align}
\widetilde{\mathcal{Z}}_{2,1}(\tau,t_{f_1},t_{f_2},m
=e^{-\widehat{\weyl }}(1-Q_\rho)\,\sum_{\widehat{\lambda}\in \widehat{P}^+_{1,1}}\,\mathcal{O}^1_{\widehat{\lambda}=[c_0,c_1,l]}(t_{f_1},t_{f_2})\,,
\end{align}
where we defined
\begin{align}
\mathcal{O}^1_{\widehat{\lambda}=[c_0,c_1,l]}(t_{f_1},t_{f_2})=\sum_{w\in \widehat{W}}(-1)^{\ell(w)}\,e^{w(\widehat{\lambda})}\,,\label{MissingNormWeyl}
\end{align}
and $\widehat{P}^+_{1,1}$ is the fundamental Weyl chamber of the affine representation generated by the weight $[1,1]$, \emph{i.e.} $\widehat{P}^+_{1,1}=\{[1,1,-l]|l\in\mathbb{N}\}$. Thus (up to a prefactor $e^{-\widehat{\weyl }}(1-Q_\rho)$), the partition function $\mathcal{Z}_{2,1}(\tau,t_{f_1},t_{f_2},m=\epsilon,\epsilon,-\epsilon)$ can be written as a sum over the states contained in $\widehat{\Gamma}_{1,1}$.

Finally, before discussing more general cases $m=n\epsilon$ with $n>1$, we remark that in the limit $Q_\rho\rightarrow 0$ we reproduce the partition function $\mathcal{Z}^{\text{line}}_{2,1}(\tau,m=\epsilon, t_{f_1},\epsilon,-\epsilon)$
\begin{align}
\lim_{Q_\rho\to 0}\widetilde{\mathcal{Z}}_{2,1}(\tau,t_{f_1},t_{f_2},m=\epsilon,\epsilon,-\epsilon)=1-Q_{f_1}\,,
\end{align}
which indeed agrees with (\ref{PartF1N2}). From the point of view of the irreducible representation $[1,1,0]$, due to (\ref{choice}), the limit $Q_\rho\rightarrow 0$ corresponds to restricting to states with grade $l=0$. Indeed, according to the weight diagram in table~\ref{AffRep110}, the partition function can thus be written as the sum of two states ($\lambda=[1,1,0]$ and $\lambda=[3,-1,0]$)
\begin{align}
\mathcal{Z}^{\text{line}}_{2,1}(\tau, t_{f_1},m=\epsilon,\epsilon,-\epsilon)&=\lim_{Q_\rho\to 0}\widetilde{\mathcal{Z}}_{2,1}(\tau,t_{f_1},t_{f_2},m=\epsilon,\epsilon,-\epsilon)=\sum_{w \in \widehat{W}} (-1)^{l(w)} e^{w(\widehat{\weyl })-\widehat{\weyl }}\nonumber\\
&=e^{-(\omega_0+\omega_1)}\sum_{k=0}^1(-1)^k\sum_{[c_0,c_1,l]=[1,1,0]-k\alpha_1}e^{c_0 \omega_0+c_1\omega_1+l\delta}=1-Q_{f_1}\,,
\end{align}
where we used the identification (\ref{choice}).
\subsubsection{Choice $\epsilon_1=-\epsilon_2=\epsilon$ and $m=n \epsilon$ for $n>1$}
For $m=n\epsilon$ with $n>1$, the partition function is an infinite sum of ratios of theta functions:
\begin{equation}
\widetilde{\mathcal{Z}}_{2,1}(\tau, \rho,t_{f_1},m=n\epsilon,\epsilon,-\epsilon)=\displaystyle \sum_{i,j}^{\infty} Q_{f_1}^i Q_{f_2}^j \prod_{r} \frac{\theta(\tau;a_r\epsilon)}{\theta(\tau;b_r\epsilon)}
\label{comp21}
\end{equation}
To illustrate this expression, we first consider in some detail the case $n=2$ and generic $n$ later.
\paragraph{n=2}
For $n=2$ the first few terms of the partition function can be written in the following suggestive form
\begin{align}
\widetilde{\mathcal{Z}}_{2,1}(\tau,\rho,t_{f_1},m=2\epsilon,\epsilon,-\epsilon)= \overbrace{e^{-4(\omega_0+\omega_1)}}^{e^{-4\widehat{\weyl }}} \Big[  &\widehat{\phi}^2_{[4,4,0]}\big( e^{4\omega_0+4\omega_1} + e^{12\omega_0-4\omega_1} + e^{-4\omega_0+12 \omega_1-4\delta} + \dots \big) \nonumber \\
 + &\widehat{\phi}^2_{[6,2,0]} \big( e^{6\omega_0+2\omega_1} + e^{10\omega_0-2\omega_1} + e^{-6 \omega_0 + 14\omega_1 - 6\delta} + \dots \big) \nonumber \\
 + & \widehat{\phi}^2_{[8,0,0]} \big( 2e^{8\omega_0} + 2e^{-8\omega_0+16 \omega_1- 8\delta} + 2e^{24 \omega_0 -16 \omega_1 - 8\delta} + \dots \big) \nonumber \\
 + &\widehat{\phi}^2_{[2,6,-1]} \big( e^{2\omega_0 + 6 \omega_1 -\delta} + e^{14\omega_0-6\omega_1 -\delta} + e^{-2\omega_0+10\omega_1 - 3\delta} + \dots  \big) \nonumber \\
 + &\widehat{\phi}^2_{[4,4,-1]} \big( e^{4\omega_0 + 4\omega_1 - \delta} + e^{12 \omega_0 - 4 \omega_1 - \delta} + e^{-4\omega_0 + 12 \omega_1 -5 \delta} + \dots \big) \nonumber \\
 + & \widehat{\phi}^2_{[8,0,-1]} \big(  2e^{8\omega_0 -\delta} + 2e^{-8\omega_0 + 16 \omega_1 - 9 \delta} + 2e^{24 \omega_0 -16 \omega_1 - 9\delta} + \dots \big) \nonumber \\
 +& \dots \Big]
\label{comp212}
\end{align}
where the notation is the same as in (\ref{WeightBaseComp2}). Indeed, the $\widehat{\phi}^2_{[c_0,c_1,l]}$ are indexed by their Dynkin labels $c_0,c_1$ and their grade $l$
\begin{align}
&\widehat{\phi}^2_{[4,4,0]}(\tau,\epsilon)=1 ,\quad
&& \widehat{\phi}^2_{[8,0,0]}(\tau,\epsilon)=2\frac{\theta_1(\tau;3\epsilon)}{\theta_1(\tau;\epsilon)}\nonumber \\
& \widehat{\phi}^2_{[4,4,-1]}(\tau,\epsilon)= \frac{\theta_1(\tau;3\epsilon)^2}{\theta_1(\tau;\epsilon)^2},\quad
&& \widehat{\phi}^2_{[8,0,-1]}(\tau,\epsilon)=-2\frac{\theta_1(\tau;5\epsilon)}{\theta_1(\tau;\epsilon)} \nonumber \\
&\widehat{\phi}^2_{[6,2,0]}(\tau,\epsilon)=\widehat{\phi}^2_{[2,6,-1]}(\tau,\epsilon)=- \frac{\theta_1(\tau;2\epsilon)^2}{\theta_1(\tau;\epsilon)^2}
\end{align}
Comparing with affine representations of $\widehat{\mathfrak{sl}}(2,\mathbb{C})$ (as given in appendix \ref{affrep}), we can write the compact partition function (\ref{comp212}) as a sum over Weyl orbits of the representatives in the fundamental Weyl chamber $\widehat{P}_{4,4}^+$ of the affine $[4,4]$ representation
\begin{equation}
\widetilde{\mathcal{Z}}_{2,1}(\tau,\rho,t_1,t_2,m=2\epsilon,\epsilon,-\epsilon)=e^{-4\widehat{\weyl }}\sum_{\widehat{\lambda}\in \widehat{P}_{4,4}^+} \widehat{\phi}^2_{[c_0,c_1,l]}(\tau,\epsilon) \mathcal{O}_{\widehat{\lambda}}^2(t_{f_1},t_{f_2})
\end{equation}
where the individual Weyl orbits are given by 
\begin{equation}
\mathcal{O}_{\widehat{\lambda}=[c_0,c_1,l]}^2(t_{f_1},t_{f_2})=\widehat{d}_{\lambda} \sum_{w\in \widehat{W}}e^{w(\widehat{\lambda})}
\end{equation}
where the normalization is given by
\begin{equation}
\widehat{d}_{\lambda=[c_0,c_1]}= \begin{cases}
\frac{1}{2} & \quad \text{if } c_0=0 \text{ or } c_1=0 \\
1 & \quad \text{else} \\
\end{cases}
\end{equation}
The weights of the affine $[4,4]$ that are in the fundamental Weyl chamber $\widehat{P}_{4,4}^+$ are those with positive Dynkin labels
\begin{equation}
\widehat{P}^+_{4,4}=\{[0,8,-l],[2,6,-l],[4,4,-l],[6,2,-l],[8,0,-l]\}_{l\in \mathbb{N}}
\end{equation}
As for the finite case there are again weights that are fixed under the action of certain elements of the Weyl group, \emph{e.g.}
\begin{equation}
s_0[0,8,-2]=[0,8,-2] \,, \quad s_0\in \widehat{\mathcal{W}}
\end{equation}
As for the non-compact cases the arguments of the $ \widehat{\phi}^2_{[c_0,c_1,l]} $ can be related to their corresponding affine weights $\widehat{\lambda}=[c_0,c_1,l]$ through
\begin{equation}
\mathcal{I}_\tau(\widehat{\phi}^2_{\lambda=[c_0,c_1,l]})= (4\widehat{\weyl} |4\widehat{\weyl} ) - (\widehat{\lambda}|\widehat{\lambda}) + 8l= 4-\frac{1}{4}c_1^2 - 8l\,,
\label{comrel}
\end{equation} 
where $(.|.)$ stands for the inner product in the affine $\widehat{\omega}_1$ basis. 

Before continuing to the case of generic $n$, we consider the decompactification limit $Q_\rho\to 0$. In this case only those weights with $\ell=0$ survive, such that (with (\ref{OmegaIntrod}))
\begin{align}
\lim_{\rho\to i\infty}\widetilde{\mathcal{Z}}_{2,1}(\tau,\rho,t_{f_1},m=2\epsilon,\epsilon,-\epsilon)&=(1+Q_{f_1}^4)\widehat{\phi}^2_{[4,4,0]}+(Q_{f_1}+Q_{f_1}^3)\widehat{\phi}^2_{[6,2,0]}+2Q_{f_1}^2\widehat{\phi}^2_{[8,0,0]}\nonumber\\
&=(1+Q_{f_1}^4)-(Q_{f_1}+Q_{f_1}^3)\frac{\theta(\tau;2\epsilon)^2}{\theta_1(\tau;\epsilon)}+2Q_{f_1}^2\,\frac{\theta_1(\tau;3\epsilon)}{\theta_1(\tau,\epsilon)}\nonumber\\
&=\mathcal{Z}^{\text{line}}_{2,1}(\tau,t_{f_1},m=2\epsilon,\epsilon,-\epsilon)\,.
\end{align}
This expression indeed agrees with (\ref{ExpZline2}) as expected, since in the limit $Q_\rho\to 0$ the brane setup corresponds to the non-compact configuration $(N,M)=(2,1)$.
\paragraph{generic n}
The above analysis can be extended for $n>2$ with a pattern arising which allows us to conjecture the structure for generic $n$: Indeed, we propose that the partition function can be written as a sum over Weyl orbits of the representatives in the fundamental Weyl chamber $\widehat{P}_{n^2,n^2}^+$ of the affine $[n^2,n^2]$ representation
\begin{equation}
\widetilde{\mathcal{Z}}_{2,1}(\tau,\rho,t_{f_1},m=n\epsilon,\epsilon,-\epsilon)= e^{-n^2\widehat{\weyl }} \sum_{\widehat{\lambda}\in \widehat{P}_{n^2,n^2}^+} \widehat{\phi}^n_{[c_0,c_1,l]}(\tau,\epsilon) \mathcal{O}_{\widehat{\lambda}}^n(t_{f_1},t_{f_2})\label{ExpandPartFct21com}
\end{equation}
where the Weyl orbits are given by\footnote{Notice that $d_\lambda=1$ for all $\lambda\in \widehat{P}^+_{1,1}$ such that no normalisation is required in (\ref{MissingNormWeyl}).}
\begin{equation}
\mathcal{O}_{\widehat{\lambda}}^n(t_{f_1},t_{f_2})= \widehat{d}_{\lambda} \sum_{w \in \widehat{W}} (-1)^{n \cdot l(w)} e^{w(\widehat{\lambda})}\,.
\end{equation}
The fundamental Weyl chamber $\widehat{P}_{n^2,n^2}^+$ is given by 
\begin{equation}
\widehat{P}_{n^2,n^2}^+= \{ [0,2n^2,-l],[2,2(n^2-1),-l],\dots,[2(n^2-1),2,-l],[2n^2,0,-l] \}_{l \in \mathbb{N}}\,.
\end{equation}
In this case the relation (\ref{comrel}) becomes
\begin{equation}
\mathcal{I}_{\tau}(\widehat{\phi}^2_{\lambda=[c_0,c_1,l]})= (n^2\widehat{\weyl} |n^2\widehat{\weyl} ) - (\widehat{\lambda}|\widehat{\lambda})+\overbrace{(c_0+c_1)}^{\text{level} \, k}l= \frac{n^4-c_1^2}{4}-kl\,.
\end{equation}
Explicit expressions for the coefficient functions $\widehat{\phi}^n_{[c_0,c_1,l]}$ are given in appendix~\ref{App:Coeffs21}. Finally, due to the fact that 
\begin{align}
\widehat{\phi}^n_{[c_0,c_1,l=0]}(\tau,\epsilon)=\phi^n_{[n^2-\tfrac{c_0-c_1}{2}]}(\tau,\epsilon)\,,
\end{align}
with $\phi^n_{[k]}$ defined in (\ref{Phin1M1}), we have in the decompactification limit
\begin{align}
&\lim_{\rho\to i\infty}\widetilde{\mathcal{Z}}_{2,1}(\tau,\rho,t_{f_1},m=n\epsilon,\epsilon,-\epsilon)=\mathcal{Z}^{\text{line}}_{2,1}(\tau,t_{f_1},m=n\epsilon,\epsilon,-\epsilon)\,,&& \forall n\geq 1\,.
\end{align}
as is expected from the point of view of the brane configurations.

\subsection{Configurations $(N,1)$ for $N>2$}
We can generalise the discussion of the previous subsection to cases $N>2$. For simplicity we restrict to $n=1$ and show that the partition function can be written as a product over simple positive roots of $\widehat{\mathfrak{a}}_{N-1}$.

The first case corresponds to $N=3$, \emph{i.e.} three M5-branes. For the partition function, this requires to sum over three different partitions $(\nu_1,\nu_2,\nu_3)$. Analysing the configurations which lead to a non-trivial contribution, we summarise the first few in the following table (with $g_{(3,1)}^{(\nu_1,\nu_2,\nu_3)}=\left(\prod_{a=1}^{3}(-Q_{f_{a}})^{|\nu_{a}|}\right)\,\prod_{a=1}^{3}\prod_{(i,j)\in \nu_{a}}\frac{\theta_{1}(\tau;z^{a}_{ij})\,\theta_{1}(\tau;v^{a}_{ij})}{\theta_1(\tau;w^{a}_{ij})\theta_1(\tau;u^{a}_{ij})}$)
\begin{center}
\begin{tabular}{|c|c|c|c|}\hline
&&&\\[-10pt]
$\nu_1$ & $\nu_2$ & $\nu_3$ & $g_{(3,1)}^{(\nu_1,\nu_2,\nu_3)}$ \\[8pt]\hline\hline
&&&\\[-10pt]
$\emptyset$ & $\emptyset$ & $\emptyset$ &$1$ \\[4pt]\hline\hline
&&&\\[-10pt]
$\parbox{0.3cm}{\ydiagram{1}}$ & $\emptyset$ & $\emptyset$ & $-Q_{f_1}$ \\[4pt]\hline
&&&\\[-10pt]
$\emptyset$ & $\parbox{0.3cm}{\ydiagram{1}}$ & $\emptyset$ & $-Q_{f_2}$ \\[4pt]\hline
&&&\\[-10pt]
$\emptyset$ & $\emptyset$ &$\parbox{0.3cm}{\ydiagram{1}}$ & $-Q_{f_3}$ \\[4pt]\hline\hline
&&&\\[-10pt]
$\parbox{0.6cm}{\ydiagram{2}}$ & $\parbox{0.3cm}{\ydiagram{1}}$ & $\emptyset$ & $Q_{f_1}^2Q_{f_2}$ \\[4pt]\hline
&&&\\[-10pt]
$\emptyset$ & $\parbox{0.6cm}{\ydiagram{2}}$ & $\parbox{0.3cm}{\ydiagram{1}}$ &  $Q_{f_2}^2Q_{f_3}$ \\[4pt]\hline
&&&\\[-10pt]
$\parbox{0.3cm}{\ydiagram{1}}$ & $\emptyset$ & $\parbox{0.6cm}{\ydiagram{2}}$ &  $Q_{f_1}Q_{f_3}^2$ \\[4pt]\hline
&&&\\[-10pt]
$\parbox{0.3cm}{\ydiagram{1,1}}$ & $\emptyset$ & $\parbox{0.3cm}{\ydiagram{1}}$ & $Q_{f_1}^2Q_{f_3}$ \\[4pt]\hline
&&&\\[-10pt]
$\parbox{0.3cm}{\ydiagram{1}}$ & $\parbox{0.3cm}{\ydiagram{1,1}}$ & $\emptyset$ & $Q_{f_1}Q_{f_2}^2$ \\[4pt]\hline
&&&\\[-10pt]
$\emptyset$ & $\parbox{0.3cm}{\ydiagram{1}}$ & $\parbox{0.3cm}{\ydiagram{1,1}}$ & $Q_{f_2}Q_{f_3}^2$ \\[4pt]\hline
\end{tabular}
\hspace{2cm}
\begin{tabular}{|c|c|c|c|}\hline
&&&\\[-10pt]
$\nu_1$ & $\nu_2$ & $\nu_3$ & $g_{(3,1)}^{(\nu_1,\nu_2,\nu_3)}$ \\[8pt]\hline\hline
&&&\\[-10pt]
$\parbox{0.3cm}{\ydiagram{1,1}}$ & $\emptyset$ & $\parbox{0.6cm}{\ydiagram{2}}$ & $-Q_{f_1}^2Q_{f_3}^2$ \\[4pt]\hline
&&&\\[-10pt]
$\emptyset$ & $\parbox{0.6cm}{\ydiagram{2}}$ & $\parbox{0.3cm}{\ydiagram{1,1}}$ & $-Q_{f_2}^2Q_{f_3}^2$ \\[4pt]\hline\hline
&&&\\[-10pt]
$\parbox{0.8cm}{\ydiagram{3,1}}$ & $\parbox{0.6cm}{\ydiagram{2}}$ & $\parbox{0.3cm}{\ydiagram{1}}$ & $-Q_{f_1}^4Q_{f_2}^2Q_{f_3}$ \\[4pt]\hline
&&&\\[-10pt]
$\parbox{0.3cm}{\ydiagram{1}}$ & $\parbox{0.8cm}{\ydiagram{3,1}}$ & $\parbox{0.6cm}{\ydiagram{2}}$ &  $-Q_{f_1}Q_{f_2}^4Q_{f_3}^2$ \\[4pt]\hline
&&&\\[-10pt]
$\parbox{0.6cm}{\ydiagram{2}}$ & $\parbox{0.3cm}{\ydiagram{1}}$ & $\parbox{0.8cm}{\ydiagram{3,1}}$ &   $-Q_{f_1}^2Q_{f_2}Q_{f_3}^4$ \\[4pt]\hline
&&&\\[-10pt]
$\parbox{0.6cm}{\ydiagram{2,1,1}}$ & $\parbox{0.3cm}{\ydiagram{1}}$ & $\parbox{0.3cm}{\ydiagram{1,1}}$ & $-Q_{f_1}^4Q_{f_2}Q_{f_3}^2$ \\[8pt]\hline
&&&\\[-10pt]
$\parbox{0.3cm}{\ydiagram{1,1}}$ & $\parbox{0.6cm}{\ydiagram{2,1,1}}$ & $\parbox{0.3cm}{\ydiagram{1}}$ &  $-Q_{f_1}^2Q_{f_2}^4Q_{f_3}$ \\[8pt]\hline
&&&\\[-10pt]
$\parbox{0.3cm}{\ydiagram{1}}$ & $\parbox{0.3cm}{\ydiagram{1,1}}$ & $\parbox{0.6cm}{\ydiagram{2,1,1}}$ &   $-Q_{f_1}Q_{f_2}^2Q_{f_3}^4$ \\[8pt]\hline
\end{tabular}
\end{center}
The first few terms in the partition function therefore take the form
\begin{align}
\widetilde{\mathcal{Z}}_{3,1}&(\tau,t_{f_1},t_{f_2},t_{f_3},m=\epsilon,\epsilon,-\epsilon)=\,1 -(Q_{f_1} + Q_{f_2} + Q_{f_3}) \nonumber\\
&+(Q_{f_1}^2 Q_{f_2} + Q_{f_1} Q_{f_2}^2 + Q_{f_1}^2 Q_{f_3} + Q_{f_2}^2 Q_{f_3} + Q_{f_1} Q_{f_3}^2 + Q_{f_2} Q_{f_3}^2) - (Q_{f_1}^2 Q_{f_2}^2 + Q_{f_1}^2 Q_{f_3}^2 + Q_{f_2}^2 Q_{f_3}^2)\nonumber\\
& - (Q_{f_1}^4 Q_{f_2}^2 Q_{f_3} +Q_{f_1}^2 Q_{f_2}^4 Q_{f_3} + Q_{f_1}^4 Q_{f_2} Q_{f_3}^2+  Q_{f_1} Q_{f_2}^4 Q_{f_3}^2 + Q_{f_1}^2 Q_{f_2} Q_{f_3}^4 + Q_{f_1} Q_{f_2}^2 Q_{f_3}^4) +\ldots
\end{align}
This expansion is matched by the expression\footnote{We have checked (\ref{Rel3Comp}) up to order $12$ in the expansion of $Q_{f_{1,2,3}}$.}
\begin{align}
&\widetilde{\mathcal{Z}}_{3,1}(\tau,t_{f_1},t_{f_2},t_{f_3},m=\epsilon,\epsilon,-\epsilon)= \prod_{k=1}^\infty \mathcal{Z}_{3,1}^{(k)}(\tau,m=\epsilon,t_{f_1},t_{f_2},t_{f_3},\epsilon,-\epsilon)\nonumber\\
&=\prod_{k=1}^\infty (1-Q_\rho^k)^2\,(1-Q_\rho^{k-1} Q_1)(1-Q_\rho^{k-1} Q_2)(1-Q_\rho^{k-1} Q_3) (1 - Q_\rho^k Q_1^{-1})(1 - Q_\rho^k Q_2^{-1}) (1-Q_\rho^k Q_3^{-1})\nonumber\\
&=\prod_{\widehat{\alpha}\in\widehat{\Delta}_+(\widehat{\mathfrak{a}}_2)}(1-e^{-\widehat{\alpha}})\,,\label{Rel3Comp}
\end{align}
where $\widehat{\Delta}_+(\widehat{\mathfrak{a}}_2)$ is the space of positive simple roots of $\widehat{\mathfrak{a}}_2$. Notice the relation
\begin{align}
\mathcal{Z}^{\text{line}}_{4,1}(\tau,t_{f_1},t_{f_2},t_{f_3},m=\epsilon,\epsilon,-\epsilon)=\frac{1}{(1-Q_\rho)(1-Q_{f_1}Q_{f_3})}\,\widetilde{\mathcal{Z}}_{3,1}^{(1)}(\tau,m=\epsilon,t_{f_1},t_{f_2},t_{f_3},\epsilon,-\epsilon)\,.
\end{align}

Repeating the computation for $N=4$, we find up to order $6$ in the expansion of $Q_{f_i}$ that the partition function can be written as
\begin{align}
\widetilde{\mathcal{Z}}_{4,1}&(\tau,t_{f_1},t_{f_2},t_{f_3},t_{f_4},m=\epsilon,\epsilon,-\epsilon)= \prod_{k=1}^\infty \widetilde{\mathcal{Z}}_{4,1}^{(k)}(\tau,m=\epsilon,t_{f_1},t_{f_2},t_{f_3},t_{f_4},\epsilon,-\epsilon)\nonumber\\
&=\prod_{k=1}^\infty (1-Q_\rho^k)^3\,(1-Q_\rho^{k-1} Q_{f_1})(1-Q_\rho^{k-1} Q_{f_2})(1-Q_\rho^{k-1} Q_{f_3}) (1-Q_\rho^{k-1} Q_{f_4}) \nonumber\\
&\hspace{1cm}\times (1 - Q_\rho^k /Q_{f_1})(1 - Q_\rho^k/ Q_{f_2}) (1-Q_\rho^k/Q_{f_3})(1-Q_\rho^k/ Q_{f_4})\nonumber\\
&\hspace{1cm}\times (1 - Q_\rho^k/(Q_{f_1}Q_{f_2}))(1 - Q_\rho^k /(Q_{f_2}Q_{f_3})) (1-Q_\rho^k /(Q_{f_3}Q_{f_4}))(1-Q_\rho^k/(Q_{f_1} Q_{f_4}))\nonumber\\
&=\prod_{\widehat{\alpha}\in\widehat{\Delta}_+(\widehat{\mathfrak{a}}_3)}(1-e^{-\widehat{\alpha}})\,,\label{Rel4Comp}
\end{align}
with $Q_\rho=Q_{f_1}Q_{f_2}Q_{f_3}Q_{f_4}$.
Notice the relation
\begin{align}
\mathcal{Z}^{\text{line}}_{5,1}(\tau,t_{f_1},t_{f_2},t_{f_3},t_{f_4},m=\epsilon,\epsilon,-\epsilon)=\frac{\widetilde{\mathcal{Z}}_{4,1}^{(1)}(\tau,m=\epsilon,t_{f_1},t_{f_2},t_{f_3}t_{f_4},\epsilon,-\epsilon)}{(1-Q_\rho)^2(1-Q_{f_1}Q_{f_4})(1-Q_{f_1}Q_{f_3}Q_{f_4})(1-Q_{f_1}Q_{f_2}Q_{f_4})}\,.
\end{align}
\subsection{Configuration $(N,M)=(2,2)$}
Finally we can similarly discuss cases $(N,M)$ with $M>1$. The simplest such case is the configuration $(2,2)$ and we shall limit ourselves to the choice $m=\epsilon_1=-\epsilon_2=\epsilon$. Analysing the partition function $\widetilde{\mathcal{Z}}_{2,2}(\mathbf{T},\mathbf{t},m=\epsilon,\epsilon,-\epsilon)$ in the same fashion as above, we can write it in the following suggestive form
\begin{align}
\widetilde{\mathcal{Z}}_{2,2}(\mathbf{T},&\mathbf{t},m=\epsilon,\epsilon,-\epsilon) = e^{-2(\widehat{\omega}_0+\widehat{\omega}_1}) \Big[\big( e^{2\widehat{\omega}_0+2\widehat{\omega}_1} + e^{6\widehat{\omega}_0 - 2\widehat{\omega}_1}+e^{-2\widehat{\omega}_0+6\widehat{\omega}_1- 2\delta}+ e^{10\widehat{\omega}_0-6\widehat{\omega}_1-2\delta}\dots \big) \nonumber \\
&-2\,\frac{\theta_1(\tau; \tilde{T}_1+\epsilon)\theta_1(\tau;\tilde{T}_1-\epsilon)}{\theta_1(\tau;\tilde{T}_1)^2}\, \big(e^{4\widehat{\omega}_0} + e^{-4\widehat{\omega}_0+ 8 \widehat{\omega}_1-4\delta} + e^{12\widehat{\omega}_0-8\widehat{\omega}_1-4\delta} + \dots  \big) \nonumber \\
&-2\,\frac{\theta_1(\tau; \tilde{T}_1+\epsilon)\theta_1(\tau;\tilde{T}_1-\epsilon)}{\theta_1(\tau;\tilde{T}_1)^2} \big( e^{4\widehat{\omega}_1-\delta} + e^{8\widehat{\omega}_0-4\widehat{\omega}_1-\delta} + e^{-8\widehat{\omega}_0 + 12 \widehat{\omega}_1-9\delta}\dots \big) \nonumber \\
&+2\,\frac{\theta_1(\tau;\tilde{T}_1+2\epsilon)\theta_1(\tau;\tilde{T}_1-2\epsilon)}{\theta_1(\tau;\tilde{T}_1)^2} \big( e^{2\widehat{\omega}_0+2\widehat{\omega}_1-\delta} + e^{6\widehat{\omega}_0-2\widehat{\omega}_1-\delta} + e^{-2\widehat{\omega}_0+6\widehat{\omega}_1-3\delta}+ \dots \big) +\dots \Big]\,, \label{Part221}
\end{align}
where we have used the same notation as in section~\ref{Sect:ConfigCompact21}. Comparing (\ref{Part221}) with the previous examples, we notice that the partition function can again be written as a sum of Weyl orbits where the affine weights of the representatives lie in the fundamental Weyl chamber of the affine representation $\widehat{P}_{2,2}^+$ with highest weight $[2,2,0]$ (see appendix~\ref{affrep})
\begin{equation}
\widetilde{\mathcal{Z}}_{2,2}(\tau, T_1,t_{f_1},t_{f_2},m=\epsilon,\epsilon,-\epsilon)= e^{-2 \widehat{\xi}} \sum_{\widehat{\lambda} \in \widehat{P}_{2,2,}^+} \widehat{\phi}^1_{[c_0,c_1,l]}(\tau,T_1,\epsilon)\, \mathcal{O}_{\widehat{\lambda}}(t_{f_1},t_{f_2})\,.\label{Part22ExpansionOrbit}
\end{equation}
Here the Weyl orbits are given by
\begin{align}
&\mathcal{O}_{\widehat{\lambda}}^1(t_{f_1},t_{f_2})= \widehat{d}_\lambda\,\sum_{w\in \widehat{\mathcal{W}}} e^{w(\widehat{\lambda})}\,,&&\text{with} &&\widehat{d}_{\lambda=[c_0,c_1,-l]}=\left\{\begin{array}{cl} \tfrac{1}{2} & \text{if } c_0=0\text{ or }c_1=0 \\ 1 &\text{else }\end{array}\right.
\end{align} 
where the Weyl reflections are explicitly given as in (\ref{AffineWeylRelections}) and 
the factor $\widehat{d}_\lambda$ takes into account the presence of fixed points in the Weyl action. The fundamental Weyl chamber is defined as
\begin{equation}
\widehat{P}_{2,2}^+= \{[0,4,-l],[2,2,-l],[4,0,-l] \}_{l\in \mathbb{N}\cup\{0\}}\,.
\end{equation}
and the $\widehat{\phi}^1_{\lambda=[c_0,c_1,-l]}$ in (\ref{Part22ExpansionOrbit}) are given by
\begin{align}
&\widehat{\phi}^1_{[0,4,-l]}(\tau,T_1,\epsilon)=\left\{\begin{array}{lcl} -2\,\frac{\theta_1(\tau;\tilde{T}_1+(2r+1)\epsilon)\theta_1(\tau;\tilde{T}_1-(2r+1)\epsilon)}{\theta_1(\tau;\tilde{T}_1)^2} & \text{if} & l=r(r+1)+1\text{ for }r\in\mathbb{N}\cup\{0\} \\[4pt] 0 &&\text{else}\end{array}\right.\label{Z22coef1}\\[10pt]
&\widehat{\phi}^1_{[2,2,-l]}(\tau,T_1,\epsilon)=\left\{\begin{array}{lcl} 1 & \text{if} & l=0 \\[4pt] 2\,\frac{\theta_1(\tau;\tilde{T}_1+2r\epsilon)\theta_1(\tau;\tilde{T}_1-2r\epsilon)}{\theta_1(\tau;\tilde{T}_1)^2} & \text{if} & l=r^2\text{ for }r\in\mathbb{N} \\[4pt] 0 &&\text{else}\end{array}\right.\label{Z22coef2}\\[10pt]
&\widehat{\phi}^1_{[4,0,-l]}(\tau,T_1,\epsilon)=\left\{\begin{array}{lcl} -2\,\frac{\theta_1(\tau;\tilde{T}_1+(2r+1)\epsilon)\theta_1(\tau;\tilde{T}_1-(2r+1)\epsilon)}{\theta_1(\tau;\tilde{T}_1)^2} & \text{if} & l=r(r+1)\text{ for }r\in\mathbb{N}\cup\{0\} \\[4pt] 0 &&\text{else}\end{array}\right.\label{Z22coef3}
\end{align}
We notice that the arguments of the $\widehat{\phi}^1_{[c_0,c_1,l]}$ are related to the affine weights by
\begin{equation}
\mathcal{I}_\tau(\widehat{\phi}^1_{[c_0,c_1,-l]})= (2\widehat{\xi}|2\widehat{\xi})-(\widehat{\lambda}|\widehat{\lambda}) + 4l =1 - \frac{1}{4}c_1^2 - 4l\,.
\end{equation}
which directly generalises the cases $M=1$ discussed above

Before closing this section we would like to make a further remark: The brane configuration $(N,M)=(2,2)$ is self-dual under the exchange of $N$ and $M$. Furthermore, the appearance of the symmetry $\widehat{\mathfrak{a}}_{N-1=1}$ is due to the expansion of $\widetilde{\mathcal{Z}}_{2,2}(\tau, T_1,\rho,t_{f_1},t_{f_2},m=\epsilon,\epsilon,-\epsilon)$ with respect to $Q_{t_{f_{1,2}}}$ and we would expect a similar structure with respect to $\bar{Q}_{1,2}$. It is therefore interesting to see whether the partition function can be written in a fashion that makes a symmetry $\widehat{\mathfrak{a}}_{N-1=1}\otimes \widehat{\mathfrak{a}}_{M-1=1}$ manifest. To this end, we first have to re-instate the normalisation factor $(W_{M=2}(\emptyset))^{N=2}\big|_{\epsilon_1=-\epsilon_2=m=\epsilon}$ in (\ref{GenDefPartFct}). The latter can be read off from (\ref{ProdRootA1p})
\begin{align}
W_2(\emptyset)\big|_{\epsilon_1=-\epsilon_2=m=\epsilon}&=\lim_{\rho\to i\infty}\widetilde{\mathcal{Z}}_{2,1}(\rho,T_1,T_2,m=\epsilon,\epsilon,-\epsilon)\nonumber\\
&=\left(\prod_{k=1}^\infty(1-\bar{Q}_1Q_\tau^k)\right)\left(\prod_{k=1}^\infty(1-Q_\tau^k/\bar{Q}_1)(1-Q_\tau)^k\right)=-i\frac{\bar{Q}_{1}^{1/2}}{Q_\tau^{1/8}}\,\theta_1(\tau;\tilde{T}_1)\,.
\end{align}
Thus, multiplying the coefficient functions (\ref{Z22coef1}) -- (\ref{Z22coef3}) with $(W_2(\emptyset))^2\big|_{\epsilon_1=-\epsilon_2=m=\epsilon}$, the non-trivial $\widehat{\phi}^1_{\lambda}$ are (up to integer coefficients) of the form 
\begin{align}
-\frac{\bar{Q}_1}{Q_\tau^{1/4}}\,\theta_1(\tau;\tilde{T}_1+k\epsilon)\theta_1(\tau;\tilde{T}_1-k\epsilon)\label{ExpansionThetaCoefs}
\end{align}
for $k\in\mathbb{N}\cup\{0\}$. Upon introducing
\begin{align}
&T_1=-2\widehat{\kappa}_0+2\widehat{\kappa}_1\,,&&Q_\tau=e^{-\mu}\,,
\end{align}
which mirror (\ref{OmegaIntrod}) and (\ref{choice}) such that
\begin{align}
&\bar{Q}_1^a \bar{Q}_2^b=e^{2(b-a)(\widehat{\kappa}_1-\widehat{\kappa}_0)-b\mu}\,,&&\forall a,b\in\mathbb{N}\,,
\end{align}
we can write for (\ref{ExpansionThetaCoefs})
\begin{align}
-\frac{\bar{Q}_1}{Q_\tau^{1/4}}\,\theta_1(\tau;\tilde{T}_1+k\epsilon)\theta_1(\tau;\tilde{T}_1-k\epsilon)= e^{-2(\widehat{\kappa}_0+\widehat{\kappa}_1)} \sum_{\widehat{\lambda} \in \widehat{P}_{2,2,}^+} \widehat{\varphi}^1_{[c_0,c_1,l]}(k,\epsilon)\, \mathcal{O}_{\widehat{\lambda}}(T_1,T_2)\,.\label{ExpandMirrorThet}
 \end{align}
where we denote $e^{-2(\widehat{\kappa}_0+\widehat{\kappa}_1)}=e^{-2\widehat{\zeta}}$ as the Weyl vector of $\widehat{a}_{M-1=1}$ and 
\begin{align}
&\widehat{\varphi}^1_{[0,4,-l]}(k,\epsilon)=\left\{\begin{array}{lcl} -(\epsilon^{-k(2s+1)}+\epsilon^{k(2s+1)}) & \text{if} & l=s(s+1)+1\text{ for }s\in\mathbb{N}\cup\{0\} \\[4pt] 0 &&\text{else}\end{array}\right.\label{Z22coef1Mirror}\\[10pt]
&\widehat{\varphi}^1_{[2,2,-l]}(k,\epsilon)=\left\{\begin{array}{lcl} 1 & \text{if} & l=0 \\[4pt] \epsilon^{2ks}+\epsilon^{-2ks} & \text{if} & l=s^2\text{ for }s\in\mathbb{N} \\[4pt] 0 &&\text{else}\end{array}\right.\label{Z22coef2Mirror}\\[10pt]
&\widehat{\varphi}^1_{[4,0,-l]}(k,\epsilon)=\left\{\begin{array}{lcl} -(\epsilon^{-k(2s+1)}+\epsilon^{k(2s+1)}) & \text{if} & l=s(s+1)\text{ for }s\in\mathbb{N}\cup\{0\} \\[4pt] 0 &&\text{else}\end{array}\right.\label{Z22coef3Mirror}
\end{align}
which exactly mirror (\ref{Z22coef1Mirror}) -- (\ref{Z22coef3Mirror}). Thus the partition function can be written in the form
\begin{align}
\mathcal{Z}_{2,2}(\mathbf{T},&\mathbf{t},m=\epsilon,\epsilon,-\epsilon) =e^{-2(\widehat{\zeta}+\widehat{\weyl})}\,\sum_{\lambda_1,\lambda_2\in \widehat{P}^+_{2,2}}\,\widehat{\mathfrak{p}}^1_{\lambda_1,\lambda_2}(\epsilon)\, \mathcal{O}_{\widehat{\lambda_1}}(t_{f_1},t_{f_2})\, \mathcal{O}_{\widehat{\lambda_2}}(T_1,T_2)\,.\label{DualPartitionFunction}
\end{align}
where the non-vanishing coefficients $\widehat{\mathfrak{p}}^1_{\lambda_1,\lambda_2}(\epsilon)$ are (with $s,s'\in \mathbb{N}\cup \{0\}$ and $r,r'\in \mathbb{N}$)
{\allowdisplaybreaks\begin{align}
&\widehat{\mathfrak{p}}^1_{[0,4,-s(s+1)-1],[c_0,c_1,-l]}(\epsilon)=\left\{\begin{array}{lcl}2\,(\epsilon^{-(2s+1)(2s'+1)}+\epsilon^{(2s+1)(2s'+1)}) & \text{if} & (c_0,c_1)=(0,4)\text{ and }l=s'(s'+1)+1  \\[4pt] -2 & \text{if} & (c_0,c_1)=(2,2)\text{ and }l=0\\[4pt] -2 (\epsilon^{-(2s+1)2r}+\epsilon^{(2s+1)2r})& \text{if} & (c_0,c_1)=(2,2)\text{ and }l=r^2 \\[4pt] 2\,(\epsilon^{-(2s+1)(2s'+1)}+\epsilon^{(2s+1)(s'+1)}) & \text{if} & (c_0,c_1)=(4,0)\text{ and }l=s'(s'+1)  \end{array}\right.\\[10pt]
&\widehat{\mathfrak{p}}^1_{[2,2,0],[c_0,c_1,-l]}(\epsilon)=\left\{\begin{array}{lcl}-2 & \text{if} & (c_0,c_1)=(0,4)\text{ and }l=s'(s'+1)+1  \\[4pt] 1 & \text{if} & (c_0,c_1)=(2,2)\text{ and }l=0\\[4pt] 2 & \text{if} & (c_0,c_1)=(2,2)\text{ and }l=r^2 \\[4pt] -2 & \text{if} & (c_0,c_1)=(4,0)\text{ and }l=s'(s'+1)  \end{array}\right.\\[10pt]
&\widehat{\mathfrak{p}}^1_{[2,2,-r^2],[c_0,c_1,-l]}(\epsilon)=\left\{\begin{array}{lcl}-2\,(\epsilon^{-2r(2s'+1)}+\epsilon^{2r(2s'+1)}) & \text{if} & (c_0,c_1)=(0,4)\text{ and }l=s'(s'+1)+1  \\[4pt] 2 & \text{if} & (c_0,c_1)=(2,2)\text{ and }l=0\\[4pt] 2 (\epsilon^{-4rs'}+\epsilon^{4rs'})& \text{if} & (c_0,c_1)=(2,2)\text{ and }l={s'}^2 \\[4pt] -2\,(\epsilon^{-2r(2s'+1)}+\epsilon^{2r(2s'+1)}) & \text{if} & (c_0,c_1)=(4,0)\text{ and }l=s'(s'+1)  \end{array}\right.\\[10pt]
&\widehat{\mathfrak{p}}^1_{[4,0,-s(s+1)],[c_0,c_1,-l]}(\epsilon)=\left\{\begin{array}{lcl}2\,(\epsilon^{-(2s+1)(2s'+1)}+\epsilon^{(2s+1)(2s'+1)}) & \text{if} & (c_0,c_1)=(0,4)\text{ and }l=s'(s'+1)+1  \\[4pt] -2 & \text{if} & (c_0,c_1)=(2,2)\text{ and }l=0\\[4pt] -2 (\epsilon^{-(2s+1)2r}+\epsilon^{(2s+1)2r})& \text{if} & (c_0,c_1)=(2,2)\text{ and }l=r^2 \\[4pt] 2\,(\epsilon^{-(2s+1)(2s'+1)}+\epsilon^{(2s+1)(2s'+1)}) & \text{if} & (c_0,c_1)=(4,0)\text{ and }l=s'(s'+1)  \end{array}\right.
\end{align}}
Notice in particular that $\widehat{\mathfrak{p}}^1_{\lambda_1,\lambda_2}(\epsilon)=\widehat{\mathfrak{p}}^1_{\lambda_2,\lambda_1}(\epsilon)$. Therefore, the form (\ref{DualPartitionFunction}) makes the duality of the partition function under the exchange $(N,M)\longleftrightarrow (M,N)$ manifest.

\section{Generic Configuration $(N,M)$ and Representations}
\label{Sect:GenericConfiguration}
After the analysis of many specific cases we compile in this section generic relations that we conjecture to hold for arbitrary $N$ , $M$ and $n$. As the non-compact case is obtained as a limit of the compact case we start with the latter. 

We propose that the normalised partition \ref{completepartitionNorm} can be written as a sum over the Weyl orbits for the representative weights in the fundamental Weyl chamber $\widehat{P}^+_{Mn^2,\dots,Mn^2}$ of the affine highest weight representation generated by $[\underbrace{Mn^2,\dots,Mn^2}_{N},0]$ of $\widehat{\mathfrak{sl}}(N,\mathbb{C})$
\begin{equation}
\widetilde{\mathcal{Z}}_{N,M}(\tau, \textbf{T},\rho,\textbf{t}_f,m=n\epsilon,\epsilon,-\epsilon) = e^{-Mn^2\widehat{\weyl }} \sum_{\widehat{\lambda} \in \widehat{P}^+_{Mn^2,\dots,Mn^2}} \widehat{\phi}^{n,M}_{[c_0,\dots,c_{N-1},l]}(\tau,\textbf{T},\epsilon)\mathcal{O}_{\widehat{\lambda}}^n(\textbf{t}_f)
\end{equation} 
where $\widehat{\weyl }=\widehat{\omega}_0 + \dots + \widehat{\omega}_{N-1}$ denotes the affine Weyl vector for  $\widehat{\mathfrak{sl}}(N,\mathbb{C})$ defined in terms of the fundamental weights $\widehat{\omega}_i$ and the Weyl orbits are given by
\begin{equation}
\mathcal{O}_{\widehat{\lambda}}^n(\textbf{t}_f)= \sum_{w \in \widehat{\mathcal{W}}} (-1)^{Mn \cdot l(w)}e^{w(\widehat{\lambda})}\,.
\end{equation}
The $\widehat{\phi}^{n,M}_{\lambda}$ are given by ratios of Jacobi theta functions. They transform under $SL(2,\mathbb{Z})_\tau$ transformations as
\begin{equation}
\widehat{\phi}^{n,M}_{[c_0,\ldots,c_{N-1},l]}\left(-\tfrac{1}{\tau},\tfrac{\mathbf{T}}{\tau},\tfrac{\epsilon}{\tau}\right)=e^{2 i \pi \mathcal{I}_\tau\,\epsilon^2} \widehat{\phi}^{n,M}_{[c_0,\ldots,c_{N-1},l]}(\tau,\mathbf{T},\epsilon)\,.\label{SLtrafoGen}
\end{equation}
where the index $\mathcal{I}_\tau$ is related to the Dynkin labels by 
\begin{equation}
\mathcal{I}_\tau\left(\widehat{\phi}^{n,M}_{\lambda=[c_0,\ldots,c_{N-1},l]}\right)= (Mn^2\widehat{\weyl} |Mn^2\widehat{\weyl} ) - (\widehat{\lambda}|\widehat{\lambda}) + kl
\label{Index}
\end{equation}
where $(.|.)$ stands for the inner product in the basis of affine fundamental weights $(\widehat{\omega}_1,\dots,\widehat{\omega}_{N-1})$. 

The partition functions of non-compact brane configurations are obtained by taking the limit
\begin{equation}
t_{f_N} \to  \infty \quad \iff \quad \delta \to \infty  \,.
\end{equation}
From the point of view of affine representations this means that we only keep the weights with grade $l=0$ as $e^{-l\delta} \to 0$. The remaining states fall into the corresponding $\mathfrak{sl}(N,\mathbb{C})$ representations with the non-affine counterpart of $\widehat{\lambda}$ as highest weight vector
\begin{equation}
\widehat{\lambda}=(\lambda,k,l) \quad \longrightarrow \quad \lambda\,.
\end{equation}
The affine Weyl group $\widehat{\mathcal{W}}$ reduces to the finite one $\mathcal{W}$. Futhermore the $\widehat{\phi}$'s at grade 0 are identified with their non-affine counterparts in the following way
\begin{equation}
\widehat{\phi}^{n,M}_{[c_0,c_1,\dots,c_{N-1},0]}(\mathbf{T},\epsilon) = \phi^{n,M}_{[c_1,\dots,c_{N-1}]}(\mathbf{T},\epsilon)\,.
\end{equation}
After taking the limit we are thus left with a sum over the Weyl orbits fot he representative weights in the fundamental Weyl chamber $P_{Mn^2,\dots,Mn^2}$ of the irreducible highest weight representation $\Gamma_{Mn^2,\dots,Mn^2}$ of $\mathfrak{sl}(N,\mathbb{C})$
\begin{equation}
\mathcal{Z}_{N,M}^{\text{line}}(\textbf{T},t_{f_1},\dots,t_{f_{N-1}},m=n\epsilon,\epsilon,-\epsilon)=e^{-Mn^2\weyl} \sum_{\lambda \in P_{Mn^2,\dots,Mn^2}^+} \phi_{[c_1,\dots,c_{N-1}]}(\textbf{T},\epsilon) \mathcal{O}_{\lambda}^n(t_{f_1},\dots,t_{N-1})\,,
\end{equation}
with the finite Weyl orbits given by
\begin{equation}
\mathcal{O}_{\lambda}^n(t_{f_1},\dots,t_{f_{N-1}})=\sum_{w\in \mathcal{W}}(-1)^{Mn\cdot l(w)}e^{w(\lambda)}\,.
\end{equation}
The index (\ref{Index}) reduces to\footnote{Notice that the transformation (\ref{SLtrafoGen}) is compatible with the decompactification limit.} 
\begin{equation}
\mathcal{I}=(Mn^2\weyl,Mn^2\weyl)-(\lambda,\lambda)\,,
\end{equation}
where $(.,.)$ denotes the inner product in the basis of fundamental weights $(\omega_1,\dots,\omega_{N-1})$.
\section{Conclusions}\label{Sect:Conclusions}
In this paper we have studied the BPS partition functions of $N$ parallel M5-branes probing a transverse $\text{ALE}_{A_{M-1}}$ space. We have distinguished cases of M5-branes separated along $\mathbb{S}^1$ (with partition function $\mathcal{Z}_{N,M}$ defined in (\ref{GenDefPartFct})) and along $\mathbb{R}$ (with partition function $\mathcal{Z}^{\text{line}}_{N,M}$ defined in (\ref{GenDefPartFctNon}). The latter can be obtained from the former through the decompactification limit that sends one of the distances $t_{f_a}$ of the branes to infinity.

To regularise the BPS partition functions, a set of deformation parameters, denoted by $(m,\epsilon_1,\epsilon_2)$ needs to be introduced. For simplicity, we have chosen to work in the so-called unrefined limit $\epsilon_1=-\epsilon_2=\epsilon$. Furthermore, motivated by studying the holonomy structure of the supercharges (from the point of view of the M-string world-sheet theory), we have imposed $m=n\epsilon$ for $n\in\mathbb{N}$. We have demonstrated in a large series of examples (and conjecture that our results hold for generic values of $N,M,n\in\mathbb{N}$) that this limit exhibits an $\mathfrak{a}_{N-1}$ (or affine $\widehat{\mathfrak{a}}_{N-1}$) symmetry of the BPS counting function. Indeed, in the case of non-compact brane configurations, (after a suitable normalisation) $\mathcal{Z}_{N,M}^{\text{line}}$ depends only polynomially on $Q_{f_a}=e^{-t_{f_a}}$. Upon identifying the latter with the roots of $\mathfrak{a}_{N-1}$, the partition function can be organised as a sum of orbits of $S_{N}$ which is the Weyl group of $\mathfrak{a}_{N-1}$. Furthermore, the representatives of each orbit fall into the fundamental Weyl chamber $P^+_{Mn^2,\ldots,Mn^2}$ of the irreducible representation $\Gamma_{Mn^2,\ldots,Mn^2}$ of $\mathfrak{a}_{N-1}$.

For compact brane configurations, the (suitably normalised) partition function $\mathcal{Z}_{N,M}(m=n\epsilon)$ is no-longer polynomial in the $Q_{f_a}$. Nevertheless, it can be arranged in a similar fashion as a sum over weights that form a single integrable representation of the affine Lie algebra $\widehat{\mathfrak{a}}_{N}$ with highest weight $[Mn^2,\ldots,Mn^2,0]$. We have again demonstrated this behaviour explicitly for a large number of examples and based on the emergent pattern conjecture that it holds in general.

Finally, compact brane configurations enjoy the duality $(N,M)\longleftrightarrow (M,N)$. For the case $(N,M)=(2,2)$ we have made this duality manifest in the full partition function $\mathcal{Z}_{2,2}$ by writing it as a double sum of weights in the fundamental domain $\widehat{P}^+_{2,2}$ of $\widehat{\mathfrak{a}}_1$. This presentation of the partition function also makes the structure of $X_{2,2}\cong X_{1,1}/(\mathbb{Z}_2\times \mathbb{Z}_2)$ more tangible, which is dual to the M5-brane configuration. 

These results make the algebraic properties of the BPS counting functions of specific M-brane configurations very tangible: indeed, in certain regions of the parameter space, the partition functions fall into the form of single highest weight representations of (affine) Lie algebras that are related to the geometric backgrounds of the M-brane configurations. While the results presented here are specific to the choice $m=n\epsilon$, the $\mathfrak{a}_N$ symmetries are expected to be unbroken for generic deformations as well: indeed the dual Calabi-Yau manifolds $X_{N,M}$ can be understood as elliptic fibrations over $A_{N-1}$. Thus, our results have highlighted a region in the moduli space in which the latter are very manifest.

In view of the many other physical systems that are dual to the M-brane configurations that we have studied here, we expect our results to have many applications in the future: one of them is the study of little string theories (LSTs) \cite{Witten:1995zh,Aspinwall:1997ye,Seiberg:1997zk,Intriligator:1997dh,Hanany:1997gh,Brunner:1997gf} (see also \cite{Aharony:1999ks,Kutasov:2001uf} for reviews). Indeed, the compact brane configurations $(N,M)$ are related to a particular class of LSTs \cite{Bhardwaj:2015oru,Hohenegger:2015btj,Hohenegger:2016eqy} with $\mathcal{N}=(1,0)$ supersymmetry. It will be interesting in the future to find further regions in the moduli spaces of LSTs which make more of their symmetries manifest or possibly reveal new ones. Furthermore, our findings may also turn out useful to study algebraic properties of double-quantised Seiberg-Witten geometry related to the topological string partition function of $X_{N,M}$ and the definition of $qq$-characters (see \cite{Kimura:2015rgi,Kimura:2016dys,Kimura:2017auj,Kimura:2017hez} and \cite{Nekrasov:2015wsu,Nekrasov:2016qym,Nekrasov:2016ydq} for recent progress respectively). . 

Finally, an interesting open question remains why in the limits we have discussed in this work, the partition function is governed by a single irreducible/integrable representation. While we have argued, based on the structure of the preserved supercharges, that the choice (\ref{UnrefinedChoice}) and (\ref{choicemasseps}) leads to cancellations among different states in the partition function (and thus to massive simplifications) it does not fully explain why the remaining contributions have the structure of a single representation. As was pointed out to us by A.~Iqbal, it would be interesting to study these results from the point of view of Chern Simons theory (see \emph{e.g.} \cite{Aganagic:2003db, Hollowood:2003cv}) to see if one can find an interpretation from this side. We leave this possibility for future work.

\section*{Acknowledgements}
We would like to thank A.~Iqbal for many stimulating discussions and very useful comments on the current manuscript. Furthermore,  SH would like to thank S.J.~Rey for many enlightening discussions.
\appendix
\section{Affine Lie Algebras}\label{app:AffExt}
\subsection{Central Extension of Simple Lie Algebras}
In this appendix we follow \cite{Kac} and \cite{DiFrancesco}. Reviews of this material can also be found in \cite{Persson:2010ms,Fuchs}. 

The affine Lie algebra $\widehat{\mathfrak{g}}$ has the following decomposition
\begin{equation}
\widehat{\mathfrak{g}}=\underbrace{\mathfrak{g} \otimes \mathbb{C}[t,t^{-1}]}_{\tilde{\mathfrak{g}}} \oplus \mathbb{C}\widehat{k} \oplus \mathbb{C}L_0
\end{equation}
where $\tilde{\mathfrak{g}}$ corresponds to the so called loop algebra. For a generator $J^a \in \mathfrak{g}$ the corresponding elements of the loop algebra take the form
\begin{equation}
J^a\otimes t^l = J^a_l \in \tilde{\mathfrak{g}} \, ,\quad l\in \mathbb{Z}
\end{equation}
The loop algebra is then centrally extended in a non-trivial way\footnote{All central extensions for simple Lie algebras turn out to be trivial. \cite{Fuchs}} by the addition of $\widehat{k}$ with the property that it commutes with all the generators
\begin{equation}
[J^a_l,\widehat{k}]=0
\end{equation}
It can be shown \cite{Fuchs} that this extension is unique.\footnote{Notice that the abelian subalgebra $\{ H^1_0,...,H^r_0,\widehat{k}\}$ is not maximally abelian. To define the  Cartan subalgebra $L_0$ needs to be included.} $L_0$ is the so called grading operator defined as a differential operator in $t$, whose eigenvalue $l$ in the sense
\begin{equation}
[L_0,J^a_l]=-lJ^a_l
\end{equation}
is called the \emph{grade} of $J^a_l$. The eigenvectors under the action of $ad(H^i_0)$ and $ad(\widehat{k})$ on the generators $E^{\alpha}_l$ are infinitely degenerate. It is thus necessary to introduce $L_0$ so that $\{H^1_0,...,H^r_0,\widehat{k},L_0 \}$ forms a Cartan subalgebra. \\ \\
An affine weight $\widehat{\lambda}$ can thus be denoted by its eigenvalues under the Cartan subalgebra
\begin{equation}
\widehat{\lambda}=(\lambda;k;l)
\label{affLab}
\end{equation}
where $\lambda$ is the corresponding weight in the finite Lie algebra $\mathfrak{g}$. The inner product between affine weights is defined as
\begin{equation}
(\widehat{\lambda}|\widehat{\mu})=(\lambda|\mu) + k_{\lambda}l_{\mu}+k_{\mu}l_{\lambda}
\end{equation}
where the first term on the right hand side is the inner product between finite weights. \\ \\
At the level of the root system the construction can be seen as follows. The root system $\Delta$ of any finite dimensional Lie algebra $\mathfrak{g}$ (whose basis is given by the simple positive roots $\alpha_i$) contains a highest root $\theta\in\Delta$, such that
\begin{align}
&\theta+\alpha_i\notin \Delta\,,&&\forall\,i=1,\ldots,r\,.
\end{align} 
We can use $\theta$ to extend the root lattice $\Lambda_{\mathfrak{g}}$. To this end, we introduce the lattice $\Pi^{1,1}$ spanned by $\{\beta_1,\beta_2\}$ whose inner product satisfies
\begin{align}
&(\beta_1|\beta_2)=1\,,&&(\beta_1|\beta_1)=(\beta_2|\beta_2)=0\,,&&(\beta_1|\alpha_i)=(\beta_2|\alpha_i)=0&&\forall i=1,\ldots,r\,.\label{affLab2}
\end{align}
We now define the root lattice $\Lambda_{\widehat{\mathfrak{g}}}$ of the new algebra $\widehat{\mathfrak{g}}$ by
\begin{align}
\Lambda_{\widehat{\mathfrak{g}}}=\sum_{a=0}^r\mathbb{Z}\,\widehat{\alpha}_a\subset\Lambda_{\mathfrak{g}}\oplus\Pi^{1,1}\,,
\end{align}
which is spanned by the new set of simple affine roots
\begin{align}
\{\widehat{\alpha}_0=\beta_1-\theta,\widehat{\alpha}_{i=1,\ldots,r}\}\,.
\end{align}
In complete analogy to the finite simple Lie algebra $\mathfrak{g}$ the affine Weyl group $\widehat{\mathcal{W}}$ is defined to be the group generated by reflections with respect to the affine roots. As there is an infinity of the latter the Weyl group is infinite as well. We will give further details for the specific case $\widehat{\alpha}_1$.

The examples that we will mostly deal with in this work is the affine extension of $\mathfrak{a}_1$, which we shall briefly discuss below.
\subsection{Lie Algebra $\widehat{\mathfrak{a}}_1$}
\label{AffA1}
The affine counterpart to the highest root of $\mathfrak{a}_1$ is
\begin{align}
\theta=\widehat{\alpha}_1=(\alpha_1;0;0)
\end{align}
The null root $\delta$ is defined  as 
\begin{equation}
\delta=\beta_1=(0;0;1)
\end{equation}
The term null root comes from the fact $(\delta|\delta)=0$. Thus, the simple positive roots of $\widehat{\mathfrak{a}}_1$ are
\begin{align}
&\widehat{\alpha}_0=\delta-\widehat{\alpha}_1=(-\alpha_1;0;1)\,,&&\text{and} &&\widehat{\alpha}_1=(\alpha_1;0;0)\,.
\end{align}
The root system of $\widehat{\mathfrak{a}}_1$ contains inifnitely many (imaginary) roots and the explicit expression can be found in \cite{Persson:2010ms}
\begin{align}
\widehat{\Delta}=\{\pm\widehat{\alpha}_1+n\delta\,\big|\, n\in\mathbb{Z}\}\cup\{k\delta \,\big|\, k\in\mathbb{Z}\setminus \{ 0\}\}\,,
\end{align}
such that the positive roots are
\begin{align}
\widehat{\Delta}_+=\big\{\widehat{\alpha}_1+n\delta\,\big|\, n\in\mathbb{N}\cup\{0\}\big\}\cup\big\{-\widehat{\alpha}_1+n\delta\,\big|\, n\in\mathbb{N}\big\}\cup\big\{k\delta \,\big|\, k\in\mathbb{N}\big\}\,.\label{PosRootA1p}
\end{align}
In analogy to the finite Lie algebras one can introduce the fundamental weights. In the case of $\widehat{\mathfrak{a}}_1$ they are given by
\begin{align}
&\widehat{\omega}_0= (0;1;0)\,,&&\text{and} && \widehat{\omega}_1=(1;2;0)\,.
\end{align}
Every affine weight $\widehat{\lambda}$ can be decomposed as
\begin{equation}
\widehat{\lambda}=\lambda_0 \widehat{\omega}_0 + \lambda_1 \widehat{\omega}_1 + l\delta\,,  \quad \lambda_0,\lambda_1,l \in \mathbb{Z}
\end{equation}
where $\lambda_0,\lambda_1$ are the so called Dynkin labels. $\lambda_1$ corresponds to the finite Dynkin label corresponding to the associated finite weight $\lambda$. $\lambda_0$ is related to the level eigenvalue $k$ by
\begin{equation}
\lambda_0=k-\lambda_1
\end{equation}
Alternatively to (\ref{affLab}) we can label the affine weights by their Dynkin labels and by their grade
\begin{equation}
\widehat{\lambda}= [ \lambda_0,\lambda_1 ,l]\,,
\end{equation}
which is the notation we will use in the main part of this work. In the affine case the Weyl vector cannot be defined in terms of the positive roots as their is an infinity of them. The definition as the sum of the fundamental weights is still valid
\begin{equation}
\widehat{\weyl }= \widehat{\omega}_0 + \widehat{\omega}_1 =[1,0,0]+[0,1,0]= [1,1,0]
\end{equation}
The Weyl group $\widehat{\mathcal{W}}(\widehat{\mathfrak{a}}_1)$ is generated by two elements $s_0,s_1$ which correspond to the reflections with respect to $\widehat{\alpha}_0$ and $\widehat{\alpha}_1$. Their action on affine weights is given as follows
\begin{align}
&s_0[\lambda_0,\lambda_1,l]=[-\lambda_0,\lambda_1+2\lambda_0,l-\lambda_0] \nonumber \\
&s_1[\lambda_0,\lambda_1,l]=[\lambda_0+2\lambda_1,-\lambda_1,l]
\label{AffWeylActionA1}
\end{align}
From this we immediately see that the action of $s_0$ changes the grade whereas the action of $s_1$ does not affect it.
\section{Representation Theory of $\mathfrak{sl}(2,\mathbb{C})$ and $\mathfrak{sl}(3,\mathbb{C})$}\label{App:Irreps}
In this section we review representations of $\mathfrak{sl}(2,\mathbb{C})$ and $\mathfrak{sl}(3,\mathbb{C})$ which are relevant for the discussion in section~\ref{Sect:ExamplesNonCompact}. Our notation mainly follows~\cite{FultonHarris} (see also \cite{ConwaySloane}).
\subsection{Irreducible Representations of $\mathfrak{sl}(2,\mathbb{C})$}\label{App:IrrepSl2}
We recall that the Lie algebra $\mathfrak{sl}(2,\mathbb{C})$ is generated by $(H,X,Y)$ which satisfy the commutation relations
\begin{align}
&[H,X]=2X\,,&&[H,Y]=-2Y\,,&&[X,Y]=H\,.
\end{align}
As explained in \cite{FultonHarris}, irreducible representation $\Gamma_n$ of $\mathfrak{sl}(2,\mathbb{C})$ (with $n\in\mathbb{N}$) can be decomposed as
\begin{align}
\Gamma_n=\bigoplus_{m=0}^n V_{n-2m}\,.
\end{align}
Here the one-dimensional eigenspaces $V_{\alpha}$ are eigenspaces of H with weight $\alpha$, \emph{i.e.} 
\begin{align}
&H\cdot v=\alpha\,v\,,&&\forall v\in V_\alpha\,,
\end{align}
while the operators $X$ and $Y$ map from one eigenspace to another
\begin{align}
&X:\,V_\alpha\longrightarrow V_{\alpha+2}\,,&&\text{and} &&Y:\,V_\alpha\longrightarrow V_{\alpha-2}\,,
\end{align}
as well as 
\begin{align}
&X\cdot v=0\,,&&\forall v\in V_n\,,\nonumber\\
&Y\cdot w=0\,,&&\forall w\in V_{-n}\,.
\end{align}
Graphically, the structure of $V$ (and the action of all generators) can be represented as follows
\begin{center}
\begin{tikzpicture}
\node at (-7.3,0) {$0$};
\node[red] at (-5,0) {$V_{-n}$};
\node[red] at (-2.5,0) {$V_{-n+2}$};
\node at (0,0) {$\cdots$};
\node[red] at (2.5,0) {$V_{n-2}$};
\node[red] at (5,0) {$V_{n}$};
\node at (7.5,0) {$0$};
\draw[->, ultra thick] (4.5,0) -- (3.2,0);
\node at (3.9,-0.3) {\footnotesize $Y$};
\draw[->, ultra thick] (1.8,0) -- (0.5,0);
\node at (1.2,-0.3) {\footnotesize $Y$};
\draw[->, ultra thick] (-0.5,0) -- (-1.8,0);
\node at (-1.2,-0.3) {\footnotesize $Y$};
\draw[->, ultra thick] (-3.2,0) -- (-4.5,0);
\node at (-3.7,-0.3) {\footnotesize $Y$};
\draw[->, ultra thick] (-5.5,0) -- (-7,0);
\node at (-6.2,-0.3) {\footnotesize $Y$};
\draw[ultra thick, ->,blue] (-4.9,0.3) to [out=45,in=180] (-3.8,0.7) to [out=0,in=135] (-2.6,0.3);
\node[blue] at (-3.7,1) {\footnotesize $X$};
\draw[ultra thick, ->,blue] (-2.4,0.3) to [out=45,in=180] (-1.2,0.7) to [out=0,in=135] (-0.1,0.3);
\node[blue] at (-1.2,1) {\footnotesize $X$};
\draw[ultra thick, ->,blue] (0.1,0.3) to [out=45,in=180] (1.2,0.7) to [out=0,in=135] (2.4,0.3);
\node[blue] at (1.2,1) {\footnotesize $X$};
\draw[ultra thick, ->,blue] (2.6,0.3) to [out=45,in=180] (3.9,0.7) to [out=0,in=135] (4.9,0.3);
\node[blue] at (3.8,1) {\footnotesize $X$};
\draw[ultra thick, ->,blue] (5.1,0.3) to [out=45,in=180] (6.4,0.7) to [out=0,in=135] (7.4,0.3);
\node[blue] at (6.4,1) {\footnotesize $X$};
\draw[ultra thick,->] (4.9,-0.3) to [out=225,in=90] (4.7,-0.6) to [out=270,in=180] (5,-0.8) to [out=0,in=270] (5.3,-0.6) to [out=90,in=315] (5.1,-0.3);
\node at (5,-1.1) {\footnotesize $H$};
\draw[ultra thick,->,xshift=-2.5cm] (4.9,-0.3) to [out=225,in=90] (4.7,-0.6) to [out=270,in=180] (5,-0.8) to [out=0,in=270] (5.3,-0.6) to [out=90,in=315] (5.1,-0.3);
\node at (2.5,-1.1) {\footnotesize $H$};
\draw[ultra thick,->,xshift=-7.5cm] (4.9,-0.3) to [out=225,in=90] (4.7,-0.6) to [out=270,in=180] (5,-0.8) to [out=0,in=270] (5.3,-0.6) to [out=90,in=315] (5.1,-0.3);
\node at (-2.5,-1.1) {\footnotesize $H$};
\draw[ultra thick,->,xshift=-10cm] (4.9,-0.3) to [out=225,in=90] (4.7,-0.6) to [out=270,in=180] (5,-0.8) to [out=0,in=270] (5.3,-0.6) to [out=90,in=315] (5.1,-0.3);
\node at (-5,-1.1) {\footnotesize $H$};
\end{tikzpicture}
\end{center}
Furthermore, for given $n\in\mathbb{N}$ the irreducible representation $\Gamma_n$ can be written as 
\begin{align}
&\Gamma_n=\text{Sym}^nV\,,&&\text{with} &&V\cong \mathbb{C}^2\,.
\end{align}
Explicitly, apart from the trivial representation $\Gamma_0$, we present the first few irreducible representations by specifying the weights of the underlying subspaces
\begin{align}
&\Gamma_1=V_{-1}\oplus V_{1}&&\hspace{3cm}\parbox{4cm}{\begin{tikzpicture}
\draw[->] (-2,0) -- (2,0);
\node at (0,0) {\footnotesize $|$};
\node at (0,-0.4) {\footnotesize $0$};
\node at (2,-0.3) {\footnotesize $n$};
\node[red] at (-1,0) {$\bullet$};
\node at (-1,0.3) {\footnotesize $-1$};
\node[red] at (1,0) {$\bullet$};
\node at (1,0.3) {\footnotesize $1$};
\end{tikzpicture}}\nonumber\\[10pt]
&\Gamma_2=V_{-2}\oplus V_0\oplus V_{2}&&\hspace{2cm}\parbox{6cm}{\begin{tikzpicture}
\draw[->] (-3,0) -- (3,0);
\node[red] at (0,0) {$\bullet$};
\node at (0,0.3) {\footnotesize $0$};
\node at (3,-0.3) {\footnotesize $n$};
\node[red] at (-2,0) {$\bullet$};
\node at (-2,0.3) {\footnotesize $-2$};
\node[red] at (2,0) {$\bullet$};
\node at (2,0.3) {\footnotesize $2$};
\end{tikzpicture}}\nonumber\\[10pt]
&\Gamma_3=V_{-3}\oplus V_{-1}\oplus V_{1}\oplus V_{3}&&\hspace{1cm}\parbox{8cm}{\begin{tikzpicture}
\draw[->] (-4,0) -- (4,0);
\node at (0,0) {\footnotesize $|$};
\node at (0,-0.4) {\footnotesize $0$};
\node at (4,-0.3) {\footnotesize $n$};
\node[red] at (-3,0) {$\bullet$};
\node at (-3,0.3) {\footnotesize $-3$};
\node[red] at (-1,0) {$\bullet$};
\node at (-1,0.3) {\footnotesize $-1$};
\node[red] at (1,0) {$\bullet$};
\node at (1,0.3) {\footnotesize $1$};
\node[red] at (3,0) {$\bullet$};
\node at (3,0.3) {\footnotesize $3$};
\end{tikzpicture}}\nonumber\\[10pt]
&\Gamma_4=V_{-4}\oplus V_{-2}\oplus V_0\oplus V_{2}\oplus V_4&&\parbox{10cm}{\begin{tikzpicture}
\draw[->] (-5,0) -- (5,0);
\node[red] at (0,0) {$\bullet$};
\node at (0,0.3) {\footnotesize $0$};
\node at (5,-0.3) {\footnotesize $n$};
\node[red] at (-4,0) {$\bullet$};
\node at (-4,0.3) {\footnotesize $-4$};
\node[red] at (-2,0) {$\bullet$};
\node at (-2,0.3) {\footnotesize $-2$};
\node[red] at (2,0) {$\bullet$};
\node at (2,0.3) {\footnotesize $2$};
\node[red] at (4,0) {$\bullet$};
\node at (4,0.3) {\footnotesize $4$};
\end{tikzpicture}}\nonumber
\end{align}
\subsection{Irreducible Representations of $\mathfrak{sl}(3,\mathbb{C})$}\label{App:IrrepsSL3}
Following \cite{FultonHarris}, in order to describe the structure of representations of $\mathfrak{sl}(3,\mathbb{C})$, we first recall the Cartan-Weyl decomposition
\begin{align}
\mathfrak{sl}(3,\mathbb{C})\cong \mathfrak{h}\oplus\left(\bigoplus_{\alpha\in S} \mathfrak{g}_\alpha\right)
\end{align}
where $\mathfrak{h}$ is the Cartan subalgebra, which is defined as
\begin{align}
\mathfrak{h}=\left\{\left(\begin{array}{ccc} c_1 & 0 & 0 \\ 0 & c_2 & 0 \\ 0 & 0 & c_3\end{array}\right)\bigg| c_{1,2,3}\in\mathbb{C}\text{ and }c_1+c_2+c_3=0\right\}\,,
\end{align} 
along with its dual (with $i=1,2,3$)
\begin{align}
&\mathfrak{h}^*=\text{Span}_{\mathbb{C}}(L_1,L_2,L_3)/\{L_1+L_2+L_3=0\}\,,&&\text{with} &&L_i\,\left(\begin{array}{ccc} c_1 & 0 & 0 \\ 0 & c_2 & 0 \\ 0 & 0 & c_3\end{array}\right)=c_i\,.\label{DefSl3Dual}
\end{align}
Furthermore we have
\begin{align}
S=\{L_i-L_j|i,j=1,2,3\text{ and }i\neq j\}\subset \mathfrak{h}^*\,
\end{align}
and the (one-dimensional) root-space $\mathfrak{g}_{L_i-L_j}$ is generated by the $3\times 3$ matrix $E_{ij}$ whose component $(i,j)=1$, while all other entries are zero. 

While each $H\subset \mathfrak{h}$ maps each of the $\mathfrak{g}_\alpha$ into itself, we have for the adjoint action $\text{ad}(X)(Y)=[X,Y]$  (with $X\in\mathfrak{g}_\alpha$ and $Y\in\mathfrak{g}_\beta$)
\begin{align}
\text{ad}(\mathfrak{g}_\alpha):\,\mathfrak{g}_\beta\longrightarrow \mathfrak{g}_{\alpha+\beta}\,.
\end{align}
As in the case of $\mathfrak{sl}(2,\mathbb{C})$, this action can be represented graphically in the form of 'translations' \cite{FultonHarris}. Indeed, while the subspaces $\mathfrak{g}_\alpha$ can be graphically represented on a two-dimensional (hexagonal) lattice, the adjoint action of a given $X\in\mathfrak{g}_\alpha$ acts through translation, \emph{e.g.} for $X\in \mathfrak{g}_{L_1-L_3}$ we have schematically
\begin{center}
\scalebox{0.63}{\parbox{16.3cm}{\begin{tikzpicture}[scale=1.2]
\draw[fill=black] (2,-3.4641) circle (0.15);
\draw[fill=black] (-2,-3.4641) circle (0.15);
\draw[fill=black] (4,0) circle (0.15);
\draw[fill=black] (2,3.4641) circle (0.15);
\draw[fill=black] (-2,3.4641) circle (0.15);
\draw[fill=black] (-4,0) circle (0.15);
\draw (-4,0) -- (4,0);
\draw (-2,3.4641) -- (2,3.4641);
\draw (-2,-3.4641) -- (2,-3.4641);
\draw (-2,-3.4641) -- (-4,0);
\draw (2,-3.4641) -- (-2,3.4641);
\draw (4,0) -- (2,3.4641);
\draw (2,-3.4641) -- (4,0);
\draw (-2,-3.4641) -- (2,3.4641);
\draw (-4,0) -- (-2,3.4641);
\draw[dashed] (0,-4) -- (0,4);
\draw[dashed] (-2,3.4641) -- (4,0);
\draw[dashed] (2,3.4641) -- (-4,0);
\draw[dashed] (-2,-3.4641) -- (4,0);
\draw[dashed] (2,-3.4641) -- (-4,0);
\draw[dashed] (3.4641,2) -- (-3.4641,-2);
\draw[dashed] (3.4641,-2) -- (-3.4641,2);
\draw[red, ultra thick] (0,0) -- (0,2.309);
\draw[red, ultra thick] (0,0) -- (2,-1.155);
\draw[red, ultra thick] (0,0) -- (-2,-1.155);
\draw[fill=black] (0,0) circle (0.03);
\node[red] at (0,2.309) {$\bullet$};
\node[red] at (2,-1.155) {$\bullet$};
\node[red] at (-2,-1.155) {$\bullet$};
\node[red] at (0.3,1.9) {\footnotesize $L_2$};
\node[red] at (2,-0.8) {\footnotesize $L_1$};
\node[red] at (-2,-0.8) {\footnotesize $L_3$};
\node at (0.3,-0.2) {\footnotesize $0$};
\node at (2.9,3.4641) {\footnotesize $L_2-L_3$};
\node at (-2.9,3.4641) {\footnotesize $L_2-L_1$};
\node at (-4.9,0) {\footnotesize $L_3-L_1$};
\node at (4.9,0) {\footnotesize $L_1-L_3$};
\node at (-2.9,-3.4641) {\footnotesize $L_3-L_2$};
\node at (2.9,-3.4641) {\footnotesize $L_1-L_2$};
\draw[ultra thick, ->, blue] (0.3,0.3) to [out=25,in=180] (2.1,0.7) to [out=0,in=155] (3.7,0.3);
\draw[ultra thick, ->, blue,xshift=-4cm] (0.3,0.3) to [out=25,in=180] (2.1,0.7) to [out=0,in=155] (3.7,0.3);
\draw[ultra thick, ->, blue,xshift=-2cm,yshift=3.4641cm] (0.3,0.3) to [out=25,in=180] (2.1,0.7) to [out=0,in=155] (3.7,0.3);
\draw[ultra thick, ->, blue,xshift=-2cm,yshift=-3.4641cm] (0.3,0.3) to [out=25,in=180] (2.1,0.7) to [out=0,in=155] (3.7,0.3);
\draw[ultra thick, ->, blue,xshift=4cm] (0.3,0.3) to [out=25,in=180] (2.1,0.7) to [out=0,in=155] (3.7,0.3);
\draw[ultra thick, ->, blue,xshift=2cm,yshift=3.4641cm] (0.3,0.3) to [out=25,in=180] (2.1,0.7) to [out=0,in=155] (3.7,0.3);
\draw[ultra thick, ->, blue,xshift=2cm,yshift=-3.4641cm] (0.3,0.3) to [out=25,in=180] (2.1,0.7) to [out=0,in=155] (3.7,0.3);
\node at (7.9,0) {\footnotesize $0$};
\node at (5.9,3.4641) {\footnotesize $0$};
\node at (5.9,-3.4641) {\footnotesize $0$};
\end{tikzpicture}}}
\end{center}
Irreducible representations of $\mathfrak{sl}(3,\mathbb{C})$ follow a similar pattern: Indeed, as explained in \cite{FultonHarris}, for any two integers $n,m\in\mathbb{N}$ there exists a finite dimensional irreducible representation $V_{n,m}$ which enjoys a weight decomposition $V_{n,m}=\bigoplus V_\alpha$. The (one-dimensional) subspaces $V_\alpha$ are characterised through their weights and are created from the heighest weight subspace $V_{nL_1-mL_3}$ through application of the generators $E_{2,1}$, $E_{3,1}$ and $E_{3,2}$.

Apart from the trivial representation ($m=n=0$), we have the following weight diagrams for $\Gamma_{1,0}\cong \mathbb{C}^3$ and its dual $\Gamma_{0,1}$
\begin{align}
\Gamma_{1,0}:\hspace{0.3cm}\parbox{5cm}{
\scalebox{0.63}{\parbox{5.8cm}{\begin{tikzpicture}[scale=1.2]
\draw (-2,0) -- (2,0);
\draw (-1,1.732) -- (1,1.732);
\draw (-1,-1.732) -- (1,-1.732);
\draw (-1,1.732) -- (1,-1.732);
\draw (-1,-1.732) -- (1,1.732);
\draw (-2,0) -- (-1,1.732);
\draw (-2,0) -- (-1,-1.732);
\draw (2,0) -- (1,1.732);
\draw (2,0) -- (1,-1.732);
\draw[fill=black] (2,0) circle (0.15);
\draw[fill=black] (-1,-1.732) circle (0.15);
\draw[fill=black] (-1,1.732) circle (0.15);
\node at (2.5,0) {\footnotesize $L_1$};
\node at (-1.3,2.1) {\footnotesize $L_2$};
\node at (-1.3,-2.1) {\footnotesize $L_3$};
\node at (0.3,0.2) {\footnotesize $0$};
\end{tikzpicture}}}
}
&&\text{and} &&
\Gamma_{0,1}:\hspace{0.3cm}\parbox{5cm}{
\scalebox{0.63}{\parbox{5.8cm}{\begin{tikzpicture}[scale=1.2]
\draw (-2,0) -- (2,0);
\draw (-1,1.732) -- (1,1.732);
\draw (-1,-1.732) -- (1,-1.732);
\draw (-1,1.732) -- (1,-1.732);
\draw (-1,-1.732) -- (1,1.732);
\draw (-2,0) -- (-1,1.732);
\draw (-2,0) -- (-1,-1.732);
\draw (2,0) -- (1,1.732);
\draw (2,0) -- (1,-1.732);
\draw[fill=black] (-2,0) circle (0.15);
\draw[fill=black] (1,-1.732) circle (0.15);
\draw[fill=black] (1,1.732) circle (0.15);
\node at (2.5,0) {\footnotesize $L_1$};
\node at (-1.3,2.1) {\footnotesize $L_2$};
\node at (-1.3,-2.1) {\footnotesize $L_3$};
\node at (0.3,0.2) {\footnotesize $0$};
\end{tikzpicture}}}
}
\end{align}
More generally, \emph{e.g.} the weight diagram of a representation $\Gamma_{m,n}$ for generic $(m,n)$ consist of hexagons and triangles that are concentric to the origin. The hexagons have vertices at $(m-i)L_1-(n-i)L_3$ for $i=0,\ldots \text{min}(m,n)-1$ and the triangles have vertices at $(m-n-3j)L_1$ for $j=0,\ldots \lfloor(m-n)/3\rfloor$, \emph{e.g.} for $(m,n)=(2,4)$ we have
\begin{center}
\scalebox{0.63}{\parbox{18.8cm}{\begin{tikzpicture}[scale=1.2]
\draw (-4,3.464) -- (4,3.464);
\draw (4,3.464) -- (6,0);
\draw (-4,3.464) -- (-6,0);
\draw (-6,0) -- (-2,-6.928);
\draw (-2,-6.928) -- (2,-6.928);
\draw (2,-6.928) -- (6,0);
\draw (-5,1.732) -- (5,1.732);
\draw (-6,0) -- (6,0);
\draw (-5,-1.732) -- (5,-1.732);
\draw (-4,-3.464) -- (4,-3.464);
\draw (-3,-5.196) -- (3,-5.196);
\draw (-5,-1.732) -- (-2,3.464);
\draw (-4,-3.464) -- (0,3.464);
\draw (-3,-5.196) -- (2,3.464);
\draw (-2,-6.928) -- (4,3.464);
\draw (0,-6.928) -- (5,1.732);
\draw (-5,1.732) -- (0,-6.928);
\draw (-4,3.464) -- (2,-6.928);
\draw (-2,3.464) -- (3,-5.196);
\draw (0,3.464) -- (4,-3.464);
\draw (2,3.464) -- (5,-1.732);
\draw[dashed] (0,-1.154) -- (0,3.464);
\draw[dashed] (0,-1.154) -- (-4,-3.464);
\draw[dashed] (0,-1.154) -- (4,-3.464);
\draw[fill=black] (-4,3.464) circle (0.15);
\draw[fill=black] (-2,3.464) circle (0.15);
\draw[fill=black] (0,3.464) circle (0.15);
\draw[fill=black] (2,3.464) circle (0.15);
\draw[fill=black] (4,3.464) circle (0.15);
\draw[fill=black] (5,1.732) circle (0.15);
\draw[fill=black] (3,1.732) circle (0.15);
\draw (3,1.732) circle (0.25);
\draw[fill=black] (1,1.732) circle (0.15);
\draw (1,1.732) circle (0.25);
\draw[fill=black] (-1,1.732) circle (0.15);
\draw (-1,1.732) circle (0.25);
\draw[fill=black] (-3,1.732) circle (0.15);
\draw (-3,1.732) circle (0.25);
\draw[fill=black] (-5,1.732) circle (0.15);
\draw[fill=black] (-6,0) circle (0.15);
\draw[fill=black] (-4,0) circle (0.15);
\draw (-4,0) circle (0.25);
\draw[fill=black] (-2,0) circle (0.15);
\draw (-2,0) circle (0.25);
\draw[fill=black] (0,0) circle (0.15);
\draw (0,0) circle (0.25);
\draw[fill=black] (2,0) circle (0.15);
\draw (2,0) circle (0.25);
\draw[fill=black] (4,0) circle (0.15);
\draw (4,0) circle (0.25);
\draw[fill=black] (6,0) circle (0.15);
\draw[fill=black] (-5,-1.732) circle (0.15);
\draw[fill=black] (-3,-1.732) circle (0.15);
\draw (-3,-1.732) circle (0.25);
\draw[fill=black] (-1,-1.732) circle (0.15);
\draw (-1,-1.732) circle (0.25);
\draw[fill=black] (1,-1.732) circle (0.15);
\draw (1,-1.732) circle (0.25);
\draw[fill=black] (3,-1.732) circle (0.15);
\draw (3,-1.732) circle (0.25);
\draw[fill=black] (5,-1.732) circle (0.15);
\draw[fill=black] (-4,-3.464) circle (0.15);
\draw[fill=black] (-2,-3.464) circle (0.15);
\draw (-2,-3.464) circle (0.25);
\draw[fill=black] (0,-3.464) circle (0.15);
\draw (0,-3.464) circle (0.25);
\draw[fill=black] (2,-3.464) circle (0.15);
\draw (2,-3.464) circle (0.25);
\draw[fill=black] (4,-3.464) circle (0.15);
\draw[fill=black] (-3,-5.196) circle (0.15);
\draw[fill=black] (-1,-5.196) circle (0.15);
\draw (-1,-5.196) circle (0.25);
\draw[fill=black] (1,-5.196) circle (0.15);
\draw (1,-5.196) circle (0.25);
\draw[fill=black] (3,-5.196) circle (0.15);
\draw[fill=black] (-2,-6.928) circle (0.15);
\draw[fill=black] (0,-6.928) circle (0.15);
\draw[fill=black] (2,-6.928) circle (0.15);
\node at (0,3.9) {\footnotesize $4L_2$};
\node at (0,-7.5) {\footnotesize $-5L_2$};
\node at (7.1,0) {\footnotesize $2L_1-4L_3$};
\node at (-7.1,0) {\footnotesize $2L_3-4L_1$};
\node at (2.9,-7.5) {\footnotesize $2L_1-4L_2$};
\node at (-2.9,-7.5) {\footnotesize $2L_3-4L_1$};
\node at (4.5,-3.7) {\footnotesize $4L_1$};
\node at (-4.5,-3.7) {\footnotesize $4L_3$};
\node at (4.8,3.9) {\footnotesize $2L_2-4L_3$};
\node at (-4.8,3.9) {\footnotesize $2L_2-4L_1$};
\node at (0.55,0.35) {\footnotesize $L_2$};
\node at (1,-1.15) {\footnotesize $L_1$};
\node at (-1,-1.15) {\footnotesize $L_3$};
\node at (0,-1.5) {\footnotesize $0$};
\draw[style={line width=2.6pt}] (-4,3.464) -- (4,3.464) -- (6,0) -- (2,-6.928) -- (-2,-6.928) -- (-6,0) -- (-4,3.464);
\draw[style={line width=2.6pt}] (-3,1.732) -- (3,1.732) -- (4,0) -- (1,-5.196) -- (-1,-5.196) -- (-4,0) -- (-3,1.732);
\draw[style={line width=2.6pt}] (-2,0) -- (2,0) -- (0,-3.464) -- (-2,0);
\end{tikzpicture}}}
\end{center}
The multiplicity (\emph{i.e.} the dimension of the corresponding subspace of $\Gamma_{m,n}$) is $(i+1)$ for the $i$th hexagon and $\text{min}(m,n)$ for the triangles in the weight diagram. In the above picture we have indicated the double multiplicity of certain weights by \parbox{0.5cm}{\begin{tikzpicture}[scale=0.7]\draw[fill=black] (0,0) circle (0.15);\draw (0,0) circle (0.25);\end{tikzpicture}}.

In the case of $m=n$ (which is the most important for us) the diagram consists of concentric regular hexagons (while for $m=0$ or $n=0$ it consists of equilateral triangles), \emph{e.g.}
{\allowdisplaybreaks
\begin{align}
&\Gamma_{1,1}:\hspace{1cm}\scalebox{0.63}{\parbox{8.6cm}{\begin{tikzpicture}[scale=1.2]
\draw[style={line width=2.6pt}] (-2,0) -- (-1,1.732) -- (1,1.732) -- (2,0) -- (1,-1.732) -- (-1,-1.732) -- (-2,0);
\draw[fill=black] (1,-1.732) circle (0.15);
\draw[fill=black] (-1,-1.732) circle (0.15);
\draw[fill=black] (2,0) circle (0.15);
\draw[fill=black] (1,1.732) circle (0.15);
\draw[fill=black] (-1,1.732) circle (0.15);
\draw[fill=black] (-2,0) circle (0.15);
\draw (-2,0) -- (2,0);
\draw (-1,1.732) -- (1,-1.732);
\draw (1,1.732) -- (-1,-1.732);
\draw[fill=black] (0,0) circle (0.15);
\node at (3,0) {\footnotesize $L_1-L_3$};
\node at (-3,0) {\footnotesize $L_3-L_1$};
\node at (1.7,2.1) {\footnotesize $L_2-L_3$};
\node at (-1.7,2.1) {\footnotesize $L_2-L_1$};
\node at (1.7,-2.1) {\footnotesize $L_1-L_2$};
\node at (-1.7,-2.1) {\footnotesize $L_3-L_2$};
\draw[dashed] (0,0) -- (0,1.732);
\draw[dashed] (0,0) -- (-1.5,-0.866);
\draw[dashed] (0,0) -- (1.5,-0.866);
\node at (-0.866,-0.5) {\footnotesize $+$};
\node at (0.866,-0.5) {\footnotesize $+$};
\node at (0,1) {\footnotesize $+$};
\node at (0.25,1.2) {\tiny $L_2$};
\node at (1.1,-0.3) {\tiny $L_1$};
\node at (-1.1,-0.3) {\tiny $L_3$};
\node at (0,-0.4) {\footnotesize $0$};
\end{tikzpicture}}}\nonumber\\[40pt]
&
\Gamma_{2,2}:\hspace{1cm}\scalebox{0.63}{\parbox{14.2cm}{\begin{tikzpicture}[scale=1.2]
\draw[fill=black] (2,3.464) circle (0.15);
\draw[fill=black] (3,1.732) circle (0.15);
\draw[fill=black] (4,0) circle (0.15);
\draw[fill=black] (3,-1.732) circle (0.15);
\draw[fill=black] (2,-3.464) circle (0.15);
\draw[fill=black] (0,-3.464) circle (0.15);
\draw[fill=black] (0,3.464) circle (0.15);
\draw[fill=black] (-2,3.464) circle (0.15);
\draw[fill=black] (-3,1.732) circle (0.15);
\draw[fill=black] (-4,0) circle (0.15);
\draw[fill=black] (-3,-1.732) circle (0.15);
\draw[fill=black] (-2,-3.464) circle (0.15);
\draw[fill=black] (1,-1.732) circle (0.15);
\draw (1,-1.732) circle (0.25);
\draw[fill=black] (-1,-1.732) circle (0.15);
\draw (-1,-1.732) circle (0.25);
\draw[fill=black] (2,0) circle (0.15);
\draw (2,0) circle (0.25);
\draw[fill=black] (1,1.732) circle (0.15);
\draw (1,1.732) circle (0.25);
\draw[fill=black] (-1,1.732) circle (0.15);
\draw (-1,1.732) circle (0.25);
\draw[fill=black] (-2,0) circle (0.15);
\draw (-2,0) circle (0.25);
\draw[fill=black] (0,0) circle (0.15);
\draw (0,0) circle (0.25);
\draw (-4,0) -- (4,0);
\draw (-3,1.732) -- (3,1.732);
\draw (-3,-1.732) -- (3,-1.732);
\draw (-2,3.4641) -- (2,3.4641);
\draw (-2,-3.4641) -- (2,-3.4641);
\draw (-2,-3.4641) -- (-4,0);
\draw (0,-3.4641) --  (-3,1.732);
\draw (2,-3.4641) -- (-2,3.4641);
\draw (3,-1.732) -- (0,3.4641);
\draw (4,0) -- (2,3.4641);
\draw (2,-3.4641) -- (4,0);
\draw (0,-3.4641) --  (3,1.732);
\draw (-2,-3.4641) -- (2,3.4641);
\draw (-3,-1.732) -- (0,3.4641);
\draw (-4,0) -- (-2,3.4641);
\draw[style={line width=2.6pt}] (-2,0) -- (-1,1.732) -- (1,1.732) -- (2,0) -- (1,-1.732) -- (-1,-1.732) -- (-2,0);
\draw[style={line width=2.6pt}] (-4,0) -- (-2,3.4641) -- (2,3.4641) -- (4,0) -- (2,-3.4641) -- (-2,-3.4641) -- (-4,0);
\node at (0,-0.5) {\footnotesize $0$};
\node at (5.2,0) {\footnotesize $2L_1-2L_3$};
\node at (-5.2,0) {\footnotesize $2L_3-2L_1$};
\node at (2.5,3.85) {\footnotesize $2L_2-2L_3$};
\node at (-2.5,3.85) {\footnotesize $2L_2-2L_1$};
\node at (2.5,-3.9) {\footnotesize $2L_1-2L_2$};
\node at (-2.5,-3.9) {\footnotesize $2L_3-2L_2$};
\node at (0,3.85) {\footnotesize $3L_2$};
\node at (0,-3.9) {\footnotesize $-3L_2$};
\node at (3.55,2.05) {\footnotesize $-3L_3$};
\node at (-3.5,-2.0) {\footnotesize $3L_3$};
\node at (3.5,-2.0) {\footnotesize $3L_1$};
\node at (-3.55,2.05) {\footnotesize $-3L_1$};
\draw[dashed] (0,0) -- (0,3.4641);
\draw[dashed] (0,0) -- (-3,-1.732);
\draw[dashed] (0,0) -- (3,-1.732);
\node at (-0.866,-0.5) {\footnotesize $+$};
\node at (0.866,-0.5) {\footnotesize $+$};
\node at (0,1) {\footnotesize $+$};
\node at (0.25,1.2) {\tiny $L_2$};
\node at (1.1,-0.3) {\tiny $L_1$};
\node at (-1.1,-0.3) {\tiny $L_3$};
\end{tikzpicture}}}\nonumber
\end{align}}

\subsection{Integrable Representations of $\mathfrak{\widehat{sl}}(2,\mathbb{C})$}
\label{affrep}
In order to describe the partition functions of compact M-brane configurations we also need (certain) irreducible representations of $\widehat{\mathfrak{sl}}(2,\mathbb{C})$ (\emph{i.e.} the affine extension of $\mathfrak{sl}(2,\mathbb{C})$). In this appendix we give a brief review of the specific representations required and we refer the reader to \cite{Fuchs} for a more rigorous and complete discussion of the representation theory of $\widehat{\mathfrak{sl}}(2,\mathbb{C})$.

The idea of constructing irreducible representation of affine algebras is similar to their finite counterparts: we start with a highest weight state $\widehat{\lambda}$ and repeatedly subtract the positive roots, which act as ladder operators. In the notation introduced in appendix~\ref{AffA1}, the latter can be written in the form
\begin{align}
&\widehat{\alpha}_0=[2,-2,1]\,,&&\text{and} && \widehat{\alpha}_1=[-2,2,0]\,.
\end{align}
However, we have to take care that the action of a single one of the two roots (\emph{i.e.} either $\alpha_0$ or $\alpha_1$) creates a (finite dimensional) irreducible representation of $\mathfrak{sl}(2,\mathbb{C})$ of the type explained in appendix~\ref{App:IrrepSl2} and in particular truncates after a finite number of steps. Starting with the highest weight state\footnote{Notice that the grade of the highest weight state has been chosen to be zero for convenience.} 
\begin{align}
&[n,n,0] && \text{for} && n\in\mathbb{N}\,,
\end{align}
(which is the case relevant for the discussion of partition functions of compact M-brane configurations in section~\ref{Sect:CompactBraneConfigs}) we obtain the following weights at grade $0$
\begin{align}
[n,n,0]\,\stackrel{-\widehat{\alpha}_1}{\longrightarrow} \,[n+2,n-2,0]\,\stackrel{-\widehat{\alpha}_1}{\longrightarrow} \,[n+4,n-4,0]\,\stackrel{-\widehat{\alpha}_1}{\longrightarrow} \,\ldots \,\stackrel{-\widehat{\alpha}_1}{\longrightarrow} \,[3n,-n,0]\,\stackrel{-\widehat{\alpha}_1}{\longrightarrow} \,0
\end{align}
Notice that the weights $\left\{[n+2r,n-2r,0]|r=0,\ldots, n\right\}$ indeed form the irreducible representation $\Gamma_n$ of $\mathfrak{sl}(2,\mathbb{C})$. Similarly, acting with the root $\widehat{\alpha}_0$ yields
\begin{align}
[n,n,0]\,\stackrel{-\widehat{\alpha}_0}{\longrightarrow} \,[n-2,n+2,-1]\,\stackrel{-\widehat{\alpha}_0}{\longrightarrow} \,[n-4,n+4,-2]\,\stackrel{-\widehat{\alpha}_0}{\longrightarrow} \,\ldots \,\stackrel{-\widehat{\alpha}_0}{\longrightarrow} \,[-n,3n,-n]\,\stackrel{-\widehat{\alpha}_0}{\longrightarrow} \,0\,,
\end{align} 
which equally forms the irreducible representation $\Gamma_n$. Acting with combinations of both roots, generates all states of the $[n,n]$ highest weight weight representation. In contrast to the irreducible representations of $\mathfrak{sl}(2,\mathbb{C})$, the highest weight representations $[n,n]$ is infinite dimensional.

Specifically, for $n=1$ we obtain the weights shown in table~\ref{AffRep110}.
\begin{table}[h!tbp]
\centering
\begin{tabular}{ccccccccccccl}
\multicolumn{1}{l}{} & \multicolumn{1}{l}{} & \multicolumn{1}{l}{}     & \multicolumn{1}{l}{} & \multicolumn{1}{l}{}     & \multicolumn{1}{l}{} & \multicolumn{1}{l}{}     & \multicolumn{1}{l}{} & \multicolumn{1}{l}{}     & \multicolumn{1}{l}{} & \multicolumn{1}{l}{}     & \multicolumn{1}{l}{} & \multicolumn{1}{l}{}                 \\ 
                     &                      &                          &                      &                          & \textcolor{red}{{[}1,1,0{]}}          & $\xrightarrow{-\widehat{\alpha}_1}$ & \textcolor{red}{{[}3,-1,0{]}}         &                          &                      &                          &                      & \multicolumn{1}{c}{}                    \\
                     &                      &                          &                      & $\myarrow[25]$           &                      & $\myarrow[25]$           &                      &                          &                      &                          &                      & \multicolumn{1}{l}{}  \\
                     &                      &                          & \textcolor{red}{{[}-1,3,-1{]}}         & $\xrightarrow{-\widehat{\alpha}_1}$ & {[}1,1,-1{]}          & $\xrightarrow{-\widehat{\alpha}_1}$ & {[}3,-1,-1{]}         & $\xrightarrow{-\widehat{\alpha}_1}$ & \textcolor{red}{{[}5,-3,-1{]}}         &                          &                      & \multicolumn{1}{c}{}                     \\
                     &                      &                          &                      & $\myarrow[25]$           &                      & $\myarrow[25]$           &                      & $\myarrow[25]$           &                      &                          &                      & \multicolumn{1}{l}{}  \\
                     &                      &                          & {[}-1,3,-2{]}         & $\xrightarrow{-\widehat{\alpha}_1}$ & {[}1,1,-2{]}          & $\xrightarrow{-\widehat{\alpha}_1}$ & {[}3,-1,-2{]}         & $\xrightarrow{-\widehat{\alpha}_1}$ & {[}5,-3,-2{]}         &                          &                      & \multicolumn{1}{c}{}                     \\
                     &                      & $\myarrow[25]$           &                      & $\myarrow[25]$           &                      & $\myarrow[25]$           &                      & $\myarrow[25]$           &                      &                          &                      & \multicolumn{1}{l}{}  \\
                     & \textcolor{red}{{[}-3,5,-3{]}}         & $\xrightarrow{-\widehat{\alpha}_1}$ & {[}-1,3,-3{]}         & $\xrightarrow{-\widehat{\alpha}_1}$ & {[}1,1,-3{]}          & $\xrightarrow{-\widehat{\alpha}_1}$ & {[}3,-1,-3{]}         & $\xrightarrow{-\widehat{\alpha}_1}$ & {[}5,-3,-3{]}         & $\xrightarrow{-\widehat{\alpha}_1}$ & \textcolor{red}{{[}7,-5,-3{]}}         & \multicolumn{1}{c}{}                     \\
                     &                      & $\myarrow[25]$           &                      & $\myarrow[25]$           &                      & $\myarrow[25]$           &                      & $\myarrow[25]$           &                      & $\myarrow[25]$           &                      & \multicolumn{1}{l}{}  \\
                     & {[}-3,5,-4{]}         & $\xrightarrow{-\widehat{\alpha}_1}$ & {[}-1,3,-4{]}         & $\xrightarrow{-\widehat{\alpha}_1}$ & {[}1,1,-4{]}          & $\xrightarrow{-\widehat{\alpha}_1}$ & {[}3,-1,-4{]}         & $\xrightarrow{-\widehat{\alpha}_1}$ & {[}5,-3,-4{]}         & $\xrightarrow{-\widehat{\alpha}_1}$ & {[}7,-5,-4{]}         & \multicolumn{1}{c}{}                     \\
\multicolumn{1}{l}{} & \multicolumn{1}{l}{} & \multicolumn{1}{l}{}     & \multicolumn{1}{l}{} & \multicolumn{1}{l}{}     & \multicolumn{1}{l}{} & $\vdots$                     & \multicolumn{1}{l}{} & \multicolumn{1}{l}{}     & \multicolumn{1}{l}{} & \multicolumn{1}{l}{}     & \multicolumn{1}{l}{} &                       
\end{tabular}
\caption{Weights of the affine highest state representation $[1,1]$}
\label{AffRep110}
\end{table}
Here, the Weyl orbit of the weight $[1,1,0]$ is coloured in red. For the highest weight state $[2,2,0]$ we find table~\ref{Tab:AffRep220}.
\begin{table}[h!tbp]
\begin{center}
\scalebox{0.63}{\parbox{30cm}{
\begin{tabular}{ccccccccccccccccccc}
&&&&&&&{\footnotesize $[2,2,0]$} & $\xrightarrow{-\widehat{\alpha}_1}$   & {\footnotesize $[4,0,0]$} & $\xrightarrow{-\widehat{\alpha}_1}$  & {\footnotesize $[6,-2,0]$} & &&&&&&\\
&&&&&& $\myarrow[25]$ && $\myarrow[25]$  && $\myarrow[25]$ && & &&\\
&& & &  &{\footnotesize$[0,4,-1]$} & $\xrightarrow{-\widehat{\alpha}_1}$ & {\footnotesize$[2,2,-1]$} & $\xrightarrow{-\widehat{\alpha}_1}$  & {\footnotesize$[4,0,-1]$} &$\xrightarrow{-\widehat{\alpha}_1}$ & {\footnotesize$[6,-2,-1]$} & $\xrightarrow{-\widehat{\alpha}_1}$ & {\footnotesize$[8,-4,-1]$}&  &  &\\  
&&&&$\myarrow[25]$ && $\myarrow[25]$ && $\myarrow[25]$ && $\myarrow[25]$ && $\myarrow[25]$ && && &\\
&&  & {\footnotesize$[-2,6,-2]$} & $\xrightarrow{-\widehat{\alpha}_1}$ & {\footnotesize$[0,4,-2]$} & $\xrightarrow{-\widehat{\alpha}_1}$ & {\footnotesize$[2,2,-2]$} & $\xrightarrow{-\widehat{\alpha}_1}$ & {\footnotesize$[4,0,-2]$} & $\xrightarrow{-\widehat{\alpha}_1}$ & {\footnotesize$[6,-2,-2]$} & $\xrightarrow{-\widehat{\alpha}_1}$ & {\footnotesize$[8,-4,-2]$} & $\xrightarrow{-\widehat{\alpha}_1}$ & {\footnotesize$[10,-6,-2]$} &   \\
&&&&$\myarrow[25]$ && $\myarrow[25]$ && $\myarrow[25]$ && $\myarrow[25]$ && $\myarrow[25]$ && && &\\
&&  & {\footnotesize$[-2,6,-3]$} & $\xrightarrow{-\widehat{\alpha}_1}$ & {\footnotesize$[0,4,-3]$} & $\xrightarrow{-\widehat{\alpha}_1}$ & {\footnotesize$[2,2,-3]$} & $\xrightarrow{-\widehat{\alpha}_1}$ & {\footnotesize$[4,0,-3]$} & $\xrightarrow{-\widehat{\alpha}_1}$ & {\footnotesize$[6,-2,-3]$} & $\xrightarrow{-\widehat{\alpha}_1}$ & {\footnotesize$[8,-4,-3]$} & $\xrightarrow{-\widehat{\alpha}_1}$ & {\footnotesize$[10,-6,-3]$} &   \\
&&$\myarrow[25]$&&$\myarrow[25]$ && $\myarrow[25]$ && $\myarrow[25]$ && $\myarrow[25]$ && $\myarrow[25]$ && && &\\
&{\footnotesize$[-4,8,-4]$} & $\xrightarrow{-\widehat{\alpha}_1}$  & {\footnotesize$[-2,6,-4]$} & $\xrightarrow{-\widehat{\alpha}_1}$ & {\footnotesize$[0,4,-4]$} & $\xrightarrow{-\widehat{\alpha}_1}$ & {\footnotesize$[2,2,-4]$} & $\xrightarrow{-\widehat{\alpha}_1}$ & {\footnotesize$[4,0,-4]$} & $\xrightarrow{-\widehat{\alpha}_1}$ & {\footnotesize$[6,-2,-4]$} & $\xrightarrow{-\widehat{\alpha}_1}$ & {\footnotesize$[8,-4,-4]$} & $\xrightarrow{-\widehat{\alpha}_1}$ & {\footnotesize$[10,-6,-4]$}  & $\xrightarrow{-\widehat{\alpha}_1}$ & {\footnotesize$[12,-8,-4]$}   \\
%
%
%
& &  &  & & & & & & $\vdots$ & & & & & & & \\
%
%
\end{tabular}
}}
\end{center}
\caption{Representation with highest state $[2,2,0]$}\label{Tab:AffRep220}
\end{table}
Finally, repeating the analysis for the highest weight state $[4,4,0]$ we find \ref{Tab:AffRep440}
\begin{table}[h!tbp]
\begin{center}
\scalebox{0.63}{\parbox{30cm}{
\begin{tabular}{ccccccccccccccccccc}
&&&&{\footnotesize $[4,4,0]$} & $\xrightarrow{-\widehat{\alpha}_1}$ & {\footnotesize $[6,2,0]$} & $\xrightarrow{-\widehat{\alpha}_1}$  & {\footnotesize $[8,0,0]$} & $\xrightarrow{-\widehat{\alpha}_1}$ & {\footnotesize $[10,-2,0]$} & $\xrightarrow{-\widehat{\alpha}_1}$ & {\footnotesize $[12,-4,0]$} &&&&&&\\
&&&$\myarrow[25]$&& $\myarrow[25]$ && $\myarrow[25]$  && $\myarrow[25]$ && $\myarrow[25]$ && &&\\
&& {\footnotesize$[2,6,-1]$} & $\xrightarrow{-\widehat{\alpha}_1}$ &{\footnotesize$[4,4,-1]$} & $\xrightarrow{-\widehat{\alpha}_1}$ & {\footnotesize$[6,2,-1]$} & $\xrightarrow{-\widehat{\alpha}_1}$  & {\footnotesize$[8,0,-1]$} &$\xrightarrow{-\widehat{\alpha}_1}$ & {\footnotesize$[10,-2,-1]$} & $\xrightarrow{-\widehat{\alpha}_1}$ & {\footnotesize$[12,-4,-1]$}& $\xrightarrow{-\widehat{\alpha}_1}$ & {\footnotesize$[14,-6,-1]$} &&\\  
&$\myarrow[25]$&&$\myarrow[25]$ && $\myarrow[25]$ && $\myarrow[25]$ && $\myarrow[25]$ && $\myarrow[25]$ && $\myarrow[25]$ && &&\\
{\footnotesize$[0,8,-2]$} & $\xrightarrow{-\widehat{\alpha}_1}$ & {\footnotesize$[2,6,-2]$} & $\xrightarrow{-\widehat{\alpha}_1}$ & {\footnotesize$[4,4,-2]$} & $\xrightarrow{-\widehat{\alpha}_1}$ & {\footnotesize$[6,2,-2]$} & $\xrightarrow{-\widehat{\alpha}_1}$ & {\footnotesize$[8,0,-2]$} & $\xrightarrow{-\widehat{\alpha}_1}$ & {\footnotesize$[10,-2,-2]$} & $\xrightarrow{-\widehat{\alpha}_1}$ & {\footnotesize$[12,-4,-2]$} & $\xrightarrow{-\widehat{\alpha}_1}$ & {\footnotesize$[14,-6,-2]$} & $\xrightarrow{-\widehat{\alpha}_1}$ & {\footnotesize$[16,-8,-2]$} \\
%
%
%
& &  &  & & & & & $\vdots$ & & & & & & & & \\
%
%
\end{tabular}
}}
\caption{Representation with highest state $[4,4,0]$}\label{Tab:AffRep440}
\end{center}
\end{table}
\section{Recursive Relation for the Configuration $(N,M)=(2,1)$}\label{App:RecursionN2}
In this appendix we provide more details on the recursive relation allowing to determine the coefficients $c_{k}^{(n)}(\tau,\epsilon)$ (introduced in (\ref{Coefsckn})) from $c_{k-1}^{(n)}(\tau,\epsilon)$ through the action of an operator $R_+$ as in (\ref{OperationRp}). We also supply as an example the explicit coefficients for $k=1,\ldots,5$ for generic $n$.

We only discuss the action of $R_+$ on $c^{(n)}_{k}$ for $k\leq \lceil\tfrac{n^2+1}{2}\rceil$, since all other cases are determined through (\ref{MirroringN2}). As explained in section~\ref{Sect:N21ng1}, the coefficients $c_{k}^{(n)}(\tau,\epsilon)$ are essentially determined through a set of partitions of integers $\mu(k,n)=(\mu_1(k,n),\ldots,\mu_{\ell}(k,n))$ of length $0\leq \ell\leq k$ with
\begin{align}
&\mu_a(k,n)\geq \mu_{a+1}(k,n)\,,&&\text{and} &&\mu_1(k,n)+2\sum_{a=2}^\ell \mu_a=2k\,,
\end{align}
for which $c(\mu(k,n))\neq 0$. These partitions $\mu(n,k)$ can be obtained iteratively in $k$: indeed to obtain the former we begin with $\mu(k-1,n)$ (for which $c(\mu(k-1,n))\neq 0$) and increase the $\mu_a(k-1,n)$ by either 1 or 2 in one of the following fashions
\begin{itemize}
\item increase $\mu_1(k-1,n)$ by two
\begin{align}
\mu(n,k)=(\mu_1(k-1,n)+2\,,\mu_2(k-1,n)\,,\ldots\,,\mu_\ell(k-1,n))\,.\label{IncreaseTwo}
\end{align}
\item increase one of the $\mu_a(k-1,n)$ (for $a>1$) by $1$
\begin{align}
\mu(n,k)=(\mu_1(k-1,n)\,,\mu_2(k-1,n)\,,\ldots\,,\mu_a(k-1,n)+1\,,\ldots\,,\mu_\ell(k-1,n))\,.\label{IncreaseOnea}
\end{align}
\item add $1$ at the end of $\mu(k-1,n)$
\begin{align}
&\mu(n,k)=(\mu_1(k-1,n)\,,\mu_2(k-1,n)\,,\ldots\,,\mu_\ell(k-1,n)\,,1)\,,\label{IncreaseOneb}
\end{align}
\end{itemize}
For each of the resulting $\mu(k,n)=(\mu_1(k,n)\,,\ldots\,,\mu_\ell(k,n))$, the coefficients $c(\mu(k,n))$ are computed as follows
\begin{align}
c(\mu(k,n))=\left\{\begin{array}{lcl} 
2 & \text{if} & \begin{array}{l}\mu_a(k,n)-\mu_{a+1}(k,n)\leq 2\hspace{0.2cm}\forall a=1,\ldots, \ell-1 \text{ and } \\ \mu_\ell(k,n)\leq 2 \text{ and }\\
\left\{\begin{array}{l} \mu_a(k,n)-\mu_{a+1}(k,n) =1 \text{ for at least one }a\in \{1,\ldots,\ell-1\}\text{ or }\\
\mu_{\ell}(n,k)=1\end{array}\right. \end{array}\\[40pt]
1 & \text{if} & \begin{array}{l}\mu_a(k,n)-\mu_{a+1}(k,n)= 0\text{ or }2\hspace{0.2cm}\forall a=1,\ldots, \ell-1\text{ and } \\ \mu_\ell(k,n)=2 \end{array}\\[20pt]
0 & & \text{else}
\end{array}\right.
\end{align}
To illustrate this procedure we can compute explicitly the first few steps of this iteration:
\begin{itemize}
\item $k=0$: For $k=0$ the length of the partition is restricted by $0\leq \ell\leq 0$, thus the only partition which may contribute is $\mu(0,n)=\emptyset$ for which $c(\emptyset)=1$, thus
\begin{align}
c_{0}^{(n)}(\tau,\epsilon)=1\,.
\end{align} 
\item $k=1$: Starting from $\mu(0,n)=\emptyset$ following (\ref{IncreaseTwo}) we have the only partition $\mu(1,n)=(2)$, for which $c((2))=1$. Furthermore, in order to satisfy the condition (\ref{ConditionPadding}) we need to choose $r_1=2$, such that
\begin{align}
c_{1}^{(n)}(\tau,\epsilon)=\frac{\theta(n)^2}{\theta(1)^2}\,.
\end{align}
\item $k=2$: Starting from $\mu(1,n)=(2)$, applying (\ref{IncreaseTwo}) and (\ref{IncreaseOneb}) we find two new partitions for $k=1$
\begin{align}
\mu(2,n)=\left\{\begin{array}{lcl} (4) & &  \ydiagram{4} \\[4pt] (2,1) & & \parbox{0.5cm}{\ydiagram{2,1}} \end{array}\right.
\end{align}
However, for the first possibility we have $c((4))=0$ since $\mu_1(2,n)=4>2$, while for the second possibility we have $c((2,1))=2$. Finally, in order to compute $c_2^{(n)}(\tau,\epsilon)$ we need to find non-negative even integers $r_{1,2}$ ($\leq 4$) that satisfy (\ref{ConditionPadding}), \emph{i.e.}
\begin{align}
&2n^2+2(n^2+1)-r_1-4r_2=4n^2-8&&\Longrightarrow&&r_1=r_2=2\,.
\end{align}
Therefore we have
\begin{align}
c_{2}^{(n)}(\tau,\epsilon)=2\,\frac{\theta(n-1)\theta(n)^2\theta(n+1)}{\theta(1)^2\theta(2)^2}\,.
\end{align}
\item $k=3$: Starting from the partition $\mu(2,n)=(2,1)$ we find with (\ref{IncreaseTwo}), (\ref{IncreaseOnea}) and (\ref{IncreaseOneb}) three new partitions
\begin{align}
\mu(3,n)=\left\{\begin{array}{lclccl} (4,1) & & \parbox{0.9cm}{{\ydiagram{4,1}}}  & & \text{with} & c((4,1))=0\,, \\[12pt] (2,2) & & \parbox{0.5cm}{{\ydiagram{2,2}} } & & \text{with} & c((2,2))=1\,, \\[12pt] (2,1,1) & & \parbox{0.5cm}{{\ydiagram{2,1,1}} } & & \text{with} & c((2,1,1))=2\,, \end{array} \right.
\end{align}
Here the coefficient $c((4,1))=0$ since $\mu_1-\mu_2=4-1>2$. Finally, in order to calculate the coefficients $c_r^{(n)}(\epsilon,\tau)$, we still need to supplement each of the two remaining partitions by suitable non-negative even integers $(r_1,r_2,r_3)$ ($\leq 4$) which need to satisfy (\ref{ConditionPadding})
\begin{align}
&(2,2):&&2n^2+4(n^2+1)-r_1-4r_2-9r_3=6(n^2-3)&&\Longrightarrow &&(r_1,r_2,r_3)=(4,2)\,,\nonumber\\ 
&(2,1,1):&&2n^2+2(n^2+1)+2(n^2+4)-r_1-4r_2-9r_3=6(n^2-3)&&\Longrightarrow &&(r_1,r_2,r_3)=(2,2,2)\,.\nonumber
\end{align} 
Therefore, we find 
\begin{align}
c_{3}^{(n)}(\tau,\epsilon)=\frac{\theta(n-1)^2\theta(n)^2\theta(n+1)^2}{\theta(1)^4\theta(3)^2}+2\frac{\theta(n-2)\theta(n-1)\theta(n)^2\theta(n+1)\theta(n+2)}{\theta(1)^2\theta(2)^2\theta(3)^2}\,.
\end{align}
\end{itemize}
We can tabulate the partitions in the following manner
\begin{center}
\begin{tabular}{|c|c|c|c|c|}\hline
&&&&\\[-10pt]
$k$ & $\mu(k,n)$ & Young diagram of $\mu(k,n)$ & $c(\mu(k,n))$ & $r_i$ \\[2pt]\hline
&&&&\\[-10pt]
0 & $\emptyset$ & --- & $1$ & $(0)$\\[2pt]\hline
&&&&\\[-10pt]
1 & $(1)$ & $\ydiagram{1}$ & $1$ & $(2)$\\[2pt]\hline
&&&&\\[-8pt]
2 & $(2,1)$ & $\parbox{0.5cm}{\ydiagram{2,1}}$ & $2$ & $(2,2)$\\[6pt]\hline
&&&&\\[-8pt]
3 & $(2,2)$ & $\parbox{0.5cm}{\ydiagram{2,2}}$ & $1$ & $(4,2)$\\[8pt]
 & $(2,1,1)$ & $\parbox{0.5cm}{\ydiagram{2,1,1}}$ & $2$ & $(2,2,2)$\\[10pt]\hline
&&&&\\[-8pt]
4 & $(4,2)$ & $\parbox{0.9cm}{\ydiagram{4,2}}$ & $1$ & $(2,4,2,0)$\\[8pt]
 & $(2,2,1)$ & $\parbox{0.5cm}{\ydiagram{2,2,1}}$ & $2$ & $(4,2,0,2)$\\[10pt]
 & $(2,1,1,1)$ & $\parbox{0.5cm}{\ydiagram{2,1,1,1}}$ & $2$ & $(2,2,2,2)$\\[12pt]\hline
&&&&\\[-8pt]
5 & $(2,2,2)$ & $\parbox{0.5cm}{\ydiagram{2,2,2}}$ & $1$ & $(4,4,0,0,2)$\\[12pt]
 & $(4,2,1)$ & $\parbox{0.9cm}{\ydiagram{4,2,1}}$ & $2$ & $(4,2,2,4,0)$\\[12pt]
 & $(2,2,1,1)$ & $\parbox{0.5cm}{\ydiagram{2,2,1,1}}$ & $2$ & $(4,2,2,0,2)$\\[16pt]
 & $(2,1,1,1,1)$ & $\parbox{0.5cm}{\ydiagram{2,1,1,1,1}}$ & $2$ & $(2,2,2,2,2)$\\[18pt]\hline
\end{tabular}
\end{center}  
which give rise to the coefficients $c_{k}^{(n)}(\epsilon,\tau)$ in (\ref{Coef1fin}). We have furthermore checked, that the algorithm described above correctly reproduces all coefficients $c_{k}^{(n)}(\epsilon,\tau)$ up to $k=8$. and we therefore conjecture that it holds for generic $k\in\mathbb{N}$.
\section{Expansion Coefficients}
In this appendix we tabulate some of the expansion coefficients that appear for various partition functions.
\subsection{Non-compact Brane Configuration $(N,M)=(3,1)$}\label{App:Coeffs31}
We list the first few coefficients $\phi^n_{[c_1,c_2]}$ appearing in the expansion~(\ref{PartFctGamnn}) of $\mathcal{Z}^{\text{line}}_{3,1}(\tau,t_{f_1},t_{f_2},m=n\epsilon,\epsilon,-\epsilon)$
{\allowdisplaybreaks\begin{align}
&\phi^n_{[n^2,n^2]}=1\,,\nonumber\\
&\phi^n_{[n^2-2,n^2+1]}=\phi^n_{[n^2+1,n^2-2]}=\frac{\theta(n)^2}{\theta(1)^2}\,,\nonumber\\
&\phi^n_{[n^2-4,n^2+2]}=\phi^n_{[n^2+2,n^2-4]}=2\,\frac{\theta(n-1)\theta(n)^2\theta(n+1)}{\theta(1)^2\theta(2)^2}\,,\nonumber\\
&\phi^n_{[n^2-6,n^2+3]}=\phi^n_{[n^2+3,n^2-6]}=\frac{\theta(n-1)^2\theta(n)^2\theta(n+1)^2}{\theta(1)^4\theta(3)^2}+2\,\frac{\theta(n-2)\theta(n-1)\theta(n)^2\theta(n+1)\theta(n+2)}{\theta(1)^2\theta(2)^2\theta(3)^2}\,,\nonumber\\
&\phi^n_{[n^2-1,n^2-1]}=\frac{\theta(n-1)\theta(n)^2\theta(n+1)}{\theta(1)^4}\,,\nonumber\\
&\phi^n_{[n^2-3,n^2]}=\phi^n_{[n^2,n^2-3]}=\frac{\theta(n-2)\theta(n)^3\theta(n+1)^2+\theta(n-1)^2\theta(n)^3\theta(n+2)}{\theta(1)^4\theta(2)^2}\,,\nonumber\\
&\phi^n_{[n^2-5,n^2+1]}=\phi^n_{[n^2+1,n^2-5]}=\frac{\theta(n-2)\theta(n-1)\theta(n)^4\theta(n+1)\theta(n+2)}{\theta(1)^6\theta(3)^2}\nonumber\\
&\hspace{1cm}+\frac{\theta(n-3)\theta(n-1)\theta(n)^3\theta(n+1)^2\theta(n+2)+\theta(n-2)\theta(n-1)^2\theta(n)^3\theta(n+1)\theta(n+3)}{\theta(1)^4\theta(2)^2\theta(3)^2}\,,\nonumber\\
&\phi^n_{[n^2-7,n^2+2]}=\phi^n_{[n^2+2,n^2-7]}=\frac{\theta(n-2)\theta(n-1)^2\theta(n)^4\theta(n+1)^2\theta(n+2)}{\theta(1)^4\theta(2)^4\theta(3)^2}\nonumber\\
&\hspace{1cm}+\frac{\theta(n-3)\theta(n-1)^2\theta(n)^4\theta(n+1)\theta(n+2)^2+\theta(n-2)^2\theta(n-1)\theta(n)^4\theta(n+1)^2\theta(n+3)}{\theta(1)^6\theta(2)^2\theta(4)^2}\nonumber\\
&\hspace{1cm}+\frac{\theta(n-4)\theta(n-2)\theta(n-1)\theta(n)^3\theta(n+1)^2\theta(n+2)\theta(n+3)}{\theta(1)^4\theta(2)^2\theta(3)^2\theta(4)^2}\nonumber\\
&\hspace{1cm}+\frac{\theta(n-3)\theta(n-2)\theta(n-1)^2\theta(n)^3\theta(n+1)\theta(n+2)\theta(n+4)}{\theta(1)^4\theta(2)^2\theta(3)^2\theta(4)^2}\,,\nonumber\\
&\phi^n_{[n^2-9,n^2+3]}=\phi^n_{[n^2+3,n^2-9]}=\frac{\theta(n-3)\theta(n-2)\theta(n-1)^2\theta(n)^4\theta(n+1)^2\theta(n+2)\theta(n+3)}{\theta(1)^6\theta(2)^4\theta(5)^2}\nonumber\\
&\hspace{1cm}+\frac{\theta(n-3)\theta(n-1)^3\theta(n)^4\theta(n+1)^2\theta(n+2)^2+\theta(n-2)^2\theta(n-1)^2\theta(n)^4\theta(n+1)^3\theta(n+3)}{\theta(1)^6\theta(2)^2\theta(3)^2\theta(4)^2}\nonumber\\
&\hspace{1cm}+\frac{\theta(n-4)\theta(n-2)\theta(n-1)^2\theta(n)^4\theta(n+1)\theta(n+2)^2\theta(n+3)}{\theta(1)^6\theta(2)^2\theta(3)^2\theta(5)^2}\nonumber\\
&\hspace{1cm}+\frac{\theta(n-3)\theta(n-2)^2\theta(n-1)\theta(n)^4\theta(n+1)^2\theta(n+2)\theta(n+4)}{\theta(1)^6\theta(2)^2\theta(3)^2\theta(5)^2}\nonumber\\
&\hspace{1cm}+\frac{\theta(n-5)\theta(n-3)\theta(n-2)\theta(n-1)\theta(n)^3\theta(n+1)^2\theta(n+2)\theta(n+3)\theta(n+4)}{\theta(1)^4\theta(2)^2\theta(3)^2\theta(4)^2\theta(5)^2}\nonumber\\
&\hspace{1cm}+\frac{\theta(n-4)\theta(n-3)\theta(n-2)\theta(n-1)^2\theta(n)^3\theta(n+1)\theta(n+2)\theta(n+3)\theta(n+5)}{\theta(1)^4\theta(2)^2\theta(3)^2\theta(4)^2\theta(5)^2}\,,\nonumber\\
&\phi^n_{[n^2-2,n^2-2]}=\frac{\theta(n-3)\theta(n)^4\theta(n+1)^3+\theta(n-1)^3\theta(n)^4\theta(n+3)}{\theta(1)^4\theta(2)^4}\nonumber\\
&\hspace{1cm}+2\frac{\theta(n-2)\theta(n-1)^2\theta(n)^2\theta(n+1)^2\theta(n+2)}{\theta(1)^4\theta(2)^4}\,,\nonumber\\
&\phi^n_{[n^2-4,n^2-1]}=\phi^n_{[n^2-1,n^2-4]}=\frac{\theta(n-3)\theta(n-2)\theta(n-1)\theta(n)^3\theta(n+1)^2\theta(n+2)^2}{\theta(1)^4\theta(2)^4\theta(3)^2}\nonumber\\
&\hspace{1cm}+\frac{\theta(n-2)^2\theta(n-1)^2\theta(n)^3\theta(n+1)\theta(n+2)\theta(n+3)}{\theta(1)^4\theta(2)^4\theta(3)^2}\nonumber\\
&\hspace{1cm}+\frac{\theta(n-4)\theta(n-1)\theta(n)^4\theta(n+1)^3\theta(n+2)+\theta(n-2)\theta(n-1)^3\theta(n)^4\theta(n+1)\theta(n+4)}{\theta(1)^4\theta(2)^4\theta(3)^2}\nonumber\\
&\hspace{1cm}+\frac{\theta(n-3)\theta(n-1)^2\theta(n)^3\theta(n+1)^3\theta(n+2)+\theta(n-2)\theta(n-1)^3\theta(n)^3\theta(n+1)^2\theta(n+3)}{\theta(1)^6\theta(2)^2\theta(3)^2}\,,\nonumber\\
&\phi^n_{[n^2-3,n^2-3]}=\frac{\theta(n-3)\theta(n-1)^4\theta(n)^2\theta(n+1)^4\theta(n+3)}{\theta(1)^8\theta(3)^4}\nonumber\\
&\hspace{1cm}+2\frac{\theta(n-3)\theta(n-2)^2\theta(n-1)^2\theta(n)^2\theta(n+1)^2\theta(n+2)^2\theta(n+3)}{\theta(1)^4\theta(2)^4\theta(3)^4}\nonumber\\
&\hspace{1cm}+\frac{\theta(n-5)\theta(n-1)^2\theta(n)^4\theta(n+1)^3\theta(n+2)^2+\theta(n-2)^2\theta(n-1)^3\theta(n)^4\theta(n+1)^2\theta(n+5)}{\theta(1)^4\theta(2)^4\theta(3)^4}\nonumber\\
&\hspace{1cm}+2\frac{\theta(n-4)\theta(n-2)\theta(n-1)\theta(n)^4\theta(n+1)^3\theta(n+2)^2}{\theta(1)^6\theta(2)^2\theta(3)^4}\nonumber\\
&\hspace{1cm}+2\frac{\theta(n-2)^2\theta(n-1)^3\theta(n)^4\theta(n+1)\theta(n+2)\theta(n+4)}{\theta(1)^6\theta(2)^2\theta(3)^4}\,.\nonumber\\
&\phi^n_{[n^2-4,n^2-4]}=\frac{\theta(n-3)\theta(n-2)^2\theta(n-1)^3\theta(n)^4\theta(n+1)^3\theta(n+2)^2\theta(n+3)}{\theta(1)^4\theta(2)^8\theta(3)^4}\nonumber\\
&\hspace{1cm}+2\,\frac{\theta(n-4)\theta(n-3)\theta(n-1)^3\theta(n)^5\theta(n+1)^3\theta(n+2)^2\theta(n+3)}{\theta(1)^6\theta(2)^6\theta(3)^2\theta(4)^2}\nonumber\\
&\hspace{1cm}+2\,\frac{\theta(n-3)\theta(n-2)^2\theta(n-1)^3\theta(n)^5\theta(n+1)^3\theta(n+3)\theta(n+4)}{\theta(1)^6\theta(2)^6\theta(3)^2\theta(4)^2}\nonumber\\
&\hspace{1cm}+\frac{\theta(n-5)\theta(n-2)^2\theta(n-1)^2\theta(n)^4\theta(n+1)^4\theta(n+2)^2\theta(n+3)}{\theta(1)^8\theta(2)^4\theta(4)^4}\nonumber\\
&\hspace{1cm}+\frac{\theta(n-3)\theta(n-2)^2\theta(n-1)^4\theta(n)^4\theta(n+1)^2\theta(n+2)^2\theta(n+5)}{\theta(1)^8\theta(2)^4\theta(4)^4}\nonumber\\
&\hspace{1cm}+2\,\frac{\theta(n-5)\theta(n-4)\theta(n-1)^2\theta(n)^5\theta(n+1)^4\theta(n+2)^2\theta(n+3)}{\theta(1)^4\theta(2)^6\theta(3)^4\theta(4)^2}\nonumber\\
&\hspace{1cm}+2\,\frac{\theta(n-3)\theta(n-2)^2\theta(n-1)^4\theta(n)^5\theta(n+1)^2\theta(n+4)\theta(n+5)}{\theta(1)^4\theta(2)^6\theta(3)^4\theta(4)^2}\nonumber\\
&\hspace{1cm}+2\,\frac{\theta(n-6)\theta(n-3)\theta(n-1)^3\theta(n)^4\theta(n+1)^3\theta(n+2)^3\theta(n+3)}{\theta(1)^6\theta(2)^4\theta(3)^2\theta(4)^4}\nonumber\\
&\hspace{1cm}+2\,\frac{\theta(n-3)\theta(n-2)^3\theta(n-1)^3\theta(n)^4\theta(n+1)^3\theta(n+3)\theta(n+6)}{\theta(1)^6\theta(2)^4\theta(3)^2\theta(4)^4}\nonumber\\
&\hspace{1cm}+2\,\frac{\theta(n-5)\theta(n-3)\theta(n-2)^2\theta(n-1)\theta(n)^4\theta(n+1)^3\theta(n+2)^2\theta(n+3)^2}{\theta(1)^6\theta(2)^4\theta(3)^2\theta(4)^4}\nonumber\\
&\hspace{1cm}+2\,\frac{\theta(n-3)^2\theta(n-2)^2\theta(n-1)^3\theta(n)^4\theta(n+1)\theta(n+2)^2\theta(n+3)\theta(n+5)}{\theta(1)^6\theta(2)^4\theta(3)^2\theta(4)^4}\nonumber\\
&\hspace{1cm}+\frac{\theta(n-7)\theta(n-2)^2\theta(n-1)^2\theta(n)^4\theta(n+1)^3\theta(n+2)^2\theta(n+3)^2}{\theta(1)^4\theta(2)^4\theta(3)^4\theta(4)^4}\nonumber\\
&\hspace{1cm}+\frac{\theta(n-3)^2\theta(n-2)^2\theta(n-1)^3\theta(n)^4\theta(n+1)^2\theta(n+2)^2\theta(n+7)}{\theta(1)^4\theta(2)^4\theta(3)^4\theta(4)^4}\nonumber\\
&\hspace{1cm}+2\,\frac{\theta(n-4)\theta(n-2)^2\theta(n-1)^4\theta(n)^2\theta(n+1)^4\theta(n+2)^2\theta(n+4)}{\theta(1)^8\theta(2)^4\theta(4)^4}\nonumber\\
&\hspace{1cm}+2\,\frac{\theta(n-4)\theta(n-3)^2\theta(n-2)^2\theta(n-1)^2\theta(n)^2\theta(n+1)^2\theta(n+2)^2\theta(n+3)^2\theta(n+4)}{\theta(1)^4\theta(2)^4\theta(3)^4\theta(4)^4}\,.\label{PhinGen}
\end{align}}
\subsection{Compact Brane Configuration $(N,M)=(2,1)$}\label{App:Coeffs21}
We list the first few coefficients $\phi^n_{[c_1,c_2,l]}$ appearing in the expansion~(\ref{ExpandPartFct21com}) of $\mathcal{Z}_{2,1}(\tau,t_{f_1},t_{f_2},m=n\epsilon,\epsilon,-\epsilon)$
{\allowdisplaybreaks\begin{align}
&\widehat{\phi}^n_{[n^2,n^2,0]}=1\,,\nonumber\\
&\widehat{\phi}^n_{[n^2+2,n^2-2,0]}=\widehat{\phi}^n_{[n^2-2,n^2+2,-1]}=\frac{\theta(n)^2}{\theta(1)^2}\,,\nonumber\\
&\widehat{\phi}^n_{[n^2+4,n^2-4,0]}=\widehat{\phi}^n_{[n^2-4,n^2+4,-2]}=2\frac{\theta(n-1)\theta(n)^2\theta(n+1)}{\theta(1)^2\theta(2)^2}\,,\nonumber\\
&\widehat{\phi}^n_{[n^2+6,n^2-6,0]}=\widehat{\phi}^n_{[n^2-6,n^2+6,-3]}=\frac{\theta(n-1)^2\theta(n)^2\theta(n+1)^2}{\theta(1)^4\theta(3)^2}+2\frac{\theta(n-2)\theta(n-1)\theta(n)^2\theta(n+1)\theta(n+2)}{\theta(1)^2\theta(2)^2\theta(3)^2}\,,\nonumber\\
&\widehat{\phi}^n_{[n^2+8,n^2-8,0]}=\widehat{\phi}^n_{[n^2-8,n^2+8,-4]}=\frac{\theta(n-1)^2\theta(n)^4\theta(n+1)^2}{\theta(1)^2\theta(2)^4\theta(3)^2}+2\frac{\theta(n-2)\theta(n-1)^2\theta(n)^2\theta(n+1)^2\theta(n+2)}{\theta(1)^4\theta(2)^2\theta(4)^2}\nonumber\\
&\hspace{1cm}+2\frac{\theta(n-3)\theta(n-2)\theta(n-1)\theta(n)^2\theta(n+1)\theta(n+2)\theta(n+3)}{\theta(1)^2\theta(2)^2\theta(3)^2\theta(4)^2}\,,\nonumber\\
&\widehat{\phi}^n_{[n^2+10,n^2-10,0]}=\frac{\theta(n-2)^2\theta(n-1)^2\theta(n)^2\theta(n+1)^2\theta(n+2)^2}{\theta(1)^4\theta(2)^4\theta(5)^2}\nonumber\\
&\hspace{1cm}+2\frac{\theta(n-2)\theta(n-1)^2\theta(n)^4\theta(n+1)^2\theta(n+2)}{\theta(1)^4\theta(2)^2\theta(3)^2\theta(4)^2}\nonumber\\
&\hspace{1cm}+2\frac{\theta(n-3)\theta(n-2)\theta(n-1)^2\theta(n)^2\theta(n+1)^2\theta(n+2)\theta(n+3)}{\theta(1)^4\theta(2)^2\theta(3)^2\theta(5)^2}\nonumber\\
&\hspace{1cm}+2\frac{\theta(n-4)\theta(n-3)\theta(n-2)\theta(n-1)\theta(n)^2\theta(n+1)\theta(n+2)\theta(n+3)\theta(n+4)}{\theta(1)^2\theta(2)^2\theta(3)^2\theta(4)^2\theta(5)^2}\,,\nonumber\\
&\widehat{\phi}^n_{[n^2,n^2,-1]}=\frac{\theta(n-1)^2\theta(n+1)^2}{\theta(1)^4}\,,\nonumber\\
&\widehat{\phi}^n_{[n^2+2,n^2-2,-1]}=\widehat{\phi}^n_{[n^2-2,n^2+2,-2]}=2\frac{\theta(n-2)\theta(n-1)\theta(n)^2\theta(n+1)\theta(n+2)}{\theta(1)^4\theta(2)^2}\nonumber\\
&\widehat{\phi}^n_{[n^2+4,n^2-4,-1]}=\widehat{\phi}^n_{[n^2-4,n^2+4,-3]}=\frac{\theta(n-2)^2\theta(n)^4\theta(n+2)^2}{\theta(1)^6\theta(3)^2}+2\frac{\theta(n-3)\theta(n-1)^2\theta(n)^2\theta(n+1)^2\theta(n+3)}{\theta(1)^4\theta(2)^2\theta(3)^2}\,,\nonumber\\
&\widehat{\phi}^n_{[n^2+6,n^2-6,-1]}=\widehat{\phi}^n_{[n^2-6,n^2+6,-4]}=\frac{\theta(n-2)^2\theta(n-1)^2\theta(n)^2\theta(n+1)^2\theta(n+2)^2}{\theta(1)^4\theta(2)^4\theta(3)^2}\nonumber\\
&\hspace{1cm}+2\frac{\theta(n-3)\theta(n-2)\theta(n-1)\theta(n)^4\theta(n+1)\theta(n+2)\theta(n+3)}{\theta(1)^6\theta(2)^2\theta(4)^2}\nonumber\\
&\hspace{1cm}+2\frac{\theta(n-4)\theta(n-2)\theta(n-1)^2\theta(n)^2\theta(n+1)^2\theta(n+2)\theta(n+4)}{\theta(1)^4\theta(2)^2\theta(3)^2\theta(4)^2}\,,\nonumber\\
&\widehat{\phi}^n_{[n^2,n^2,-2]}=2\frac{\theta(n-2)^2\theta(n-1)^2\theta(n+1)^2\theta(n+2)^2}{\theta(1)^4\theta(2)^4}+2\frac{\theta(n-3)\theta(n-1)\theta(n)^4\theta(n+1)\theta(n+3)}{\theta(1)^4\theta(2)^4}\,,\nonumber\\
&\widehat{\phi}^n_{[n^2+2,n^2-2,-2]}=\widehat{\phi}^n_{[n^2-2,n^2+2,-3]}=2\frac{\theta(n-4)\theta(n-1)^2\theta(n)^4\theta(n+1)^2\theta(n-4)}{\theta(1)^4\theta(2)^4\theta(3)^2}\nonumber\\
&\hspace{1cm}+2\frac{\theta(n-3)\theta(n-2)\theta(n-1)^2\theta(n)^2\theta(n+1)^2\theta(n+2)\theta(n+3)}{\theta(1)^6\theta(2)^2\theta(3)^2}\nonumber\\
&\hspace{1cm}+2\frac{\theta(n-3)\theta(n-2)^2\theta(n-1)\theta(n)^2\theta(n+1)\theta(n+2)^2\theta(n+3)}{\theta(1)^4\theta(2)^4\theta(3)^2)}\,,\nonumber\\
&\widehat{\phi}^n_{[n^2+4,n^2-4,-2]}=\widehat{\phi}^n_{[n^2-4,n^2+4,-4]}=2\frac{\theta(n-3)\theta(n-2)^2\theta(n-1)^2\theta(n)^2\theta(n+1)^2\theta(n+2)^2\theta(n+3)}{\theta(1)^4\theta(2)^6\theta(3)^2}\nonumber\\
&\hspace{1cm}+2\frac{\theta(n-3)^2\theta(n-2)\theta(n-1)\theta(n)^4\theta(n+1)\theta(n+2)\theta(n+3)^2}{\theta(1)^6\theta(2)^4\theta(4)^2}\nonumber\\
&\hspace{1cm}+2\frac{\theta(n-4)\theta(n-2)\theta(n-1)^3\theta(n)^2\theta(n+1)^3\theta(n+2)\theta(n+4)}{\theta(1)^6\theta(2)^4\theta(4)^2}\nonumber\\
&\hspace{1cm}+2\frac{\theta(n-4)\theta(n-3)\theta(n-2)\theta(n-1)^2\theta(n)^2\theta(n+1)^2\theta(n+2)\theta(n+3)\theta(n+4)}{\theta(1)^4\theta(2)^4\theta(3)^2\theta(4)^2}\nonumber\\
&\hspace{1cm}+2\frac{\theta(n-5)\theta(n-2)\theta(n-1)^2\theta(n)^4\theta(n+1)^2\theta(n+2)\theta(n+5)}{\theta(1)^4\theta(2)^4\theta(3)^2\theta(4)^2}\,,
\end{align}}

\section{Contributing Partitions for $(N,M)=(3,1)$ and $m=\epsilon$}\label{App:NM31m1}
To analyse the restrictions on the sum over partitions in the case $N=3$, we consider two generic Young diagrams $(\nu_1,\nu_2)$ and try to restrict their forms by analysing their contributions to the partition function (\ref{DefZline})
\begin{itemize}
\item contribution of the $(\nu_{2,1}-1)$th box in the first row of $\nu_2$
\begin{align}
&\parbox{1.6cm}{\ydiagram{6, 5, 3,2}}\hspace{1cm}\parbox{1.8cm}{\ydiagram{7, 5, 4,4,3,1} *[*(black)]{5+1,0,0,0,0,0}} && i_2=1\,, &&j_2=\nu_{2,1}-1
\end{align}
For this box, we have $z^{(2)}_{1,\nu_{2,1}-1}=\epsilon(\nu_{2,1}-1-(\nu_{2,1}-1))=0$, therefore, the Young diagram $\nu_2$ is not allowed to have a second column, but is restricted to consist of a single column.
\item contribution of the last box in the second row of $\nu_1$
\begin{align}
&\parbox{1.6cm}{\ydiagram{6, 5, 3,2}*[*(black)]{0,4+1,0,0}}\hspace{1cm}\parbox{0.3cm}{\ydiagram{1, 1, 1,1,1,1}} && i_1=2\,, &&j_1=\nu_{1,2}
\end{align}
such that $v^{(1)}_{2,\nu_{1,2}}=-\epsilon(\nu_{1,2}-2-\nu_{1,2}+2)=0$, therefore, the Young diagram $\nu_1$ cannot have a second row and is restricted to consist of a single row. 
\item restrictions on the form of $\nu_1$\\
Here we have to distinguish four different possibilities
\begin{itemize}
\item $\nu_1=\emptyset$: in this case we have no restriction on the form of $\nu_2$
\item $\nu_1=\ydiagram{1}$\,: in this case we have $z^{(1)}_{1,1}=\epsilon(1+\nu_{2,1}^t-1-1)$, which restricts $\nu_2$ to be 
\begin{align}
&\nu_1=\ydiagram{1}\,,&&\nu_2\in\left\{\emptyset,\parbox{0.3cm}{\ydiagram{1,1}},\parbox{0.3cm}{\ydiagram{1,1,1}},\ldots\right\}\,,
\end{align}
and in particular excludes $(\nu_1,\nu_2)=(\parbox{0.3cm}{\ydiagram{1}},\parbox{0.3cm}{\ydiagram{1}})$.
\item $\nu_1=\ydiagram{2}$\,: in this case we have $z^{(1)}_{1,1}=\epsilon(2+\nu_{2,1}^t-1-1)$ and $z_{1,2}^{(1)}=\epsilon(2-1-2)\neq 0$, which restricts the form of $\nu_2$
\begin{align}
&\nu_1=\ydiagram{2}\,,&&\nu_2\in\left\{\parbox{0.3cm}{\ydiagram{1}},\parbox{0.3cm}{\ydiagram{1,1}},\parbox{0.3cm}{\ydiagram{1,1,1}},\ldots\right\}\,,
\end{align}
which particularly excludes $(\nu_1,\nu_2)=(\parbox{0.6cm}{\ydiagram{2}},\emptyset)$
\item $\nu_1=\ydiagram{5}$\,, with length $\nu_{1,1}\geq 3$: in this case we consider the contribution of the box $\nu_{1,1}-1$ in the first row of $\nu_1$
\begin{align}
&\parbox{1.4cm}{\ydiagram{5}*[*(black)]{3+1}}\hspace{1cm}\parbox{0.3cm}{\ydiagram{1, 1, 1,1,1,1}} && i_1=1\,, &&j_1=\nu_{1,1}-1\geq 2\,,
\end{align}
for which we have $z^{(1)}_{1,\nu-{1,1}-1}=\epsilon(\nu_{1,1}-1-(\nu_{1,1}-1))=0$. Therefore, the length $\nu_{1,1}<3$ is restricted.
\end{itemize}
Summarising, we are left with the following three sets of configurations
\begin{align}
&(\nu_1,\nu_2)\in\left\{(\emptyset,\emptyset)\,,\left(\emptyset,\parbox{0.3cm}{\ydiagram{1}}\right)\,,\left(\emptyset,\parbox{0.3cm}{\ydiagram{1,1}}\right)\,,\left(\emptyset,\parbox{0.3cm}{\ydiagram{1,1,1}}\right)\,,\ldots \right\}\,,\label{N2class1}\\[2pt]
&(\nu_1,\nu_2)\in\left\{(\ydiagram{1}\,,\emptyset)\,,\left(\ydiagram{1}\,,\parbox{0.3cm}{\ydiagram{1,1}}\right)\,,\left(\ydiagram{1}\,,\parbox{0.3cm}{\ydiagram{1,1,1}}\right)\,,\ldots \right\}\,,\label{N2class2}\\[2pt]
&(\nu_1,\nu_2)\in\left\{(\ydiagram{2}\,,\parbox{0.3cm}{\ydiagram{1}})\,,\left(\ydiagram{2}\,,\parbox{0.3cm}{\ydiagram{1,1}}\right)\,,\left(\ydiagram{2}\,,\parbox{0.3cm}{\ydiagram{1,1,1}}\right)\,,\ldots \right\}\,.\label{N2class3}
\end{align}
\item restrictions on the form of $\nu_2$\\
We can further constrain the three classes of contributions (\ref{N2class1}) -- (\ref{N2class2}):
\begin{itemize}
\item $\nu_1=\emptyset$ (see (\ref{N2class1})): in this case we consider the first box in the second row of $\nu_2$\\
\begin{align}
&\emptyset\hspace{1cm}\parbox{0.3cm}{\ydiagram{1, 1, 1,1,1,1}*[*(black)]{0,0+1,0,0}} && i_2=2\,, &&j_2=1\,,
\end{align}
which yields $v^{(2)}_{2,1}=-\epsilon(\nu_{2,1}+0-2-1+2)=\nu_{2,1}-1$ and therefore only $\nu_{2,1}=0$ contributes, which restricts (\ref{N2class1}) to the following cases
\begin{align}
&(\nu_1,\nu_2)\in\left\{(\emptyset,\emptyset)\,,\left(\emptyset,\parbox{0.3cm}{\ydiagram{1}}\right)\right\}\,,
\end{align}
\item $\nu_1=\ydiagram{1}$ (see (\ref{N2class2})): the case $(\ydiagram{1}\,,\emptyset)$ contributes to the partition function, while for the cases $\nu_{2,2}\neq 0$ we consider the first box in the third row of $\nu_2$\\
\begin{align}
&\parbox{0.3cm}{\ydiagram{1}}\hspace{1cm}\parbox{0.3cm}{\ydiagram{1, 1, 1,1,1,1}*[*(black)]{0,0,0+1,0}} && i_2=3\,, &&j_2=1\,,
\end{align}
which yields $v^{(2)}_{3,1}=-\epsilon(\nu_{2,1}+1-3-1+2)=-\epsilon(\nu_{2,1}-1)$
and therefore only $\nu^2_{3,1}=0$ contributes, which restricts (\ref{N2class2}) to the following contributions
\begin{align}
(\nu_1,\nu_2)\in\left\{(\ydiagram{1}\,,\emptyset)\,,\left(\ydiagram{1}\,,\parbox{0.3cm}{\ydiagram{1,1}}\right) \right\}\,.
\end{align}
\item $\nu_1=\ydiagram{2}$\,(see (\ref{N2class3})): in this case we consider the first box in the third row of $\nu_2$
\begin{align}
&\parbox{0.6cm}{\ydiagram{2}}\hspace{1cm}\parbox{0.3cm}{\ydiagram{1, 1, 1,1,1,1}*[*(black)]{0,0,0+1,0}} && i_2=3\,, &&j_2=1\,,
\end{align}
which yields $v^{(2)}_{3,1}=-\epsilon(\nu_{2,1}+1-3-1+2)=-\epsilon(\nu_{2,1}-1)$ and therefore only $\nu_{2,1}=0$ contributes, which restricts (\ref{N2class2}) to the following contributions
\begin{align}
(\nu_1,\nu_2)\in\left\{(\ydiagram{2}\,,\parbox{0.3cm}{\ydiagram{1}})\,,\left(\ydiagram{2}\,,\parbox{0.3cm}{\ydiagram{1,1}}\right)\right\}
\end{align}

\end{itemize}
\end{itemize}



\end{document}